\documentclass[preprint,12pt]{elsarticle}

\usepackage{graphicx}
\usepackage{latexsym}
\usepackage{dcolumn}
\usepackage[utf8]{inputenc}
\usepackage[T1]{fontenc}
\usepackage{tabularx}
\usepackage{hyperref}
\usepackage{color}
\usepackage{placeins}
\usepackage{float}
\usepackage{amssymb}
\usepackage{amsmath}
\usepackage{gensymb}
\usepackage[separate-uncertainty=true]{siunitx}
\usepackage{orcidlink}
\usepackage{soul}

\usepackage{cleveref}
\usepackage{subfig}
\usepackage{multirow}
\usepackage{subcaption}
\usepackage{array}
\usepackage{longtable}
\usepackage{rotating}
\usepackage{booktabs}
\usepackage[version=4]{mhchem}
\usepackage{subfiles}
\newcolumntype{?}{!{\vrule width 1.5pt}}
\newcolumntype{P}[1]{>{\centering\arraybackslash}p{#1}}
\newcolumntype{M}[1]{>{\centering\arraybackslash}m{#1}}

\makeatletter
\def\hlinewd#1{%
\noalign{\ifnum0=`}\fi\hrule \@height #1 %
\futurelet\reserved@a\@xhline}
\makeatother

\DeclareSIUnit\angstrom{\text {Å}}
\DeclareSIUnit\mips{\text {mips}}
\renewcommand{\citet}[1]{Ref.~\cite{#1}}

\newcommand{\SiGe}{Si$_{1-x}$Ge$_{x}$}
\newcommand{\CB}{C$_{1-x}$B$_{x}$}

\journal{Diamond and Related Materials}

\biboptions{sort&compress}

\begin{document}

\begin{frontmatter}



\title{Structural effects of boron doping in diamond crystals for gamma-ray light-source applications: Insights from molecular dynamics simulations}


\author[kent_phys]{Matthew D. Dickers\corref{cor1}\orcidlink{0000-0001-9615-9101}} 
\ead{M.D.Dickers@kent.ac.uk}
\cortext[cor1]{Corresponding author}

\author[kent_chem]{Felipe Fantuzzi\orcidlink{0000-0002-8200-8262}}

\author[kent_phys]{Nigel J. Mason\orcidlink{0000-0002-4468-8324}}

\author[MBN]{Gennady B. Sushko}

\author[MBN]{Andrei V. Korol\orcidlink{0000-0002-6807-5194}}

\author[MBN]{Andrey V. Solov'yov}

\address[kent_phys]{Physics and Astronomy, School of Engineering, Mathematics and Physics, University of Kent, Park Wood Rd, Canterbury CT2 7NH, UK}

\address[kent_chem]{Chemistry and Forensic Science, School of Natural Sciences, University of Kent, Park Wood Rd, Canterbury CT2 7NH, UK}

\address[MBN]{MBN Research Center, Altenh\"oferallee 3, 60438 Frankfurt am Main, Germany}

\begin{abstract}
\noindent Boron-doped diamond crystals (BDD, \CB) exhibit exceptional mechanical strength, electronic tunability, and resistance to radiation damage. This makes them promising materials for use in gamma-ray crystal-based light sources. To better understand and quantify the structural distortions introduced by doping, which are critical for maintaining channelling efficiency, we perform atomistic-level molecular dynamics simulations on periodic \CB \, systems of various sizes. These simulations allow the influence of boron concentration on the lattice constant and the (1\;1\;0) and (1\;0\;0) inter-planar distances to be evaluated over the concentration range from pure diamond (0\%) to 5\% boron at room temperature (\SI{300}{\kelvin}).  Linear relationships between both lattice constant and inter-planar distance with increasing dopant concentration are observed, with a deviation from Vegard's Law. This deviation is larger than that reported by other theoretical and computational studies; however, this may be attributed to an enhanced crystal quality over these studies, a vital aspect when considering gamma-ray crystal light source design. The methodology presented here incorporates several refinements to closely reflect the conditions of microwave plasma chemical vapour deposition (MPCVD) crystal growth. These results enable reliable atomistic modelling of doped diamond crystals and support their use in the design and fabrication of periodically bent structures for next-generation gamma-ray light source technologies.
\end{abstract}



\begin{keyword}
Boron-doped Diamond \sep Doped Crystals \sep Inter-planar Distance \sep Molecular Dynamics
\end{keyword}

\end{frontmatter}

\section{Introduction} \label{sec:Intro}
Boron-doped diamond (BDD) has emerged as a versatile material with applications spanning electronics, sensing, and radiation technologies, owing to its exceptional mechanical hardness, thermal conductivity, and chemical stability. Incorporation of boron atoms into the diamond lattice introduces p-type semiconducting behaviour while preserving the intrinsic robustness of the host structure. This combination of structural and electronic properties makes BDD a valuable basis for both fundamental research and applied technologies, including electrochemical devices \cite{Muzyka_2019, Einaga_2022}, quantum applications \cite{Alkahtani_2022, Bhattacharya_2025}, and photodetectors \cite{Thaiyotin_2000, Aksenova_2017}. Notably, the material has attracted recent interest in the development of crystal-based light sources (CLSs) operating in the gamma-ray regime, where its structural integrity and resistance to radiation damage are key for the controlled manipulation of particle beams \cite{Korol2014Book, Connell_2015, Tran_2017, Korol2020, Korol2022Book, KOROL2023}.

Gamma-ray CLSs offer a novel method for generating high-brilliance, sub-angstrom wavelength electromagnetic radiation \cite{Korol2014Book, Korol2020, Korol2022Book, Sushko_2022}. Such devices operate by directing beams of ultra-relativistic charged particles through the crystalline planes of oriented crystals where they undergo a process known as \textit{channelling} \cite{Lindhard_1965}. The operation of these devices is based upon the emission of \textit{channelling radiation}, which is emitted as charged particles oscillate within the potential wells formed by the crystalline planes \cite{Kumakhov_1976}. The focus of this work concerns the structural properties of BDD crystals, which may be used to manufacture gamma-ray CLSs. A discussion of the theoretical foundations and physical mechanisms governing these processes is not provided here; these may be found in Refs.~\cite{Korol2014Book, Korol2020, Korol2022Book, Sushko_2022, Uggerhoj_2005}, which provide extensive overviews on this topic. 

The properties of the emitted radiation are heavily dependent on the crystal’s internal structure, with different internal geometries leading to the emission of different types of radiation. One particularly advantageous geometry is the periodically bent crystal, in which the crystalline planes are bent to follow a sinusoidal profile. The movement of charged particles along this periodic path results in the emission of \textit{undulator radiation} \cite{Korol_1999}, in addition to channelling radiation. This mechanism is closely analogous to that employed in modern free-electron lasers. This crystal geometry provides a high degree of tunability; by varying the bending amplitude and period, one can precisely control the spectral and angular properties of the emitted radiation.

Although numerous methods for the fabrication of bent crystals exist \cite{Guidi_2007, Guidi_2011, Malagutti_2025, Bellucci_2003, Guidi_2005, Bagli_2014, Camattari_2017, Balling_2009, Bellucci_2015, Baryshevsky_1980, Ikezi_1984, Dedkov_1994, Korol_1998, Korol_1999, Wagner_2011, Kaleris_2025}, the present study focuses on the approach of introducing spatially modulated dopant atoms into the crystal structure \cite{Breese_1997, Mikkelsen_2000, Avakian_2003}. The incorporation of atoms with different covalent radii from those of the host lattice atoms leads to a local lattice mismatch. This mismatch induces an internal strain and leads to a change in the local geometry, displacing neighbouring atoms from their ideal crystallographic positions. Such impurity-induced distortions have been described in detail for doped \SiGe\ crystals by \citet{Breese_1997}, and the same physical principles apply to BDD systems. As the dopant concentration increases, the cumulative effect of these local distortions leads to a measurable change in the inter-planar distances. In the context of channelling in periodically bent crystals, where the (1\;1\;0) crystallographic planes are of most interest, a spatially varying dopant concentration along the axial length of the crystal (usually along the [1\;0\;0] direction) produces a periodic variation in the (1\;1\;0) inter-planar distances, and thus a bent channel through which charged particles may propagate. These dopant-induced lattice distortions directly influence the trajectories of channelled particles, and thus the production of undulator radiation. For an in-depth description of this mechanism, we direct the reader to Refs.~\cite{Breese_1997, Mikkelsen_2000, Krause_2002}; this bending mechanism is illustrated in Figure 4 in \citet{Krause_2002}. In the case of BDD, boron atoms substitute directly for carbon atoms in the lattice; acting as substitutional point defects, they induce local strain due to their larger covalent radius (\SI{0.88}{\angstrom} vs \SI{0.77}{\angstrom} for carbon \cite{goldschmidinterstitial}), resulting in a net expansion of inter-planar distances.

BDD crystals may be fabricated using microwave plasma chemical vapour deposition (MPCVD) \cite{Ashfold_1994, Brunet_1998, Connell_2015}, in which a gas mixture of carbon-containing precursor molecules, such as methane (\ce{CH4}), is exposed to a microwave-induced plasma. The plasma decomposes the precursor molecule, allowing atomic carbon to deposit onto a substrate, typically high pressure high temperature (HPHT) synthetic type Ib diamond, chosen for its common availability and low cost. Depending on the substrate and growth conditions, this process produces either thin polycrystalline or homoepitaxial diamond films. For doped crystals, a controlled concentration of dopant precursor molecules is introduced during growth. In the case of boron doping, diborane gas (\ce{B2H6}) is commonly used, with boron atoms incorporating substitutionally into the diamond lattice. To fabricate the periodically bent crystals used in gamma-ray CLS applications, the concentration of boron precursors in the gas mixture and the gas flow rate are periodically varied during the growth process to produce a spatially graded boron concentration along the length of the crystal. To provide a consistent designation of the boron content, the doped crystal is denoted as \CB, where $x$ represents the fractional concentration of boron atoms: $x = 0$ corresponds to undoped diamond, and $x = 1$ to a hypothetical lattice composed entirely of boron atoms. This notation will be used throughout this work, interchangeably with BDD.

The development of effective gamma-ray CLSs requires high-quality crystals with minimal structural defects. Low-quality crystals, characterised by point defects and dislocations \cite{Boer_2020}, disrupt the periodic potential experienced by channelled particles and lead to dechannelling \cite{Lindhard_1965}: the premature exit of particles from the channelled trajectory. In this context, the quality of a CLS crystal can be quantified by the dechannelling length \cite{Lindhard_1965}, the average distance a channelled particle travels before being dechannelled. Crystals of higher quality have fewer defects, and thus longer dechannelling lengths, leading to greater radiation intensity and improved CLS performance. Among the factors affecting crystal quality, dopant concentration plays a critical role: excessive doping introduces lattice strain and increases the likelihood of defect formation, which can degrade or eliminate channelling paths entirely. Careful control of dopant levels is therefore essential in crystal fabrication. In addition, the quality of the substrate significantly influences the quality of the grown crystal. Studies have demonstrated that type IIa HPHT diamond substrates yield higher-quality crystals compared to their type Ib counterparts \cite{Tallaire_2017}. Structural defects commonly present in type Ib substrates can propagate into the growing crystal, thereby reducing its crystalline quality. Consequently, the use of high-quality substrates is essential to achieve high-quality crystal growth.

A key advantage of \CB\ crystals is their demonstrated resistance to defect formation even at relatively high boron concentrations ($\sim$\SI{e21}{\text{atom}\;\centi\meter^{-3}}) \cite{Brunet_1998, Tran_2017}. Moreover, the strong C–C bonding in diamond provides exceptional lattice stability under high-energy particle irradiation \cite{Campbell_2000, Campbell_2002, Uggerhoj_2005}, offering potentially longer operational lifetimes in CLSs compared to other systems such as \SiGe. These properties make \CB\ an especially promising material for the next generation of gamma-ray light sources.

Herein, we present a follow-up to our recent study \cite{Si-Ge_Paper} on the atomistic-level effects of germanium concentration on the structure of \SiGe\ crystals. The present work investigates the influence of boron doping on the crystalline structure of diamond, incorporating several key improvements to the previous methodology of \citet{Si-Ge_Paper}. Specifically, the impact of boron atoms is examined across three distinct regions of the \CB\ crystal: near the substrate, within the bulk, and close to the free surface. The effects of different dopant concentration and crystal sizes on both the lattice constant and inter-planar distances are evaluated in each region. This analysis focuses on the (1\;1\;0) and (1\;0\;0) crystalline planes, which are particularly relevant to the doping strategies employed in the fabrication of CLS crystals. Across each simulation the dopant concentration is kept fixed, and crystals with spatially graded dopant concentrations that are typically used to fabricate periodically bent crystals are not modelled. This work instead focuses on the direct lattice response to dopant incorporation and does not model the MPCVD growth process. These results may be used to inform models for the design of periodically bent crystals for gamma-ray CLS applications.

Molecular dynamics (MD) simulations are performed using the \textsc{MBN Explorer} \cite{MBNExplorer} and \textsc{MBN Studio} \cite{MBNStudio} software packages, enabling a detailed investigation of dopant-induced structural changes. These results are directly relevant to the design and practical realisation of gamma-ray CLSs, as reliable fabrication of periodically bent crystals requires a detailed understanding of manufacturing tolerances, especially concerning the maximum dopant concentration that can be introduced without compromising crystal quality.

The structure of this work is as follows. In \Cref{subsec:Method_Gen}, the computational methodology for generating and modelling \CB\ crystals is introduced, along with the range of crystal sizes and dopant concentrations considered. A geometric correction is presented in \Cref{subsec:Geometric_Correction}, that may be applied to the results to enable direct comparison with experimental and theoretical studies. \Cref{subsec:overview} provides an overview of the structural metrics considered in this work, namely the lattice constant and inter-planar distances, with the corresponding simulation results reported in Sections \ref{sec:LC} and \ref{subsec:IP}, respectively. Finally, in \Cref{sec:Conclusions}, the implications of these results for the design and practical realisation of gamma-ray CLSs are discussed, along with prospects for methodological improvements and future research directions.

\section{Methodology} \label{sec:Method}
\subsection{Crystal Structures and Classification} \label{subsec:Method_Gen}

\begin{figure*}[t!]
    \centering
    \includegraphics[width=\textwidth]{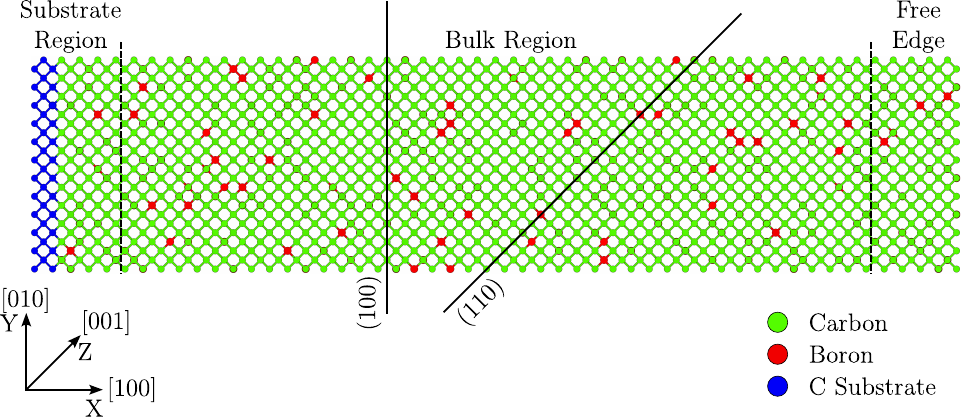}
    \caption{Representative diagram of the crystal configuration used in the simulations. Carbon atoms are shown in green, boron atoms in red, and substrate carbon atoms in blue. The orientations of the (1\;1\;0) and (1\;0\;0) crystallographic planes are indicated by solid black lines, while the boundaries between the three analysis regions are marked by black dashed lines. Further details on these regions are provided in the text.}
    \label{fig:Crystal_Regions}
\end{figure*}

This section outlines the computational methodology used to generate defect-free doped \CB\ crystals, and their subsequent atomistic-level analysis. All simulations were carried out using the \textsc{MBN Explorer} software package \cite{MBNExplorer}, designed for advanced multiscale modelling of complex molecular structures and dynamics. \textsc{MBN Studio} \cite{MBNStudio}, a multi-task toolkit and dedicated graphical user interface for \textsc{MBN Explorer}, was utilised to construct the systems, prepare input files, and analyse simulation outputs. The methods used in this work are based on that presented in our previous study \cite{Si-Ge_Paper}, with a number of notable improvements. Unlike the prior study, the present simulations are conducted under periodic boundary conditions at a fixed temperature of \SI{300}{\kelvin}, corresponding to the canonical (NVT) ensemble. A range of crystal sizes with varying dimensions, denoted by $L_X$, $L_Y$, and $L_Z$, are considered. The specific crystal sizes and the rationale for their selection are outlined later in this section. To avoid confusion between spatial coordinates and dopant concentration, the coordinate directions are denoted as $X$, $Y$, and $Z$, while $x$ is reserved for the dopant concentration. The structural configuration of the generated crystal is illustrated in \Cref{fig:Crystal_Regions}: the system is periodic in the $Y$ and $Z$ directions, which serve as the lateral (transverse) directions, while the $X$ direction is axial. A three-atom-thick carbon substrate is fixed at the base along the negative $X$-axis, constraining motion in that direction, while the positive $X$-axis is exposed to vacuum, allowing free expansion. This configuration enables subdivision of the crystal into three analysis regions, as shown in \Cref{fig:Crystal_Regions}, which are described in detail later in this section.

BDD crystals were generated by first defining a single diamond unit cell in the \textsc{MBN Explorer} input file, which was then replicated along each axis to produce a crystal of a given dimension. In all simulations, $Y$ and $Z$ simulation box directions were fixed to match the corresponding crystal size. To ensure correct periodicity, the entire structure was offset by half the standard C–C bond length (\SI{0.77}{\angstrom}) from the simulation-box edges. Because periodicity was required only along $Y$ and $Z$, and \textsc{MBN Explorer} permits a single boundary condition per simulation, the box length along $X$ was extended by \SI{15}{\angstrom} beyond the initial crystal length to allow free axial expansion and prevent interactions with periodic images. Although larger than the expected expansion, this buffer avoided artefacts and had a negligible impact on simulation time.

A pre-determined fraction of carbon atoms was then randomly selected and substituted with boron atoms, corresponding to target boron concentrations between $x=0.00$ (0\%) and $x=0.05$ (5\%, \SI{8.82e21}{\centi\meter^{-3}}) in increments of 0.01. In BDD crystals, dopant concentrations are usually small, in the region of $\lesssim 1\%$ \cite{Connell_2015, Tran_2017} due to the limited solubility of boron in diamond (typically $\sim$1--2\%) \cite{Solozhenko_2009}. Higher concentrations were included here to (i) reflect ranges considered in previous studies and (ii) evaluate structural effects at levels that may become achievable with future technologies. Because carbon atoms were selected at random for substitution by boron, the precise target concentration cannot be guaranteed. Calculations of the ‘true’ dopant concentrations show minimal deviation from the target values. A full breakdown for each target concentration and crystal size is provided in Tables S1–S5 of the Supplementary Information (SI), as well as a dedicated statistical. This random substitution method also assumes that boron atoms only occupy substitutional sites, and does not explicitly account for boron located at interstitial positions. As such, the resulting crystals correspond to the idealised case of fully substitutional BDD.

The substrate was generated to match each crystal's $L_Y$ and $L_Z$ dimensions, but truncated along $L_X$ to a thickness of three atoms to minimise system size. In MPCVD growth, an HPHT synthetic type Ib diamond substrate is typically employed as the growth surface. In this work, the HPHT substrate is not explicitly modelled; instead, its structural support is mimicked by fixing all substrate atoms. Thermal vibrations were neglected during substrate generation, and the atoms were held fixed throughout the simulation, preventing expansion along the negative $X$ direction. The substrate was positioned to the left of each crystal, with the interface aligned to preserve the diamond lattice. Atomic interactions were modelled using the multi-component Tersoff potential \cite{TersoffOriginal, Tersoff_C}, with the specific parameters and interacting atoms summarised in \Cref{tab:TersoffTable}.

\begin{table*}[]
    \centering
    \caption{Table of multi-component Tersoff potential \cite{TersoffOriginal, Tersoff_C} parameters for C-B. Parameters for C from \citet{Tersoff_C} and parameters for B from \citet{Tersoff_B}. Note that the parameters $\beta$, $n$, $c$, $d$, $h$, $\chi_{i-j}$, and $\omega_{i-j}$ are dimensionless.}
    \begin{tabular}{c|c|c}
        Elements & C \cite{Tersoff_C} & B \cite{Tersoff_B} \\ \hline
        A (\unit{\electronvolt}) & \SI{1.3936e3}{} & \SI{2.7702e2}{} \\
        B (\unit{\electronvolt}) & \SI{3.467e2}{} & \SI{1.8349e2}{} \\
        $\lambda$ (\unit{\angstrom^{-1}}) & 3.4879 & 1.9922 \\
        $\mu$ (\unit{\angstrom^{-1}}) & 2.2119 & 1.5856 \\
        $\beta$ & \SI{1.5724e-7}{} & \SI{1.6000e-6}{} \\
        $n$ & 0.72751 & 3.9929 \\
        $c$ & \SI{3.8049e4}{} & \SI{5.2629e-1}{} \\
        $d$ & 4.384 & \SI{1.5870e-3}{} \\
        $h$ & -0.57058 & 0.5000 \\
        $R_{\text{min}}$ (\unit{\angstrom}) & 1.8 & 1.8 \\
        $R_{\text{max}}$ (\unit{\angstrom}) & 2.1 & 2.1 \\ \hlinewd{1.5pt}
        Interactions $(i-j)$ & \multicolumn{2}{c}{C-B \cite{Tersoff_B}} \\ \hline
        $\chi_{i-j}$ & \multicolumn{2}{c}{1.0025} \\
        $\omega_{i-j}$ & \multicolumn{2}{c}{1.0000}
    \end{tabular}
    \label{tab:TersoffTable}
\end{table*}

A range of crystal sizes, varying in $L_X$, $L_Y$, and $L_Z$, was considered in this work to optimise and simplify the simulations while minimising the total number of atoms for computational efficiency. Lateral dimensions were analysed to elucidate their effect on the results, thereby allowing the lateral size to be optimised. The axial dimension was treated differently due to the use of non-periodic boundary conditions along this axis. The lateral dimensions were then kept fixed while $L_X$ was varied to determine whether the crystal structure is sensitive to axial length. In all cases, the lateral dimensions were chosen as integer multiples of the diamond lattice constant, \SI{3.5669\pm0.0001}{\angstrom} \cite{Saotome_1998}, to make the system fully periodic laterally, effectively simulating a laterally infinite structure.

\begin{table*}[]
    \centering
    \caption{Summary of all crystal configurations used in this study. Each structure is identified by a label and defined by its $X$, $Y$, and $Z$ dimensions in \si{\angstrom}, the total number of atoms in the crystal $N_{\text{crys}}$, the total number of atoms in the substrate $N_{\text{sub}}$, and the corresponding figure where the structure is shown. Figures denoted `S' are located in the SI. In cases where two figure references are given the configuration appears in both.}
    \begin{tabular}{c|c|c|c|c}
        Label & $L_X \times L_Y \times L_Z$ (\unit{\angstrom}) & $N_{\text{cry}}$ & $N_{\text{sub}}$ & Figure Ref. \\ \hline
        C1  & 21.402 $\times$ 21.402 $\times$ 21.402  & 1800  & 216 & \Cref{fig:Crystals}/S1 \\
        C2  & 24.969 $\times$ 24.969 $\times$ 24.969  & 2842  & 294 & Figure S1 \\
        C3  & 28.536 $\times$ 28.536 $\times$ 28.536  & 4224  & 384 & Figure S1\\ \hline
        E1A & 89.175 $\times$ 21.402 $\times$ 21.402  & 7272  & 216 & \Cref{fig:Crystals}/S2 \\
        E1B & 131.979 $\times$ 21.402 $\times$ 21.402 & 10728 & 216 & \Cref{fig:Crystals}/S2 \\
        E1C & 178.350 $\times$ 21.402 $\times$ 21.402 & 14472 & 216 & \Cref{fig:Crystals}/S2 \\ \hline
        E2A & 89.175 $\times$ 24.969 $\times$ 24.969  & 9898 & 294 & Figure S3 \\
        E2B & 131.979 $\times$ 24.969 $\times$ 24.969 & 14602 & 294 & Figure S3 \\
        E2C & 178.35 $\times$ 24.969 $\times$ 24.969  & 19698 & 294 & Figure S3 \\ \hline
        E3A & 89.175 $\times$ 28.536 $\times$ 28.536  & 12928 & 384 & Figure S4 \\
        E3B & 131.979 $\times$ 28.536 $\times$ 28.536 & 19072 & 384 & Figure S4 \\
        E3C & 178.350 $\times$ 28.536 $\times$ 28.536 & 25728 & 384 & Figure S4 \\ \hline
        S1  & 35.670 $\times$ 71.340 $\times$ 28.536  & 13120 & 960 & Figure S5 \\
        S2  & 35.670 $\times$ 71.340 $\times$ 71.340  & 32800 & 2400 & Figure S5 \\
    \end{tabular}
    \label{tab:Crystals}
\end{table*}

\begin{figure*}[t!]
    \centering
    \includegraphics[width=\textwidth]{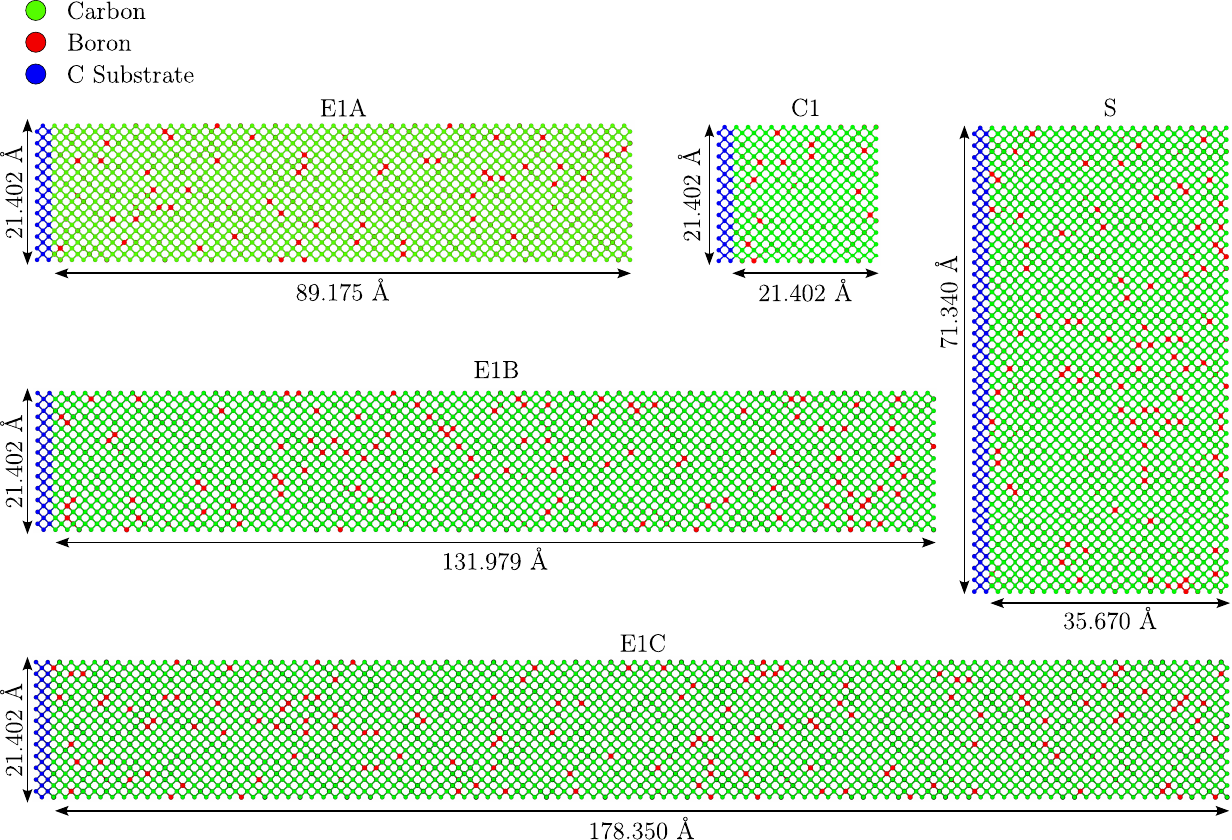}
    \caption{Representative diagrams of selected crystal structures used in this study, including the cubic crystal C1, slab crystals S1 and S2, and extended crystals E1A, E1B, and E1C. Carbon atoms are shown in green, boron dopants in red, and substrate carbon atoms in blue. Each structure shows a random dopant distribution corresponding to a boron concentration of $x = 0.05$ (5\%). The complete set of crystal structures listed in \Cref{tab:Crystals} is provided in the SI.}
    \label{fig:Crystals}
\end{figure*}

The resulting crystals fall into three categories: \textit{cubic} crystals, where $L_X = L_Y = L_Z$; \textit{extended} crystals, where $L_Y = L_Z < L_X$; and \textit{slab} crystals, where $L_Y > L_X$ and $L_Z > L_X$. The lateral dimensions are always equal and take one of three values: \SI{21.402}{\angstrom}, \SI{24.969}{\angstrom}, or \SI{28.536}{\angstrom} (equivalent to 6, 7, and 8 unit cells). Three axial dimensions were used: \SI{89.175}{\angstrom}, \SI{131.979}{\angstrom}, and \SI{178.350}{\angstrom} (equivalent to 25, 37, and 50 unit cells). To systematically label these configurations, the following nomenclature is used: \textit{C} for cubic, \textit{E} for extended, and \textit{S} for slab, followed by a numerical index indicating the lateral dimensions: \textit{1} for \SI{21.402}{\angstrom}, \textit{2} for \SI{24.969}{\angstrom}, and \textit{3} for \SI{28.536}{\angstrom}. For E-category crystals, an additional letter denotes the axial length: \textit{A} for \SI{89.175}{\angstrom}, \textit{B} for \SI{131.979}{\angstrom}, and \textit{C} for \SI{178.350}{\angstrom}. In S-category crystals, the numerical index distinguishes $L_Y > L_Z$ (\textit{1}) from $L_Z = L_Y$ (\textit{2}), with a single axial length being used in both cases. The complete set of configurations is listed in \Cref{tab:Crystals}, and representative examples are shown in \Cref{fig:Crystals}. Visualisations of all crystal sizes are provided in the SI.

For each dopant concentration, between 50 and 200 unique crystalline structures were generated, each featuring a unique distribution of boron atoms. The number of structures generated was limited by the available CPU time; smaller crystal sizes, being faster to simulate, allowed for more statistical repeats. The total number of simulations for each crystal size is listed in the SI. Generating multiple unique dopant distributions provides a robust and statistically meaningful analysis. All structures initially underwent geometry optimisation using the \textsc{MBN Explorer} velocity quenching algorithm over \num{10000} optimisation steps, ensuring that each crystal reached a local energy minimum. This was performed with the simulation box dimensions $L_X$, $L_Y$, and $L_Z$ held fixed, such that only atomic positions were relaxed and no optimisation of the lattice or simulation box dimensions was performed. These optimised configurations were used as initial geometries for MD simulations, which were conducted at \SI{300}{\kelvin} using a Langevin thermostat with a damping time of \SI{0.2}{\pico\second} for a period of \SI{1}{\nano\second}. This process allowed for full thermalisation of the system within the canonical (NVT) ensemble, with the number of particles, simulation volume, and temperature kept constant throughout. It should be noted that this approach neglects temperature gradients that are present during MPCVD growth; however, such gradients primarily affect the growth kinetics, and can lead to the formation of defect. This simplifying assumption may lead to underestimations of the lattice distortion, and therefore the simulations represent the idealised, post-growth equilibrium properties of the crystal.

The crystal generation methodology described here offers three advantages: (i) it enables the use of periodic boundary conditions to simulate an effectively infinite crystal while maintaining a manageable number of atoms; (ii) it eliminates the majority of edge effects observed in our previous \SiGe\ study \cite{Si-Ge_Paper}; (iii) it enables the analysis of structural behaviour in different regions of the crystal. The system is divided into three zones based on distance from the fixed substrate: (i) the region adjacent to the substrate, where mechanical support is strongest; (ii) the central ``bulk-like” region, distant from direct constraint; and (iii) the free surface region, where atoms are free to expand into vacuum at the top of the simulation box. These are shown in \Cref{fig:Crystal_Regions}. Separate analysis of each of these regions allows the response of the local crystal structure to dopant atoms to be assessed. For structural analysis, the final ten frames of each trajectory, corresponding to the last \SI{10}{\pico\second}, were averaged to account for thermal vibrations. This timescale was chosen to capture multiple vibration periods while excluding larger-scale structural changes occurring over longer durations. Atoms belonging to the fixed substrate were excluded from all analyses.

\subsection{Geometric Correction} \label{subsec:Geometric_Correction}
The use of periodic boundary conditions introduces limitations when comparing simulation results with empirical models of dopant-induced lattice expansion, such as Vegard’s Law \cite{Vegard_1921}, which predicts a linear relationship between lattice constant and dopant concentration for alloys of identical crystal structure. Vegard’s Law assumes isotropic three-dimensional expansion, whereas the systems here exhibit anisotropic behaviour: expansion in the $Y$ and $Z$ directions is suppressed by the periodic boundaries, while expansion is permitted only along the positive $X$ direction and constrained by the fixed substrate along the negative $X$ direction. The system therefore undergoes effectively one-dimensional expansion. This anisotropy does not influence the analysis of the (1\;0\;0) inter-planar distances but alters that of the (1\;1\;0) planes. To enable meaningful comparison with Vegard’s Law and experimental data, a geometric correction must be applied to account for the restricted lateral expansion.

The distance between the crystalline planes $(i\;j\;k)$ may be denoted as $d_{ijk}$. In an isotropically expanding crystal, the distance between neighbouring (1\;1\;0) planes, $d_{110}$, is related to that of the (1\;0\;0) and (0\;1\;0) planes ($d_{100}$ and $d_{010}$, respectively). Since $d_{100} \equiv d_{010}$ in isotropic geometry, we obtain $d_{110} = \sqrt{d_{100}^2 + d_{010}^2} = \sqrt{2}d_{100}$. 

In anisotropic geometry, $d_{010}$ remains fixed, while $d_{100}$ increases by a small amount $\Delta X$. The expanded inter-planar distance is denoted by $\tilde{d}_{ijk}$, such that $\tilde{d}_{100} = d_{100} + \Delta X$, with $d_{100} \gg \Delta X$. Therefore, the isotropic and anisotropic expansion of the (1\;1\;0) planes, $\tilde{d}_{110,i}$ and $\tilde{d}_{110,a}$, are given by:

\begin{gather}
    \tilde{d}_{110,i} = \sqrt{2}(d_{100} + \Delta X) \\
    \tilde{d}_{110,a} = \sqrt{(d_{100} + \Delta X)^2 + d_{010}^2} \approx \tilde{d}_{110,i}\left(1 - \frac{\Delta X}{2d_{100}}\right) 
\end{gather}

\noindent Using these, a geometric correction factor $C$, which rescales the anisotropic expansion to reflect the isotropic scenario, can be obtained:

\begin{equation}
    C = \frac{\tilde{d}_{110,i}}{\tilde{d}_{110,a}} \approx 1 + \frac{\Delta X}{2d_{100}}
\end{equation}

\noindent The values of $d_{010}$ and $\Delta X$ are obtained from the simulations. The correction factor $C$ is computed independently for each crystal size, dopant concentration, and spatial region, and is applied by multiplying the determined $d_{110}$ by the corresponding $C$. While this approach substantially reduces the bias associated with constrained expansion, it does not account for cases in which multiple crystal facets such as (1\;1\;0) and (1\;1\;1) contribute simultaneously to the overall strain response.

\section{Results and Discussion} \label{sec:R&D}
In this section, the structural properties of \CB\ crystals are analysed, including the lattice constant and inter-planar distances as functions of boron concentration.

\subsection{Overview of Structural Metrics} \label{subsec:overview}
The nominal values for the diamond lattice constant $a_{\text{C}}$ and the inter-planar spacings used here are \SI{3.5669\pm0.0001}{\angstrom} \cite{Saotome_1998} for $a_{\text{C}}$, \SI{1.2611}{\angstrom} for $d_{110}$, and \SI{0.8917}{\angstrom} for $d_{100}$ at \SI{300}{\kelvin}. Due to the cubic structure of the crystals, the lattice constant is equal in all crystallographic directions. The $d_{110}$ inter-planar distances are of most interest in this work, as it is through these planes that particles channel in CLSs; see \Cref{fig:Crystal_Regions}. In comparison, the $d_{100}$ inter-planar distances provide a measure of the axial expansion or contraction.

Inter-planar distances were determined by projecting all atoms onto the $(1,1,0)$ and $(1,0,0)$ planes. Atoms belonging to each plane were grouped, and the plane positions were defined as the average coordinates of their constituent atoms. Inter-planar distances were calculated as the separation between successive plane positions and averaged over all planes of the same type within each crystal, yielding characteristic $d_{110}$ and $d_{100}$ values. To obtain statistically representative values, ensemble averages across all crystals of a given size and dopant concentration were taken. Relative changes in lattice constant and inter-planar distance, $\Delta a_{CB}/a_{\text{C}}$, $\Delta d_{110}/d_{110}$, and $\Delta d_{100}/d_{100}$, were also calculated, as these quantities are widely reported in the context of lattice expansion and crystal characterisation \cite{Brunet_1998, Connell_2015}.

Vegard’s Law \cite{Vegard_1921} provides a useful point of comparison for the calculated lattice constants and inter-planar distances. This empirical relationship predicts a linear dependence of the lattice constant $a_{\text{AB}}$ on the dopant concentration $x$ for binary systems composed of two components, A and B, that share the same crystalline structure, and is given by \Cref{eqn:Vegards_Law}. For \CB\ systems, its direct application is complicated by the fact that elemental boron does not naturally crystallise into a stable cubic structure. At room temperature, boron is most stable in the $\beta$-rhombohedral phase \cite{Hayami_2024}. In BDD, however, boron atoms act as substitutional impurities occupying carbon lattice sites. As a result, the lattice constant $a_{\text{B}}$ required for \Cref{eqn:Vegards_Law} cannot be defined in the same way as for conventional substitutional alloys such as \SiGe. To address this limitation, two modifications have been proposed: one based on covalent radii \cite{Brunet_1998} and another on atomic volumes \cite{Brazhkin_2006}. Both take the simplified form of \Cref{eqn:Modified_VL}:

\begin{gather}
    a_{\text{AB}} = (1-x)a_{\text{A}} + a_{\text{B}}x \label{eqn:Vegards_Law} \\
    a_{\text{CB}} = a_{\text{C}}(\kappa x + 1) \label{eqn:Modified_VL}
\end{gather}

\noindent where $\kappa$ is a coefficient determined by the chosen modification, $x$ is the fractional concentration of boron, and $a_{\text{C}}$ is the lattice constant of pure diamond. The covalent radii of carbon and boron ($r_{\text{C}} = \SI{0.77}{\angstrom}$ and $r_{\text{B}} = \SI{0.88}{\angstrom}$ \cite{goldschmidinterstitial}) are used to determine the coefficient $\kappa_r$ (see below), which is adopted here as the reference form of Vegard’s Law for \CB\ systems. The atomic volumes $V_{\text{C}} = \SI{5.67}{\angstrom^3/\text{atom}}$ \cite{goldschmidinterstitial} for diamond and $V_{\text{B}} = \SI{7.28}{\angstrom^3/\text{atom}}$\,\footnote{\label{fn:Boron_Volume}The origin of this value is not specified in \citet{Brazhkin_2006}, but it is most likely derived from the density of $\beta$-rhombohedral boron.} \citet{Brazhkin_2006}, derived from crystallographic measurements of unit cell volumes and densities, are used to determine $\kappa_V$:

\begin{equation}
    \kappa = 
    \begin{cases}
        \kappa_r = 0.14286 & \text{(covalent radii) \citet{Brunet_1998}} \\
        \kappa_V = 0.09465 & \text{(atomic volumes) \citet{Brazhkin_2006}}
    \end{cases}
    \label{eqn:k_vals}
\end{equation}

\begin{figure}[t!]
    \centering
    \includegraphics[width=\columnwidth]{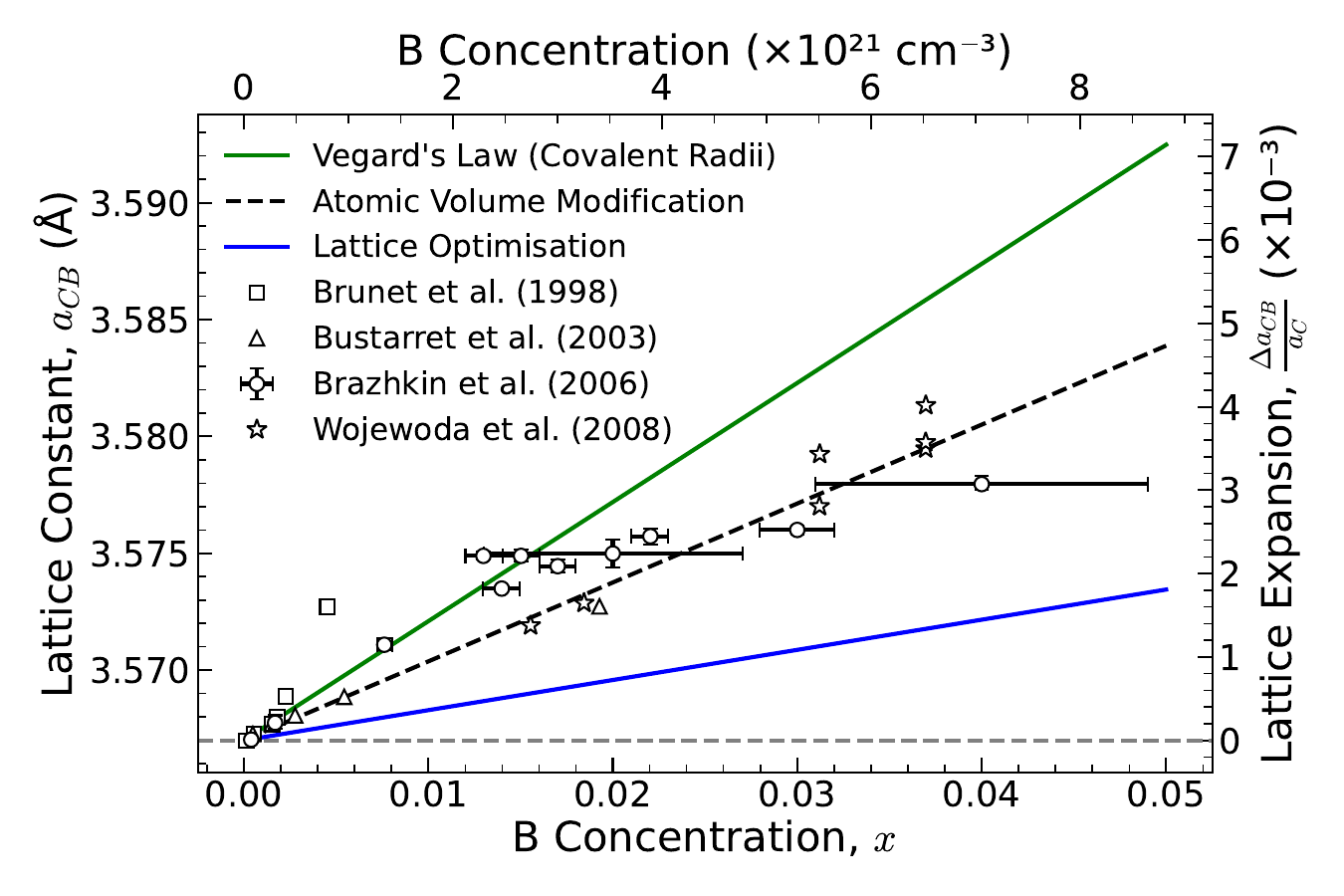}
    \caption{Compilation of literature data showing the lattice constant $a_{\text{CB}}$ and relative lattice expansion $\Delta a_{\text{CB}}/a_{\text{C}}$ of \CB\ crystals as a function of boron concentration $x$. The solid green line represents the covalent radius modification to Vegard's Law \cite{Brunet_1998}, while the dashed black line shows the atomic volume modification \citet{Brazhkin_2006}. The solid blue line shows the results of lattice optimisation, which are discussed in \Cref{sec:LC}, but presented here as a point of comparison. Comparison data are taken from multiple sources including experimental works from \citet{Brunet_1998} (open squares), \citet{Bustarret_2003} (open triangles), and \citet{Brazhkin_2006} (open circles), and \textit{ab initio} calculations from \citet{Wojewoda_2008} (open stars).}
    \label{fig:Lattice_Constant}
\end{figure}

These relationships are shown in \Cref{fig:Lattice_Constant}, with the green line representing the covalent radius modification, and the black dashed line the atomic volume modification, together with experimental data \cite{Brunet_1998, Bustarret_2003, Brazhkin_2006} and computational results \cite{Wojewoda_2008}. The large uncertainties in dopant concentration reported by \citet{Brazhkin_2006} are not explained, but are likely the result of measurement uncertainties in the methods used to estimate the boron concentration of their samples. It is evident from this plot that the atomic volume modification provides a better estimate of the lattice expansion at higher dopant concentrations ($x \gtrsim 0.015$), though both relationships are valid at low dopant concentrations ($x \lesssim 0.005$).

Although widely used, the atomic-volume modification raises questions: it appears to rely on the atomic volume of $\beta$-rhombohedral boron\footref{fn:Boron_Volume}\textemdash whose structure differs fundamentally from cubic diamond\textemdash yet the precise method is not specified. \citet{Wojewoda_2008} instead proposed a three-component relationship based on covalent radii, the fraction of boron atoms forming B$-$B pairs (\ce{B2} dimers), and the concentration of free charge carriers:

\begin{equation}
    a_{CB} = a_C(\kappa_rx + \kappa_{\text{pair}}x_{BB} + \kappa_fx + 1),
    \label{eqn:Three_Vegard}
\end{equation}

\noindent where $\kappa_{\text{pair}} = 0.021824$ is a coefficient accounting for the presence of \ce{B2} dimers (derived from \textit{ab initio} calculations), $x_{BB}$ is the fraction of \ce{B2}, and $\kappa_f = \kappa_V - \kappa_r = -0.04821$ is the free-carrier coefficient, evaluated using the values in \Cref{eqn:k_vals}. This relationship is exactly equivalent to the atomic-volume modification derived by \citet{Brazhkin_2006} and is therefore omitted from \Cref{fig:Lattice_Constant} for clarity.

Since Vegard’s Law assumes isotropic, linear expansion, the inter-planar distances are expected to scale proportionally. The relation in \Cref{eqn:Modified_VL} can therefore be applied to $d_{ijk}$ by substituting $a_{\text{CB}} \to d_{\text{CB}}$ and $a_{\text{C}} \to d_{\text{C}}$, where $d_{\text{C}}$ denotes the inter-planar spacing in pure diamond for a given crystalline plane. As both lattice constants and inter-planar distances scale linearly, the relative change $\Delta d_{ijk}/d_{ijk}$ is equivalent to $\Delta a_{CB}/a_{\text{C}}$.

\subsection{Variation of Lattice Constants} \label{sec:LC}

\begin{figure*}[t!]
 \subfloat[]{\includegraphics[width=0.48\textwidth]{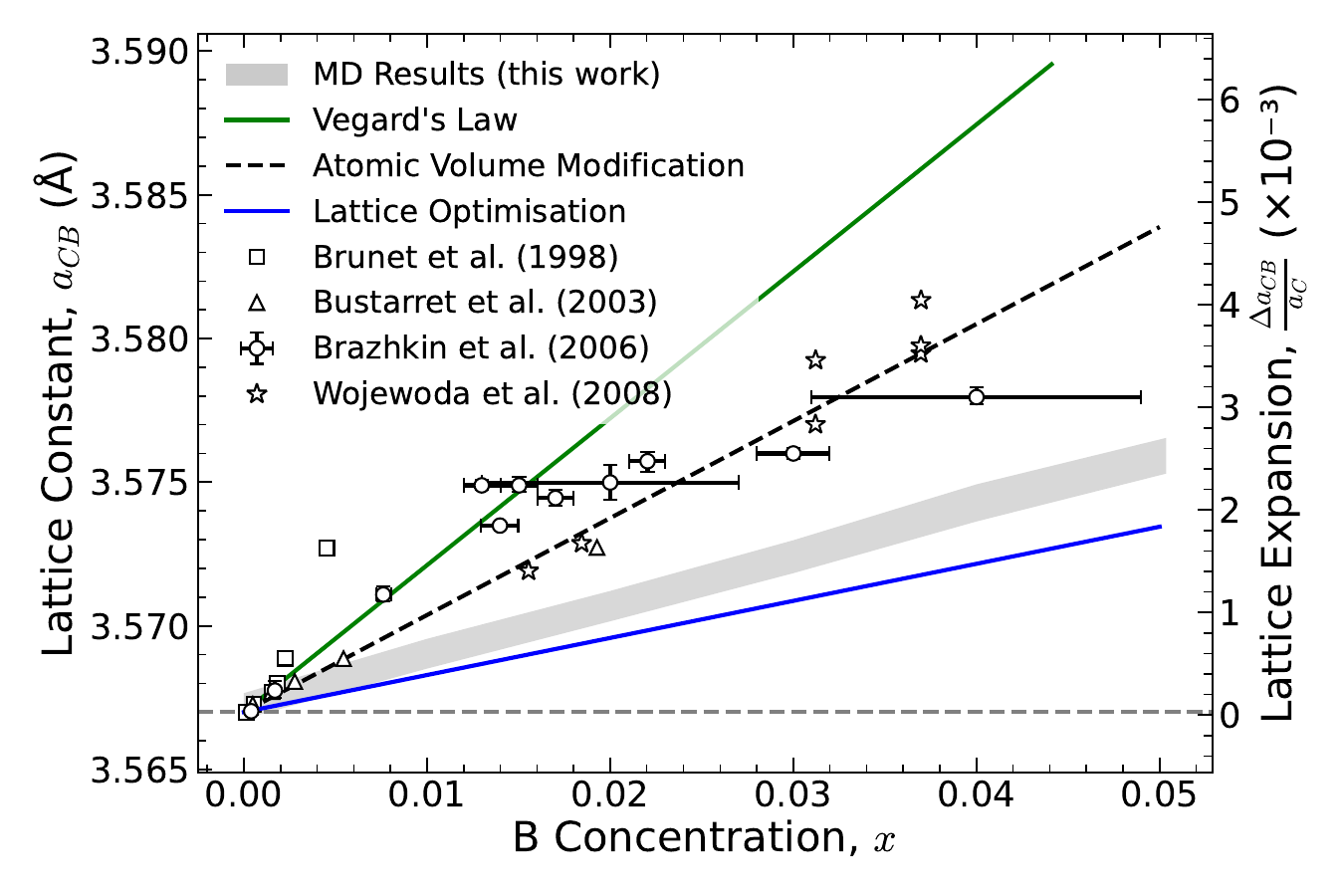}}\quad
 \subfloat[]{\includegraphics[width=0.48\textwidth]{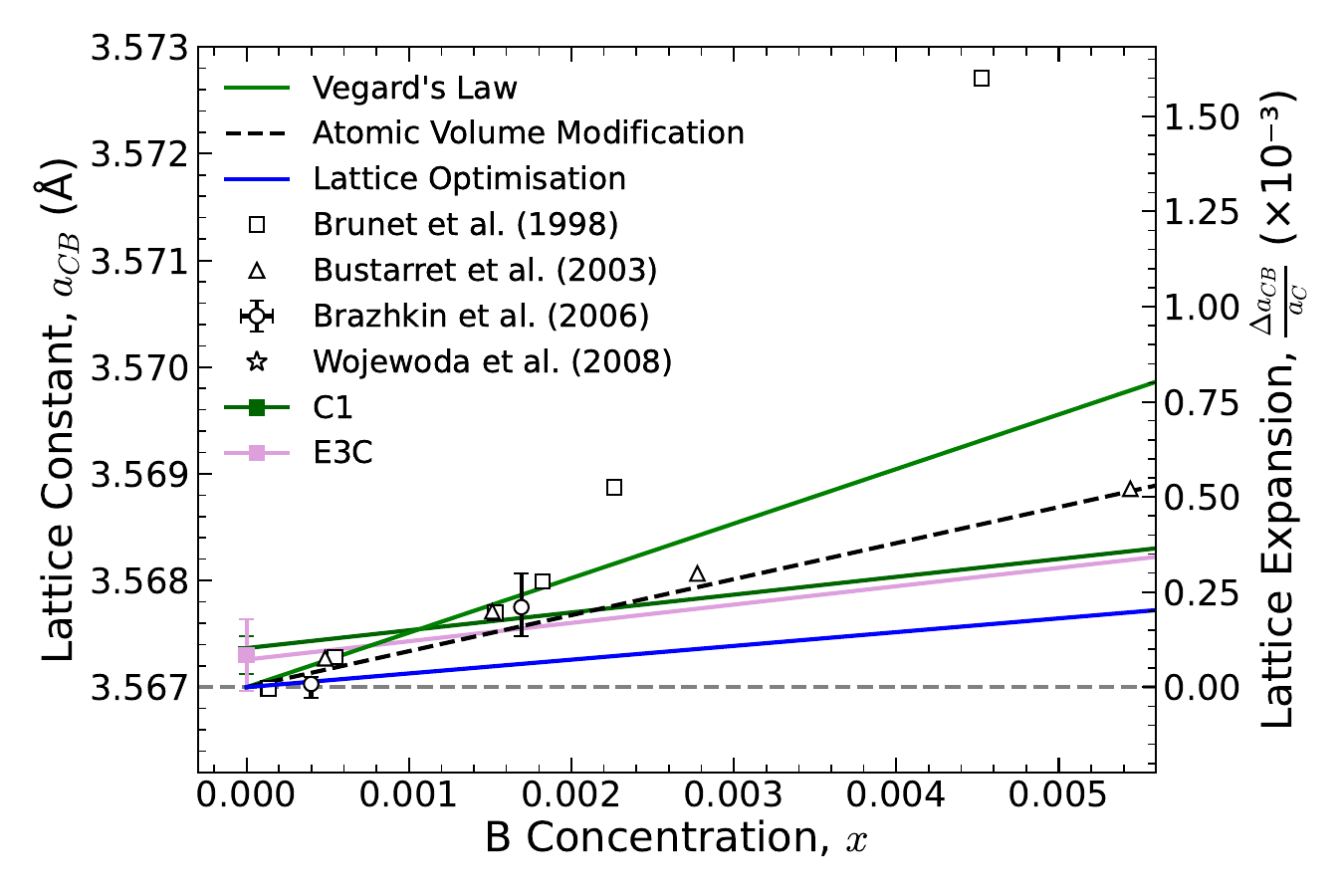}}
 \caption{Plots of the lattice constant $a_{\text{CB}}$ as a function of boron dopant concentration. The grey shaded region in panel \textbf{(a)} spans the full range of values obtained across all crystal sizes} listed in \Cref{tab:Crystals}. Lines corresponding to individual crystal sizes are available in the SI. The solid green line corresponds to Vegard's Law \cite{Brunet_1998}, while the dashed black line represents the atomic volume modification \cite{Brazhkin_2006}. The dashed grey line indicates the nominal lattice constant of pure diamond. Comparison data are taken from multiple sources including experimental works from \citet{Brunet_1998} (open squares), \citet{Bustarret_2003} (open triangles), and \citet{Brazhkin_2006} (open circles), and \textit{ab initio} calculations from \citet{Wojewoda_2008} (open stars). Panel \textbf{(b)} shows a zoom in on lower concentrations and shows only the two extreme cases (C1 and E3C) to allow easier differentiation between the two lines.
 \label{fig:LC_Plots}
\end{figure*}

The lattice constant was calculated directly from the atomic coordinates of the thermalised systems after MD simulations. The variation in the lattice constant with dopant concentration is shown in \Cref{fig:LC_Plots}a. Due to the large number of crystal geometries considered in this work, a shaded region is used to represent the full range of results across all crystal sizes at each boron concentration, rather than plotting each individual line. This approach is also used in later sections to present similar results involving many geometries. The complete data sets for all individual geometries are provided in the SI. The nominal lattice constant of pure diamond, the covalent-radius and atomic-volume modifications, and the experimental and computational data are compiled from \Cref{fig:Lattice_Constant}. The results show a consistent linear trend, in line with expectations from Vegard's Law, reaching a relative lattice expansion of $\Delta a_{\text{CB}}/a_{\text{C}} \sim \num{2.5e-3}$ across all sizes at a dopant concentration of 5\%. Although the data points have associated boron-concentration uncertainties, these have been omitted from the plots for clarity. The error bars on both axes are small and are provided explicitly in the SI.

Compared to the covalent radius and atomic volume modifications, and with literature data, discrepancies are clearly evident. However, at low dopant concentrations (below $\sim0.5\%$), there is reasonably good agreement between the simulation results and literature data, as seen in the zoomed-in view in \Cref{fig:LC_Plots}b. It is exactly these low levels of boron doping that are of most interest for the manufacture of gamma-ray CLS crystals, with candidate crystals manufactured to date not exceeding $\sim1\%$ of B \cite{Connell_2015, Tran_2017}. A slight deviation from the nominal value of the lattice constant is also observed, which may be attributed to edge effects that occur in the free-edge region of the crystal. While the area most affected has been excluded from analysis, these effects still propagate into the overall structure.

Across the full range of dopant concentrations considered, the lattice constant remains linear with concentration, consistent with the behaviour predicted by Vegard's Law. However, the slope of this linear relationship is systematically lower than Vegard's Law predicts. At higher dopant concentrations, a significant deviation from the atomic volume modification and literature data is observed. This deviation can be quantified by fitting simulation results to \Cref{eqn:Modified_VL} to obtain an average value of $\kappa_{\text{MD}} \approx 0.05093$ across all crystal sizes. It should be noted that such deviations are well documented for \CB\ crystals \cite{Brunet_1998, Bustarret_2003, Wojewoda_2008, Kawano_2010} and for other materials \cite{Jacob_2007}, although not to the extent observed in our simulations. This deviation may arise for several reasons. The first concerns the structural quality of the crystals used in this work. The methods herein utilised generate crystals very close to the ideal diamond structure, with direct, random substitution of boron atoms. In contrast, experimental methods such as MPCVD grow crystals layer by layer, introducing boron via the gas phase \cite{Ashfold_1994, Brunet_1998, Connell_2015}. The chemical and kinetic processes associated with this growth and incorporation can inherently lead to the formation of point defects, such as dopant atom clusters, vacancies, atoms located at interstitial positions, as well as dislocations, and residual stresses, all of which can alter the effective lattice parameter and structure. This represents a fundamental limitation of the methodology implemented in this work, and thus the results must be considered to show the equilibrium structural properties of the system at the upper bound of crystal quality.

\begin{figure*}[t!]
 \subfloat[]{\includegraphics[width=\textwidth]{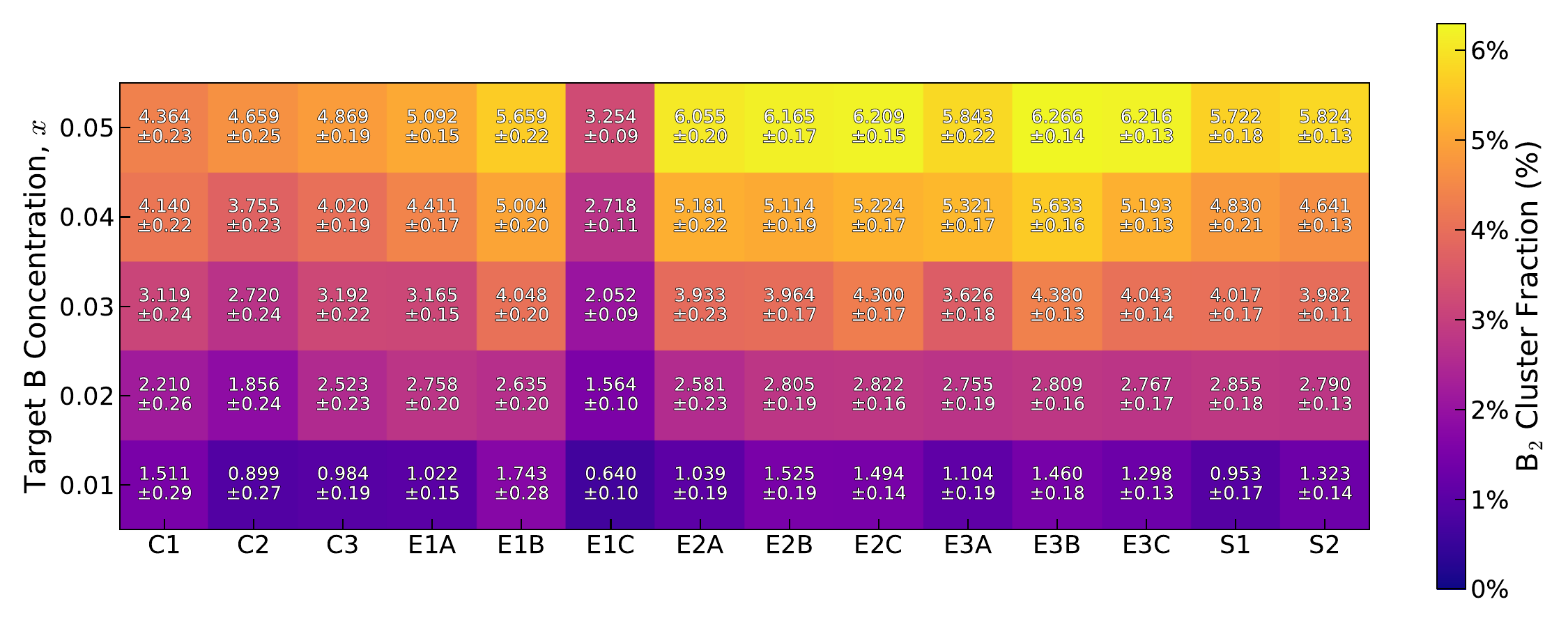}}\\[-2ex]
 \subfloat[]{\includegraphics[width=\textwidth]{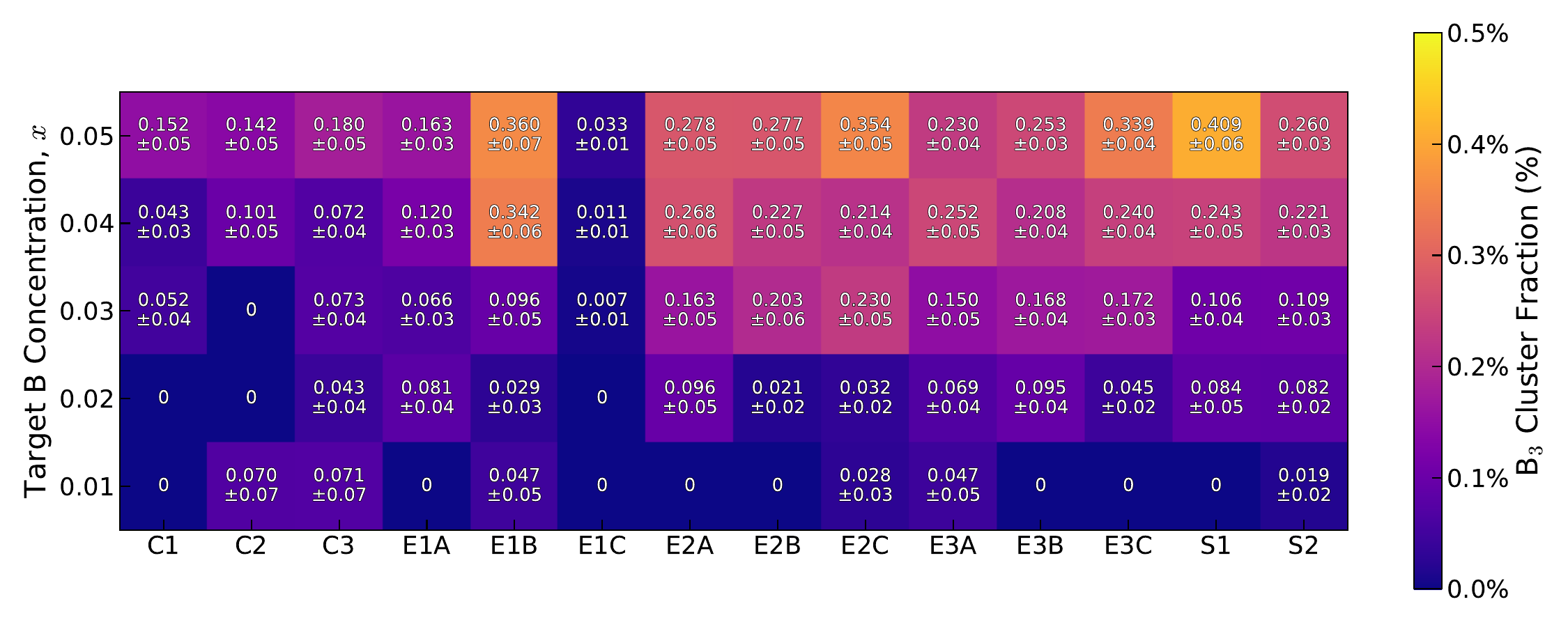}}
 \caption{Heatmaps showing the average fraction of B$-$B atoms pairs \textbf{(a)} and B$_{3}$ clusters \textbf{(b)} relative to isolated B atoms as a function of boron concentration and crystal size. Each box reports the average percentage of B$-$B atom pairs out of all B atoms for the corresponding crystal size and dopant concentration.}
 \label{fig:B-B_Pairs}
\end{figure*}

Another possible origin of the discrepancy observed in \Cref{fig:LC_Plots} is the formation of boron clusters. The three-component form of Vegard's Law \cite{Wojewoda_2008} given in \Cref{eqn:Three_Vegard} explicitly accounts for contributions from \ce{B2} dimers. At higher boron concentrations, the formation of \ce{B2} and larger clusters (\ce{B3} and \ce{B4}) becomes energetically favourable and has been shown to impact the overall lattice structure \cite{Bourgeois_2006, Wojewoda_2008}. Incorporation of these clusters leads to lattice expansion due to the larger covalent radius of boron. Figures \ref{fig:B-B_Pairs}a and \ref{fig:B-B_Pairs}b show heatmaps of the fractions of boron atoms found in \ce{B2} and \ce{B3} clusters, compared to isolated dopants, as functions of concentration and crystal size. The fraction of \ce{B2} clusters increases with dopant concentration, as expected, but remains uniform across crystal sizes. An exception is the C-category crystals which show a lower fraction of \ce{B2} compared to the larger E and S-category crystals, most likely due to the smaller overall number of atoms in these systems 

Overall, the proportion of \ce{B2} clusters observed in the simulation results is very small: at 5\% doping, at most $\sim6.2\%$ of all boron atoms exist in \ce{B2} clusters, corresponding to just $\sim0.31\%$ of all atoms in the crystal. This low cluster fraction likely reflects the small system sizes relative to experimental MPCVD-grown crystals. To our knowledge, no experimental study has quantitatively measured \ce{B2} (or larger) clusters in MPCVD-grown crystals; nevertheless, clustering is expected, with several studies suggesting significant \ce{B2} dimer formation \cite{Gross_2003, Goss_2006, Goss_2008, Watanabe_2018}, and \citet{Bourgeois_2006} explicitly stating that ``a significant fraction of boron atoms may form B dimers''. Thus, while present in the crystals analysed in this work, the impact of \ce{B2} clusters is likely negligible.

A similar analysis of \ce{B3} clusters (\Cref{fig:B-B_Pairs}b) shows a much lower concentration than for \ce{B2}, with similar trends: a higher fraction of \ce{B3} clusters in larger E- and S-category crystals at concentrations above $x=0.03$, and almost none in C-category crystals. Analyses of \ce{B4} clusters were also conducted; however, across the 1940 unique crystal structures generated here, only 17 \ce{B4} clusters were identified, compared with 674 \ce{B3} and 3444 \ce{B2} clusters. No clusters larger than \ce{B4} were found. This extremely low prevalence of boron clusters provides further evidence of the high-quality crystal structures generated by the present methodology and may, in part, explain the deviation from Vegard’s Law observed. As emphasised in \Cref{sec:Intro}, such high-quality, defect-free crystals are essential for gamma-ray CLS applications, while defect characterisation in MPCVD-grown crystals remains an active topic \cite{Korol2014Book, Korol2022Book}. Based on the computational methodology used, the lack of dopant clusters is not necessarily unexpected. As previously discussed, the dopant methodology of random substitution  employed here is representative of a equilibrium structure, and neglects the consideration of the growth kinetics associated with the MPCVD fabrication process. MPCVD permits cluster formation from surface diffusion of deposited atoms, allowing for both the formation of clusters and lattice vacancies. However, these processes take place over timescales much longer, and temperatures much higher than modelled here.

The three-component model of \Cref{eqn:Three_Vegard} \cite{Wojewoda_2008} also accounts for the influence of free carriers on the lattice: boron doping leads to free holes, thereby rendering the material a p-type semiconductor. These free carriers induce a negative lattice strain via hydrostatic deformation of the valence-band maximum, as quantified by \citet{Wojewoda_2008}, leading to a negative deviation from Vegard’s Law at higher dopant concentrations. The MD simulations conducted here do not capture such electronic (valence-band) effects; instead, the predicted lattice expansion arises solely from steric effects associated with substitutional boron. While this represents a limitation of the method, it is of minor consequence for this work, which focuses on evaluating the structural influence of dopant atoms to inform crystal fabrication. 

As a final test, lattice optimisation simulations were performed. In MD simulations, crystals were initially generated with the experimental lattice constant of diamond, $a_\text{C} = \SI{3.5669 \pm 0.0001}{\angstrom}$ \cite{Saotome_1998}. Lattice optimisation of pure diamond yielded an optimised lattice constant of $\SI{3.561}{\angstrom}$. Corresponding lattice optimisation of a cubic boron crystal constructed by replacing all carbon atoms with boron yielded a lattice constant of $\SI{3.690}{\angstrom}$. Using these values with Vegard's Law, \Cref{eqn:Vegards_Law}, results in a slightly smaller slope than obtained from our MD simulations ($\kappa_{\text{MD}} \approx 0.05093$), with $\kappa_{\text{opt}} = 0.03623$. These results are shown as the blue line in Figures \ref{fig:Lattice_Constant} and \ref{fig:LC_Plots}a. Atomic volumes calculated from the optimised structures are $V_\text{C} = \SI{5.64}{\angstrom^3/\text{atom}}$ and $V_\text{B} = \SI{6.28}{\angstrom^3/\text{atom}}$. While the diamond volume closely matches literature values \cite{goldschmidinterstitial}, the cubic boron volume is lower than that of $\beta$-rhombohedral boron. Using these volumes to calculate $\kappa_V$ gives $\kappa_V = 0.03760$, consistent with the $\kappa_{\text{opt}}$. Lattice optimisation simulations are conducted without a thermostat, so thermal effects are not included. Accounting for thermal expansion would increase the optimised lattice constants, owing to the increased thermal expansion of boron compared to diamond ($\sim$\SI{6e-6}{\kelvin^{-1}} \cite{Lundström_1998} vs $\sim$\SI{1e-6}{\kelvin^{-1}} \cite{Jacobson_2019}), reducing the difference relative to the MD simulation results. These lattice optimisation results are more in line with the results of MD simulations, and suggest that the atomic volume for boron quoted by \citet{Brazhkin_2006}\footref{fn:Boron_Volume} leads to an overestimation in the lattice constant by the atomic volume modification. This warrants further investigation and may form the focus of a future study.

Taken together, a combination of enhanced crystal quality, negligible \ce{B2} clustering, and overestimation of the atomic volume of boron in comparison to the results of lattice optimisation provides a reasonable explanation for the discrepancy observed between the results of MD simulations and results from literature.

\subsection{Variation of Inter-planar Distances} \label{subsec:IP}

Building on the analysis of lattice constants, the influence of boron doping on inter-planar distances can be investigated across the three spatial regions outlined in \Cref{sec:Method}. The boundaries between these regions are shown in \Cref{fig:Crystal_Regions}, and were defined based on a fixed number of atomic layers measured along the [1\;0\;0] direction, from the variation in (1\;0\;0) inter-planar distance as a function of the distance from the substrate. For consistency, the same boundary positions were used across all crystal sizes: seven atoms thick for the substrate region (excluding the fixed substrate), and ten atoms thick for the free edge. More details describing the method by which these regions were defined is available in the SI.

\begin{figure*}[t!]
 \subfloat[]{\includegraphics[width=0.5\textwidth]{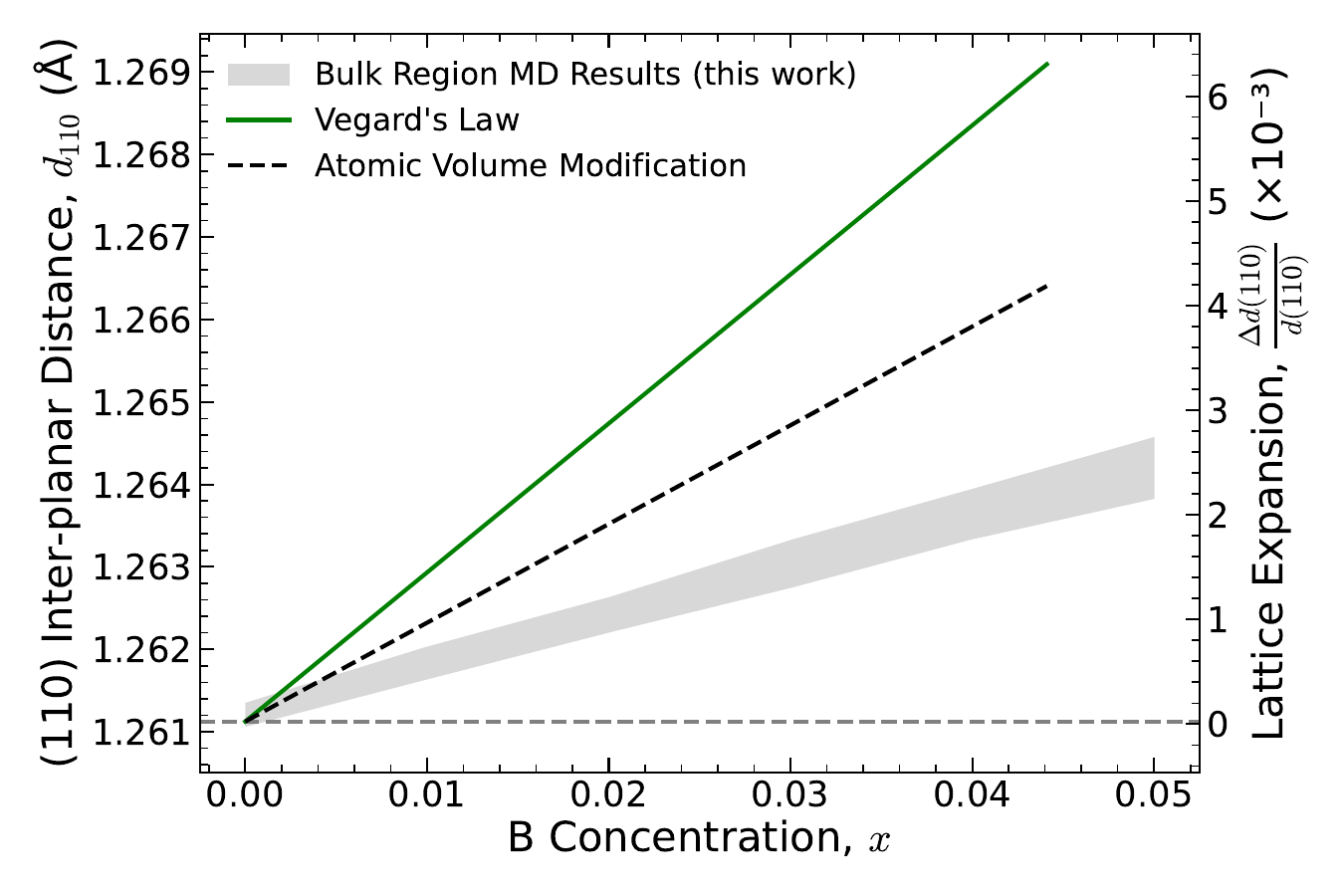}}\quad
 \subfloat[]{\includegraphics[width=0.5\textwidth]{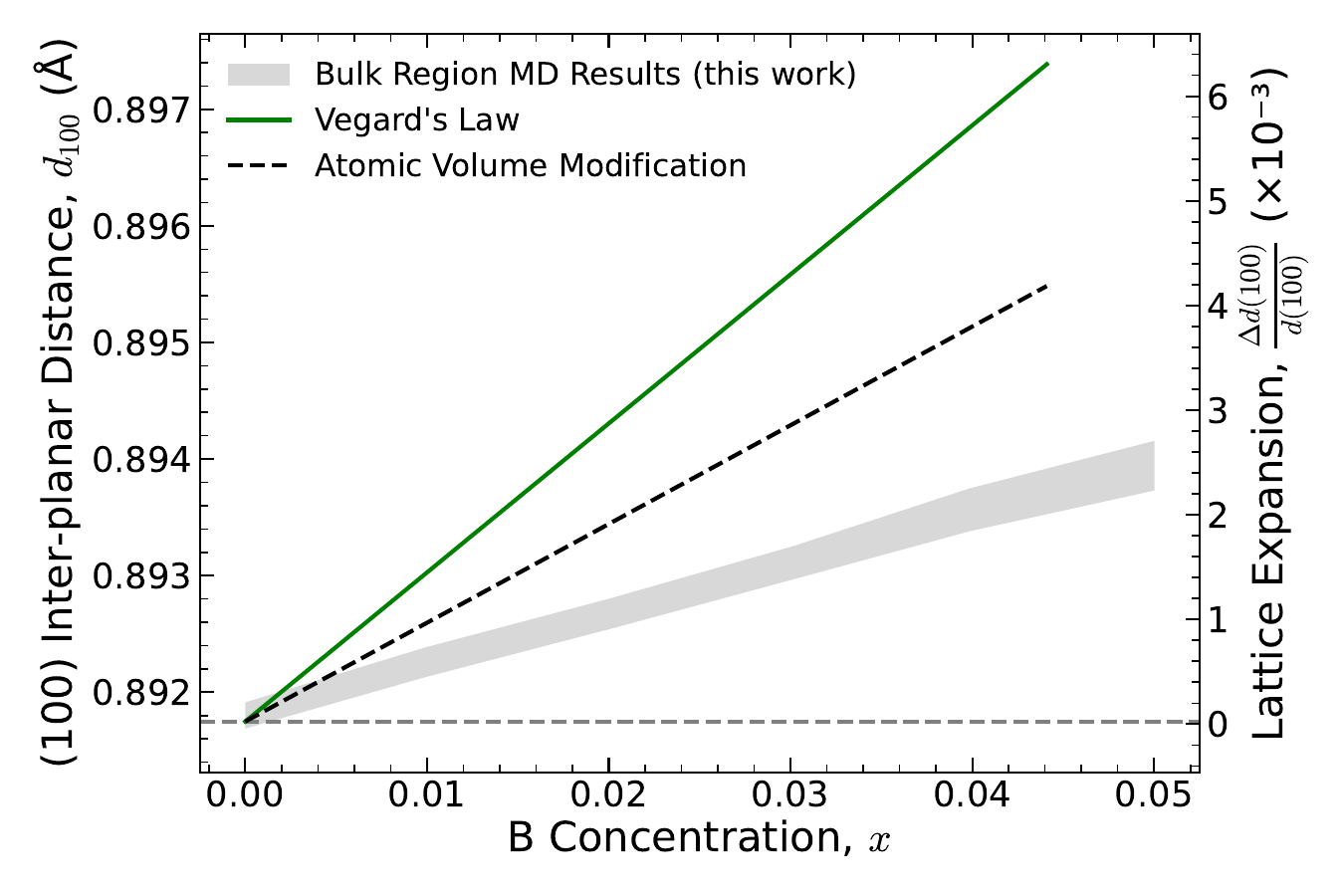}}
 \caption{(1\;1\;0) and (1\;0\;0) inter-planar distances (panels \textbf{(a)} and \textbf{(b)} respectively) as a function of boron dopant concentration. The grey shaded regions represents the full range of values obtained for the bulk region across all crystal sizes. Specific cases showing the region-by-region inter-planar distances for individual crystal sizes are available in the SI. The solid green line represents Vegard’s Law \cite{Vegard_1921}; the dashed black line shows the atomic volume interpolation correction \cite{Brazhkin_2006}; and the dashed grey line indicates the nominal inter-planar spacing in pure diamond.}
 \label{fig:IP_Plots}
\end{figure*}

\begin{figure*}[t!]
    \centering
    \includegraphics[width=\textwidth]{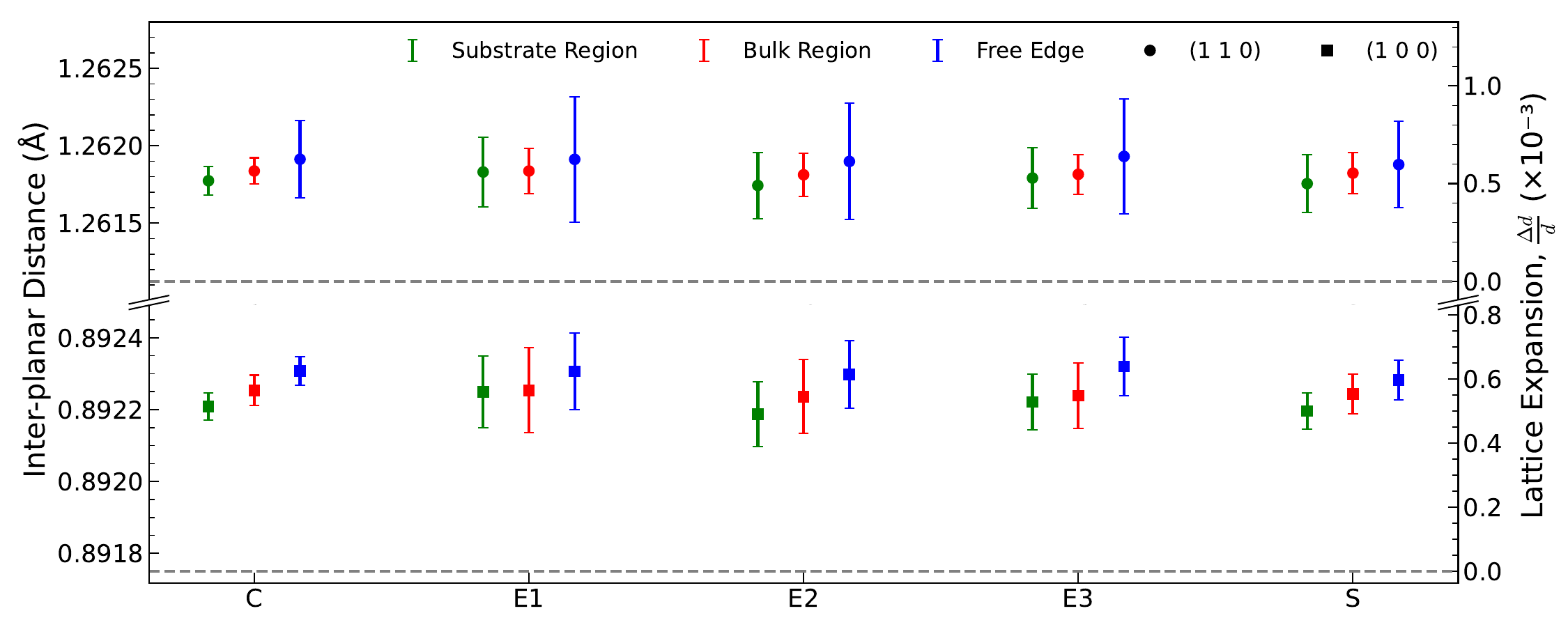}
    \caption{Calculated values of the (1\;1\;0) (circles) and (1\;0\;0) (squares) inter-planar distances across the major crystal categories C, E1, E2, E3, and S. Each point represents the average over all crystal sizes within the corresponding category and is shown for each spatial region of the crystal: substrate (green), bulk (red), and free edge (blue). A corresponding plot showing results for each individual crystal size listed in \Cref{tab:Crystals} is provided in Figure S13 of the SI.}
    \label{fig:Dist_Bars}
\end{figure*}

The variation in (1\;1\;0) and (1\;0\;0) inter-planar distances with boron dopant concentration are shown in Figures \ref{fig:IP_Plots}a and b, respectively, with the grey shaded regions representing the full range of values obtained for the bulk region across all crystal sizes. The complete set of plots for all crystal sizes and spatial regions is provided in the SI, and is summarised later in this section. To facilitate comparison between crystal sizes, \Cref{fig:Dist_Bars} shows the (1\;1\;0) and (1\;0\;0) inter-planar distances for each region, where each point represents the average over all sizes within the corresponding major crystal category (C, E1, E2, E3, or S). In all cases a dopant concentration of $x=0.01$, which is most representative of gamma-ray CLS crystals, is used. A corresponding plot showing all individual crystal sizes listed in \Cref{tab:Crystals} can be seen in Figure S13 of the SI.

\subsubsection{Substrate Region}
As outlined in \Cref{sec:Method}, the crystal generation method includes a three-atom-thick fixed diamond layer to mimic the supportive role of the substrate typically used in CVD crystal growth. The size of this region was fixed at seven atomic layers from the fixed substrate, independent of the overall crystal size. This consistent definition enables a direct comparison of the influence of bulk region size and substrate anchoring on inter-planar distances.

In the substrate region, the inter-planar distances between the (1\;1\;0) and (1\;0\;0) planes are broadly similar, as can be seen in \Cref{fig:Dist_Bars} (and Figure S13 in the SI) which show similar values for all crystal sizes and for both plane sets. This is particularly the case for the C-category crystals, where simulation statistics are highest and thus the error bars in \Cref{fig:Dist_Bars} are smallest. Larger crystals (E1 and E2) show slight variation between inter-category sizes, as seen from the larger statistical error bars and directly from Figure S13. These, however, largely remain within these statistical error bars and are therefore not considered significant. Overall, these observations indicate that the substrate is playing a supportive role in constraining this region for both the (1\;1\;0) and (1\;0\;0) crystalline planes.

\subsubsection{Bulk Region}
The bulk region constitutes the central portion of the crystal, surrounded by atoms on all sides: laterally through the periodic boundaries, and axially by the substrate and free-edge regions. As such, this configuration effectively emulates the interior of an infinitely periodic crystal. The extent of the bulk region does not correspond to a fixed number of atomic layers, but is defined by the volume not occupied by the substrate or free edge. Therefore, the number of atomic layers in this region varies from a minimum of 8 for the smallest crystal (C1) up to 184 for the largest one (E3C).

The bulk region exhibits a greater dependence on the axial crystal dimension compared to the substrate region. The full plots showing dependences for individual crystal sizes provided in Section S4 of the SI show a systematic increase in the (1\;1\;0) inter-planar distances as the axial dimensions of the crystal increase. This behaviour is most pronounced at higher dopant concentrations but is still observed across most crystal categories at $x=0.01$. This can also be seen in a direct comparison of \Cref{fig:IP_Plots}a and b, where the shaded region representing the full range of bulk region values across all crystal sizes is wider for the (1\;1\;0) inter-planar distances than for the (1\;0\;0) inter-planar distances. This can also be directly observed from the red points in Figure S13. The only exception is the S-category, where the axial dimensions are the same for S1 and S2. This effect is not observed for the (1\;0\;0) planes, where the inter-planar distances remain uniform across all crystal sizes.

This dependence of the (1\;1\;0) planes on axial dimensions may be attributed to bulk-size effects: at smaller sizes the crystals do not occupy a sufficiently bulk-like volume, so finite-size effects yield smaller increases in inter-planar distance. This effect is most pronounced along the axial direction, as this is the direction of expansion, and the (1\;1\;0) inter-planar distances have components from both the $X$ direction expansion, and the fixed $Y$ direction. In comparison, for the (1\;0\;0) planes, there is only a single expansion component, so the limited expansion of the crystal in the $Y$ and $Z$ directions does not play a role. In short, this is a geometric artefact arising from the anisotropic expansion of the system, even after applying the geometric correction, and it provides further evidence that the substrate plays a supportive role in the substrate region.

\subsubsection{Free Edge}
The free-edge region is the portion of the crystal that is exposed to vacuum. It is initially defined to include the ten atomic layers adjacent to the free edge. Of these, the three layers closest to the surface are excluded from analysis due to atypical coordination and geometry arising from the absence of periodic boundaries along the [1\;0\;0] direction. This leaves seven atomic layers included in the analysis, matching the number used for the substrate region.

In this region, the absence of periodic support along $X$ leads to surface relaxation and edge effects, producing large crystal-to-crystal deviations in inter-planar distance, even between crystals of the same size and dopant concentration. A deviation from the nominal inter-planar distance is evident from the non-zero width of the shaded region in the undoped case in Figures \ref{fig:IP_Plots}a and \ref{fig:IP_Plots}b, directly resulting from these edge effects. Similar behaviour is observed for the lattice constant in \Cref{fig:LC_Plots}b and, to a lesser extent, in the bulk and substrate regions. Although an overall linear dependence is present here, the associated uncertainties are substantially larger. The (1\;1\;0) and (1\;0\;0) planes show comparable variability across all crystal sizes (blue points in \Cref{fig:Dist_Bars}), and this region consistently exhibits the largest inter-planar distances. In general, there is no distinct size trend, although the S-category crystals show the least variation across sizes from Figure S13. This most likely reflects their larger unconstrained surface (greater $Y$ and $Z$ dimensions), which provides more atoms in the averaging region and thus better statistics. Overall, a free surface permits larger local relaxations that depart from bulk behaviour, as expected. The diamond (1\;0\;0) surface is also known to undergo a $2\times1$ dimer reconstruction to reduce the energy associated with dangling bonds \cite{Halicioglu_1992, Kern_1996}, which is not explicitly modelled in these simulations. This relaxation mainly affects the few atomic layers closest to the surface, thus the absence of explicit surface reconstruction is not expected to significantly influence the lattice expansion in this region.

Results in this region should be treated with caution, as the uncertainties are inherently large. We include the free-edge results for completeness and as a methodological demonstration, but the analysis of this region is not exhaustive; a fuller understanding would require further work beyond the scope of this study.

\section{Conclusions and Outlook} \label{sec:Conclusions}
In this study, atomistic-scale molecular dynamics simulations were performed on a range of periodically replicated \CB\ crystals with boron concentrations between $x=0.00$ and $x=0.05$. The aim was to elucidate the effects of boron atom substitution on the lattice structure for crystals of various sizes. These simulations have been conducted in the context of the design and practical realisation of crystal-based gamma-ray light sources based on periodically bent crystals. Given that the performance of such crystals depends on precise structural control, particularly regarding dopant incorporation, a detailed understanding of the effects of boron concentration is essential for establishing practical manufacturing tolerances.

A clear linear relationship has been observed between boron concentration and both the lattice constant and the inter-planar distances of the (1\;1\;0) and (1\;0\;0) planes across all crystal sizes studied. By dividing the simulation box into three distinct regions (substrate-adjacent, bulk, and free edge), the response of distinct structural environments to boron doping has been analysed. Across all crystal sizes, the lattice constant follows the same linear dependence, with only minor variations that fall within evaluated error bars. In the substrate region, the inter-planar distance remains essentially uniform across all crystal sizes, confirming the mechanical support provided by the fixed substrate. In contrast, the bulk region exhibits a geometric dependence with increasing axial crystal dimension on the (1\;1\;0) inter-planar distance. This may be attributed to the finite size effects of the crystal and artifacts originating from the anisotropic nature of the expansion in the simulations. At the free edge, increased variability in the inter-planar distances is observed, attributed to edge effects caused by the absence of periodic support.

Across all regions, the results consistently deviate from the predictions of the covalent radius \cite{Brunet_1998} and atomic volume \cite{Brazhkin_2006} modifications to Vegard's Law \cite{Vegard_1921}, and other literature data \cite{Brunet_1998, Bustarret_2003, Brazhkin_2006, Wojewoda_2008}. This deviation may be attributed to several factors, including enhanced crystal quality relative to MPCVD-grown crystals: the methodology presented here yields near-ideal structures in which boron atoms directly substitute for carbon, and interstitial atoms are excluded. By comparison, MPCVD growth is not expected to produce purely substitutional boron but rather an alloy-like structure with clusters, interstitial atoms, and defects. Indeed, our simulations show only a very small fraction of boron clusters, which may also contribute to the discrepancy.

In the context of gamma-ray CLS crystals, a defect-free structure as close to an ideal crystal as possible is desirable. Such high-quality crystals were not the focus of the experimental works of Refs.~\cite{Brunet_1998, Bustarret_2003, Brazhkin_2006}, where crystal quality was not the main focus, thus a deviation from their results is expected. The results presented here are therefore indicative of an upper-bound case for the design of gamma-ray CLSs. While crystals of such high quality are not currently obtainable in practice, these results therefore serve as a reference for defect-free, stress-free, fully substitutional crystals, providing a clear benchmark for the manufacture of high-quality BDD light-source materials. Indeed, recent reports of near defect-free crystals manufactured for gamma-ray CLS applications \cite{TECHNO-CLS_2025} suggest that this idealised limit is relevant to the cutting edge of current fabrication techniques.

Another important aspect is the boron concentration. This study considers dopant levels up to 5\%, well above those typically used in BDD crystals. At the lower concentrations of $\lesssim 1\%$ typically used in BDD crystal growth, our results show good agreement with other literature data. At higher dopant concentrations less data is available and uncertainties increase due to the formation of defects. The crystal sizes used here are small compared with typical nanometre-scale defects, which therefore cannot be analysed by the present methods. MPCVD could instead be modelled as a stochastic process via kinetic Monte Carlo (kMC) simulations. A number of works have utilised such methods to model CVD growth of diamond \cite{Netto_1998, Rodgers_2015}, though to our knowledge boron incorporation has never been explicitly included, and instead only modelled using Density Functional Theory (DFT) \cite{Cheesman_2005, Lu_2025}. Importantly, the work presented here should be viewed as a first step toward more comprehensive modelling of boron-doped diamond systems. Future studies may extend this framework by incorporating additional complexities such as interstitial dopant atoms, periodically graded dopant profiles, and realistic growth processes.

Building upon our earlier work on \SiGe\ systems \cite{Si-Ge_Paper}, this study implements an improved simulation framework incorporating both substrate effects and periodic boundary conditions in \CB\ crystals, making it more representative of real materials used in gamma-ray CLSs. This methodology offers a computationally efficient and accurate platform not only for studying lattice distortions but also for investigating other material responses relevant to CLS development, such as amorphisation and phase transitions, radiation damage, and thermal effects. Coupled with experimental data, these simulations offer a powerful route for guiding the design of next-generation crystal-based light sources.

\section*{Acknowledgements}
\noindent This work was funded by UK Research and Innovation (UKRI) under the UK government’s Horizon Europe funding Guarantee under the grant No. 10037865 in collaboration with the European Commission's Horizon Europe-EIC-Pathfinder-Open TECHNO-CLS (G.A. 101046458) project. The authors also acknowledge financial support from the H2020 RISE-NLIGHT project (G.A. 872196), and from the COST Action
CA20129 MultIChem, supported by COST (European Cooperation in Science and Technology).
The work was supported in part by Deutsche Forschungsgemeinschaft, Germany (Project No. 413220201).
The authors gratefully acknowledge the Specialist and High Performance Computing systems provided by Information Services at the University of Kent. Special thanks are extended to Dr Timothy Kinnear for HPC assistance with the Icarus cluster at the University of Kent. The authors kindly thank the four anonymous referees, whose valuable comments have improved the quality and clarity of this manuscript.

\section*{CRediT Author Contribution}
\noindent \textbf{Matthew D. Dickers:} Conceptualisation, Methodology, Software, Formal analysis, Data curation, Visualisation, Writing - original draft, Writing - review \& editing; \textbf{Felipe Fantuzzi:} Supervision, Writing - review \& editing; \textbf{Nigel J. Mason:} Supervision, Writing - review \& editing; \textbf{Gennady B. Sushko}: Methodology, Software; \textbf{Andrei V. Korol:} Conceptualisation, Methodology, Validation, Supervision, Writing - review \& editing; \textbf{Andrey V. Solov'yov:} Conceptualisation, Methodology, Software, Validation, Supervision, Writing - review \& editing

\section*{Competing Interests}
\noindent The authors do not declare any conflicts of interest, and there is no financial interest to report.

\section*{Data Availability}
\noindent The datasets generated and/or analysed during the current study are available from the corresponding author upon reasonable request. Data presented in plots of lattice constant and inter-planar distances are tabulated in the supplementary information

\bibliographystyle{elsarticle-num} 
\bibliography{bibliography.bib}

\begin{thebibliography}{10}
\expandafter\ifx\csname url\endcsname\relax
  \def\url#1{\texttt{#1}}\fi
\expandafter\ifx\csname urlprefix\endcsname\relax\def\urlprefix{URL }\fi
\expandafter\ifx\csname href\endcsname\relax
  \def\href#1#2{#2} \def\path#1{#1}\fi

\bibitem{Muzyka_2019}
K.~Muzyka, J.~Sun, T.~H. Fereja, Y.~Lan, W.~Zhang, G.~Xu, {Boron-doped diamond:
  current progress and challenges in view of electroanalytical applications},
  Anal. Methods 11 (2019) 397--414.
\newblock \href {https://doi.org/10.1039/C8AY02197J}
  {\path{doi:10.1039/C8AY02197J}}.

\bibitem{Einaga_2022}
Y.~Einaga, {Boron-Doped Diamond Electrodes: Fundamentals for Electrochemical
  Applications}, Acc. Chem. Res. 55~(24) (2022) 3605--3615.
\newblock \href {https://doi.org/10.1021/acs.accounts.2c00597}
  {\path{doi:10.1021/acs.accounts.2c00597}}.

\bibitem{Alkahtani_2022}
M.~Alkahtani, D.~K. Zharkov, A.~V. Leontyev, A.~G. Shmelev, V.~G. Nikiforov,
  P.~R. Hemmer, {Lightly Boron-Doped Nanodiamonds for Quantum Sensing
  Applications}, Nanomaterials 12~(4) (2022).
\newblock \href {https://doi.org/10.3390/nano12040601}
  {\path{doi:10.3390/nano12040601}}.

\bibitem{Bhattacharya_2025}
S.~Bhattacharya, J.~Boyd, S.~Reichardt, V.~Allard, A.~H. Talebi, N.~Maccaferri,
  O.~Shenderova, A.~L. Lereu, L.~Wirtz, G.~Strangi, R.~M. Sankaran,
  {Intervalence plasmons in boron-doped diamond}, Nat. Commun. 16~(1) (2025)
  444.
\newblock \href {https://doi.org/10.1038/s41467-024-55353-0}
  {\path{doi:10.1038/s41467-024-55353-0}}.

\bibitem{Thaiyotin_2000}
L.~Thaiyotin, T.~Phetchakul, S.~Cheirsirikul, S.~Supadech, {UV photodetectors
  from B-doped diamond film}, in: {2000 TENCON Proceedings. Intelligent Systems
  and Technologies for the New Millennium (Cat. No.00CH37119)}, Vol.~3, 2000,
  pp. 230--233 vol.3.
\newblock \href {https://doi.org/10.1109/TENCON.2000.892263}
  {\path{doi:10.1109/TENCON.2000.892263}}.

\bibitem{Aksenova_2017}
A.~S. Aksenova, A.~A. Altuhov, E.~V. Ryabeva, V.~T. Samosadnyi, V.~S.
  Feshchenko, A.~P. Chernyaev, V.~A. Shepelev, {The investigation of
  boron-doped diamond absorbance spectrum}, J. Phys. Conf. Ser. 798~(1) (2017)
  012149.
\newblock \href {https://doi.org/10.1088/1742-6596/798/1/012149}
  {\path{doi:10.1088/1742-6596/798/1/012149}}.

\bibitem{Korol2014Book}
A.~V. Korol, A.~V. Solov'yov, W.~Greiner, {Channeling and Radiation in
  Periodically Bent Crystals}, Springer Berlin, Heidelberg, 2014.
\newblock \href {https://doi.org/https://doi.org/10.1007/978-3-642-54933-5}
  {\path{doi:https://doi.org/10.1007/978-3-642-54933-5}}.

\bibitem{Connell_2015}
S.~H. Connell, J.~H{\"a}rtwig, A.~Masvaure, D.~Mavunda, T.~N. {Tran Thi},
  \href{https://events.saip.org.za/event/34/attachments/1143/1398/SAIP2014-169.pdf}{{Towards
  a crystal undulator}}, in: C.~Engelbrecht, S.~Karataglidis (Eds.),
  {Proceedings of the 59th Annual Conference of the South African Institute of
  Physics (SAIP2014)}, University of Johannesburg, Johannesburg, 2015, pp.
  169--174.
\newline\urlprefix\url{https://events.saip.org.za/event/34/attachments/1143/1398/SAIP2014-169.pdf}

\bibitem{Tran_2017}
T.~N. Tran~Thi, J.~Morse, D.~Caliste, B.~Fernandez, D.~Eon, J.~H{\"{a}}rtwig,
  C.~Barbay, C.~Mer-Calfati, N.~Tranchant, J.~C. Arnault, T.~A. Lafford,
  J.~Baruchel, {Synchrotron Bragg Diffraction Imaging Characterization of
  Synthetic Diamond Crystals for Optical and Electronic Power Device
  Applications}, J. Appl. Crystallogr. 50~(2) (2017) 561--569.
\newblock \href {https://doi.org/10.1107/S1600576717003831}
  {\path{doi:10.1107/S1600576717003831}}.

\bibitem{Korol2020}
A.~V. Korol, A.~V. Solov'yov, {Crystal-based Intensive Gamma-ray Light
  Sources}, Europ. Phys. J. D. 74~(10) (2020) 201.
\newblock \href {https://doi.org/10.1140/epjd/e2020-10239-8}
  {\path{doi:10.1140/epjd/e2020-10239-8}}.

\bibitem{Korol2022Book}
A.~Korol, A.~V. Solov'yov, {Novel Lights Sources Beyond Free Electron Lasers},
  Springer Nature, Cham, 2022.
\newblock \href {https://doi.org/https://doi.org/10.1007/978-3-031-04282-9}
  {\path{doi:https://doi.org/10.1007/978-3-031-04282-9}}.

\bibitem{KOROL2023}
A.~V. Korol, A.~V. Solov'yov, {Atomistic Modeling and Characterizaion of Light
  Sources Based on Small-amplitude Short-period Periodically Bent Crystals},
  Nucl. Instrum. Methods Phys. Res. B 537 (2023) 1--13.
\newblock \href {https://doi.org/https://doi.org/10.1016/j.nimb.2023.01.012}
  {\path{doi:https://doi.org/10.1016/j.nimb.2023.01.012}}.

\bibitem{Sushko_2022}
G.~B. Sushko, A.~V. Korol, A.~V. Solov'yov, {Extremely Brilliant Crystal-based
  Light Sources}, Europ. Phys. J. D. 76~(9) (2022) 166.
\newblock \href {https://doi.org/10.1140/epjd/s10053-022-00502-7}
  {\path{doi:10.1140/epjd/s10053-022-00502-7}}.

\bibitem{Lindhard_1965}
J.~Lindhard, {Influence Of Crystal Lattice On Motion Of Energetic Charged
  Particles}, {Kongel. Dan. Vidensk. Selsk., Mat.-Fys. Medd.} 34~(14) (1965)
  1--64.

\bibitem{Kumakhov_1976}
M.~Kumakhov, {On the Theory of Electromagnetic Radiation of Charged Particles
  in a Crystal}, Phys. Lett. A 57~(1) (1976) 17--18.
\newblock \href {https://doi.org/https://doi.org/10.1016/0375-9601(76)90438-2}
  {\path{doi:https://doi.org/10.1016/0375-9601(76)90438-2}}.

\bibitem{Uggerhoj_2005}
U.~I. Uggerh\o{}j, {The Interaction of Relativistic Particles with Strong
  Crystalline Fields}, Rev. Mod. Phys. 77~(4) (2005) 1131--1171.
\newblock \href {https://doi.org/10.1103/RevModPhys.77.1131}
  {\path{doi:10.1103/RevModPhys.77.1131}}.

\bibitem{Korol_1999}
A.~V. Korol, A.~V. Solov'yov, W.~Greiner, {Photon Emission by an
  Ultra-relativistic Particle Channeling in a Periodically Bent Crystal}, Int.
  J. Mod. Phys. E 08~(01) (1999) 49--100.
\newblock \href {https://doi.org/10.1142/S0218301399000069}
  {\path{doi:10.1142/S0218301399000069}}.

\bibitem{Guidi_2007}
V.~Guidi, L.~Lanzoni, A.~Mazzolari, G.~Martinelli, A.~Tralli, {Design of a
  Crystalline Undulator Based on Patterning by Tensile Si3N4 Strips on a Si
  Crystal}, Appl. Phys. Lett. 90~(11) (2007) 114107.
\newblock \href {https://doi.org/10.1063/1.2712510}
  {\path{doi:10.1063/1.2712510}}.

\bibitem{Guidi_2011}
V.~Guidi, L.~Lanzoni, A.~Mazzolari, {Patterning and Modeling of Mechanically
  Bent Silicon Plates Deformed through Coactive Stresses}, Thin Solid Films
  520~(3) (2011) 1074--1079.
\newblock \href {https://doi.org/https://doi.org/10.1016/j.tsf.2011.09.008}
  {\path{doi:https://doi.org/10.1016/j.tsf.2011.09.008}}.

\bibitem{Malagutti_2025}
L.~Malagutti, L.~Bandiera, F.~Bonaf\`{e}, N.~Canale, D.~{De Salvador},
  P.~Fedeli, J.~R. Garrido, V.~Guidi, L.~Lanzoni, A.~Korol, F.~Mancarella,
  R.~Negrello, G.~Patern\`{o}, M.~Romagnoni, F.~Sgarbossa, A.~Solov'yov,
  A.~Sytov, D.~Valzani, A.~Mazzolari, {From simulation to fabrication:
  Realizing silicon crystalline undulators with silicon nitride stressor layer
  patterning}, Nucl. Instrum. Meth. Phys. Res. A 1076 (2025) 170480.
\newblock \href {https://doi.org/https://doi.org/10.1016/j.nima.2025.170480}
  {\path{doi:https://doi.org/10.1016/j.nima.2025.170480}}.

\bibitem{Bellucci_2003}
S.~Bellucci, S.~Bini, V.~M. Biryukov, Y.~A. Chesnokov, S.~Dabagov, G.~Giannini,
  V.~Guidi, Y.~M. Ivanov, V.~I. Kotov, V.~A. Maisheev, C.~Malag\`u,
  G.~Martinelli, A.~A. Petrunin, V.~V. Skorobogatov, M.~Stefancich,
  D.~Vincenzi, {Experimental Study for the Feasibility of a Crystalline
  Undulator}, Phys. Rev. Lett. 90~(3) (2003) 034801.
\newblock \href {https://doi.org/10.1103/PhysRevLett.90.034801}
  {\path{doi:10.1103/PhysRevLett.90.034801}}.

\bibitem{Guidi_2005}
V.~Guidi, A.~Antonini, S.~Baricordi, F.~Logallo, C.~Malag\`{u}, E.~Milan,
  A.~Ronzoni, M.~Stefancich, G.~Martinelli, A.~Vomiero, {Tailoring of Silicon
  Crystals for Relativistic-particle Channeling}, Nucl. Instrum. Methods Phys.
  Res. B 234~(1) (2005) 40--46.
\newblock \href {https://doi.org/https://doi.org/10.1016/j.nimb.2005.01.008}
  {\path{doi:https://doi.org/10.1016/j.nimb.2005.01.008}}.

\bibitem{Bagli_2014}
E.~Bagli, L.~Bandiera, V.~Bellucci, A.~Berra, R.~Camattari, D.~De~Salvador,
  G.~Germogli, V.~Guidi, L.~Lanzoni, D.~Lietti, A.~Mazzolari, M.~Prest, V.~V.
  Tikhomirov, E.~Vallazza, {Experimental evidence of planar channeling in a
  periodically bent crystal}, Eur. Phys. J. C 74~(10) (2014) 3114.
\newblock \href {https://doi.org/10.1140/epjc/s10052-014-3114-x}
  {\path{doi:10.1140/epjc/s10052-014-3114-x}}.

\bibitem{Camattari_2017}
R.~Camattari, G.~Patern{\`{o}}, M.~Romagnoni, V.~Bellucci, A.~Mazzolari,
  V.~Guidi, {Homogeneous self-standing curved monocrystals, obtained using
  sandblasting, to be used as manipulators of hard X-rays and charged particle
  beams}, J. Appl. Crystallogr. 50~(1) (2017) 145--151.
\newblock \href {https://doi.org/10.1107/S1600576716018768}
  {\path{doi:10.1107/S1600576716018768}}.

\bibitem{Balling_2009}
P.~Balling, J.~Esberg, K.~Kirsebom, D.~Le, U.~Uggerh\o{}j, S.~Connell,
  J.~H\"{a}rtwig, F.~Masiello, A.~Rommeveaux, {Bending diamonds by femtosecond
  laser ablation}, Nucl. Instrum. Methods Phys. Res. B 267~(17) (2009)
  2952--2957.
\newblock \href {https://doi.org/https://doi.org/10.1016/j.nimb.2009.06.109}
  {\path{doi:https://doi.org/10.1016/j.nimb.2009.06.109}}.

\bibitem{Bellucci_2015}
V.~Bellucci, R.~Camattari, V.~Guidi, A.~Mazzolari, G.~Patern\`{o}, G.~Mattei,
  C.~Scian, L.~Lanzoni, {Ion implantation for manufacturing bent and
  periodically bent crystals}, Appl. Phys. Lett. 107~(6) (2015) 064102.
\newblock \href {https://doi.org/10.1063/1.4928553}
  {\path{doi:10.1063/1.4928553}}.

\bibitem{Baryshevsky_1980}
V.~Baryshevsky, I.~Dubovskaya, A.~Grubich, {Generation of gamma-quanta by
  channeled particles in the presence of a variable external field}, Phys.
  Lett. A 77~(1) (1980) 61--64.
\newblock \href {https://doi.org/https://doi.org/10.1016/0375-9601(80)90637-4}
  {\path{doi:https://doi.org/10.1016/0375-9601(80)90637-4}}.

\bibitem{Ikezi_1984}
H.~Ikezi, Y.~Lin-Liu, T.~Ohkawa, {Channeling Radiation in a Periodically
  Distorted Crystal}, Phys. Rev. B 30~(3) (1984) 1567--1569.
\newblock \href {https://doi.org/10.1103/PhysRevB.30.1567}
  {\path{doi:10.1103/PhysRevB.30.1567}}.

\bibitem{Dedkov_1994}
G.~V. Dedkov, {Channeling Radiation in a Crystal Undergoing an Action of
  Ultrasonic or Electromagnetic Waves}, Phys. Stat. Sol. 184~(2) (1994)
  535--542.
\newblock \href {https://doi.org/https://doi.org/10.1002/pssb.2221840227}
  {\path{doi:https://doi.org/10.1002/pssb.2221840227}}.

\bibitem{Korol_1998}
A.~V. Korol, A.~V. Solov'yov, W.~Greiner, {Coherent Radiation of an
  Ultrarelativistic Charged Particle Channelled in a Periodically Bent
  Crystal}, J. Phys. G 24~(5) (1998) L45.
\newblock \href {https://doi.org/10.1088/0954-3899/24/5/001}
  {\path{doi:10.1088/0954-3899/24/5/001}}.

\bibitem{Wagner_2011}
W.~Wagner, B.~Azadegan, H.~Buettig, L.~S. Grigoryan, A.~Mkrtchyan, J.~Pawelke,
  {Channeling Radiation on Quartz Stimulated by Acoustic Waves}, Nuovo Cimento
  C 34~(4) (2011) 133–140.
\newblock \href {https://doi.org/10.1393/ncc/i2011-10899-4}
  {\path{doi:10.1393/ncc/i2011-10899-4}}.

\bibitem{Kaleris_2025}
K.~Kaleris, E.~Kaselouris, V.~Dimitriou, E.~Kaniolakis-Kaloudis, M.~Bakarezos,
  M.~Tatarakis, N.~A. Papadogiannis, G.~B. Sushko, A.~V. Korol, A.~V.
  Solov'yov, {Narrowband $\ensuremath{\gamma}$-ray radiation generation by
  acoustically driven crystalline undulators}, Phys. Rev. Accel. Beams 28
  (2025) 033502.
\newblock \href {https://doi.org/10.1103/PhysRevAccelBeams.28.033502}
  {\path{doi:10.1103/PhysRevAccelBeams.28.033502}}.

\bibitem{Breese_1997}
M.~Breese, {Beam bending using graded composition strained layers}, Nucl.
  Instrum. Methods Phys. Res. B 132~(3) (1997) 540--547.
\newblock \href {https://doi.org/https://doi.org/10.1016/S0168-583X(97)00455-2}
  {\path{doi:https://doi.org/10.1016/S0168-583X(97)00455-2}}.

\bibitem{Mikkelsen_2000}
U.~Mikkelsen, E.~Uggerh\o{}j, {A Crystalline Undulator Based on Graded
  Composition Strained Layers in a Superlattice}, Nucl. Instrum. Methods Phys.
  Res. B 160~(3) (2000) 435--439.
\newblock \href {https://doi.org/https://doi.org/10.1016/S0168-583X(99)00637-0}
  {\path{doi:https://doi.org/10.1016/S0168-583X(99)00637-0}}.

\bibitem{Avakian_2003}
R.~Avakian, K.~Avetyan, K.~Ispirian, E.~Melikian, {Bent crystallographic planes
  in gradient crystals and crystalline undulators}, Nucl. Instrum. Meth. Phys.
  Res. A 508~(3) (2003) 496--499.
\newblock \href {https://doi.org/https://doi.org/10.1016/S0168-9002(03)01656-5}
  {\path{doi:https://doi.org/10.1016/S0168-9002(03)01656-5}}.

\bibitem{Krause_2002}
W.~Krause, A.~Korol, A.~Solov'yov, W.~Greiner, {Photon emission by
  ultra-relativistic positrons in crystalline undulators: the high-energy
  regime}, Nucl. Instrum. Meth. Phys. Res. A 483~(1) (2002) 455--460.
\newblock \href {https://doi.org/https://doi.org/10.1016/S0168-9002(02)00361-3}
  {\path{doi:https://doi.org/10.1016/S0168-9002(02)00361-3}}.

\bibitem{goldschmidinterstitial}
H.~J. Goldschmid, {Interstitial Alloys}, Butterworths, London, 1967.
\newblock \href {https://doi.org/https://doi.org/10.1007/978-1-4899-5880-8}
  {\path{doi:https://doi.org/10.1007/978-1-4899-5880-8}}.

\bibitem{Ashfold_1994}
M.~N.~R. Ashfold, P.~W. May, C.~A. Rego, N.~M. Everitt, {Thin film diamond by
  chemical vapour deposition methods}, Chem. Soc. Rev. 23 (1994) 21--30.
\newblock \href {https://doi.org/10.1039/CS9942300021}
  {\path{doi:10.1039/CS9942300021}}.

\bibitem{Brunet_1998}
F.~Brunet, P.~Germi, M.~Pernet, A.~Deneuville, E.~Gheeraert, F.~Laugier,
  M.~Burdin, G.~Rolland, {The effect of boron doping on the lattice parameter
  of homoepitaxial diamond films}, Diam. Relat. Mater. 7~(6) (1998) 869--873.
\newblock \href {https://doi.org/https://doi.org/10.1016/S0925-9635(97)00316-6}
  {\path{doi:https://doi.org/10.1016/S0925-9635(97)00316-6}}.

\bibitem{Boer_2020}
K.~W. B{\"o}er, U.~W. Pohl, {Crystal Defects}, Springer International
  Publishing, Cham, 2020, pp. 1--54.
\newblock \href {https://doi.org/10.1007/978-3-319-06540-3_15-4}
  {\path{doi:10.1007/978-3-319-06540-3_15-4}}.

\bibitem{Tallaire_2017}
A.~Tallaire, V.~Mille, O.~Brinza, T.~N. {Tran Thi}, J.~Brom, Y.~Loguinov,
  A.~Katrusha, A.~Koliadin, J.~Achard, {Thick CVD diamond films grown on
  high-quality type IIa HPHT diamond substrates from New Diamond Technology},
  Diam. Relat. Mater. 77 (2017) 146--152.
\newblock \href {https://doi.org/https://doi.org/10.1016/j.diamond.2017.07.002}
  {\path{doi:https://doi.org/10.1016/j.diamond.2017.07.002}}.

\bibitem{Campbell_2000}
B.~Campbell, A.~Mainwood, {Radiation Damage of Diamond by Electron and Gamma
  Irradiation}, physica status solidi (a) 181~(1) (2000) 99--107.
\newblock \href
  {https://doi.org/https://doi.org/10.1002/1521-396X(200009)181:1<99::AID-PSSA99>3.0.CO;2-5}
  {\path{doi:https://doi.org/10.1002/1521-396X(200009)181:1<99::AID-PSSA99>3.0.CO;2-5}}.

\bibitem{Campbell_2002}
B.~Campbell, W.~Choudhury, A.~Mainwood, M.~Newton, G.~Davies, {Lattice damage
  caused by the irradiation of diamond}, Nucl. Instrum. Meth. Phys. Res. A
  476~(3) (2002) 680--685.
\newblock \href {https://doi.org/https://doi.org/10.1016/S0168-9002(01)01664-3}
  {\path{doi:https://doi.org/10.1016/S0168-9002(01)01664-3}}.

\bibitem{Si-Ge_Paper}
M.~D. Dickers, G.~B. Sushko, A.~V. Korol, N.~J. Mason, F.~Fantuzzi, A.~V.
  Solov'yov, {Dopant concentration effects on Si$_{1-x}$Ge$_{x}$ crystals for
  emerging light-source technologies: a molecular dynamics study}, Europ. Phys.
  J. D. 78~(6) (2024) 77.
\newblock \href {https://doi.org/10.1140/epjd/s10053-024-00870-2}
  {\path{doi:10.1140/epjd/s10053-024-00870-2}}.

\bibitem{MBNExplorer}
I.~A. Solov'yov, A.~V. Yakubovich, P.~V. Nikolaev, I.~Volkovets, A.~V.
  Solov'yov, {MesoBioNano Explorer--a Universal Program for Multiscale Computer
  Simulations of Complex Molecular Structure and Dynamics.}, J. Comput. Chem.
  33~(30) (2012) 2412--2439.
\newblock \href {https://doi.org/10.1002/jcc.23086}
  {\path{doi:10.1002/jcc.23086}}.

\bibitem{MBNStudio}
G.~B. Sushko, I.~A. Solov'yov, A.~V. Solov'yov, {Modeling MesoBioNano Systems
  with MBN Studio Made Easy}, J. Mol. Graph. Model. 88 (2019) 247--260.
\newblock \href {https://doi.org/10.1016/j.jmgm.2019.02.003}
  {\path{doi:10.1016/j.jmgm.2019.02.003}}.

\bibitem{Solozhenko_2009}
V.~L. Solozhenko, O.~O. Kurakevych, D.~Andrault, Y.~Le~Godec, M.~Mezouar,
  {Ultimate Metastable Solubility of Boron in Diamond: Synthesis of Superhard
  Diamondlike ${\mathrm{BC}}_{5}$}, Phys. Rev. Lett. 102 (2009) 015506.
\newblock \href {https://doi.org/10.1103/PhysRevLett.102.015506}
  {\path{doi:10.1103/PhysRevLett.102.015506}}.

\bibitem{TersoffOriginal}
J.~Tersoff, {New empirical approach for the structure and energy of covalent
  systems}, Phys. Rev. B 37 (1988) 6991--7000.
\newblock \href {https://doi.org/10.1103/PhysRevB.37.6991}
  {\path{doi:10.1103/PhysRevB.37.6991}}.

\bibitem{Tersoff_C}
J.~Tersoff, {Modeling solid-state chemistry: Interatomic potentials for
  multicomponent systems}, Phys. Rev. B 39 (1989) 5566--5568.
\newblock \href {https://doi.org/10.1103/PhysRevB.39.5566}
  {\path{doi:10.1103/PhysRevB.39.5566}}.

\bibitem{Tersoff_B}
K.~Matsunaga, C.~Fisher, H.~Matsubara, {Tersoff Potential Parameters for
  Simulating Cubic Boron Carbonitrides}, Jpn. J. Appl. Phys. 39 (01 2000).
\newblock \href {https://doi.org/10.1143/JJAP.39.L48}
  {\path{doi:10.1143/JJAP.39.L48}}.

\bibitem{Saotome_1998}
T.~Saotome, K.~Ohashi, T.~Sato, H.~Maeta, K.~Haruna, F.~Ono, {Thermal expansion
  of a boron-doped diamond single crystal at low temperatures}, J. Phys.
  Condens. Matter 10 (1998) 1267--1272.

\bibitem{Vegard_1921}
L.~Vegard, {Die Konstitution der Mischkristalle und die Raumf\"{u}llung der
  Atome}, Zeitschrift f\"{u}r Physik 5~(1) (1921) 17--26.
\newblock \href {https://doi.org/10.1007/BF01349680}
  {\path{doi:10.1007/BF01349680}}.

\bibitem{Hayami_2024}
W.~Hayami, T.~Hiroto, K.~Soga, T.~Ogitsu, K.~Kimura, {Thermodynamic stability
  of elemental boron allotropes with varying numbers of interstitial atoms}, J.
  Solid State Chem. 329 (2024) 124407.
\newblock \href {https://doi.org/https://doi.org/10.1016/j.jssc.2023.124407}
  {\path{doi:https://doi.org/10.1016/j.jssc.2023.124407}}.

\bibitem{Brazhkin_2006}
V.~V. Brazhkin, E.~A. Ekimov, A.~G. Lyapin, S.~V. Popova, A.~V. Rakhmanina,
  S.~M. Stishov, V.~M. Lebedev, Y.~Katayama, K.~Kato, {Lattice parameters and
  thermal expansion of superconducting boron-doped diamonds}, Phys. Rev. B 74
  (2006) 140502.
\newblock \href {https://doi.org/10.1103/PhysRevB.74.140502}
  {\path{doi:10.1103/PhysRevB.74.140502}}.

\bibitem{Bustarret_2003}
E.~Bustarret, E.~Gheeraert, K.~Watanabe, {Optical and electronic properties of
  heavily boron-doped homo-epitaxial diamond}, Phys. Status Solidi (a) 199~(1)
  (2003) 9--18.
\newblock \href {https://doi.org/https://doi.org/10.1002/pssa.200303819}
  {\path{doi:https://doi.org/10.1002/pssa.200303819}}.

\bibitem{Wojewoda_2008}
T.~Wojewoda, P.~Achatz, L.~Ort\'{e}ga, F.~Omn\`{e}s, C.~Marcenat, E.~Bourgeois,
  X.~Blase, F.~Jomard, E.~Bustarret, {Doping-induced anisotropic lattice strain
  in homoepitaxial heavily boron-doped diamond}, Diam. Relat. Mater. 17~(7)
  (2008) 1302--1306.
\newblock \href {https://doi.org/https://doi.org/10.1016/j.diamond.2008.01.040}
  {\path{doi:https://doi.org/10.1016/j.diamond.2008.01.040}}.

\bibitem{Kawano_2010}
A.~Kawano, H.~Ishiwata, S.~Iriyama, R.~Okada, T.~Yamaguchi, Y.~Takano,
  H.~Kawarada, {Superconductor-to-insulator transition in boron-doped diamond
  films grown using chemical vapor deposition}, Phys. Rev. B 82 (2010) 085318.
\newblock \href {https://doi.org/10.1103/PhysRevB.82.085318}
  {\path{doi:10.1103/PhysRevB.82.085318}}.

\bibitem{Jacob_2007}
K.~T. Jacob, S.~Raj, L.~Rannesh, {Vegard's law: a fundamental relation or an
  approximation?}, Int. J. of Mater. Res. 98~(9) (2007) 776--779.
\newblock \href {https://doi.org/doi:10.3139/146.101545}
  {\path{doi:doi:10.3139/146.101545}}.

\bibitem{Bourgeois_2006}
E.~Bourgeois, E.~Bustarret, P.~Achatz, F.~Omn\`es, X.~Blase, {Impurity dimers
  in superconducting B-doped diamond: Experiment and first-principles
  calculations}, Phys. Rev. B 74 (2006) 094509.
\newblock \href {https://doi.org/10.1103/PhysRevB.74.094509}
  {\path{doi:10.1103/PhysRevB.74.094509}}.

\bibitem{Gross_2003}
J.~P. Goss, P.~R. Briddon, R.~Jones, Z.~Teukam, D.~Ballutaud, F.~Jomard,
  J.~Chevallier, M.~Bernard, A.~Deneuville, {Deep hydrogen traps in heavily
  B-doped diamond}, Phys. Rev. B 68 (2003) 235209.
\newblock \href {https://doi.org/10.1103/PhysRevB.68.235209}
  {\path{doi:10.1103/PhysRevB.68.235209}}.

\bibitem{Goss_2006}
J.~P. Goss, P.~R. Briddon, {Theory of boron aggregates in diamond:
  First-principles calculations}, Phys. Rev. B 73 (2006) 085204.
\newblock \href {https://doi.org/10.1103/PhysRevB.73.085204}
  {\path{doi:10.1103/PhysRevB.73.085204}}.

\bibitem{Goss_2008}
J.~P. Goss, R.~J. Eyre, P.~R. Briddon, {Bound substitutional impurity pairs in
  diamond: a density functional study}, J. Phys. Condens. Matter 20~(8) (2008)
  085217.
\newblock \href {https://doi.org/10.1088/0953-8984/20/8/085217}
  {\path{doi:10.1088/0953-8984/20/8/085217}}.

\bibitem{Watanabe_2018}
T.~Watanabe, S.~Yoshioka, T.~Yamamoto, H.~Sepehri-Amin, T.~Ohkubo,
  S.~Matsumura, Y.~Einaga, {The local structure in heavily boron-doped diamond
  and the effect this has on its electrochemical properties}, Carbon 137 (2018)
  333--342.
\newblock \href {https://doi.org/https://doi.org/10.1016/j.carbon.2018.05.026}
  {\path{doi:https://doi.org/10.1016/j.carbon.2018.05.026}}.

\bibitem{Lundström_1998}
T.~Lundström, B.~Lönnberg, J.~Bauer, Thermal expansion of
  $\beta$-rhombohedral boron, J. Alloys Compd. 267~(1) (1998) 54--58.
\newblock \href {https://doi.org/https://doi.org/10.1016/S0925-8388(97)00545-8}
  {\path{doi:https://doi.org/10.1016/S0925-8388(97)00545-8}}.

\bibitem{Jacobson_2019}
P.~Jacobson, S.~Stoupin, Thermal expansion coefficient of diamond in a wide
  temperature range, Diam. Relat. Mater. 97 (2019) 107469.
\newblock \href {https://doi.org/https://doi.org/10.1016/j.diamond.2019.107469}
  {\path{doi:https://doi.org/10.1016/j.diamond.2019.107469}}.

\bibitem{Halicioglu_1992}
T.~Halicioglu, {(2$\times$1) Reconstructed patterns of diamond (100) surface},
  Diamond and Related Materials 1~(9) (1992) 963--967.
\newblock \href {https://doi.org/https://doi.org/10.1016/0925-9635(92)90118-8}
  {\path{doi:https://doi.org/10.1016/0925-9635(92)90118-8}}.

\bibitem{Kern_1996}
G.~Kern, J.~Hafner, J.~Furthmüller, G.~Kresse, {(2$\times$1)reconstruction and
  hydrogen-induced de-reconstruction of the diamond (100) and (111) surfaces},
  Surface Science 352-354 (1996) 745--749.
\newblock \href {https://doi.org/https://doi.org/10.1016/0039-6028(95)01244-3}
  {\path{doi:https://doi.org/10.1016/0039-6028(95)01244-3}}.

\bibitem{TECHNO-CLS_2025}
{TECHNO-CLS Consortium}, {TECHNO-CLS Periodic Report: Reporting Period 3},
  Tech. rep., Horizon Europe EIC-Pathfinder Project TECHNO-CLS, internal
  report, Grant Agreement No. 101046458, unpublished (2025).

\bibitem{Netto_1998}
A.~Netto, M.~Frenklach, {Kinetic Monte Carlo Simulation of Diamond Film Growth
  with the Inclusion of Surface Migration}, MRS Online Proceedings Library
  527~(1) (1998) 383--388.
\newblock \href {https://doi.org/10.1557/PROC-527-383}
  {\path{doi:10.1557/PROC-527-383}}.

\bibitem{Rodgers_2015}
W.~J. Rodgers, P.~W. May, N.~L. Allan, J.~N. Harvey, {Three-dimensional kinetic
  Monte Carlo simulations of diamond chemical vapor deposition}, J Chem Phys
  142~(21) (2015) 214707.

\bibitem{Cheesman_2005}
A.~Cheesman, J.~N. Harvey, M.~N.~R. Ashfold, {Computational studies of
  elementary steps relating to boron doping during diamond chemical vapour
  deposition}, Phys. Chem. Chem. Phys. 7 (2005) 1121--1126.
\newblock \href {https://doi.org/10.1039/B418664H}
  {\path{doi:10.1039/B418664H}}.

\bibitem{Lu_2025}
M.~Lu, C.~Zhang, F.~Sun, {Growth mechanisms and material properties of
  boron-doped single crystal diamond synthesized by HFCVD}, Surf. Interfaces 62
  (2025) 106217.
\newblock \href {https://doi.org/10.1016/j.surfin.2025.106217}
  {\path{doi:10.1016/j.surfin.2025.106217}}.

\end{thebibliography}


\begin{thebibliography}{S99}

\bibitem[S1]{supp:Brunet_1998} F.~Brunet, P.~Germi, M.~Pernet, A.~Deneuville, E.~Gheeraert, F.~Laugier, M.~Burdin, G.~Rolland, {The effect of boron doping on the lattice parameter of homoepitaxial diamond films}, Diam. Relat. Mater. 7~(6) (1998) 869--873. \href{https://doi.org/10.1016/S0925-9635(97)00316-6}{\texttt{doi:10.1016/S0925-9635(97)00316-6.}}

\bibitem[S2]{supp:Brazhkin_2006} V.~V. Brazhkin, E.~A. Ekimov, A.~G. Lyapin, S.~V. Popova, A.~V. Rakhmanina, S.~M. Stishov, V.~M. Lebedev, Y.~Katayama, K.~Kato, {Lattice parameters and thermal expansion of superconducting boron-doped diamonds}, Phys. Rev. B 74 (2006) 140502. \href{https://doi.org/10.1103/PhysRevB.74.140502}{\texttt{doi:10.1103/PhysRevB.74.140502.}}

\bibitem[S3]{supp:Bustarret_2003} E.~Bustarret, E.~Gheeraert, K.~Watanabe, {Optical and electronic properties of heavily boron-doped homo-epitaxial diamond}, Phys. Status Solidi (a) 199~(1) (2003) 9--18. \href{https://doi.org/10.1002/pssa.200303819}{\texttt{doi:10.1002/pssa.200303819.}}

\bibitem[S4]{supp:Wojewoda_2008} T.~Wojewoda, P.~Achatz, L.~Ort\'{e}ga, F.~Omn\`{e}s, C.~Marcenat, E.~Bourgeois, X.~Blase, F.~Jomard, E.~Bustarret, {Doping-induced anisotropic lattice strain in homoepitaxial heavily boron-doped diamond}, Diam. Relat. Mater. 17~(7) (2008) 1302--1306. \href{https://doi.org/10.1016/j.diamond.2008.01.040}{\texttt{doi:10.1016/j.diamond.2008.01.040.}}

\bibitem[S5]{supp:Vegard_1921} L.~Vegard, {Die Konstitution der Mischkristalle und die Raumf\"{u}llung der Atome}, Zeitschrift f\"{u}r Physik 5~(1) (1921) 17--26. \href{https://doi.org/10.1007/BF01349680}{\texttt{doi:10.1007/BF01349680.}}

\bibitem[S6]{supp:Hartmut_2025} H.~Backe, J.~Baruchel, S.~B\'{e}nichou, R.~Dowek, D.~Eon, P.~Everaere, L.~Kirste, P.~Klag, W.~Lauth, P.~Stra\v{n}\'{a}k, T.~N.~T. Caliste, {Observation of narrow-band $\gamma$ radiation from a boron-doped diamond superlattice with an 855 MeV electron beam} (2025). \href{http://arxiv.org/abs/2504.17988}{\texttt{arXiv:2504.17988}}, \href{https://doi.org/10.48550/arXiv.2504.17988}{\texttt{doi:10.48550/arXiv.2504.17988.}}

\bibitem[S7]{supp:Zhang_2011} J.~Zhang, C.~Yan, H.~Wang, Y.~Liu, W.~Wang, M.~Saenger, {Hafnium-doped GaN with
  n-type electrical resistivity in the 10$^{-4}\;\omega\;$cm range}, Appl.
  Phys. Lett. 99~(20) (2011) 202113. \href{https://doi.org/10.1063/1.3663570}{\texttt{doi:10.1063/1.3663570.}}

\bibitem[S8]{supp:Klein_1993} C. A. Klein, G. F. Cardinale, Young’s modulus and Poisson’s ratio of CVD diamond, Diam. Relat. Mater. 2 (5) (1993) 918–923. \href{https://doi.org/10.1016/0925-9635(93)90250-6}{\texttt{doi:10.1016/0925-9635(93)90250-6.}}

\bibitem[S9]{supp:Mohr_2014} M. Mohr, A. Caron, P. Herbeck-Engel, R. Bennewitz, P. Gluche,
K. Brühne, H.-J. Fecht, Young’s modulus, fracture strength, and Poisson’s ratio of nanocrystalline diamond films, J. Appl. Phys. 116 (12) (2014) 124308. \href{https://doi.org/10.1063/1.4896729}{\texttt{doi:10.1063/1.4896729.}}

\bibitem[S10]{supp:Hess_2012} P. Hess, The mechanical properties of various chemical vapor deposition diamond structures compared to the ideal single crystal, J. Appl. Phys. 111 (5) (2012) 051101. \href{https://doi.org/10.1063/1.3683544}{\texttt{doi:10.1063/1.3683544.}}

\bibitem[S11]{supp:Clerc_2005} D. G. Clerc, H. Ledbetter, Second-order and third-order elastic properties of diamond: An ab initio study, J. Phys. Chem. Solids 66 (10) (2005) 1589–1597. \href{https://doi.org/10.1016/j.jpcs.2005.05.075}{\texttt{doi:10.1016/j.jpcs.2005.05.075.}}

\bibitem[S12]{supp:Liu_2017}X. Liu, Y.-Y. Chang, S. N. Tkachev, C. R. Bina, S. D. Jacobsen, Elastic
and mechanical softening in boron-doped diamond, Sci. Rep. 7 (1) (2017)
42921. \href{https://doi.org/10.1038/srep42921}{\texttt{doi:10.1038/srep42921.}}

\bibitem[S13]{supp:Nye_1985}X. J. F. Nye, Physical Properties of Crystals, Oxford University Press, London, England, 1985.

\end{thebibliography}

\setcounter{section}{0}
\setcounter{page}{1}
\setcounter{figure}{0}
\setcounter{equation}{0}

\renewcommand{\thesection}{S\arabic{section}}
\renewcommand{\thepage}{S\arabic{page}}
\renewcommand{\thetable}{S\arabic{table}}
\renewcommand{\thefigure}{S\arabic{figure}}
\renewcommand{\theequation}{S\arabic{equation}}

\newpage

\begin{center}
\large Supplemental Information for:\\\textbf{Structural effects of boron doping in diamond crystals for gamma-ray light-source applications: Insights from molecular dynamics simulations}
\end{center}

\section{Illustrations of Crystal Structures}
Here we provide additional details of the crystal structures modelled in this study. The following figures illustrate representative atomistic configurations of all crystal sizes at a dopant concentration of $x=0.05$ (5\%).

\subsection{C Crystals}
\begin{figure}[H]
    \centering
    \includegraphics[width=0.8\textwidth]{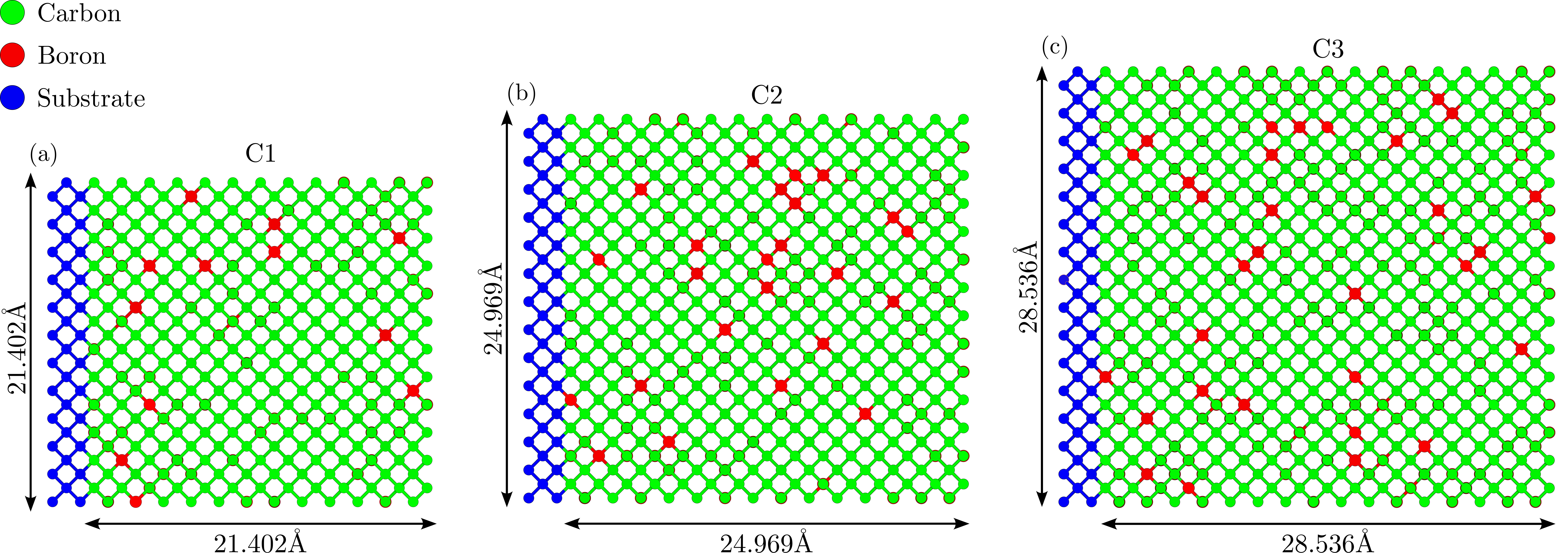}
    \caption{Representative C1-category crystal structures used in this study. Panels show \textbf{(a)}: C1, \textbf{(b)}: C2, and \textbf{(c)}: C3. The corresponding crystal dimensions are 21.402$\times$21.402$\times$21.40,  24.969$\times$24.969$\times$24.969, and 28.536$\times$28.536$\times$28.536, respectively.}
    \label{fig:C_Images}
\end{figure}

\begin{figure}[H]
    \centering
    \includegraphics[width=\textwidth]{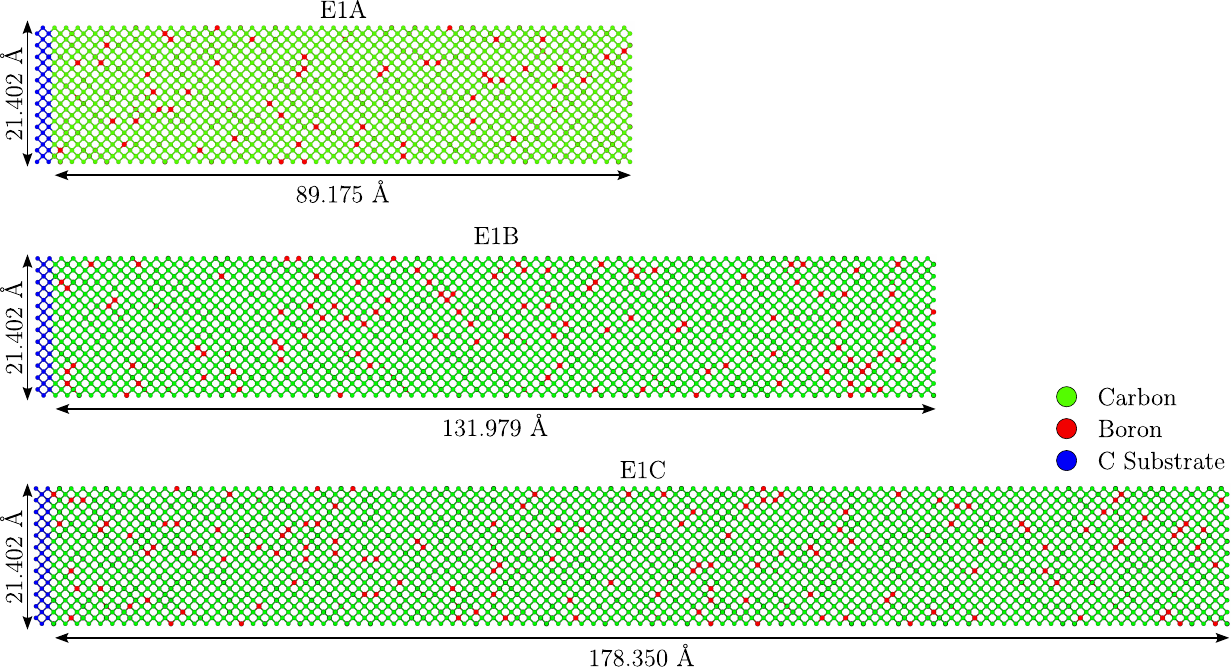}
    \caption{Representative E1-category crystal structures used in this study. Panels show \textbf{(a)}: E1A, \textbf{(b)}: E1B, and \textbf{(c)}: E1C. The corresponding crystal dimensions are 21.402$\times$21.402$\times$89.175, 21.402$\times$21.402$\times$131.979, and 21.402$\times$21.402$\times$178.350, respectively.}
    \label{fig:E1_Images}
\end{figure}

\subsection{E2 Crystals}
\begin{figure}[H]
    \centering
    \includegraphics[width=\textwidth]{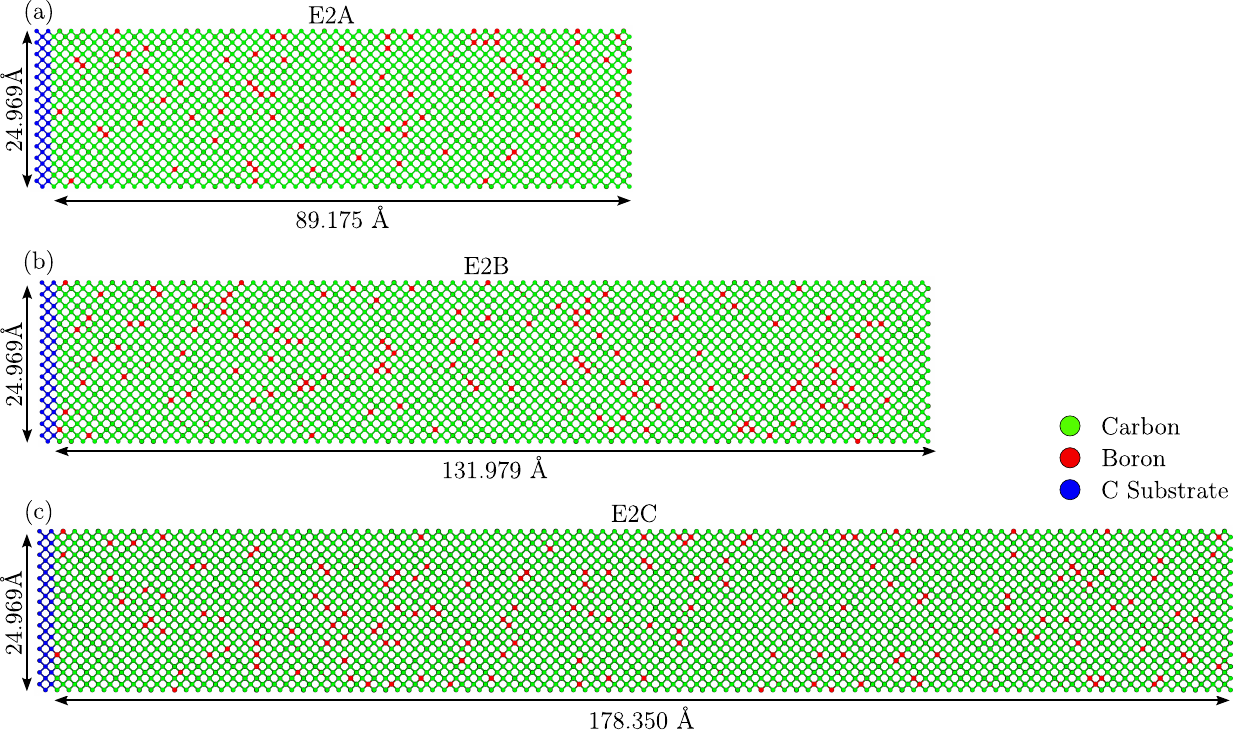}
    \caption{Representative E2-category crystal structures used in this study. Panels show \textbf{(a)}: E2A, \textbf{(b)}: E2B, and \textbf{(c)}: E2C. The corresponding crystal dimensions are  24.969$\times$24.969$\times$89.175, 24.969$\times$24.969$\times$131.979, and 24.969$\times$24.969$\times$178.35, respectively.}
    \label{fig:E2_Images}
\end{figure}

\subsection{E3 Crystals}
\begin{figure}[H]
    \centering
    \includegraphics[width=\textwidth]{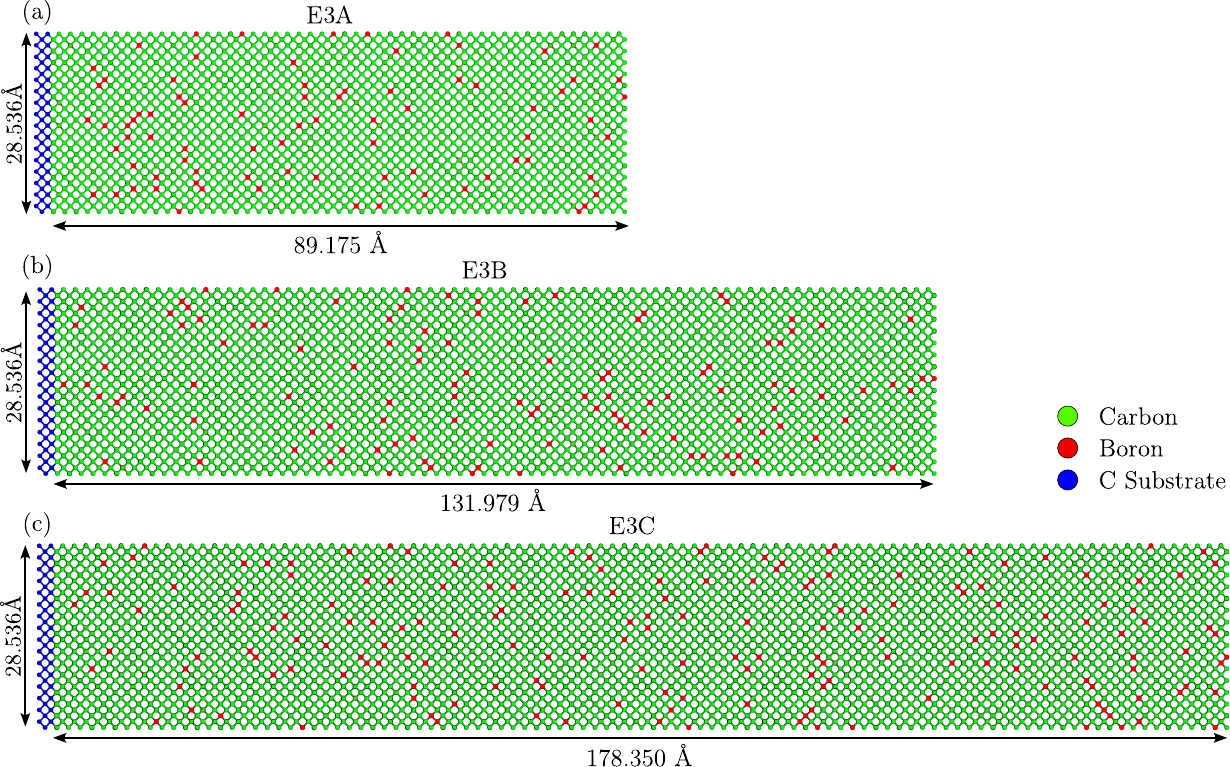}
    \caption{Representative E3-category crystal structures used in this study. Panels show \textbf{(a)}: E3A, \textbf{(b)}: E3B, and \textbf{(c)}: E3C. The corresponding crystal dimensions are  28.536$\times$28.536$\times$89.175, 28.536$\times$28.536$\times$131.979, and 28.536$\times$28.536$\times$178.350, respectively.}
    \label{fig:E3_Images}
\end{figure}

\subsection{S Crystals}
\begin{figure}[H]
    \centering
    \includegraphics[width=0.3\textwidth]{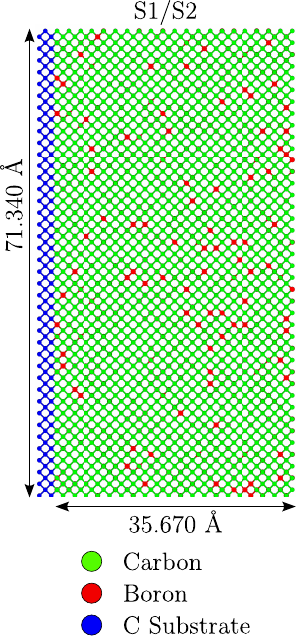}
    \caption{Representative S-category crystal structure used in this study. For brevity a single structure has been used to represent S1 and S2, as only the $Z$ dimension varies. The corresponding crystal dimensions are 35.670 $\times$ 71.340 $\times$ 28.536 for S1, and 35.670 $\times$ 71.340 $\times$ 71.340 for S2}
    \label{fig:S_Images}
\end{figure}

\section{Table of Calculated Dopant Concentrations}
Here we provide additional details on the average calculated dopant concentrations, obtained from the mean number of carbon and boron atoms in each crystal size, and compare them to the targeted dopant concentrations. The following tables summarise the values for each crystal size at a given target concentration. In addition, the total number of carbon and boron atoms are listed, alongside the number of simulations conducted for a particular crystal.

\begin{table}[H]
    \centering
    \caption{Summary of simulation statistics for crystals with a target dopant concentration of $x=0.01$ (1\%). Reported values include the average calculated dopant concentration (based on the fraction of carbon and boron atoms), the mean number of carbon atoms $N_{\text{atoms, C}}$ and boron atoms $N_{\text{atoms, B}}$ per crystal, and the total number of simulations for each crystal size $N_{\text{simulations}}$. All values are averaged over simulations of a given crystal size at the specified target concentration. Uncertainties correspond to one standard deviation ($1\sigma$).}
    \label{tab:dopant_concentrations_x01}
\resizebox{\textwidth}{!}{%
\begin{tabular}{c|c|c|c|c}
Crystal Size (\SI{}{\angstrom^3}) & Calc. Conc. (\%)  & $N_{\text{atoms, C}}$ & $N_{\text{atoms, B}}$ & $N_{\text{simulations}}$ \\ \hline
21.402 $\times$ 21.402 $\times$ 21.402  & $1.02\pm0.02$ & $1781.66\pm4.22$   & $18.34\pm4.22$   & 170 \\
24.969 $\times$ 24.969 $\times$ 24.969  & $1.00\pm0.02$ & $2813.46\pm5.20$   & $28.54\pm5.20$   & 100 \\
28.536 $\times$ 28.536 $\times$ 28.536  & $1.00\pm0.02$ & $4181.56\pm7.23$   & $42.44\pm7.23$   & 100 \\
89.175 $\times$ 21.402 $\times$ 21.402  & $1.02\pm0.01$ & $7198.17\pm9.34$   & $73.83\pm9.34$   & 100 \\
131.979 $\times$ 21.402 $\times$ 21.402 & $0.99\pm0.01$ & $10622.22\pm10.64$ & $105.78\pm10.64$ & 50  \\
178.350 $\times$ 21.402 $\times$ 21.402 & $1.00\pm0.01$ & $14178.46\pm15.63$ & $293.54\pm15.63$ & 100 \\
89.175 $\times$ 24.969 $\times$ 24.969  & $0.99\pm0.01$ & $9800.34\pm8.38$   & $97.66\pm8.98$   & 50  \\
131.979 $\times$ 24.969 $\times$ 24.969 & $1.01\pm0.01$ & $14454.28\pm12.89$ & $147.72\pm12.89$ & 50  \\
178.35 $\times$ 24.969 $\times$ 24.969  & $0.99\pm0.01$ & $19503.00\pm11.96$ & $195.00\pm11.96$ & 50  \\
89.175 $\times$ 28.536 $\times$ 28.536  & $1.01\pm0.01$ & $12797.94\pm10.57$ & $130.06\pm10.57$ & 50  \\
131.979 $\times$ 28.536 $\times$ 28.536 & $1.00\pm0.01$ & $18880.36\pm19.02$ & $376.14\pm19.02$ & 50  \\
178.350 $\times$ 28.536 $\times$ 28.536 & $1.00\pm0.01$ & $25469.88\pm1.50$  & $258.12\pm16.50$ & 50  \\
35.670 $\times$ 71.340 $\times$ 28.536  & $1.00\pm0.01$ & $12989.08\pm10.49$ & $130.92\pm10.49$ & 50  \\
35.670 $\times$ 71.340 $\times$ 71.340  & $0.99\pm0.01$ & $32476.20\pm19.03$ & $323.80\pm19.03$ & 50              
\end{tabular}
}
\end{table}

\begin{table}[H]
    \centering
    \caption{Summary of simulation statistics for crystals with a target dopant concentration of $x=0.02$ (2\%). Reported values include the average calculated dopant concentration (based on the fraction of carbon and boron atoms), the mean number of carbon atoms $N_{\text{atoms, C}}$ and boron atoms $N_{\text{atoms, B}}$ per crystal, and the total number of simulations for each crystal size $N_{\text{simulations}}$. All values are averaged over simulations of a given crystal size at the specified target concentration. Uncertainties correspond to one standard deviation ($1\sigma$).}
    \label{tab:dopant_concentrations_x02}
\resizebox{\textwidth}{!}{%
\begin{tabular}{c|c|c|c|c}
Crystal Size (\SI{}{\angstrom^3}) & Calc. Conc. (\%) & $N_{\text{atoms, C}}$ & $N_{\text{atoms, B}}$ & $N_{\text{simulations}}$ \\ \hline
21.402 $\times$ 21.402 $\times$ 21.402  & $1.97\pm0.02$ & $1764.47\pm5.32$   & $35.53\pm5.32$   & 170 \\
24.969 $\times$ 24.969 $\times$ 24.969  & $2.00\pm0.03$ & $2785.14\pm7.74$   & $56.86\pm7.74$   & 100 \\
28.536 $\times$ 28.536 $\times$ 28.536  & $2.03\pm0.02$ & $4138.35\pm8.78$   & $85.65\pm8.78$   & 100 \\
89.175 $\times$ 21.402 $\times$ 21.402  & $2.01\pm0.02$ & $7126.10\pm11.45$  & $145.90\pm11.45$ & 100 \\
131.979 $\times$ 21.402 $\times$ 21.402 & $1.97\pm0.02$ & $10516.42\pm17.18$ & $211.58\pm17.18$ & 50  \\
178.350 $\times$ 21.402 $\times$ 21.402 & $2.03\pm0.01$ & $14178.46\pm15.63$ & $293.54\pm15.63$ & 100 \\
89.175 $\times$ 24.969 $\times$ 24.969  & $2.02\pm0.02$ & $9697.82\pm14.11$  & $200.18\pm14.11$ & 50  \\
131.979 $\times$ 24.969 $\times$ 24.969 & $2.00\pm0.02$ & $14309.74\pm16.84$ & $292.26\pm16.84$ & 50  \\
178.35 $\times$ 24.969 $\times$ 24.969  & $2.00\pm0.02$ & $19304.86\pm21.26$ & $393.14\pm21.26$ & 50  \\
89.175 $\times$ 28.536 $\times$ 28.536  & $2.00\pm0.01$ & $12670.00\pm12.47$ & $258.00\pm12.47$ & 50  \\
131.979 $\times$ 28.536 $\times$ 28.536 & $1.97\pm0.01$ & $18695.86\pm19.02$ & $376.14\pm19.02$ & 50  \\
178.350 $\times$ 28.536 $\times$ 28.536 & $2.00\pm0.01$ & $25212.82\pm24.96$ & $515.18\pm24.93$ & 50  \\
35.670 $\times$ 71.340 $\times$ 28.536  & $2.01\pm0.02$ & $12855.88\pm17.55$ & $395.88\pm17.68$ & 50  \\
35.670 $\times$ 71.340 $\times$ 71.340  & $2.01\pm0.01$ & $32140.82\pm24.12$ & $659.18\pm24.12$ & 50                            
\end{tabular}
}
\end{table}

\begin{table}[H]
    \centering
    \caption{Summary of simulation statistics for crystals with a target dopant concentration of $x=0.03$ (3\%). Reported values include the average calculated dopant concentration (based on the fraction of carbon and boron atoms), the mean number of carbon atoms $N_{\text{atoms, C}}$ and boron atoms $N_{\text{atoms, B}}$ per crystal, and the total number of simulations for each crystal size $N_{\text{simulations}}$. All values are averaged over simulations of a given crystal size at the specified target concentration. Uncertainties correspond to one standard deviation ($1\sigma$).}
    \label{tab:dopant_concentrations_x03}
\resizebox{\textwidth}{!}{%
\begin{tabular}{c|c|c|c|c}
Crystal Size (\SI{}{\angstrom^3}) & Calc. Conc. (\%) & $N_{\text{atoms, C}}$ & $N_{\text{atoms, B}}$ & $N_{\text{simulations}}$ \\ \hline
21.402 $\times$ 21.402 $\times$ 21.402  & $2.98\pm0.02$ & $1746.39\pm6.72$   & $53.61\pm6.72$   & 200 \\
24.969 $\times$ 24.969 $\times$ 24.969  & $3.01\pm0.03$ & $2756.33\pm8.25$   & $85.67\pm8.25$   & 100 \\
28.536 $\times$ 28.536 $\times$ 28.536  & $3.00\pm0.03$ & $4097.34\pm10.58$  & $126.66\pm10.58$ & 100 \\
89.175 $\times$ 21.402 $\times$ 21.402  & $3.00\pm0.02$ & $7053.57\pm16.27$  & $290.70\pm16.27$ & 100 \\
131.979 $\times$ 21.402 $\times$ 21.402 & $3.06\pm0.02$ & $10400.20\pm17.94$ & $327.80\pm17.94$ & 50  \\
178.350 $\times$ 21.402 $\times$ 21.402 & $3.02\pm0.01$ & $14035.54\pm18.88$ & $436.4\pm18.88$  & 100 \\
89.175 $\times$ 24.969 $\times$ 24.969  & $3.01\pm0.02$ & $9599.58\pm27.13$  & $298.42\pm27.13$ & 50  \\
131.979 $\times$ 24.969 $\times$ 24.969 & $3.01\pm0.02$ & $14161.86\pm23.50$ & $440.14\pm23.50$ & 50  \\
178.35 $\times$ 24.969 $\times$ 24.969  & $3.02\pm0.02$ & $19104.08\pm22.76$ & $593.92\pm22.76$ & 50  \\
89.175 $\times$ 28.536 $\times$ 28.536  & $3.00\pm0.02$ & $12540.78\pm22.62$ & $387.22\pm22.62$ & 50  \\
131.979 $\times$ 28.536 $\times$ 28.536 & $3.00\pm0.01$ & $18499.72\pm18.20$ & $572.28\pm18.20$ & 50  \\
178.350 $\times$ 28.536 $\times$ 28.536 & $2.99\pm0.02$ & $24958.66\pm26.78$ & $769.34\pm26.78$ & 50  \\
35.670 $\times$ 71.340 $\times$ 28.536  & $3.02\pm0.02$ & $12724.12\pm17.68$ & $395.88\pm17.68$ & 50  \\
35.670 $\times$ 71.340 $\times$ 71.340  & $3.02\pm0.01$ & $31810.36\pm29.00$ & $989.64\pm29.00$ & 50                  
\end{tabular}
}
\end{table}

\begin{table}[H]
    \centering
    \caption{Summary of simulation statistics for crystals with a target dopant concentration of $x=0.04$ (4\%). Reported values include the average calculated dopant concentration (based on the fraction of carbon and boron atoms), the mean number of carbon atoms $N_{\text{atoms, C}}$ and boron atoms $N_{\text{atoms, B}}$ per crystal, and the total number of simulations for each crystal size $N_{\text{simulations}}$. All values are averaged over simulations of a given crystal size at the specified target concentration. Uncertainties correspond to one standard deviation ($1\sigma$).}
    \label{tab:dopant_concentrations_x04}
\resizebox{\textwidth}{!}{%
\begin{tabular}{c|c|c|c|c}
Crystal Size (\SI{}{\angstrom^3}) & Calc. Conc. (\%) & $N_{\text{atoms, C}}$ & $N_{\text{atoms, B}}$ & $N_{\text{simulations}}$ \\ \hline
21.402 $\times$ 21.402 $\times$ 21.402  & $3.96\pm0.04$ & $1728.80\pm8.92$   & $71.20\pm8.92$    & 200 \\
24.969 $\times$ 24.969 $\times$ 24.969  & $3.97\pm0.04$ & $2729.28\pm11.56$  & $112.72\pm11.56$  & 100 \\
28.536 $\times$ 28.536 $\times$ 28.536  & $4.00\pm0.03$ & $4054.90\pm12.98$  & $169.10\pm12.98$  & 100 \\
89.175 $\times$ 21.402 $\times$ 21.402  & $4.00\pm0.02$ & $6981.30\pm16.27$  & $290.70\pm16.27$  & 100 \\
131.979 $\times$ 21.402 $\times$ 21.402 & $4.06\pm0.03$ & $10291\pm20.47$    & $436.08\pm20.47$  & 50  \\
178.350 $\times$ 21.402 $\times$ 21.402 & $3.99\pm0.02$ & $13893.91\pm22.22$ & $578.09\pm22.22$  & 100 \\
89.175 $\times$ 24.969 $\times$ 24.969  & $4.00\pm0.03$ & $9501.88\pm19.50$  & $396.12\pm19.50$  & 50  \\
131.979 $\times$ 24.969 $\times$ 24.969 & $4.00\pm0.02$ & $14018.56\pm22.69$ & $583.44\pm22.69$  & 50  \\
178.35 $\times$ 24.969 $\times$ 24.969  & $4.00\pm0.02$ & $18911.06\pm28.51$ & $786.94\pm28.51$  & 50  \\
89.175 $\times$ 28.536 $\times$ 28.536  & $4.04\pm0.02$ & $12405.32\pm21.78$ & $522.68\pm21.78$  & 50  \\
131.979 $\times$ 28.536 $\times$ 28.536 & $3.99\pm0.02$ & $18310.90\pm31.42$ & $761.10\pm31.42$  & 50  \\
178.350 $\times$ 28.536 $\times$ 28.536 & $3.96\pm0.02$ & $24708.16\pm30.27$ & $1019.84\pm30.27$ & 50  \\
35.670 $\times$ 71.340 $\times$ 28.536  & $3.97\pm0.02$ & $12460.10\pm21.93$ & $521.44\pm21.93$  & 50  \\
35.670 $\times$ 71.340 $\times$ 71.340  & $3.98\pm0.01$ & $31493.04\pm31.99$ & $1306.96\pm31.99$ & 50           
\end{tabular}
}
\end{table}

\begin{table}[H]
    \centering
    \caption{Summary of simulation statistics for crystals with a target dopant concentration of $x=0.05$ (5\%). Reported values include the average calculated dopant concentration (based on the fraction of carbon and boron atoms), the mean number of carbon atoms $N_{\text{atoms, C}}$ and boron atoms $N_{\text{atoms, B}}$ per crystal, and the total number of simulations for each crystal size $N_{\text{simulations}}$. All values are averaged over simulations of a given crystal size at the specified target concentration. Uncertainties correspond to one standard deviation ($1\sigma$).}
    \label{tab:dopant_concentrations_x05}
\resizebox{\textwidth}{!}{%
\begin{tabular}{c|c|c|c|c}
Crystal Size (\SI{}{\angstrom^3}) & Calc. Conc. (\%) & $N_{\text{atoms, C}}$ & $N_{\text{atoms, B}}$ & $N_{\text{simulations}}$ \\ \hline
21.402 $\times$ 21.402 $\times$ 21.402  & $5.00\pm0.04$ & $1710.04\pm9.61$   & $89.96\pm9.61$    & 200 \\
24.969 $\times$ 24.969 $\times$ 24.969  & $5.02\pm0.04$ & $2699.38\pm10.91$  & $142.62\pm10.91$  & 100 \\
28.536 $\times$ 28.536 $\times$ 28.536  & $5.02\pm0.03$ & $4011.77\pm13.84$  & $212.23\pm13.84$  & 100 \\
89.175 $\times$ 21.402 $\times$ 21.402  & $5.02\pm0.03$ & $6907\pm19.50$     & $364.71\pm19.50$  & 100 \\
131.979 $\times$ 21.402 $\times$ 21.402 & $5.02\pm0.03$ & $10189.34\pm20.47$ & $538.66\pm20.42$  & 50  \\
178.350 $\times$ 21.402 $\times$ 21.402 & $5.02\pm0.02$ & $13745.35\pm26.45$ & $726.65\pm26.45$  & 100 \\
89.175 $\times$ 24.969 $\times$ 24.969  & $4.99\pm0.03$ & $9404.26\pm18.73$  & $493.74\pm18.73$  & 50  \\
131.979 $\times$ 24.969 $\times$ 24.969 & $4.95\pm0.03$ & $13879.74\pm27.75$ & $722.26\pm27.75$  & 50  \\
178.35 $\times$ 24.969 $\times$ 24.969  & $4.99\pm0.02$ & $18716.04\pm31.41$ & $981.96\pm31.41$  & 50  \\
89.175 $\times$ 28.536 $\times$ 28.536  & $5.00\pm0.02$ & $12281.24\pm19.13$ & $646.76\pm19.13$  & 50  \\
131.979 $\times$ 28.536 $\times$ 28.536 & $4.94\pm0.02$ & $18130.20\pm26.09$ & $941.80\pm26.09$  & 50  \\
178.350 $\times$ 28.536 $\times$ 28.536 & $5.01\pm0.02$ & $24439.22\pm30.97$ & $1288.87\pm30.97$ & 50  \\
35.670 $\times$ 71.340 $\times$ 28.536  & $5.03\pm0.03$ & $12460\pm27.87$    & $659.90\pm27.87$  & 50  \\
35.670 $\times$ 71.340 $\times$ 71.340  & $5.00\pm0.01$ & $31159.44\pm34.53$ & $1640.56\pm34.53$ & 50              
\end{tabular}
}
\end{table}

\section{Definition of Region Boundaries}

\begin{figure*}[t!]
 \subfloat[]{\includegraphics[width=0.48\textwidth]{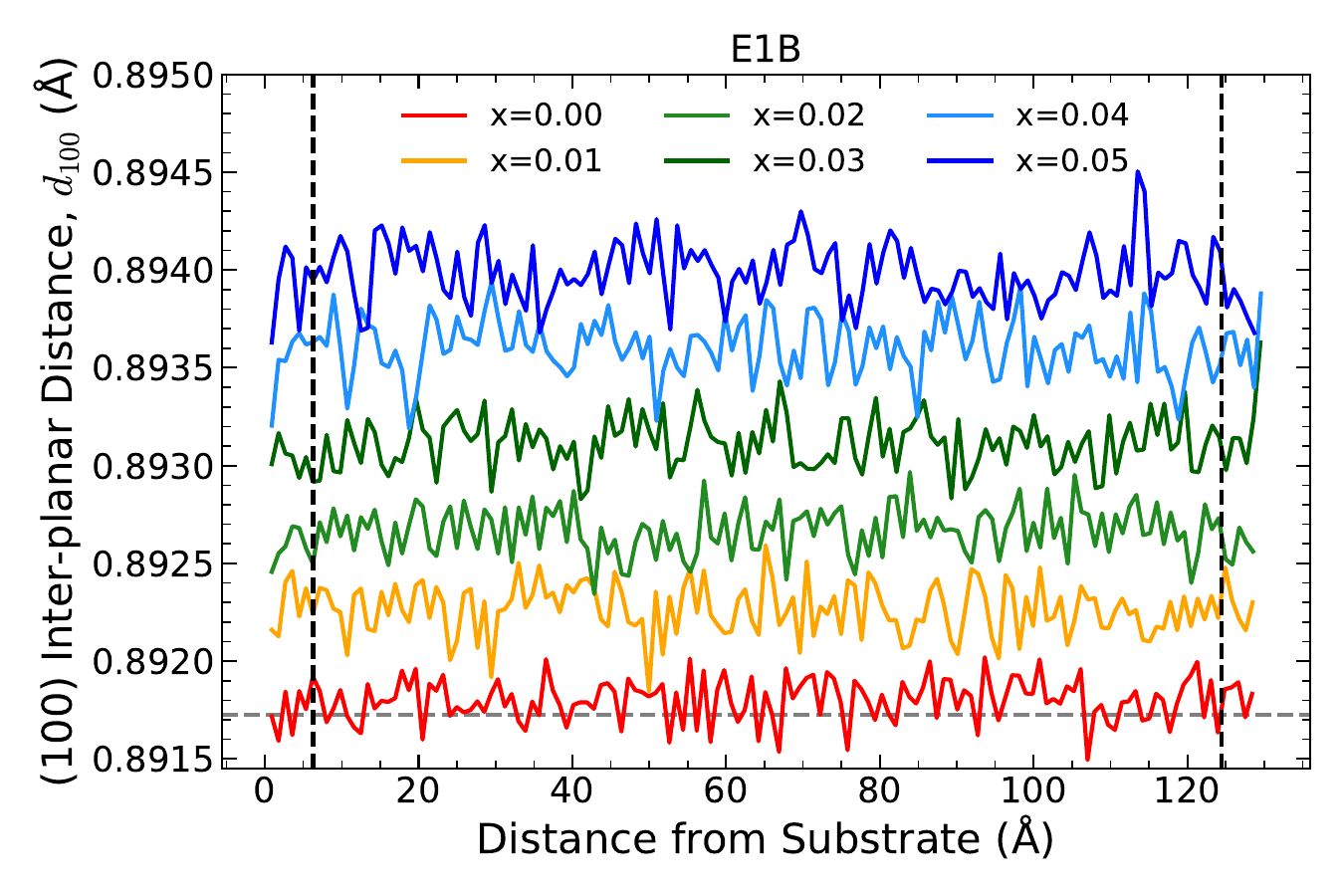}}\quad
 \subfloat[]{\includegraphics[width=0.48\textwidth]{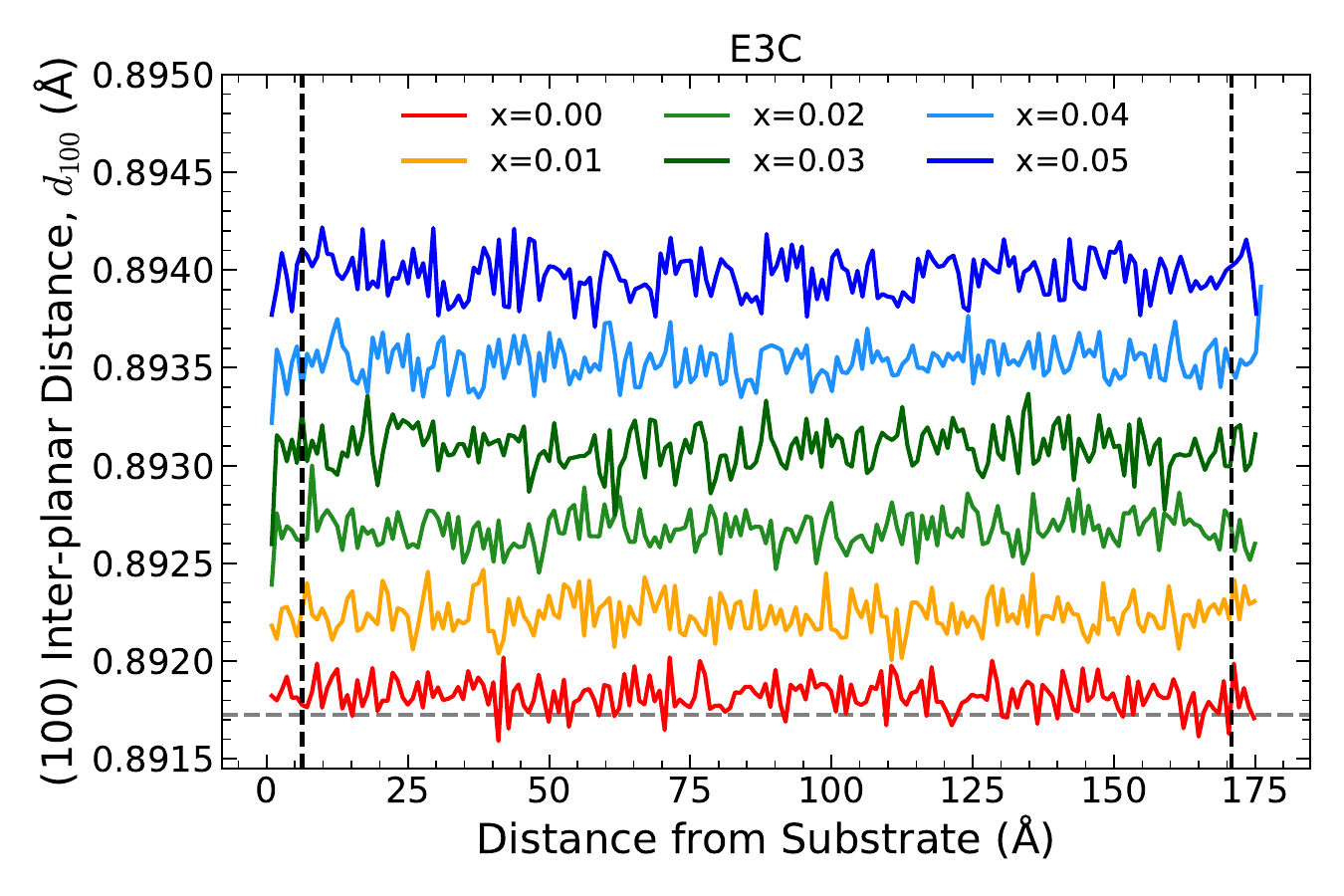}}
 \caption{Plots of the (1\;0\;0) inter-planar distance $d_{100}$ as a function of the distance from the substrate in two representative example crystals: E1C \textbf{(a)} and E3C \textbf{(b)}. Each concentration is shown with a separate line. The grey dashed line shows the nominal (100) inter-planar distance, and the black dashed lines show the region boundaries. The left line corresponds to the boundary between the substrate region and bulk region, and the right line corresponds to the boundary between the bulk region and the free edge. The last three (1\;0\;0) planes at the free edge are excluded due to surface-induced distortions.}
 \label{fig:Height_Plots}
\end{figure*}

This is illustrated for two representative crystals, E1C and E3C, in Figures \ref{fig:Height_Plots}a and \ref{fig:Height_Plots}b respectively. These cases were chosen as they most clearly represent the separation between each region. Each data point corresponds to the separation between two neighbouring planes, with increasing distance from the substrate, effectively providing a measure of how the inter-planar distance changes as the distance from the substrate increases. The data are averaged over all crystals of a given size, resulting in a degree of noise. The black dashed lines mark the boundaries between the three regions: the left line separates the substrate and bulk regions, while the right line separates the bulk region and the free edge. In these examples, slight variations in the inter-planar distance are visible with increasing distance from the substrate. Within the bulk region, the inter-planar distance remains relatively uniform. In the substrate region, the inter-planar distance increases up to the bulk region boundary. In contras, the free edge region exhibits variability between crystal size and dopant concentrations, consistent with the loss of support at the edge of this region. These plots were used to determine the most consistent positions for defining the region boundaries. For consistency, the same boundary positions were used across all crystal sizes: seven atoms thick for the substrate region (excluding the fixed substrate), and ten atoms thick for the free edge.

\section{Lattice Constant}
\begin{figure}
    \centering
    \includegraphics[width=0.75\textwidth]{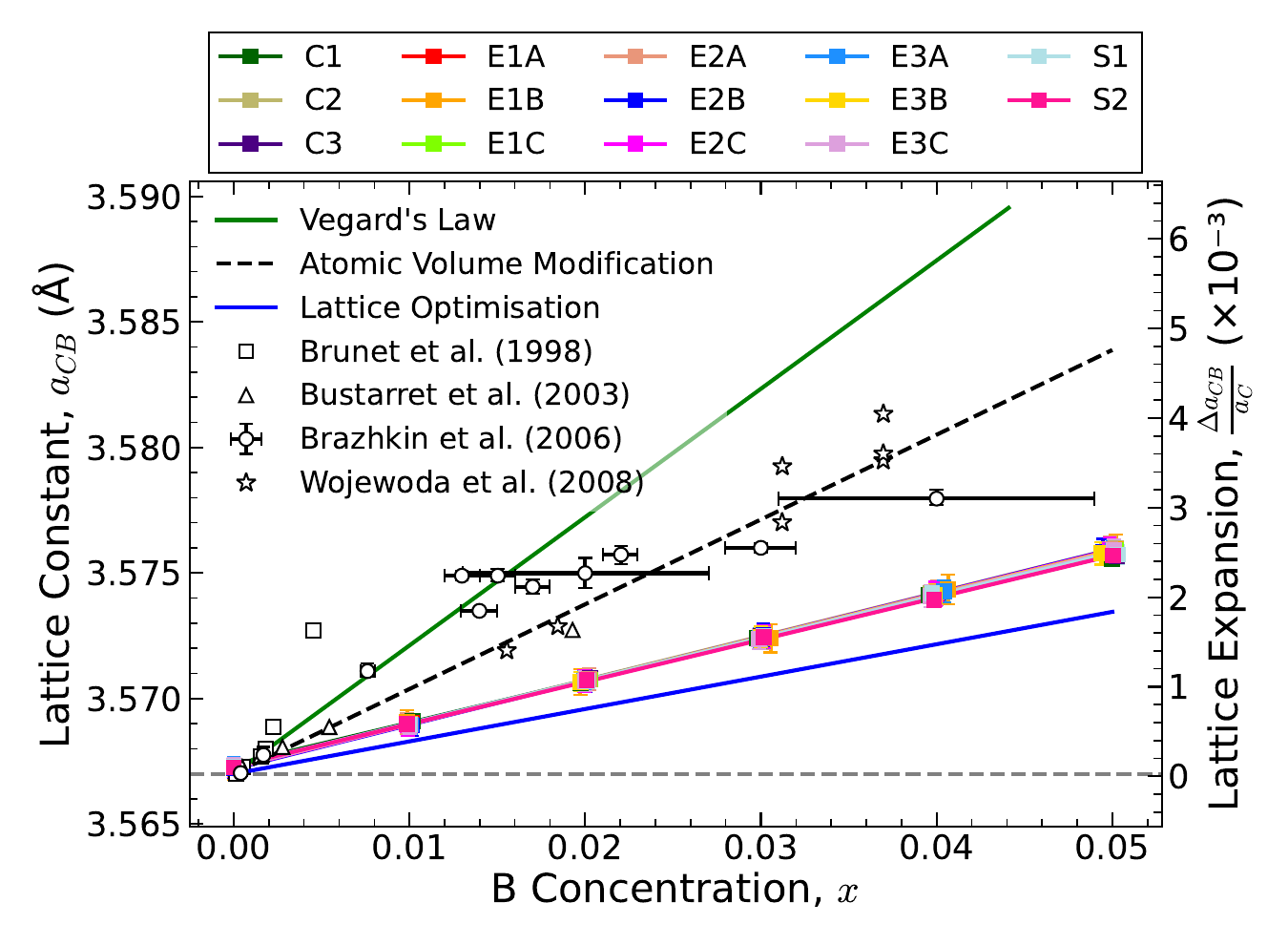}
    \caption{Plot of the lattice constant $a_{\text{CB}}$ as a function of boron dopant concentration for each crystal size considered in this work. The solid green line corresponds to Vegard's Law \cite{supp:Brunet_1998}, while the dashed black line represents the atomic volume modification \cite{supp:Brazhkin_2006}. The dashed grey line indicates the nominal lattice constant of pure diamond. Comparison data are taken from multiple sources including experimental works from \citet{supp:Brunet_1998} (open squares), \citet{supp:Bustarret_2003} (open triangles), and \citet{supp:Brazhkin_2006} (open circles), and \textit{ab initio} calculations from \citet{supp:Wojewoda_2008} (open stars).}
    \label{fig:IP_All}
\end{figure}

Here we provide additional details on the lattice constants of \CB~crystals obtained from molecular dynamics simulations. \Cref{fig:IP_All} shows the lattice constant plotted as a function of inter-planar distance for all crystal sizes. The following table reports the simulated lattice constant $a_{\text{CB}}$ and the relative lattice expansion $a_{\text{CB}}/a_{\text{C}}$ for each crystal size, with values listed as a function of dopant concentration.

\begin{longtable}{ccc}
\caption{Lattice constants and relative expansions for \CB~crystals determined from molecular dynamics simulations. For each crystal size, the table lists the simulated lattice constant $a_{\text{CB}}$ and the relative lattice expansion $\Delta a_{\text{CB}}/a_{\text{C}}$ as a function of dopant concentration. Values are averaged over all simulations of a given crystal size and concentration. Uncertainties correspond to one standard deviation ($1\sigma$)} \\

\multicolumn{1}{c|}{B Concentration, $x$} & \multicolumn{1}{c|}{\rule{0pt}{2.5ex}$a_{\text{CB}}$ (\unit{\angstrom})} & $\Delta a_{\text{CB}}/a_{\text{C}}$ (\SI{e-3}{\angstrom}) \\ \hline
\endfirsthead 

\multicolumn{1}{c|}{B Concentration, $x$} & \multicolumn{1}{c|}{\rule{0pt}{2.5ex}$a_{\text{CB}}$ (\unit{\angstrom})} & $\Delta a_{\text{CB}}/a_{\text{C}}$ (\SI{e-3}{\angstrom}) \\ \hline
\endhead 

\endfoot

\multicolumn{3}{c}{\rule{0pt}{2.5ex}\textbf{C1}} \\ \hline
\multicolumn{1}{c|}{0.00} & \multicolumn{1}{c|}{$3.5673\pm$\num{0.00018}} & $0.0851\pm0.0497$ \\
\multicolumn{1}{c|}{0.01} & \multicolumn{1}{c|}{$3.5691\pm$\num{0.00016}} & $0.5891\pm0.0456$ \\
\multicolumn{1}{c|}{0.02} & \multicolumn{1}{c|}{$3.5706\pm$\num{0.00018}} & $1.0172\pm0.0516$ \\
\multicolumn{1}{c|}{0.03} & \multicolumn{1}{c|}{$3.5724\pm$\num{0.00019}} & $1.5146\pm0.0542$ \\
\multicolumn{1}{c|}{0.04} & \multicolumn{1}{c|}{$3.5741\pm$\num{0.00021}} & $1.9957\pm0.0582$ \\
\multicolumn{1}{c|}{0.05} & \multicolumn{1}{c|}{$3.5756\pm$\num{0.00027}} & $2.4008\pm0.0744$ \\ \toprule
\multicolumn{3}{c}{\rule{0pt}{2.5ex}\textbf{C2}} \\ \hline
\multicolumn{1}{c|}{0.00} & \multicolumn{1}{c|}{$3.5673\pm$\num{0.00015}} & $0.0787\pm0.0410$ \\
\multicolumn{1}{c|}{0.01} & \multicolumn{1}{c|}{$3.5690\pm$\num{0.00018}} & $0.5589\pm0.0501$ \\
\multicolumn{1}{c|}{0.02} & \multicolumn{1}{c|}{$3.5707\pm$\num{0.00021}} & $1.0493\pm0.0597$ \\
\multicolumn{1}{c|}{0.03} & \multicolumn{1}{c|}{$3.5724\pm$\num{0.00024}} & $1.5221\pm0.0660$ \\
\multicolumn{1}{c|}{0.04} & \multicolumn{1}{c|}{$3.5742\pm$\num{0.00026}} & $2.0170\pm0.0726$ \\
\multicolumn{1}{c|}{0.05} & \multicolumn{1}{c|}{$3.5760\pm$\num{0.00027}} & $2.5189\pm0.0761$ \\ \toprule
\multicolumn{3}{c}{\rule{0pt}{2.5ex}\textbf{C3}} \\ \hline
\multicolumn{1}{c|}{0.00} & \multicolumn{1}{c|}{$3.5673\pm$\num{0.00013}} & $0.0740\pm0.0374$ \\
\multicolumn{1}{c|}{0.01} & \multicolumn{1}{c|}{$3.5690\pm$\num{0.00016}} & $0.5488\pm0.0447$ \\
\multicolumn{1}{c|}{0.02} & \multicolumn{1}{c|}{$3.5708\pm$\num{0.00018}} & $1.0647\pm0.0517$ \\
\multicolumn{1}{c|}{0.03} & \multicolumn{1}{c|}{$3.5725\pm$\num{0.00021}} & $1.5350\pm0.0581$ \\
\multicolumn{1}{c|}{0.04} & \multicolumn{1}{c|}{$3.5742\pm$\num{0.00023}} & $2.0205\pm0.0636$ \\
\multicolumn{1}{c|}{0.05} & \multicolumn{1}{c|}{$3.5757\pm$\num{0.00029}} & $2.4305\pm0.0817$ \\ \toprule
\multicolumn{3}{c}{\rule{0pt}{2.5ex}\textbf{E1A}} \\ \hline
\multicolumn{1}{c|}{0.00} & \multicolumn{1}{c|}{$3.5673\pm0.0002$} & $0.0754\pm0.0494$ \\
\multicolumn{1}{c|}{0.01} & \multicolumn{1}{c|}{$3.5690\pm0.0002$} & $0.5636\pm0.0611$ \\
\multicolumn{1}{c|}{0.02} & \multicolumn{1}{c|}{$3.5707\pm0.0003$} & $1.0398\pm0.0710$ \\
\multicolumn{1}{c|}{0.03} & \multicolumn{1}{c|}{$3.5724\pm0.0003$} & $1.5218\pm0.0790$ \\
\multicolumn{1}{c|}{0.04} & \multicolumn{1}{c|}{$3.5742\pm0.0003$} & $2.0108\pm0.0864$ \\
\multicolumn{1}{c|}{0.05} & \multicolumn{1}{c|}{$3.5759\pm0.0004$} & $2.4867\pm0.0985$ \\ \toprule
\multicolumn{3}{c}{\rule{0pt}{2.5ex}\textbf{E1B}} \\ \hline
\multicolumn{1}{c|}{0.00} & \multicolumn{1}{c|}{$3.5672\pm0.0004$} & $0.0590\pm0.1258$ \\
\multicolumn{1}{c|}{0.01} & \multicolumn{1}{c|}{$3.5691\pm0.0005$} & $0.5817\pm0.1331$ \\
\multicolumn{1}{c|}{0.02} & \multicolumn{1}{c|}{$3.5707\pm0.0005$} & $1.0274\pm0.1418$ \\
\multicolumn{1}{c|}{0.03} & \multicolumn{1}{c|}{$3.5724\pm0.0006$} & $1.5121\pm0.1571$ \\
\multicolumn{1}{c|}{0.04} & \multicolumn{1}{c|}{$3.5743\pm0.0006$} & $2.0575\pm0.1645$ \\
\multicolumn{1}{c|}{0.05} & \multicolumn{1}{c|}{$3.5759\pm0.0006$} & $2.5059\pm0.1692$ \\ \toprule
\multicolumn{3}{c}{\rule{0pt}{2.5ex}\textbf{E1C}} \\ \hline
\multicolumn{1}{c|}{0.00} & \multicolumn{1}{c|}{$3.5672\pm0.0001$} & $0.0635\pm0.0250$ \\
\multicolumn{1}{c|}{0.01} & \multicolumn{1}{c|}{$3.5690\pm0.0002$} & $0.5568\pm0.0431$ \\
\multicolumn{1}{c|}{0.02} & \multicolumn{1}{c|}{$3.5707\pm0.0002$} & $1.0506\pm0.0562$ \\
\multicolumn{1}{c|}{0.03} & \multicolumn{1}{c|}{$3.5725\pm0.0002$} & $1.5287\pm0.0658$ \\
\multicolumn{1}{c|}{0.04} & \multicolumn{1}{c|}{$3.5742\pm0.0003$} & $2.0093\pm0.0746$ \\
\multicolumn{1}{c|}{0.05} & \multicolumn{1}{c|}{$3.5759\pm0.0003$} & $2.4967\pm0.0847$ \\ \toprule
\multicolumn{3}{c}{\rule{0pt}{2.5ex}\textbf{E2A}} \\ \hline
\multicolumn{1}{c|}{0.00} & \multicolumn{1}{c|}{$3.5672\pm0.0004$} & $0.0613\pm0.1067$ \\
\multicolumn{1}{c|}{0.01} & \multicolumn{1}{c|}{$3.5690\pm0.0004$} & $0.5603\pm0.1160$ \\
\multicolumn{1}{c|}{0.02} & \multicolumn{1}{c|}{$3.5708\pm0.0004$} & $1.0567\pm0.1249$ \\
\multicolumn{1}{c|}{0.03} & \multicolumn{1}{c|}{$3.5724\pm0.0005$} & $1.5224\pm0.1309$ \\
\multicolumn{1}{c|}{0.04} & \multicolumn{1}{c|}{$3.5742\pm0.0005$} & $2.0232\pm0.1371$ \\
\multicolumn{1}{c|}{0.05} & \multicolumn{1}{c|}{$3.5759\pm0.0005$} & $2.5046\pm0.1428$ \\ \toprule
\multicolumn{3}{c}{\rule{0pt}{2.5ex}\textbf{E2B}} \\ \hline
\multicolumn{1}{c|}{0.00} & \multicolumn{1}{c|}{$3.5672\pm0.0004$} & $0.0623\pm0.1079$ \\
\multicolumn{1}{c|}{0.01} & \multicolumn{1}{c|}{$3.5689\pm0.0004$} & $0.5387\pm0.1147$ \\
\multicolumn{1}{c|}{0.02} & \multicolumn{1}{c|}{$3.5707\pm0.0004$} & $1.0390\pm0.1216$ \\
\multicolumn{1}{c|}{0.03} & \multicolumn{1}{c|}{$3.5725\pm0.0005$} & $1.5452\pm0.1313$ \\
\multicolumn{1}{c|}{0.04} & \multicolumn{1}{c|}{$3.5742\pm0.0005$} & $2.0082\pm0.1359$ \\
\multicolumn{1}{c|}{0.05} & \multicolumn{1}{c|}{$3.5758\pm0.0005$} & $2.4796\pm0.1439$ \\ \toprule
\multicolumn{3}{c}{\rule{0pt}{2.5ex}\textbf{E2C}} \\ \hline
\multicolumn{1}{c|}{0.00} & \multicolumn{1}{c|}{$3.5673\pm0.0004$} & $0.0711\pm0.1086$ \\
\multicolumn{1}{c|}{0.01} & \multicolumn{1}{c|}{$3.5689\pm0.0004$} & $0.5413\pm0.1155$ \\
\multicolumn{1}{c|}{0.02} & \multicolumn{1}{c|}{$3.5707\pm0.0004$} & $1.0335\pm0.1232$ \\
\multicolumn{1}{c|}{0.03} & \multicolumn{1}{c|}{$3.5725\pm0.0005$} & $1.5302\pm0.1302$ \\
\multicolumn{1}{c|}{0.04} & \multicolumn{1}{c|}{$3.5742\pm0.0005$} & $2.0070\pm0.1369$ \\
\multicolumn{1}{c|}{0.05} & \multicolumn{1}{c|}{$3.5759\pm0.0005$} & $2.4988\pm0.1412$ \\ \toprule
\multicolumn{3}{c}{\rule{0pt}{2.5ex}\textbf{E3A}} \\ \hline
\multicolumn{1}{c|}{0.00} & \multicolumn{1}{c|}{$3.5673\pm0.0003$} & $0.0903\pm0.0936$ \\
\multicolumn{1}{c|}{0.01} & \multicolumn{1}{c|}{$3.5690\pm0.0004$} & $0.5513\pm0.1019$ \\
\multicolumn{1}{c|}{0.02} & \multicolumn{1}{c|}{$3.5707\pm0.0004$} & $1.0425\pm0.1056$ \\
\multicolumn{1}{c|}{0.03} & \multicolumn{1}{c|}{$3.5724\pm0.0004$} & $1.5155\pm0.1135$ \\
\multicolumn{1}{c|}{0.04} & \multicolumn{1}{c|}{$3.5743\pm0.0004$} & $2.0371\pm0.1191$ \\
\multicolumn{1}{c|}{0.05} & \multicolumn{1}{c|}{$3.5759\pm0.0004$} & $2.4833\pm0.1235$ \\ \toprule
\multicolumn{3}{c}{\rule{0pt}{2.5ex}\textbf{E3B}} \\ \hline
\multicolumn{1}{c|}{0.00} & \multicolumn{1}{c|}{$3.5673\pm0.0003$} & $0.0709\pm0.0924$ \\
\multicolumn{1}{c|}{0.01} & \multicolumn{1}{c|}{$3.5690\pm0.0004$} & $0.5571\pm0.1002$ \\
\multicolumn{1}{c|}{0.02} & \multicolumn{1}{c|}{$3.5707\pm0.0004$} & $1.0279\pm0.1079$ \\
\multicolumn{1}{c|}{0.03} & \multicolumn{1}{c|}{$3.5725\pm0.0004$} & $1.5306\pm0.1128$ \\
\multicolumn{1}{c|}{0.04} & \multicolumn{1}{c|}{$3.5741\pm0.0004$} & $1.9969\pm0.1204$ \\
\multicolumn{1}{c|}{0.05} & \multicolumn{1}{c|}{$3.5758\pm0.0004$} & $2.4621\pm0.1260$ \\ \toprule
\multicolumn{3}{c}{\rule{0pt}{2.5ex}\textbf{E3C}} \\ \hline
\multicolumn{1}{c|}{0.00} & \multicolumn{1}{c|}{$3.5673\pm0.0003$} & $0.0841\pm0.0939$ \\
\multicolumn{1}{c|}{0.01} & \multicolumn{1}{c|}{$3.5690\pm0.0004$} & $0.5485\pm0.1014$ \\
\multicolumn{1}{c|}{0.02} & \multicolumn{1}{c|}{$3.5707\pm0.0004$} & $1.0345\pm0.1076$ \\
\multicolumn{1}{c|}{0.03} & \multicolumn{1}{c|}{$3.5724\pm0.0004$} & $1.5046\pm0.1142$ \\
\multicolumn{1}{c|}{0.04} & \multicolumn{1}{c|}{$3.5741\pm0.0004$} & $1.9825\pm0.1229$ \\
\multicolumn{1}{c|}{0.05} & \multicolumn{1}{c|}{$3.5759\pm0.0004$} & $2.4951\pm0.1239$ \\ \toprule
\multicolumn{3}{c}{\rule{0pt}{2.5ex}\textbf{S1}} \\ \hline
\multicolumn{1}{c|}{0.00} & \multicolumn{1}{c|}{$3.5673\pm0.0002$} & $0.0818\pm0.0577$ \\
\multicolumn{1}{c|}{0.01} & \multicolumn{1}{c|}{$3.5690\pm0.0002$} & $0.5518\pm0.0638$ \\
\multicolumn{1}{c|}{0.02} & \multicolumn{1}{c|}{$3.5707\pm0.0002$} & $1.0387\pm0.0665$ \\
\multicolumn{1}{c|}{0.03} & \multicolumn{1}{c|}{$3.5725\pm0.0003$} & $1.5365\pm0.0713$ \\
\multicolumn{1}{c|}{0.04} & \multicolumn{1}{c|}{$3.5742\pm0.0003$} & $2.0151\pm0.0718$ \\
\multicolumn{1}{c|}{0.05} & \multicolumn{1}{c|}{$3.5757\pm0.0004$} & $2.4462\pm0.1024$ \\ \toprule
\multicolumn{3}{c}{\rule{0pt}{2.5ex}\textbf{S2}} \\ \hline
\multicolumn{1}{c|}{0.00} & \multicolumn{1}{c|}{$3.5672\pm0.0001$} & $0.0694\pm0.0374$ \\
\multicolumn{1}{c|}{0.01} & \multicolumn{1}{c|}{$3.5690\pm0.0001$} & $0.5528\pm0.0407$ \\
\multicolumn{1}{c|}{0.02} & \multicolumn{1}{c|}{$3.5707\pm0.0002$} & $1.0466\pm0.0434$ \\
\multicolumn{1}{c|}{0.03} & \multicolumn{1}{c|}{$3.5724\pm0.0002$} & $1.5279\pm0.0449$ \\
\multicolumn{1}{c|}{0.04} & \multicolumn{1}{c|}{$3.5739\pm0.0003$} & $1.9380\pm0.0779$ \\
\multicolumn{1}{c|}{0.05} & \multicolumn{1}{c|}{$3.5757\pm0.0003$} & $2.4358\pm0.0785$
\end{longtable}

\section{(1\;0\;0) and (1\;1\;0) Region-by-region Inter-planar Distances}
Here we provide additional details on the region-by-region inter-planar distances of \CB~crystals as a function of boron dopant concentration. The following subsections present results for each crystal type (C, E1, E2, E3, and S). For each crystal, the region-by-region inter-planar distances are plotted with left panels showing the (1\;1\;0) inter-planar distances, and the right panels showing the (1\;0\;0) inter-planar distances. Within each panel, the top, middle, and bottom rows correspond to the substrate, bulk, and free-edge regions, respectively. Each plot includes comparisons to the covalent radius \cite{supp:Brunet_1998} and atomic volume \cite{supp:Brazhkin_2006} modifications to Vegard's Law \cite{supp:Vegard_1921}, with the relative lattice expansion $\Delta d_{ijk}/d_{ijk}$, where $(i\;j\;k)$ are a given set of planes, plotted on the right-hand $y$ axis. Following the figures, tables are provided for each plane , first (1\;1\;0) then (1\;0\;0), listing the dopant concentration, inter-planar distance, and relative lattice expansion for each region.

\subsection{C Crystals}
\begin{figure}[H]
 \subfloat[]{\includegraphics[width=0.48\textwidth]{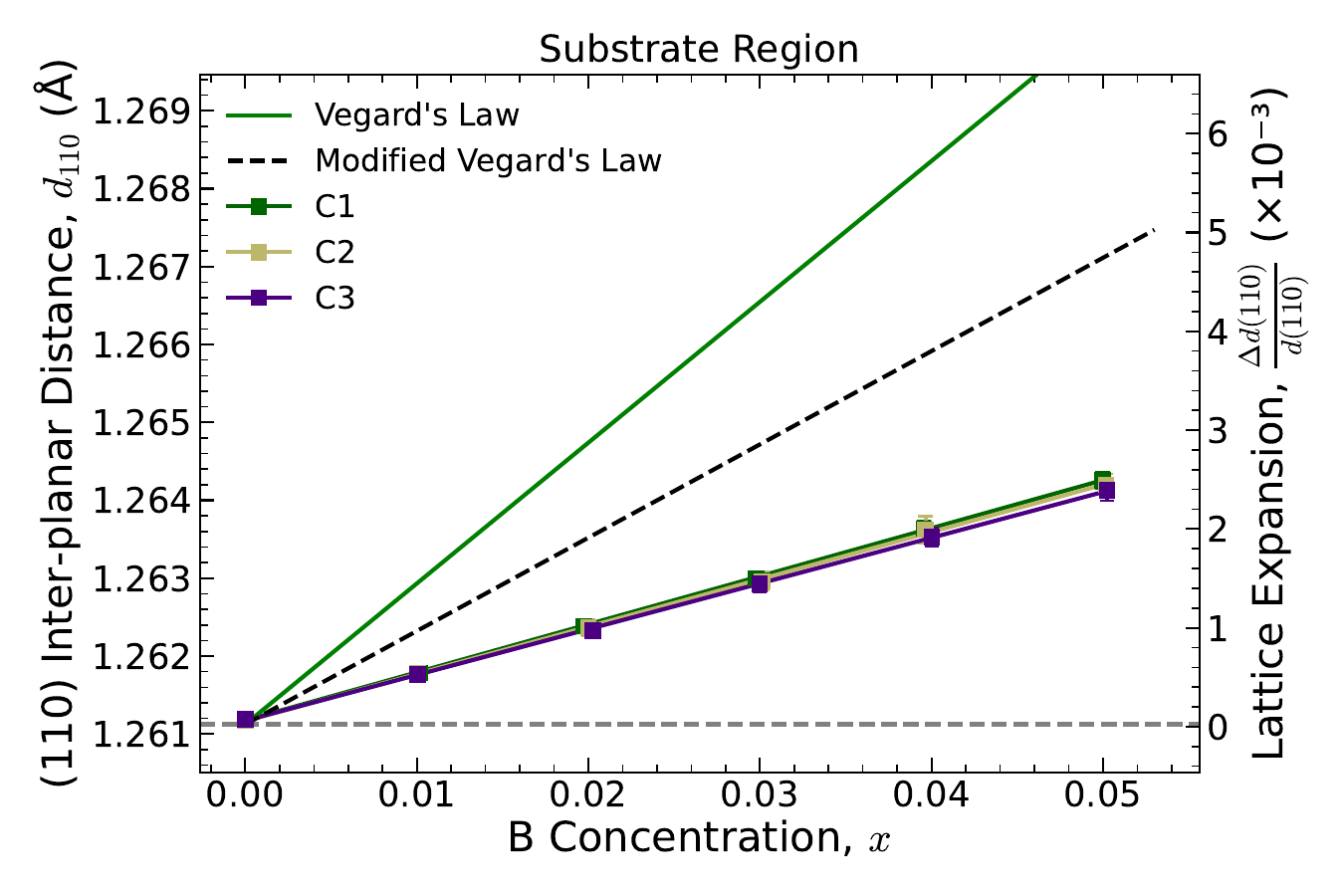}}\hfill
 \setcounter{subfigure}{3}
 \subfloat[]{\includegraphics[width=0.48\textwidth]{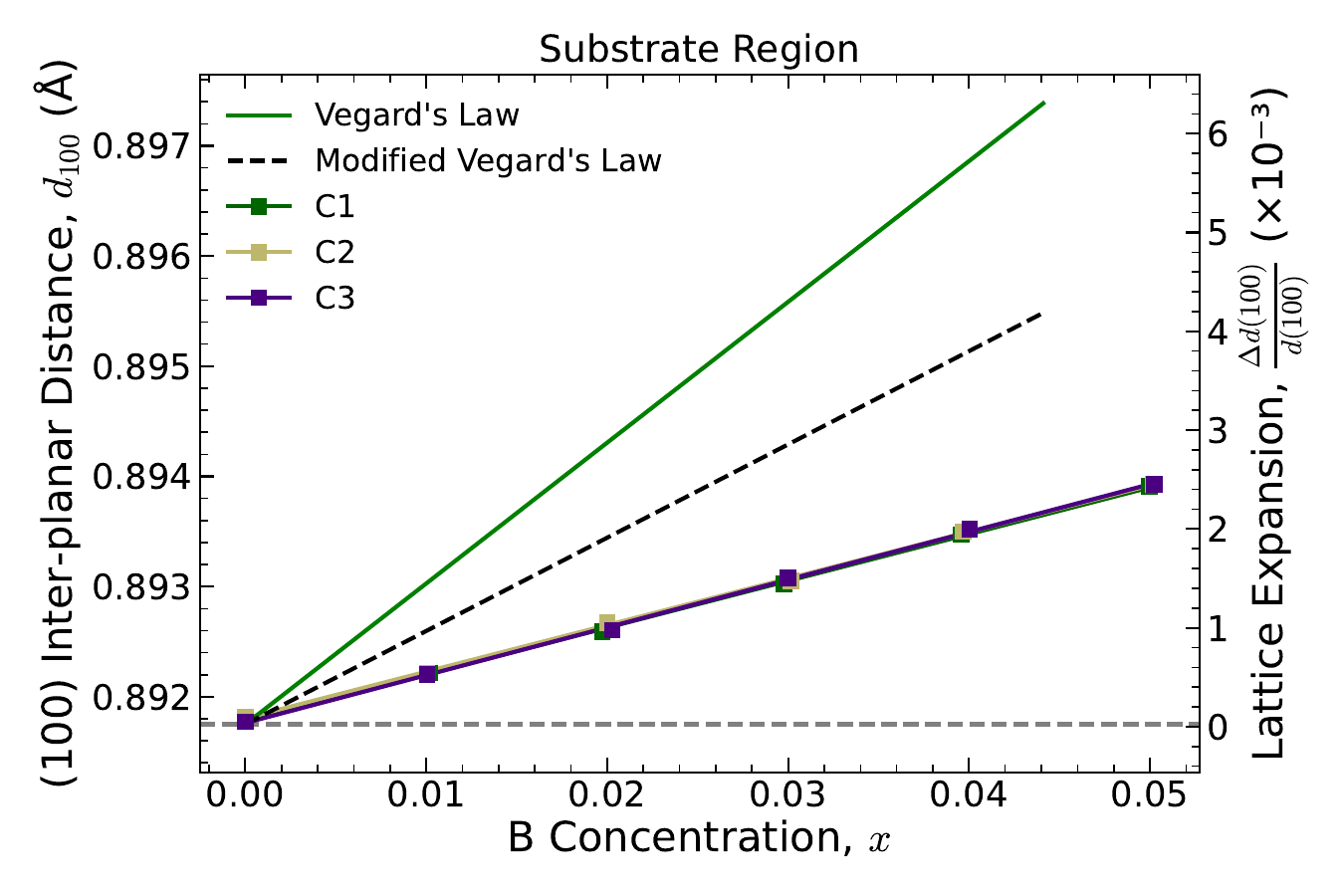}}\\[-2ex]
 \setcounter{subfigure}{1}
 \subfloat[]{\includegraphics[width=0.48\textwidth]{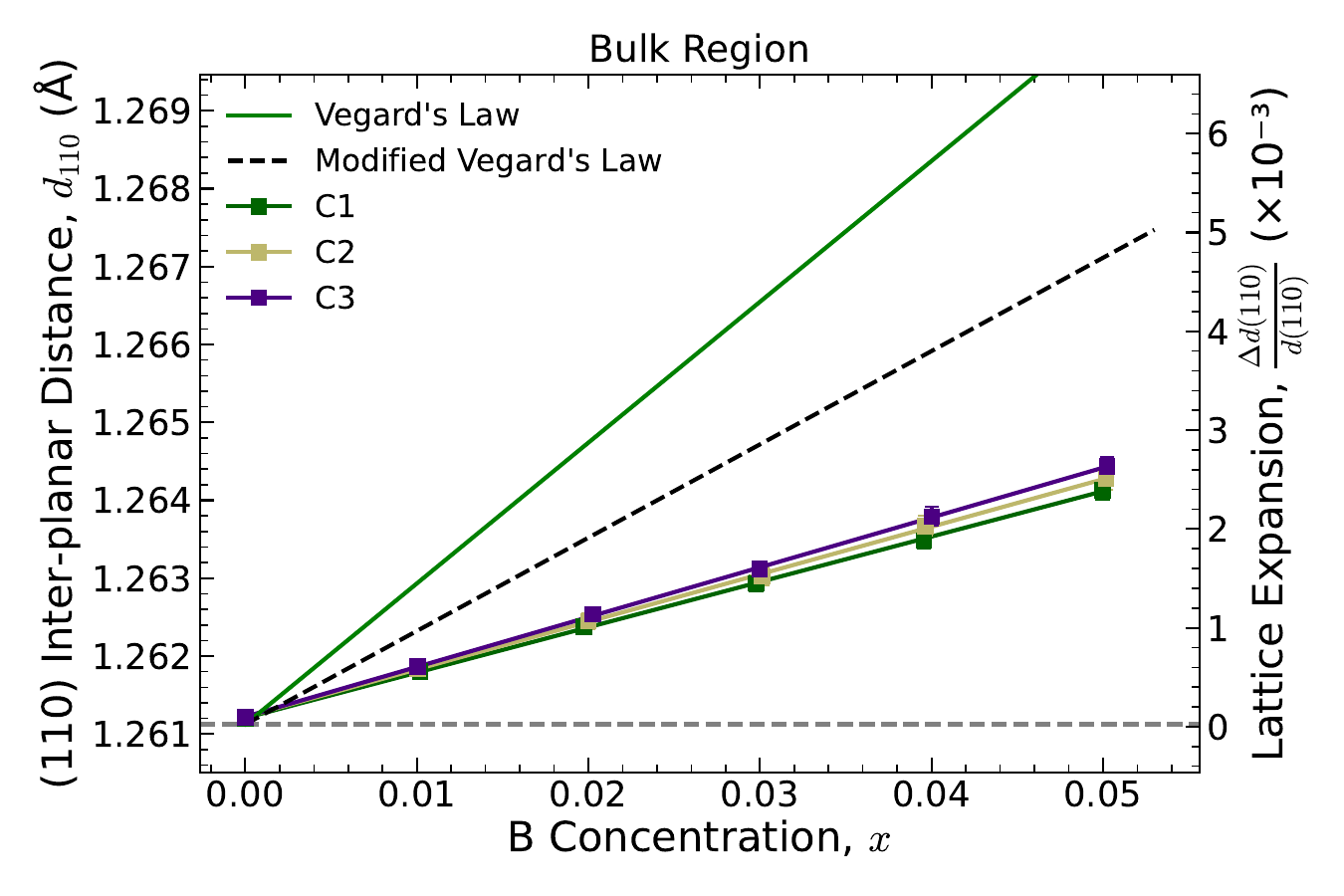}}\hfill
 \setcounter{subfigure}{4}
 \subfloat[]{\includegraphics[width=0.48\textwidth]{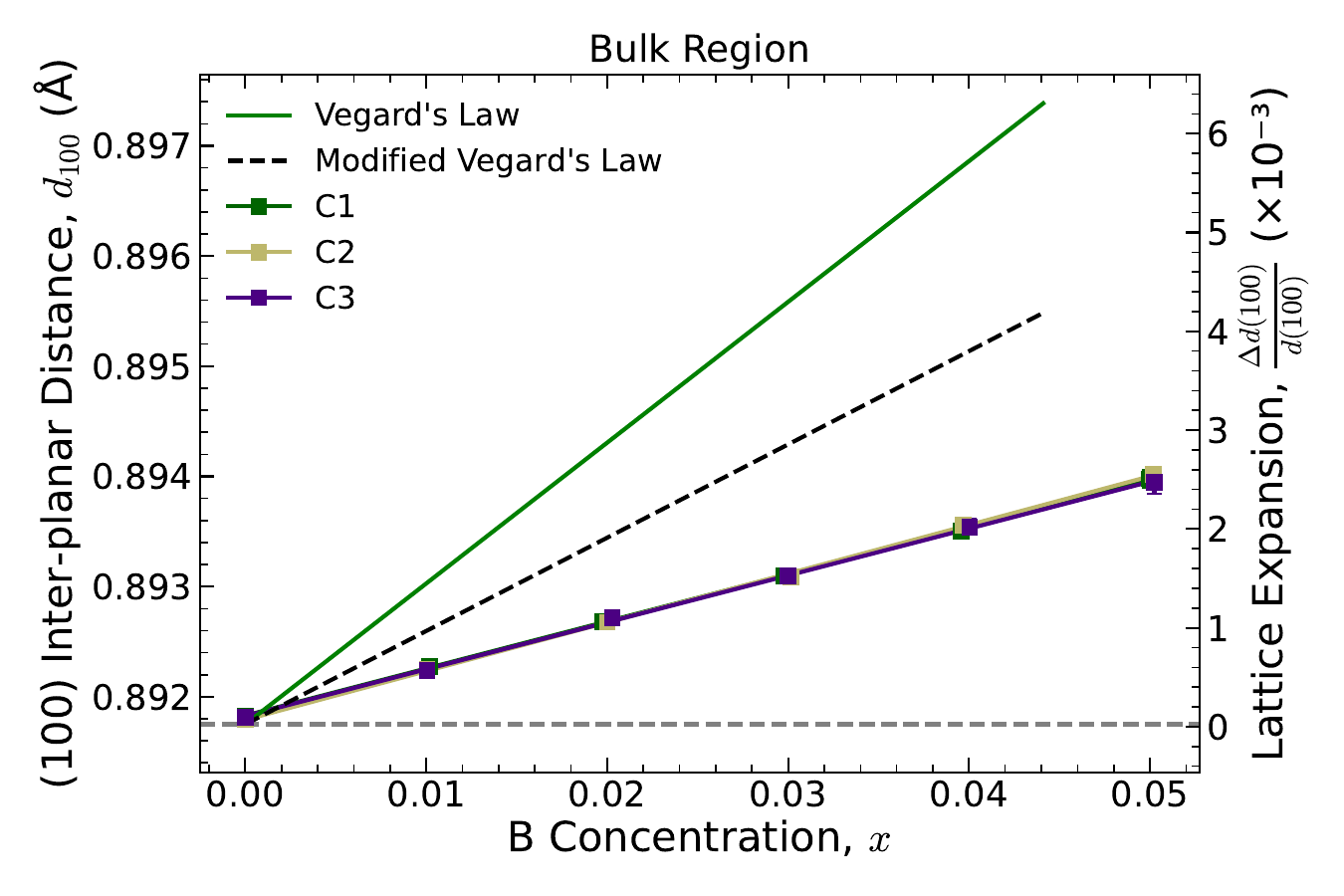}}\\[-2ex]
 \setcounter{subfigure}{2}
 \subfloat[]{\includegraphics[width=0.48\textwidth]{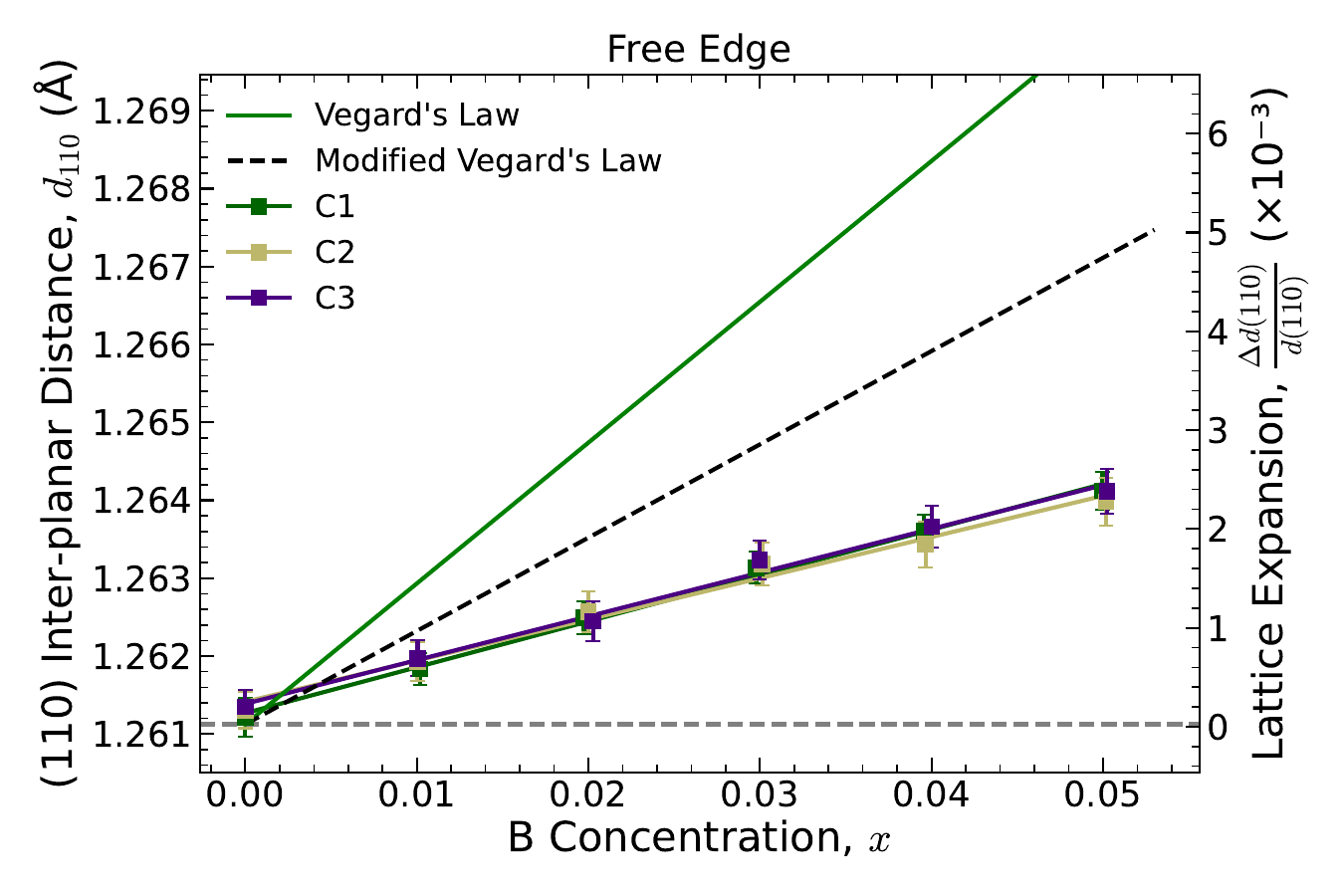}}\hfill
 \setcounter{subfigure}{5}
 \subfloat[]{\includegraphics[width=0.48\textwidth]{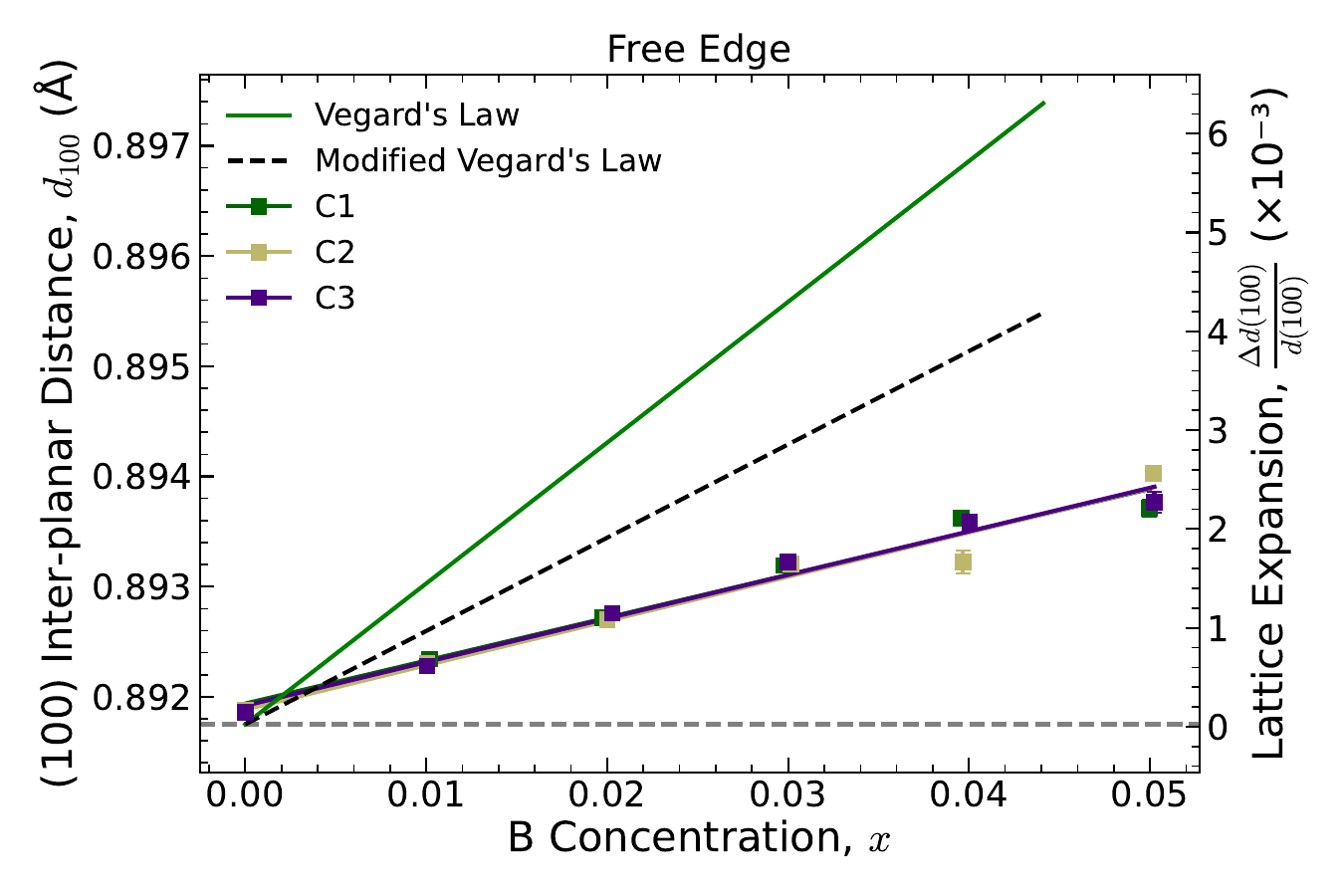}}\\[-2ex]
 \caption{Region-by-region inter-planar distances of C-category crystals as a function of boron concentration, with left/right panels for (1\;1\;0)/(1\;0\;0) planes and top/middle/bottom rows for substrate, bulk, and free-edge regions.}
 \label{fig:IP_Plots_C}
\end{figure}

\begin{sidewaystable}[]
\centering
    \caption{(1\;1\;0) inter-planar distance $d_{110}$ and lattice expansion $\Delta d_{110}/d_{110}$ as a function of dopant concentration for the substrate, bulk, and free-edge regions of the C1, C2, and C3 crystals.}
    \resizebox{\textheight}{!}{%
    \begin{tabular}{ccccccc}
    \multicolumn{1}{c|}{\multirow{2}{*}{B Concentration, $x$}} & \multicolumn{2}{c|}{Substrate Region} & \multicolumn{2}{c|}{Bulk Region} & \multicolumn{2}{c}{Free Edge} \\
    \cline{2-7} \multicolumn{1}{c|}{} & \multicolumn{1}{c|}{\rule{0pt}{2.5ex}$d_{110}$ (\unit{\angstrom})} & \multicolumn{1}{c|}{$\Delta d_{110}/d_{110}$ (\SI{e-3}{\angstrom})} & \multicolumn{1}{c|}{\rule{0pt}{2.5ex}$d_{110}$ (\unit{\angstrom})} & \multicolumn{1}{c|}{$\Delta d_{110}/d_{110}$ (\SI{e-3}{\angstrom})} & \multicolumn{1}{c|}{\rule{0pt}{2.5ex}$d_{110}$ (\unit{\angstrom})} & $\Delta d_{110}/d_{110}$ (\SI{e-3}{\angstrom}) \\ \hline
    \multicolumn{7}{c}{\rule{0pt}{2.5ex}\textbf{C1}} \\ \hline
    \multicolumn{1}{c|}{0.00} & \multicolumn{1}{c|}{$1.2611\pm$\num{8.6e-05}} & \multicolumn{1}{c|}{$0.0567\pm0.0678$} & \multicolumn{1}{c|}{$1.2611\pm$\num{8.6e-05}} & \multicolumn{1}{c|}{$0.0579\pm0.0680$} & \multicolumn{1}{c|}{$1.2603\pm$\num{0.00025}} & $0.0712\pm0.1949$ \\
    \multicolumn{1}{c|}{0.01} & \multicolumn{1}{c|}{$1.2612\pm$\num{7.7e-05}} & \multicolumn{1}{c|}{$0.5218\pm0.0611$} & \multicolumn{1}{c|}{$1.2613\pm$\num{8.1e-05}} & \multicolumn{1}{c|}{$0.5359\pm0.0640$} & \multicolumn{1}{c|}{$1.2605\pm$\num{0.0002}} & $0.5620\pm0.1608$ \\
    \multicolumn{1}{c|}{0.02} & \multicolumn{1}{c|}{$1.2614\pm$\num{8.7e-05}} & \multicolumn{1}{c|}{$0.9998\pm0.0694$} & \multicolumn{1}{c|}{$1.2614\pm$\num{9.3e-05}} & \multicolumn{1}{c|}{$0.9947\pm0.0739$} & \multicolumn{1}{c|}{$1.2606\pm$\num{0.00021}} & $1.0812\pm0.1680$ \\
    \multicolumn{1}{c|}{0.03} & \multicolumn{1}{c|}{$1.2615\pm$\num{9e-05}} & \multicolumn{1}{c|}{$1.4853\pm0.0712$} & \multicolumn{1}{c|}{$1.2616\pm$\num{9.7e-05}} & \multicolumn{1}{c|}{$1.4338\pm0.0770$} & \multicolumn{1}{c|}{$1.2607\pm$\num{0.00021}} & $1.5930\pm0.1627$ \\
    \multicolumn{1}{c|}{0.04} & \multicolumn{1}{c|}{$1.2616\pm$\num{0.0001}} & \multicolumn{1}{c|}{$1.9908\pm0.0794$} & \multicolumn{1}{c|}{$1.2617\pm$\num{0.0001}} & \multicolumn{1}{c|}{$1.8711\pm0.0813$} & \multicolumn{1}{c|}{$1.2609\pm$\num{0.00021}} & $1.9627\pm0.1698$ \\
    \multicolumn{1}{c|}{0.05} & \multicolumn{1}{c|}{$1.2617\pm$\num{0.00011}} & \multicolumn{1}{c|}{$2.4827\pm0.0864$} & \multicolumn{1}{c|}{$1.2619\pm$\num{0.00012}} & \multicolumn{1}{c|}{$2.3810\pm0.0913$} & \multicolumn{1}{c|}{$1.2611\pm$\num{0.00024}} & $2.3774\pm0.1896$ \\ \toprule
    \multicolumn{7}{c}{\rule{0pt}{2.5ex}\textbf{C2}} \\ \hline
    \multicolumn{1}{c|}{0.00} & \multicolumn{1}{c|}{$1.2611\pm$\num{8e-05}} & \multicolumn{1}{c|}{$0.0383\pm0.0633$} & \multicolumn{1}{c|}{$1.2612\pm$\num{7.4e-05}} & \multicolumn{1}{c|}{$0.0729\pm0.0588$} & \multicolumn{1}{c|}{$1.2604\pm$\num{0.00023}} & $0.1453\pm0.1852$ \\
    \multicolumn{1}{c|}{0.01} & \multicolumn{1}{c|}{$1.2612\pm$\num{9.1e-05}} & \multicolumn{1}{c|}{$0.5170\pm0.0719$} & \multicolumn{1}{c|}{$1.2613\pm$\num{8.5e-05}} & \multicolumn{1}{c|}{$0.5680\pm0.0676$} & \multicolumn{1}{c|}{$1.2606\pm$\num{0.00025}} & $0.6411\pm0.1973$ \\
    \multicolumn{1}{c|}{0.02} & \multicolumn{1}{c|}{$1.2613\pm$\num{0.00011}} & \multicolumn{1}{c|}{$0.9811\pm0.0845$} & \multicolumn{1}{c|}{$1.2615\pm$\num{0.0001}} & \multicolumn{1}{c|}{$1.0504\pm0.0792$} & \multicolumn{1}{c|}{$1.2607\pm$\num{0.00026}} & $1.1472\pm0.2086$ \\
    \multicolumn{1}{c|}{0.03} & \multicolumn{1}{c|}{$1.2614\pm$\num{0.00012}} & \multicolumn{1}{c|}{$1.4551\pm0.0961$} & \multicolumn{1}{c|}{$1.2617\pm$\num{0.00011}} & \multicolumn{1}{c|}{$1.5034\pm0.0887$} & \multicolumn{1}{c|}{$1.2608\pm$\num{0.00028}} & $1.6330\pm0.2202$ \\
    \multicolumn{1}{c|}{0.04} & \multicolumn{1}{c|}{$1.2616\pm$\num{0.00017}} & \multicolumn{1}{c|}{$1.9808\pm0.1369$} & \multicolumn{1}{c|}{$1.2619\pm$\num{0.00013}} & \multicolumn{1}{c|}{$2.0182\pm0.1011$} & \multicolumn{1}{c|}{$1.2607\pm$\num{0.00029}} & $1.8311\pm0.2312$ \\
    \multicolumn{1}{c|}{0.05} & \multicolumn{1}{c|}{$1.2617\pm$\num{0.00014}} & \multicolumn{1}{c|}{$2.4363\pm0.1148$} & \multicolumn{1}{c|}{$1.2620\pm$\num{0.00014}} & \multicolumn{1}{c|}{$2.4978\pm0.1088$} & \multicolumn{1}{c|}{$1.2609\pm$\num{0.00031}} & $2.2665\pm0.2446$ \\ \toprule
    \multicolumn{7}{c}{\rule{0pt}{2.5ex}\textbf{C3}} \\ \hline
    \multicolumn{1}{c|}{0.00} & \multicolumn{1}{c|}{$1.2611\pm$\num{7.7e-05}} & \multicolumn{1}{c|}{$0.0531\pm0.0611$} & \multicolumn{1}{c|}{$1.2612\pm$\num{6.3e-05}} & \multicolumn{1}{c|}{$0.0713\pm0.0496$} & \multicolumn{1}{c|}{$1.2605\pm$\num{0.00022}} & $0.1764\pm0.1754$ \\
    \multicolumn{1}{c|}{0.01} & \multicolumn{1}{c|}{$1.2612\pm$\num{8.5e-05}} & \multicolumn{1}{c|}{$0.5047\pm0.0671$} & \multicolumn{1}{c|}{$1.2614\pm$\num{8e-05}} & \multicolumn{1}{c|}{$0.5900\pm0.0634$} & \multicolumn{1}{c|}{$1.2606\pm$\num{0.00023}} & $0.6718\pm0.1831$ \\
    \multicolumn{1}{c|}{0.02} & \multicolumn{1}{c|}{$1.2613\pm$\num{9.7e-05}} & \multicolumn{1}{c|}{$0.9516\pm0.0766$} & \multicolumn{1}{c|}{$1.2616\pm$\num{9.1e-05}} & \multicolumn{1}{c|}{$1.1194\pm0.0718$} & \multicolumn{1}{c|}{$1.2605\pm$\num{0.00025}} & $1.0476\pm0.2014$ \\
    \multicolumn{1}{c|}{0.03} & \multicolumn{1}{c|}{$1.2614\pm$\num{0.00011}} & \multicolumn{1}{c|}{$1.4315\pm0.0894$} & \multicolumn{1}{c|}{$1.2618\pm$\num{0.0001}} & \multicolumn{1}{c|}{$1.5789\pm0.0820$} & \multicolumn{1}{c|}{$1.2608\pm$\num{0.00025}} & $1.6725\pm0.2007$ \\
    \multicolumn{1}{c|}{0.04} & \multicolumn{1}{c|}{$1.2615\pm$\num{0.00012}} & \multicolumn{1}{c|}{$1.9068\pm0.0949$} & \multicolumn{1}{c|}{$1.2620\pm$\num{0.00012}} & \multicolumn{1}{c|}{$2.1142\pm0.0957$} & \multicolumn{1}{c|}{$1.2610\pm$\num{0.00027}} & $2.0113\pm0.2150$ \\
    \multicolumn{1}{c|}{0.05} & \multicolumn{1}{c|}{$1.2616\pm$\num{0.00013}} & \multicolumn{1}{c|}{$2.3817\pm0.1056$} & \multicolumn{1}{c|}{$1.2622\pm$\num{0.00013}} & \multicolumn{1}{c|}{$2.6213\pm0.1017$} & \multicolumn{1}{c|}{$1.2610\pm$\num{0.00029}} & $2.3729\pm0.2304$
    \end{tabular}
    }
    \label{tab:IP_table_C_110}
\end{sidewaystable}

\begin{sidewaystable*}[]
\centering
    \caption{(1\;0\;0) inter-planar distance $d_{110}$ and lattice expansion $\Delta d_{110}/d_{110}$ as a function of dopant concentration for the substrate, bulk, and free-edge regions of the C1, C2, and C3 crystals.}
    \resizebox{\textheight}{!}{%
    \begin{tabular}{ccccccc}
    \multicolumn{1}{c|}{\multirow{2}{*}{B Concentration, $x$}} & \multicolumn{2}{c|}{Substrate Region} & \multicolumn{2}{c|}{Bulk Region} & \multicolumn{2}{c}{Free Edge} \\
    \cline{2-7} \multicolumn{1}{c|}{} & \multicolumn{1}{c|}{\rule{0pt}{2.5ex}$d_{100}$ (\unit{\angstrom})} & \multicolumn{1}{c|}{$\Delta d_{100}/d_{100}$ (\SI{e-3}{\angstrom})} & \multicolumn{1}{c|}{\rule{0pt}{2.5ex}$d_{100}$ (\unit{\angstrom})} & \multicolumn{1}{c|}{$\Delta d_{100}/d_{100}$ (\SI{e-3}{\angstrom})} & \multicolumn{1}{c|}{\rule{0pt}{2.5ex}$d_{100}$ (\unit{\angstrom})} & $\Delta d_{100}/d_{100}$ (\SI{e-3}{\angstrom}) \\ \hline
    \multicolumn{7}{c}{\rule{0pt}{2.5ex}\textbf{C1}} \\ \hline
    \multicolumn{1}{c|}{0.00} & \multicolumn{1}{c|}{$0.8918\pm$\num{3.8e-05}} & \multicolumn{1}{c|}{$0.0514\pm0.0431$} & \multicolumn{1}{c|}{$0.8918\pm$\num{4e-05}} & \multicolumn{1}{c|}{$0.0813\pm0.0448$} & \multicolumn{1}{c|}{$0.8919\pm$\num{4.2e-05}} & $0.1255\pm0.0474$ \\
    \multicolumn{1}{c|}{0.01} & \multicolumn{1}{c|}{$0.8922\pm$\num{3.6e-05}} & \multicolumn{1}{c|}{$0.5230\pm0.0403$} & \multicolumn{1}{c|}{$0.8923\pm$\num{3.6e-05}} & \multicolumn{1}{c|}{$0.5893\pm0.0405$} & \multicolumn{1}{c|}{$0.8923\pm$\num{3.7e-05}} & $0.6633\pm0.0415$ \\
    \multicolumn{1}{c|}{0.02} & \multicolumn{1}{c|}{$0.8926\pm$\num{3.9e-05}} & \multicolumn{1}{c|}{$0.9399\pm0.0432$} & \multicolumn{1}{c|}{$0.8927\pm$\num{4.2e-05}} & \multicolumn{1}{c|}{$1.0410\pm0.0467$} & \multicolumn{1}{c|}{$0.8927\pm$\num{4e-05}} & $1.0827\pm0.0453$ \\
    \multicolumn{1}{c|}{0.03} & \multicolumn{1}{c|}{$0.8930\pm$\num{4.2e-05}} & \multicolumn{1}{c|}{$1.4261\pm0.0474$} & \multicolumn{1}{c|}{$0.8931\pm$\num{4.2e-05}} & \multicolumn{1}{c|}{$1.5093\pm0.0475$} & \multicolumn{1}{c|}{$0.8932\pm$\num{4.2e-05}} & $1.6148\pm0.0471$ \\
    \multicolumn{1}{c|}{0.04} & \multicolumn{1}{c|}{$0.8935\pm$\num{4.4e-05}} & \multicolumn{1}{c|}{$1.9316\pm0.0497$} & \multicolumn{1}{c|}{$0.8935\pm$\num{4.6e-05}} & \multicolumn{1}{c|}{$1.9659\pm0.0515$} & \multicolumn{1}{c|}{$0.8936\pm$\num{4.5e-05}} & $2.0994\pm0.0504$ \\
    \multicolumn{1}{c|}{0.05} & \multicolumn{1}{c|}{$0.8939\pm$\num{4.8e-05}} & \multicolumn{1}{c|}{$2.4202\pm0.0539$} & \multicolumn{1}{c|}{$0.8940\pm$\num{7.7e-05}} & \multicolumn{1}{c|}{$2.4981\pm0.0866$} & \multicolumn{1}{c|}{$0.8937\pm$\num{7.6e-05}} & $2.2023\pm0.0853$ \\ \toprule
    \multicolumn{7}{c}{\rule{0pt}{2.5ex}\textbf{C2}} \\ \hline
    \multicolumn{1}{c|}{0.00} & \multicolumn{1}{c|}{$0.8918\pm$\num{3.1e-05}} & \multicolumn{1}{c|}{$0.0726\pm0.0346$} & \multicolumn{1}{c|}{$0.8918\pm$\num{3.5e-05}} & \multicolumn{1}{c|}{$0.0480\pm0.0390$} & \multicolumn{1}{c|}{$0.8919\pm$\num{3.5e-05}} & $0.1424\pm0.0395$ \\
    \multicolumn{1}{c|}{0.01} & \multicolumn{1}{c|}{$0.8922\pm$\num{3.8e-05}} & \multicolumn{1}{c|}{$0.5154\pm0.0424$} & \multicolumn{1}{c|}{$0.8922\pm$\num{4.2e-05}} & \multicolumn{1}{c|}{$0.5545\pm0.0473$} & \multicolumn{1}{c|}{$0.8923\pm$\num{4e-05}} & $0.6177\pm0.0450$ \\
    \multicolumn{1}{c|}{0.02} & \multicolumn{1}{c|}{$0.8927\pm$\num{4.5e-05}} & \multicolumn{1}{c|}{$1.0358\pm0.0507$} & \multicolumn{1}{c|}{$0.8927\pm$\num{4.9e-05}} & \multicolumn{1}{c|}{$1.0402\pm0.0554$} & \multicolumn{1}{c|}{$0.8927\pm$\num{4.8e-05}} & $1.0622\pm0.0537$ \\
    \multicolumn{1}{c|}{0.03} & \multicolumn{1}{c|}{$0.8930\pm$\num{5.1e-05}} & \multicolumn{1}{c|}{$1.4567\pm0.0576$} & \multicolumn{1}{c|}{$0.8931\pm$\num{5.4e-05}} & \multicolumn{1}{c|}{$1.4982\pm0.0605$} & \multicolumn{1}{c|}{$0.8932\pm$\num{5.3e-05}} & $1.6324\pm0.0595$ \\
    \multicolumn{1}{c|}{0.04} & \multicolumn{1}{c|}{$0.8935\pm$\num{5.6e-05}} & \multicolumn{1}{c|}{$1.9615\pm0.0624$} & \multicolumn{1}{c|}{$0.8936\pm$\num{6e-05}} & \multicolumn{1}{c|}{$2.0279\pm0.0672$} & \multicolumn{1}{c|}{$0.8932\pm$\num{0.0001}} & $1.6492\pm0.1169$ \\
    \multicolumn{1}{c|}{0.05} & \multicolumn{1}{c|}{$0.8939\pm$\num{5.7e-05}} & \multicolumn{1}{c|}{$2.4399\pm0.0635$} & \multicolumn{1}{c|}{$0.8940\pm$\num{6.2e-05}} & \multicolumn{1}{c|}{$2.5419\pm0.0696$} & \multicolumn{1}{c|}{$0.8940\pm$\num{6e-05}} & $2.5568\pm0.0673$ \\ \toprule
    \multicolumn{7}{c}{\rule{0pt}{2.5ex}\textbf{C3}} \\ \hline
    \multicolumn{1}{c|}{0.00} & \multicolumn{1}{c|}{$0.8918\pm$\num{2.9e-05}} & \multicolumn{1}{c|}{$0.0240\pm0.0330$} & \multicolumn{1}{c|}{$0.8918\pm$\num{3.1e-05}} & \multicolumn{1}{c|}{$0.0728\pm0.0345$} & \multicolumn{1}{c|}{$0.8919\pm$\num{3.3e-05}} & $0.1251\pm0.0367$ \\
    \multicolumn{1}{c|}{0.01} & \multicolumn{1}{c|}{$0.8922\pm$\num{3.2e-05}} & \multicolumn{1}{c|}{$0.5051\pm0.0361$} & \multicolumn{1}{c|}{$0.8922\pm$\num{3.8e-05}} & \multicolumn{1}{c|}{$0.5501\pm0.0431$} & \multicolumn{1}{c|}{$0.8923\pm$\num{3.7e-05}} & $0.5938\pm0.0415$ \\
    \multicolumn{1}{c|}{0.02} & \multicolumn{1}{c|}{$0.8926\pm$\num{3.8e-05}} & \multicolumn{1}{c|}{$0.9568\pm0.0431$} & \multicolumn{1}{c|}{$0.8927\pm$\num{4.4e-05}} & \multicolumn{1}{c|}{$1.0834\pm0.0496$} & \multicolumn{1}{c|}{$0.8928\pm$\num{4e-05}} & $1.1311\pm0.0451$ \\
    \multicolumn{1}{c|}{0.03} & \multicolumn{1}{c|}{$0.8931\pm$\num{4.2e-05}} & \multicolumn{1}{c|}{$1.4891\pm0.0466$} & \multicolumn{1}{c|}{$0.8931\pm$\num{4.9e-05}} & \multicolumn{1}{c|}{$1.5087\pm0.0547$} & \multicolumn{1}{c|}{$0.8932\pm$\num{4.6e-05}} & $1.6513\pm0.0518$ \\
    \multicolumn{1}{c|}{0.04} & \multicolumn{1}{c|}{$0.8935\pm$\num{4.6e-05}} & \multicolumn{1}{c|}{$1.9853\pm0.0518$} & \multicolumn{1}{c|}{$0.8935\pm$\num{7.2e-05}} & \multicolumn{1}{c|}{$2.0097\pm0.0806$} & \multicolumn{1}{c|}{$0.8936\pm$\num{5e-05}} & $2.0565\pm0.0562$ \\
    \multicolumn{1}{c|}{0.05} & \multicolumn{1}{c|}{$0.8939\pm$\num{4.7e-05}} & \multicolumn{1}{c|}{$2.4406\pm0.0530$} & \multicolumn{1}{c|}{$0.8939\pm$\num{0.0001}} & \multicolumn{1}{c|}{$2.4601\pm0.1124$} & \multicolumn{1}{c|}{$0.8938\pm$\num{9.5e-05}} & $2.2576\pm0.1060$
    \end{tabular}
    }
    \label{tab:IP_table_C_100}
\end{sidewaystable*}

\subsection{E1 Crystals}
\begin{figure}[H]
 \subfloat[]{\includegraphics[width=0.48\textwidth]{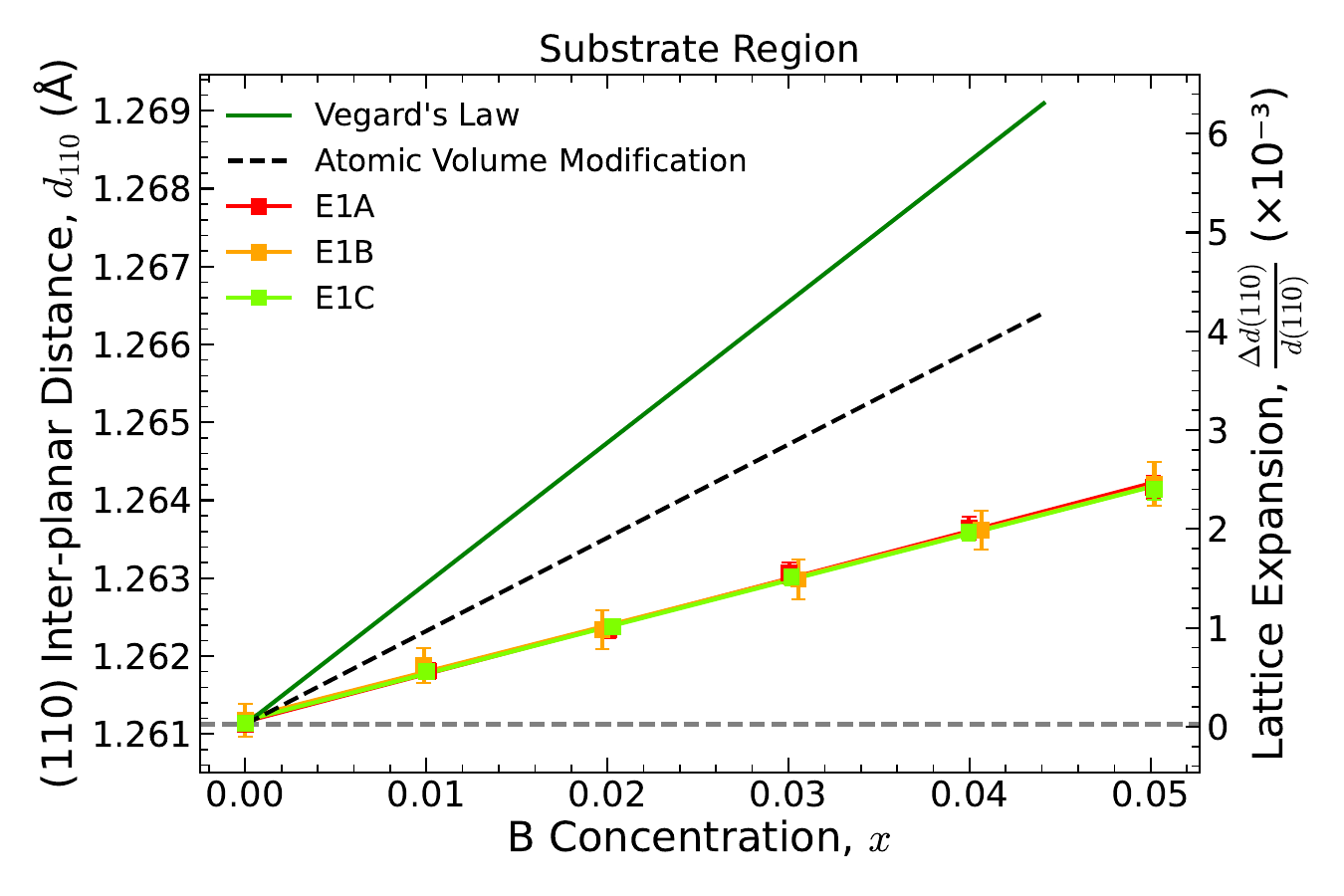}}\hfill
 \setcounter{subfigure}{3}
 \subfloat[]{\includegraphics[width=0.48\textwidth]{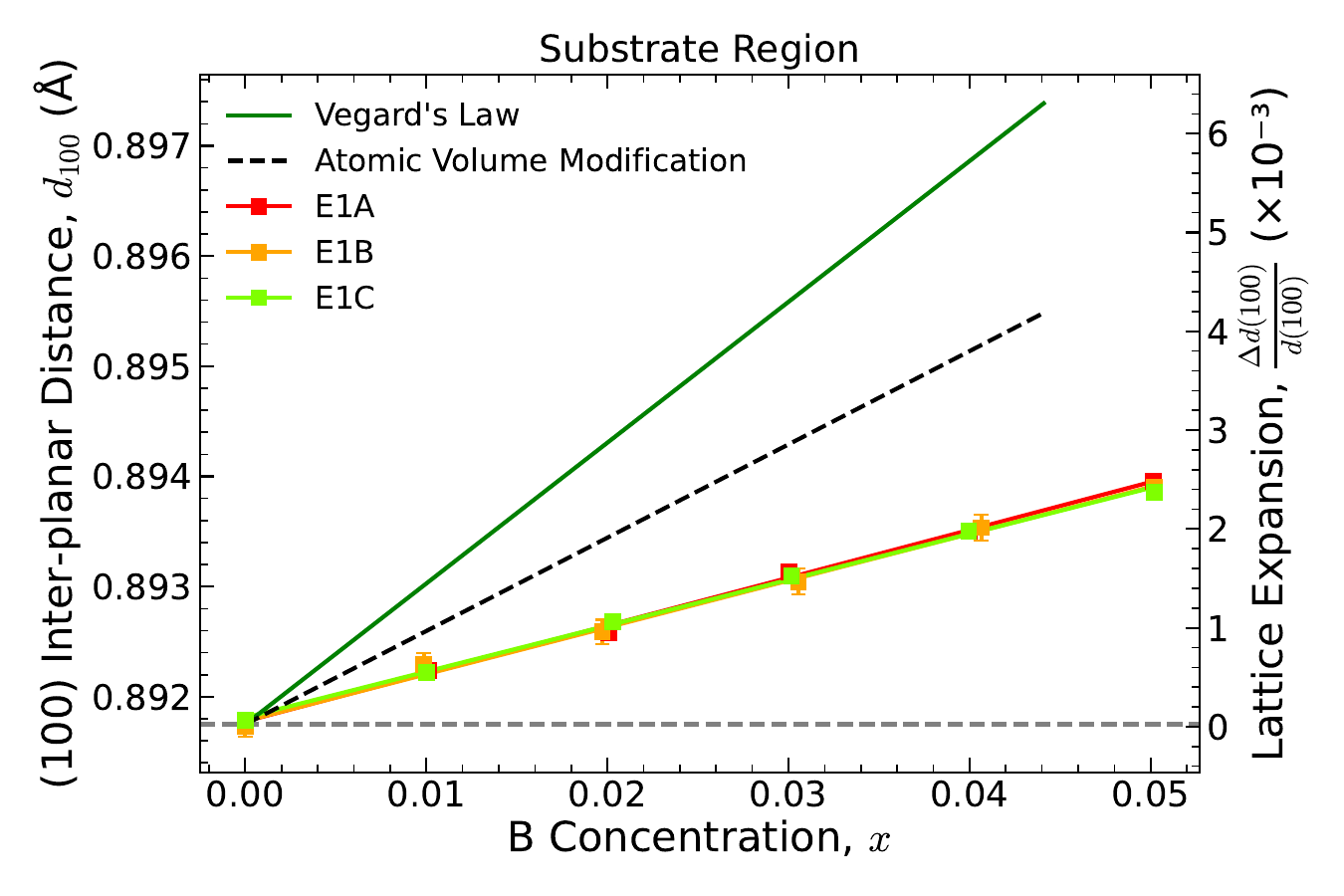}}\\[-2ex]
 \setcounter{subfigure}{1}
 \subfloat[]{\includegraphics[width=0.48\textwidth]{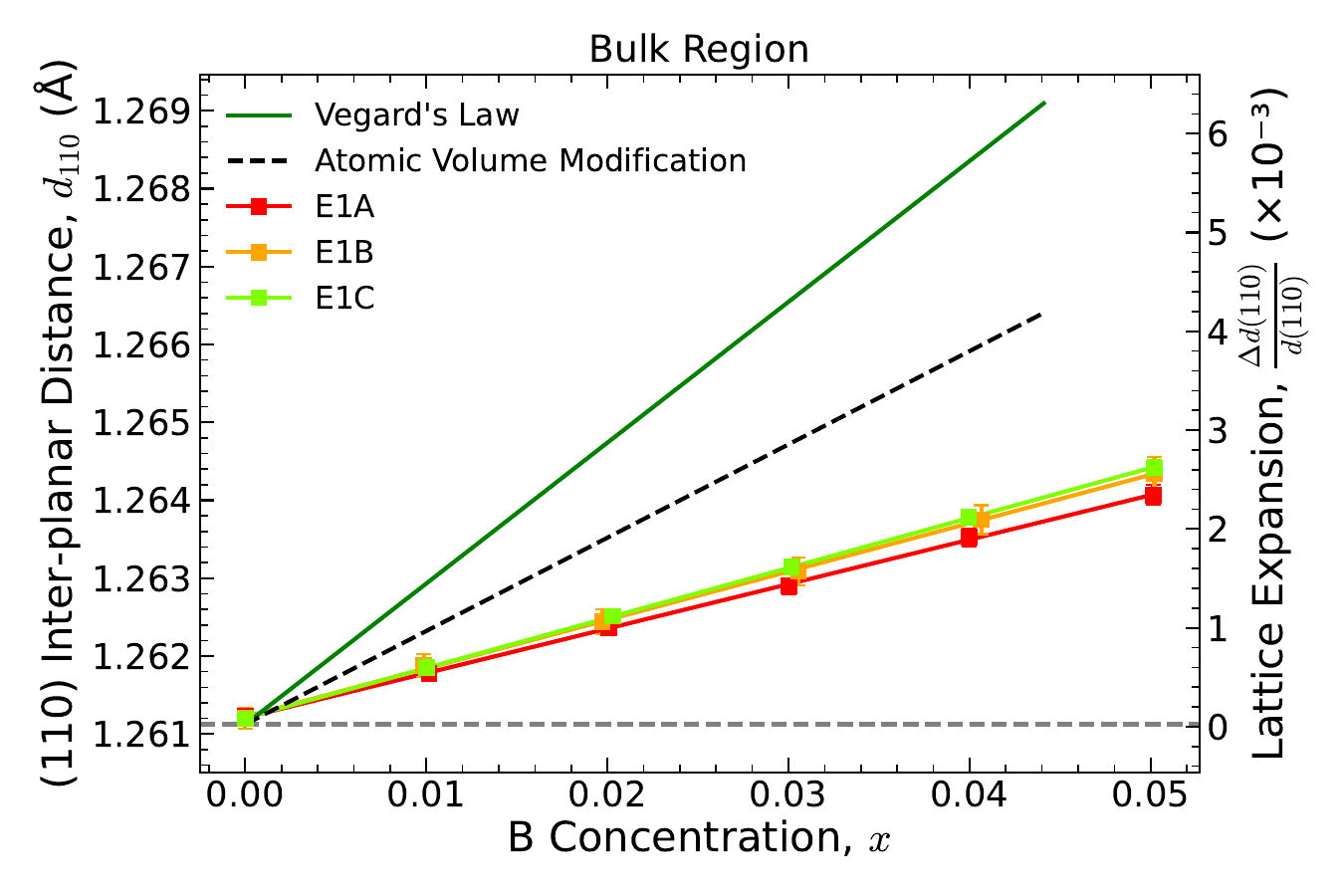}}\hfill
 \setcounter{subfigure}{4}
 \subfloat[]{\includegraphics[width=0.48\textwidth]{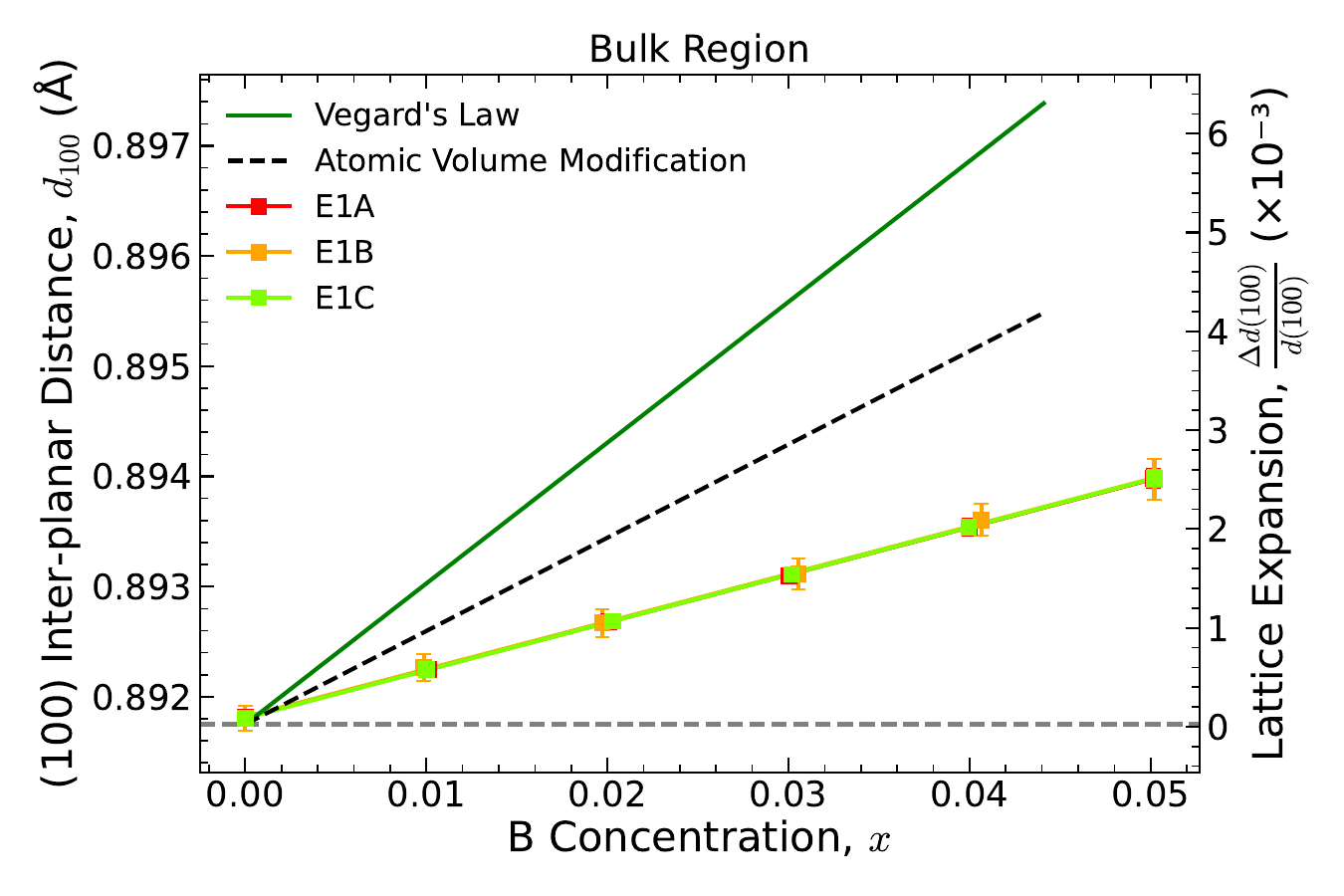}}\\[-2ex]
 \setcounter{subfigure}{2}
 \subfloat[]{\includegraphics[width=0.48\textwidth]{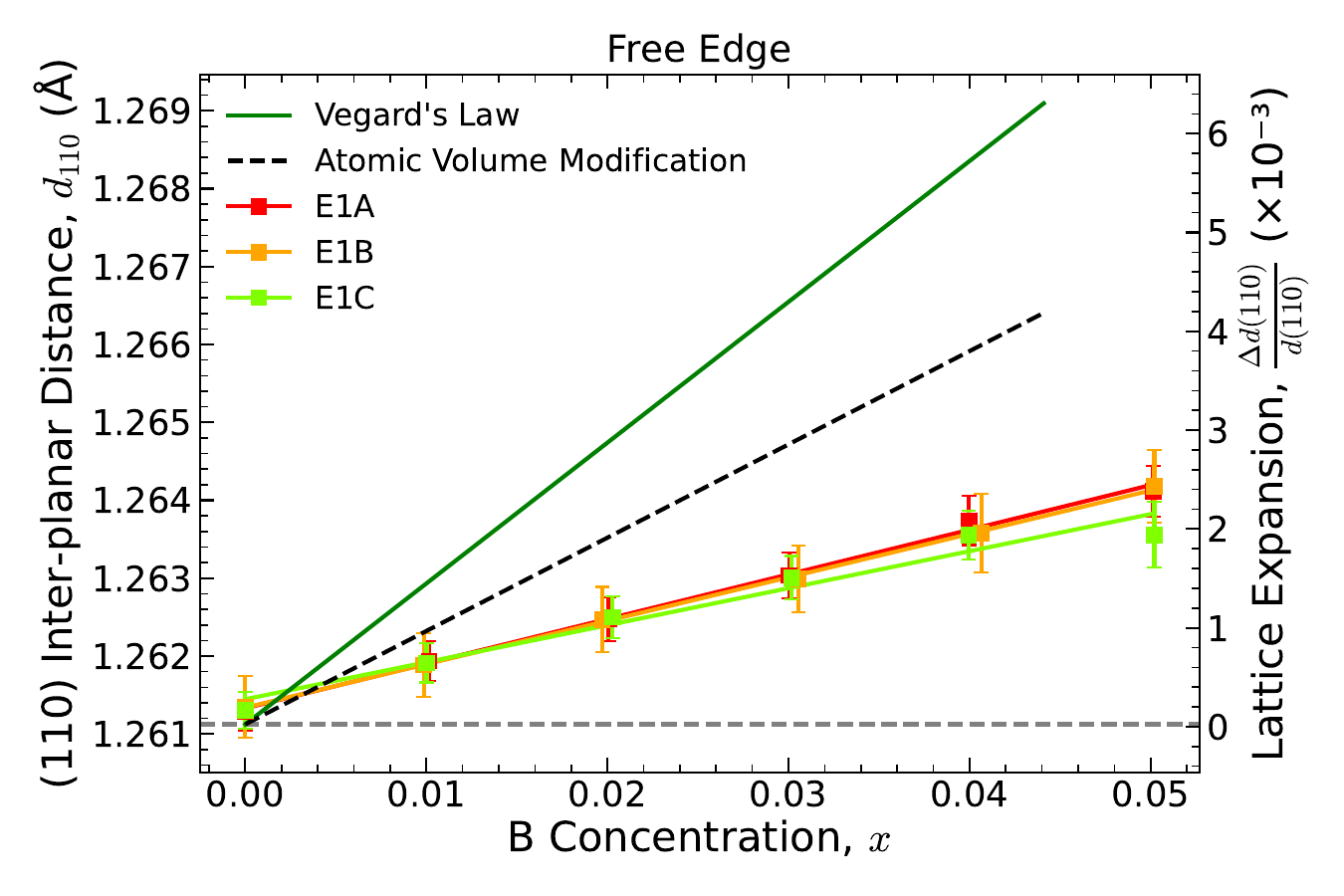}}\hfill
 \setcounter{subfigure}{5}
 \subfloat[]{\includegraphics[width=0.48\textwidth]{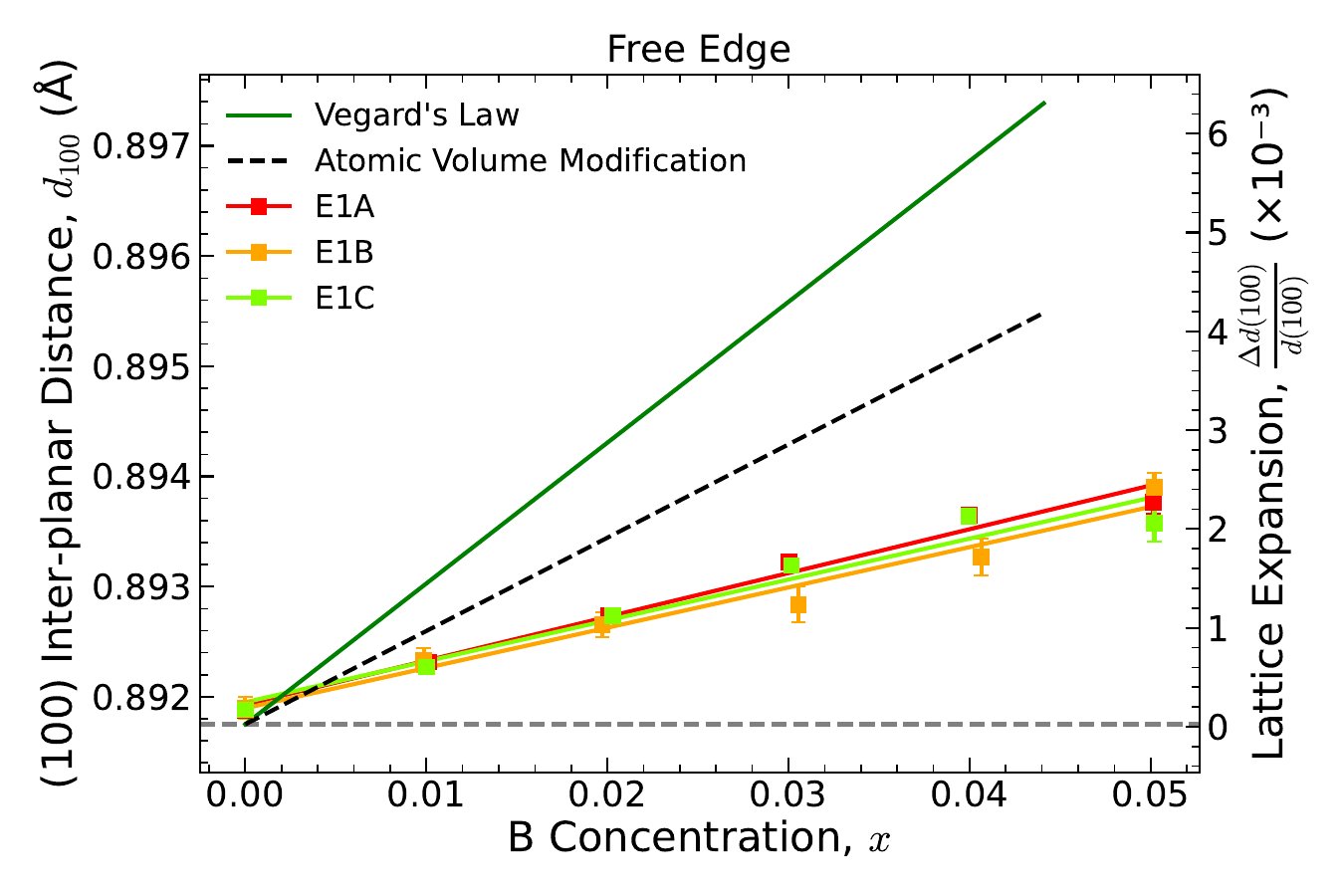}}\\[-2ex]
 \caption{Region-by-region inter-planar distances of E1-category crystals as a function of boron concentration, with left/right panels for (1\;1\;0)/(1\;0\;0) planes and top/middle/bottom rows for substrate, bulk, and free-edge regions.}
 \label{fig:IP_Plots_E1}
\end{figure}

\begin{sidewaystable*}[]
\centering
    \caption{(1\;1\;0) inter-planar distance $d_{110}$ and lattice expansion $\Delta d_{110}/d_{110}$ as a function of dopant concentration for the substrate, bulk, and free-edge regions of the E1A, E1B, and E1C crystals.}
    \resizebox{\textheight}{!}{%
    \begin{tabular}{ccccccc}
    \multicolumn{1}{c|}{\multirow{2}{*}{B Concentration, $x$}} & \multicolumn{2}{c|}{Substrate Region} & \multicolumn{2}{c|}{Bulk Region} & \multicolumn{2}{c}{Free Edge} \\
    \cline{2-7} \multicolumn{1}{c|}{} & \multicolumn{1}{c|}{\rule{0pt}{2.5ex}$d_{110}$ (\unit{\angstrom})} & \multicolumn{1}{c|}{$\Delta d_{110}/d_{110}$ (\SI{e-3}{\angstrom})} & \multicolumn{1}{c|}{\rule{0pt}{2.5ex}$d_{110}$ (\unit{\angstrom})} & \multicolumn{1}{c|}{$\Delta d_{110}/d_{110_{100}}$ (\SI{e-3}{\angstrom})} & \multicolumn{1}{c|}{\rule{0pt}{2.5ex}$d_{110}$ (\unit{\angstrom})} & $\Delta d_{110}/d_{110}$ (\SI{e-3}{\angstrom}) \\ \hline
    \multicolumn{7}{c}{\rule{0pt}{2.5ex}\textbf{E1A}} \\ \hline
    \multicolumn{1}{c|}{0.00} & \multicolumn{1}{c|}{$1.2611\pm$\num{8.7e-05}} & \multicolumn{1}{c|}{$0.0030\pm0.0686$} & \multicolumn{1}{c|}{$1.2612\pm$\num{6e-05}} & \multicolumn{1}{c|}{$0.0782\pm0.0474$} & \multicolumn{1}{c|}{$1.2603\pm$\num{0.00025}} & $0.1344\pm0.1972$ \\
    \multicolumn{1}{c|}{0.01} & \multicolumn{1}{c|}{$1.2612\pm$\num{9.9e-05}} & \multicolumn{1}{c|}{$0.5417\pm0.0785$} & \multicolumn{1}{c|}{$1.2617\pm$\num{6.9e-05}} & \multicolumn{1}{c|}{$0.5169\pm0.0550$} & \multicolumn{1}{c|}{$1.2605\pm$\num{0.00026}} & $0.6452\pm0.2052$ \\
    \multicolumn{1}{c|}{0.02} & \multicolumn{1}{c|}{$1.2614\pm$\num{0.00011}} & \multicolumn{1}{c|}{$0.9639\pm0.0896$} & \multicolumn{1}{c|}{$1.2621\pm$\num{8.4e-05}} & \multicolumn{1}{c|}{$0.9788\pm0.0667$} & \multicolumn{1}{c|}{$1.2606\pm$\num{0.00028}} & $1.0696\pm0.2230$ \\
    \multicolumn{1}{c|}{0.03} & \multicolumn{1}{c|}{$1.2615\pm$\num{0.00013}} & \multicolumn{1}{c|}{$1.5406\pm0.1055$} & \multicolumn{1}{c|}{$1.2626\pm$\num{0.0001}} & \multicolumn{1}{c|}{$1.4067\pm0.0799$} & \multicolumn{1}{c|}{$1.2608\pm$\num{0.00029}} & $1.5158\pm0.2305$ \\
    \multicolumn{1}{c|}{0.04} & \multicolumn{1}{c|}{$1.2616\pm$\num{0.00014}} & \multicolumn{1}{c|}{$1.9962\pm0.1137$} & \multicolumn{1}{c|}{$1.2631\pm$\num{0.00011}} & \multicolumn{1}{c|}{$1.9031\pm0.0889$} & \multicolumn{1}{c|}{$1.2609\pm$\num{0.00031}} & $2.0732\pm0.2493$ \\
    \multicolumn{1}{c|}{0.05} & \multicolumn{1}{c|}{$1.2617\pm$\num{0.00015}} & \multicolumn{1}{c|}{$2.4110\pm0.1194$} & \multicolumn{1}{c|}{$1.2636\pm$\num{0.00013}} & \multicolumn{1}{c|}{$2.3331\pm0.1002$} & \multicolumn{1}{c|}{$1.2609\pm$\num{0.00032}} & $2.3700\pm0.2575$ \\
     \toprule
    \multicolumn{7}{c}{\rule{0pt}{2.5ex}\textbf{E1B}} \\ \hline
    \multicolumn{1}{c|}{0.00} & \multicolumn{1}{c|}{$1.2612\pm$\num{0.00021}} & \multicolumn{1}{c|}{$0.0402\pm0.1654$} & \multicolumn{1}{c|}{$1.2612\pm$\num{0.00013}} & \multicolumn{1}{c|}{$0.0586\pm0.1043$} & \multicolumn{1}{c|}{$1.2603\pm$\num{0.00039}} & $0.1772\pm0.3110$ \\
    \multicolumn{1}{c|}{0.01} & \multicolumn{1}{c|}{$1.2613\pm$\num{0.00022}} & \multicolumn{1}{c|}{$0.5961\pm0.1778$} & \multicolumn{1}{c|}{$1.2618\pm$\num{0.00014}} & \multicolumn{1}{c|}{$0.5985\pm0.1146$} & \multicolumn{1}{c|}{$1.2604\pm$\num{0.0004}} & $0.6025\pm0.3207$ \\
    \multicolumn{1}{c|}{0.02} & \multicolumn{1}{c|}{$1.2614\pm$\num{0.00025}} & \multicolumn{1}{c|}{$0.9635\pm0.1975$} & \multicolumn{1}{c|}{$1.2622\pm$\num{0.00016}} & \multicolumn{1}{c|}{$1.0444\pm0.1257$} & \multicolumn{1}{c|}{$1.2606\pm$\num{0.00042}} & $1.0661\pm0.3322$ \\
    \multicolumn{1}{c|}{0.03} & \multicolumn{1}{c|}{$1.2615\pm$\num{0.00025}} & \multicolumn{1}{c|}{$1.4760\pm0.1999$} & \multicolumn{1}{c|}{$1.2628\pm$\num{0.00018}} & \multicolumn{1}{c|}{$1.5567\pm0.1413$} & \multicolumn{1}{c|}{$1.2607\pm$\num{0.00043}} & $1.4764\pm0.3404$ \\
    \multicolumn{1}{c|}{0.04} & \multicolumn{1}{c|}{$1.2616\pm$\num{0.00025}} & \multicolumn{1}{c|}{$1.9754\pm0.1993$} & \multicolumn{1}{c|}{$1.2633\pm$\num{0.00018}} & \multicolumn{1}{c|}{$2.0825\pm0.1465$} & \multicolumn{1}{c|}{$1.2607\pm$\num{0.00051}} & $1.9490\pm0.4007$ \\
    \multicolumn{1}{c|}{0.05} & \multicolumn{1}{c|}{$1.2618\pm$\num{0.00028}} & \multicolumn{1}{c|}{$2.4480\pm0.2220$} & \multicolumn{1}{c|}{$1.2639\pm$\num{0.0002}} & \multicolumn{1}{c|}{$2.5560\pm0.1610$} & \multicolumn{1}{c|}{$1.2609\pm$\num{0.00047}} & $2.4195\pm0.3714$ \\ \toprule
    \multicolumn{7}{c}{\rule{0pt}{2.5ex}\textbf{E1C}} \\ \hline
    \multicolumn{1}{c|}{0.00} & \multicolumn{1}{c|}{$1.2611\pm$\num{4.5e-05}} & \multicolumn{1}{c|}{$0.0123\pm0.0353$} & \multicolumn{1}{c|}{$1.2612\pm$\num{2.6e-05}} & \multicolumn{1}{c|}{$0.0567\pm0.0203$} & \multicolumn{1}{c|}{$1.2603\pm$\num{0.00024}} & $0.1452\pm0.1876$ \\
    \multicolumn{1}{c|}{0.01} & \multicolumn{1}{c|}{$1.2612\pm$\num{6.7e-05}} & \multicolumn{1}{c|}{$0.5404\pm0.0529$} & \multicolumn{1}{c|}{$1.2618\pm$\num{4.6e-05}} & \multicolumn{1}{c|}{$0.5782\pm0.0366$} & \multicolumn{1}{c|}{$1.2604\pm$\num{0.00025}} & $0.6248\pm0.1988$ \\
    \multicolumn{1}{c|}{0.02} & \multicolumn{1}{c|}{$1.2614\pm$\num{8.7e-05}} & \multicolumn{1}{c|}{$0.9927\pm0.0691$} & \multicolumn{1}{c|}{$1.2623\pm$\num{6.2e-05}} & \multicolumn{1}{c|}{$1.1003\pm0.0491$} & \multicolumn{1}{c|}{$1.2606\pm$\num{0.00027}} & $1.0896\pm0.2129$ \\
    \multicolumn{1}{c|}{0.03} & \multicolumn{1}{c|}{$1.2615\pm$\num{0.0001}} & \multicolumn{1}{c|}{$1.4997\pm0.0814$} & \multicolumn{1}{c|}{$1.2628\pm$\num{7.7e-05}} & \multicolumn{1}{c|}{$1.5991\pm0.0611$} & \multicolumn{1}{c|}{$1.2608\pm$\num{0.00028}} & $1.4926\pm0.2203$ \\
    \multicolumn{1}{c|}{0.04} & \multicolumn{1}{c|}{$1.2616\pm$\num{0.00011}} & \multicolumn{1}{c|}{$1.9607\pm0.0899$} & \multicolumn{1}{c|}{$1.2634\pm$\num{9.1e-05}} & \multicolumn{1}{c|}{$2.1090\pm0.0719$} & \multicolumn{1}{c|}{$1.2607\pm$\num{0.00031}} & $1.9238\pm0.2475$ \\
    \multicolumn{1}{c|}{0.05} & \multicolumn{1}{c|}{$1.2617\pm$\num{0.00013}} & \multicolumn{1}{c|}{$2.3900\pm0.1069$} & \multicolumn{1}{c|}{$1.2639\pm$\num{0.00011}} & \multicolumn{1}{c|}{$2.6086\pm0.0840$} & \multicolumn{1}{c|}{$1.2603\pm$\num{0.00042}} & $1.9282\pm0.3326$ 
    \end{tabular}
    }
    \label{tab:IP_table_E1_110}
\end{sidewaystable*}

\begin{sidewaystable*}[]
\centering
    \caption{(1\;0\;0) inter-planar distance $d_{100}$ and lattice expansion $\Delta d_{100}/d_{100}$ as a function of dopant concentration for the substrate, bulk, and free-edge regions of the E1A, E1B, and E1C crystals}
    \resizebox{\textheight}{!}{%
    \begin{tabular}{ccccccc}
    \multicolumn{1}{c|}{\multirow{2}{*}{B Concentration, $x$}} & \multicolumn{2}{c|}{Substrate Region} & \multicolumn{2}{c|}{Bulk Region} & \multicolumn{2}{c}{Free Edge} \\
    \cline{2-7} \multicolumn{1}{c|}{} & \multicolumn{1}{c|}{\rule{0pt}{2.5ex}$d_{100}$ (\unit{\angstrom})} & \multicolumn{1}{c|}{$\Delta d_{100}/d_{100}$ (\SI{e-3}{\angstrom})} & \multicolumn{1}{c|}{\rule{0pt}{2.5ex}$d_{100}$ (\unit{\angstrom})} & \multicolumn{1}{c|}{$\Delta d_{100}/d_{100}$ (\SI{e-3}{\angstrom})} & \multicolumn{1}{c|}{\rule{0pt}{2.5ex}$d_{100}$ (\unit{\angstrom})} & $\Delta d_{100}/d_{100}$ (\SI{e-3}{\angstrom}) \\ \hline
    \multicolumn{7}{c}{\rule{0pt}{2.5ex}\textbf{E1A}} \\ \hline
    \multicolumn{1}{c|}{0.00} & \multicolumn{1}{c|}{$0.8918\pm$\num{3.9e-05}} & \multicolumn{1}{c|}{$0.0371\pm0.0439$} & \multicolumn{1}{c|}{$0.8918\pm$\num{4.4e-05}} & \multicolumn{1}{c|}{$0.0742\pm0.0491$} & \multicolumn{1}{c|}{$0.8919\pm$\num{4e-05}} & $0.1414\pm0.0448$ \\
    \multicolumn{1}{c|}{0.01} & \multicolumn{1}{c|}{$0.8922\pm$\num{4.9e-05}} & \multicolumn{1}{c|}{$0.5444\pm0.0553$} & \multicolumn{1}{c|}{$0.8922\pm$\num{5.4e-05}} & \multicolumn{1}{c|}{$0.5587\pm0.0604$} & \multicolumn{1}{c|}{$0.8923\pm$\num{4.8e-05}} & $0.6342\pm0.0540$ \\
    \multicolumn{1}{c|}{0.02} & \multicolumn{1}{c|}{$0.8926\pm$\num{5.3e-05}} & \multicolumn{1}{c|}{$0.9310\pm0.0591$} & \multicolumn{1}{c|}{$0.8927\pm$\num{6.3e-05}} & \multicolumn{1}{c|}{$1.0419\pm0.0707$} & \multicolumn{1}{c|}{$0.8927\pm$\num{5.1e-05}} & $1.1020\pm0.0570$ \\
    \multicolumn{1}{c|}{0.03} & \multicolumn{1}{c|}{$0.8931\pm$\num{5.8e-05}} & \multicolumn{1}{c|}{$1.5537\pm0.0645$} & \multicolumn{1}{c|}{$0.8931\pm$\num{7e-05}} & \multicolumn{1}{c|}{$1.5088\pm0.0785$} & \multicolumn{1}{c|}{$0.8932\pm$\num{6.2e-05}} & $1.6514\pm0.0693$ \\
    \multicolumn{1}{c|}{0.04} & \multicolumn{1}{c|}{$0.8935\pm$\num{6.4e-05}} & \multicolumn{1}{c|}{$1.9661\pm0.0722$} & \multicolumn{1}{c|}{$0.8935\pm$\num{7.7e-05}} & \multicolumn{1}{c|}{$2.0073\pm0.0858$} & \multicolumn{1}{c|}{$0.8936\pm$\num{6.5e-05}} & $2.1265\pm0.0724$ \\
    \multicolumn{1}{c|}{0.05} & \multicolumn{1}{c|}{$0.8940\pm$\num{6.5e-05}} & \multicolumn{1}{c|}{$2.4711\pm0.0730$} & \multicolumn{1}{c|}{$0.8940\pm$\num{8.8e-05}} & \multicolumn{1}{c|}{$2.5042\pm0.0989$} & \multicolumn{1}{c|}{$0.8938\pm$\num{0.00011}} & $2.2599\pm0.1194$ \\ \toprule
    \multicolumn{7}{c}{\rule{0pt}{2.5ex}\textbf{E1B}} \\ \hline
    \multicolumn{1}{c|}{0.00} & \multicolumn{1}{c|}{$0.8917\pm$\num{8.9e-05}} & \multicolumn{1}{c|}{$-0.0229\pm0.0996$} & \multicolumn{1}{c|}{$0.8918\pm$\num{0.00011}} & \multicolumn{1}{c|}{$0.0582\pm0.1260$} & \multicolumn{1}{c|}{$0.8919\pm$\num{0.0001}} & $0.1685\pm0.1127$ \\
    \multicolumn{1}{c|}{0.01} & \multicolumn{1}{c|}{$0.8923\pm$\num{0.0001}} & \multicolumn{1}{c|}{$0.6081\pm0.1121$} & \multicolumn{1}{c|}{$0.8923\pm$\num{0.00012}} & \multicolumn{1}{c|}{$0.5779\pm0.1325$} & \multicolumn{1}{c|}{$0.8923\pm$\num{0.00011}} & $0.6549\pm0.1202$ \\
    \multicolumn{1}{c|}{0.02} & \multicolumn{1}{c|}{$0.8926\pm$\num{0.00011}} & \multicolumn{1}{c|}{$0.9425\pm0.1221$} & \multicolumn{1}{c|}{$0.8927\pm$\num{0.00013}} & \multicolumn{1}{c|}{$1.0327\pm0.1417$} & \multicolumn{1}{c|}{$0.8927\pm$\num{0.00011}} & $1.0162\pm0.1249$ \\
    \multicolumn{1}{c|}{0.03} & \multicolumn{1}{c|}{$0.8930\pm$\num{0.00012}} & \multicolumn{1}{c|}{$1.4504\pm0.1312$} & \multicolumn{1}{c|}{$0.8931\pm$\num{0.00014}} & \multicolumn{1}{c|}{$1.5281\pm0.1593$} & \multicolumn{1}{c|}{$0.8928\pm$\num{0.00016}} & $1.2177\pm0.1820$ \\
    \multicolumn{1}{c|}{0.04} & \multicolumn{1}{c|}{$0.8935\pm$\num{0.00012}} & \multicolumn{1}{c|}{$2.0014\pm0.1298$} & \multicolumn{1}{c|}{$0.8936\pm$\num{0.00014}} & \multicolumn{1}{c|}{$2.0794\pm0.1616$} & \multicolumn{1}{c|}{$0.8933\pm$\num{0.00017}} & $1.6994\pm0.1867$ \\
    \multicolumn{1}{c|}{0.05} & \multicolumn{1}{c|}{$0.8939\pm$\num{0.00012}} & \multicolumn{1}{c|}{$2.4147\pm0.1307$} & \multicolumn{1}{c|}{$0.8940\pm$\num{0.00019}} & \multicolumn{1}{c|}{$2.4896\pm0.2088$} & \multicolumn{1}{c|}{$0.8939\pm$\num{0.00013}} & $2.4132\pm0.1410$ \\ \toprule
    \multicolumn{7}{c}{\rule{0pt}{2.5ex}\textbf{E1C}} \\ \hline
    \multicolumn{1}{c|}{0.00} & \multicolumn{1}{c|}{$0.8918\pm$\num{2e-05}} & \multicolumn{1}{c|}{$0.0413\pm0.0229$} & \multicolumn{1}{c|}{$0.8918\pm$\num{2.2e-05}} & \multicolumn{1}{c|}{$0.0612\pm0.0247$} & \multicolumn{1}{c|}{$0.8919\pm$\num{2.3e-05}} & $0.1469\pm0.0259$ \\
    \multicolumn{1}{c|}{0.01} & \multicolumn{1}{c|}{$0.8922\pm$\num{3.1e-05}} & \multicolumn{1}{c|}{$0.5257\pm0.0345$} & \multicolumn{1}{c|}{$0.8922\pm$\num{3.8e-05}} & \multicolumn{1}{c|}{$0.5570\pm0.0430$} & \multicolumn{1}{c|}{$0.8923\pm$\num{3.3e-05}} & $0.5834\pm0.0375$ \\
    \multicolumn{1}{c|}{0.02} & \multicolumn{1}{c|}{$0.8927\pm$\num{3.7e-05}} & \multicolumn{1}{c|}{$1.0465\pm0.0419$} & \multicolumn{1}{c|}{$0.8927\pm$\num{5e-05}} & \multicolumn{1}{c|}{$1.0489\pm0.0561$} & \multicolumn{1}{c|}{$0.8927\pm$\num{4.1e-05}} & $1.1071\pm0.0461$ \\
    \multicolumn{1}{c|}{0.03} & \multicolumn{1}{c|}{$0.8931\pm$\num{4.4e-05}} & \multicolumn{1}{c|}{$1.5121\pm0.0494$} & \multicolumn{1}{c|}{$0.8931\pm$\num{5.8e-05}} & \multicolumn{1}{c|}{$1.5257\pm0.0656$} & \multicolumn{1}{c|}{$0.8932\pm$\num{5e-05}} & $1.6158\pm0.0558$ \\
    \multicolumn{1}{c|}{0.04} & \multicolumn{1}{c|}{$0.8935\pm$\num{5.1e-05}} & \multicolumn{1}{c|}{$1.9647\pm0.0572$} & \multicolumn{1}{c|}{$0.8935\pm$\num{6.6e-05}} & \multicolumn{1}{c|}{$2.0080\pm0.0746$} & \multicolumn{1}{c|}{$0.8936\pm$\num{5.2e-05}} & $2.1203\pm0.0578$ \\
    \multicolumn{1}{c|}{0.05} & \multicolumn{1}{c|}{$0.8939\pm$\num{5.5e-05}} & \multicolumn{1}{c|}{$2.3632\pm0.0618$} & \multicolumn{1}{c|}{$0.8940\pm$\num{7.9e-05}} & \multicolumn{1}{c|}{$2.5039\pm0.0886$} & \multicolumn{1}{c|}{$0.8936\pm$\num{0.00017}} & $2.0445\pm0.1882$
    \end{tabular}
    }
    \label{tab:IP_table_E1_100}
\end{sidewaystable*}

\subsection{E2 Crystals}
\begin{figure}[H]
 \subfloat[]{\includegraphics[width=0.48\textwidth]{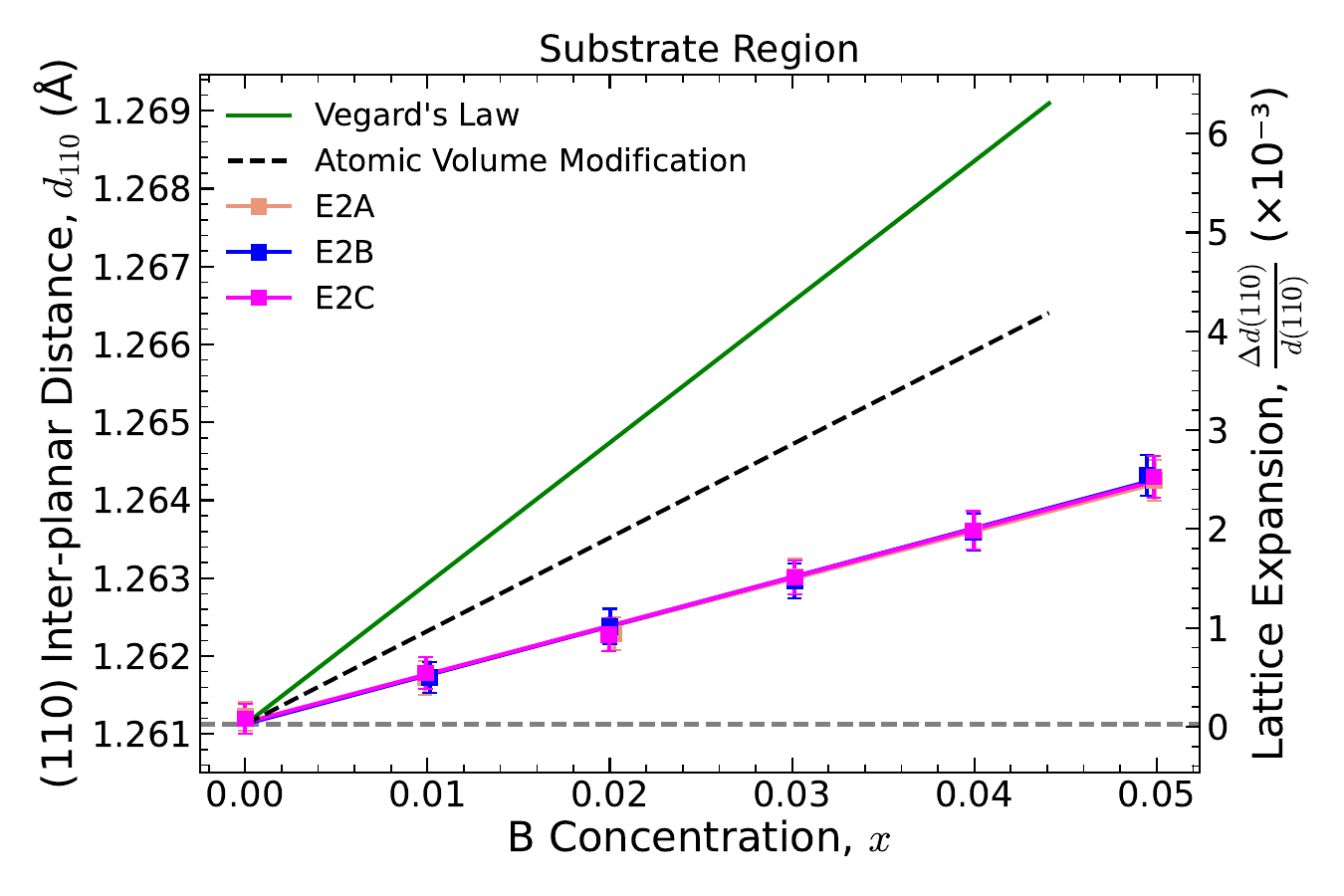}}\hfill
 \setcounter{subfigure}{3}
 \subfloat[]{\includegraphics[width=0.48\textwidth]{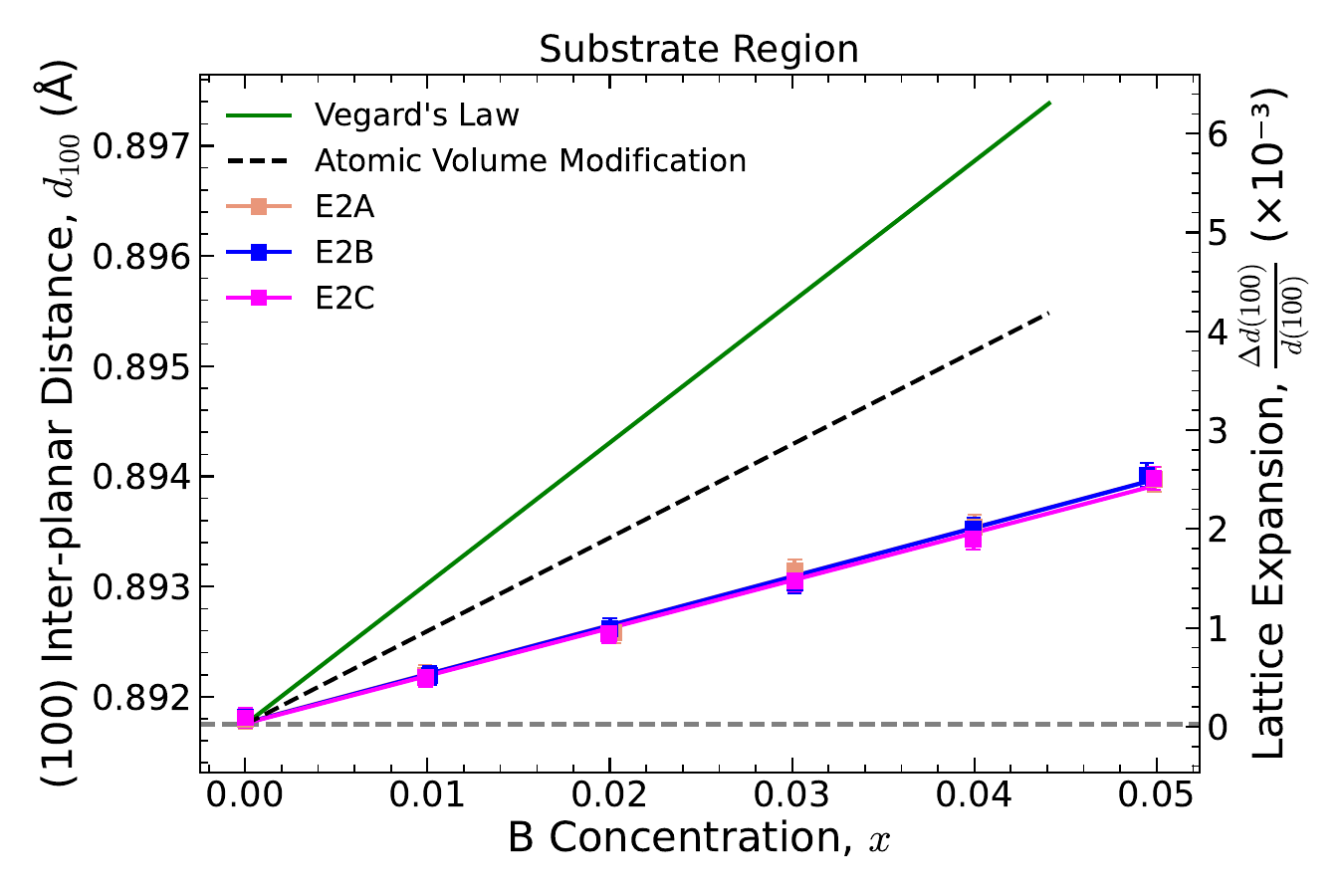}}\\[-2ex]
 \setcounter{subfigure}{1}
 \subfloat[]{\includegraphics[width=0.48\textwidth]{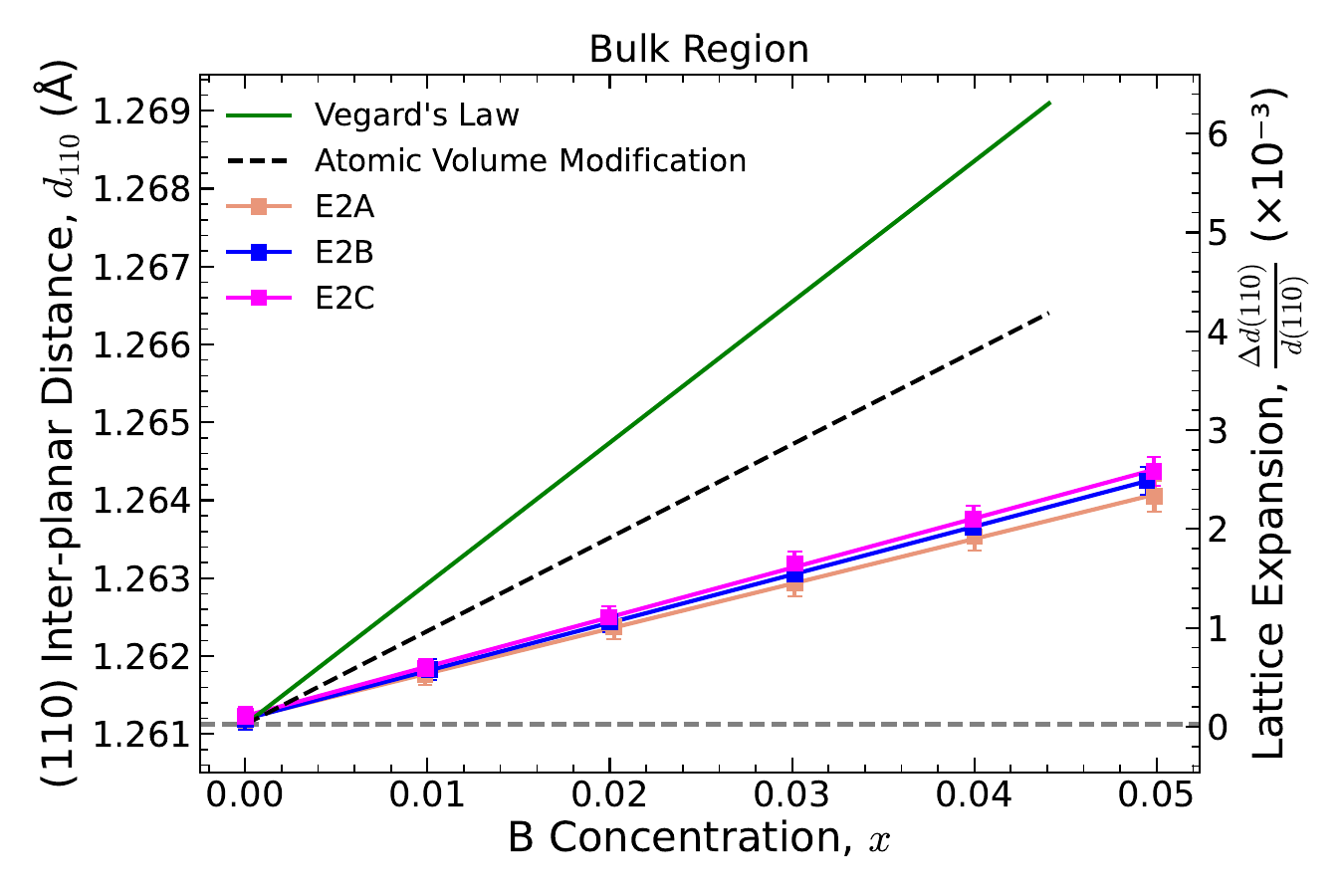}}\hfill
 \setcounter{subfigure}{4}
 \subfloat[]{\includegraphics[width=0.48\textwidth]{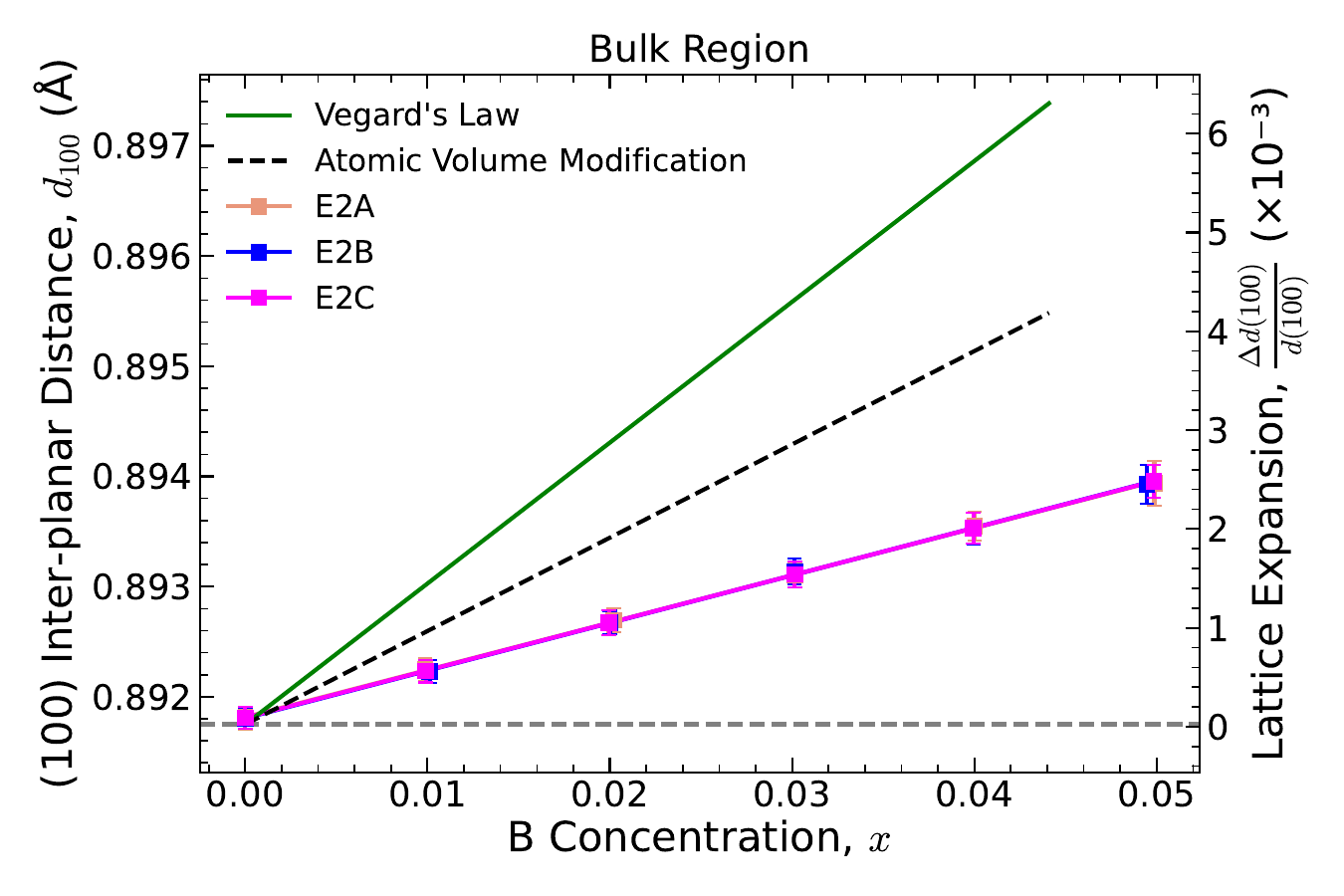}}\\[-2ex]
 \setcounter{subfigure}{2}
 \subfloat[]{\includegraphics[width=0.48\textwidth]{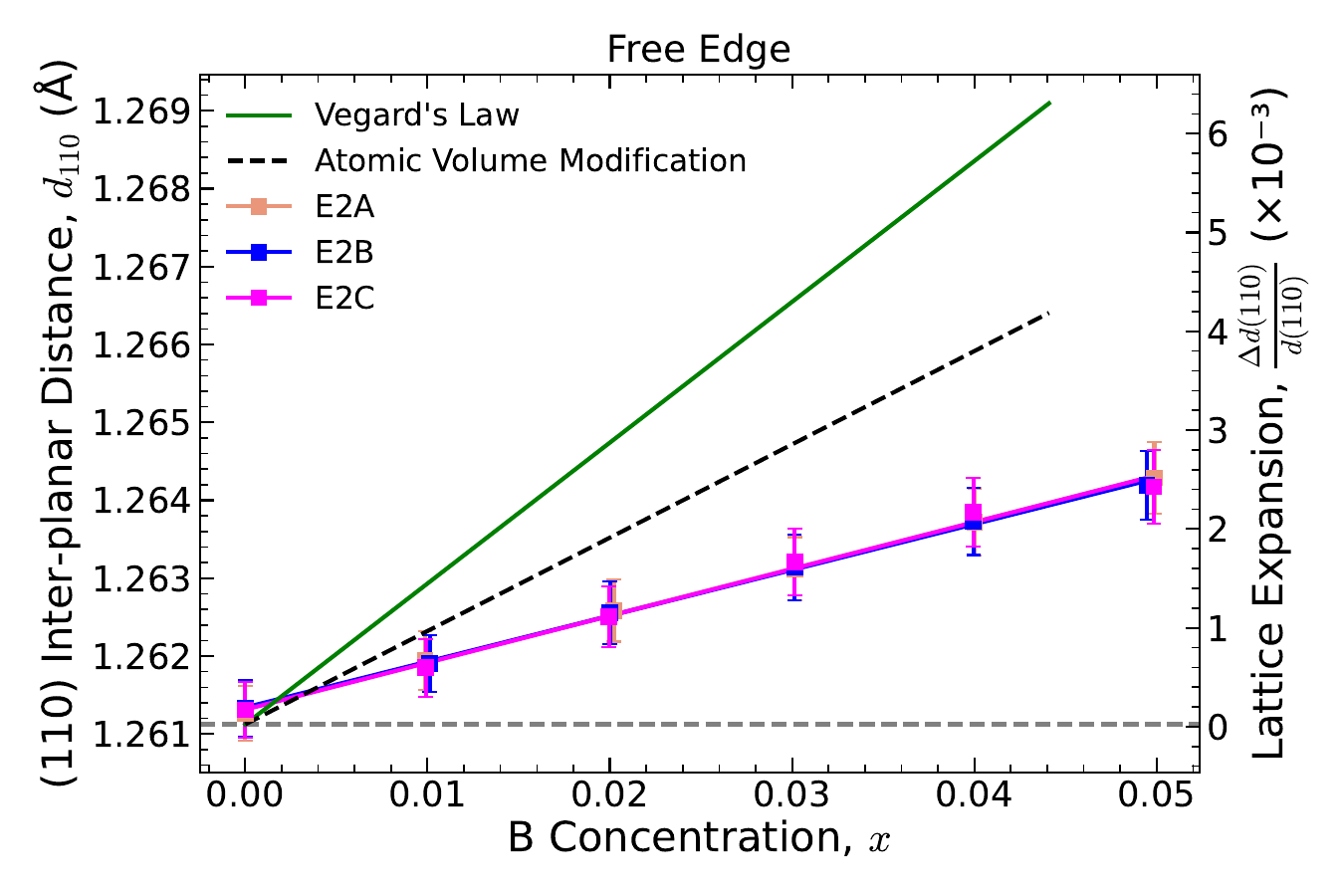}}\hfill
 \setcounter{subfigure}{5}
 \subfloat[]{\includegraphics[width=0.48\textwidth]{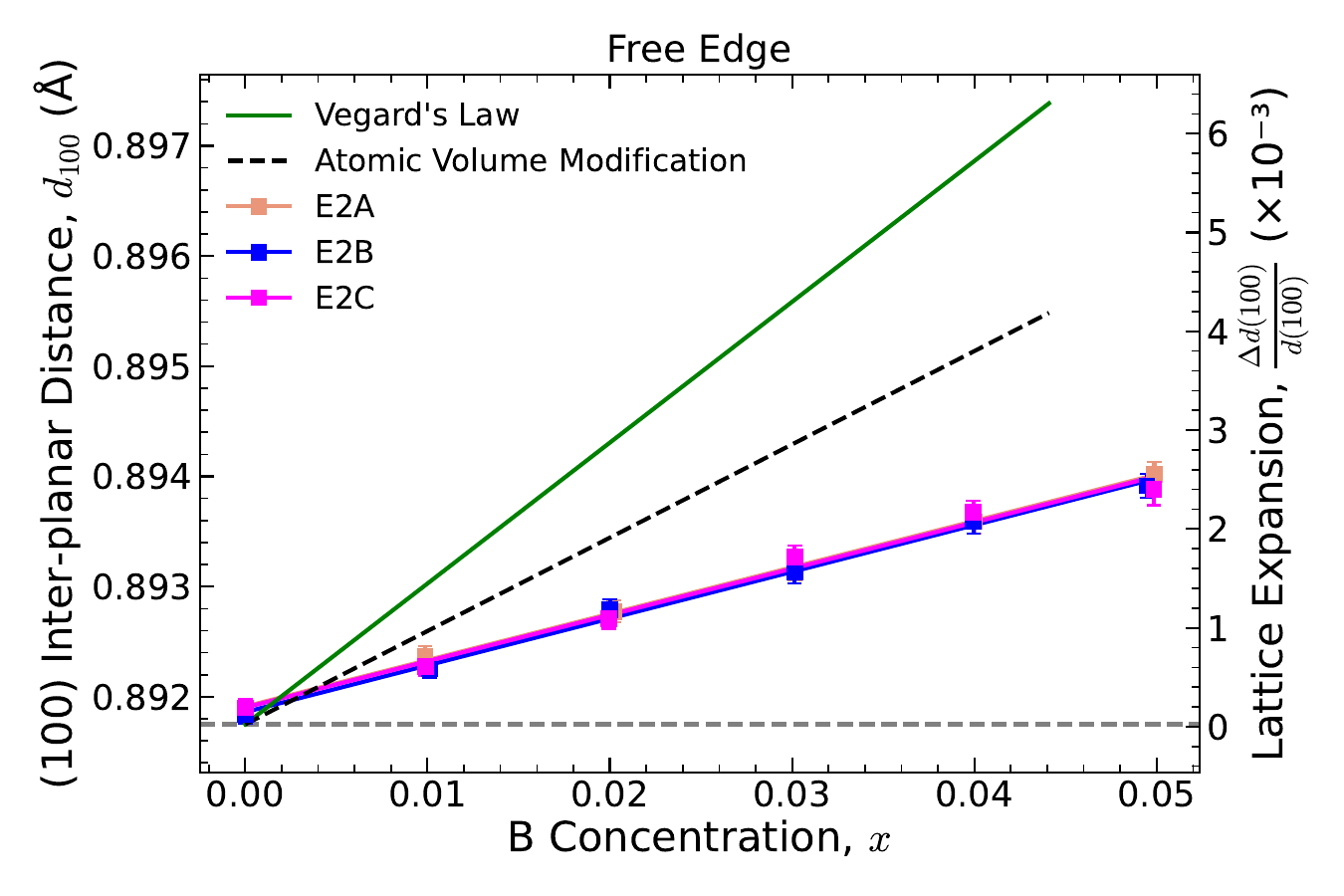}}\\[-2ex]
 \caption{Region-by-region inter-planar distances of E2-category crystals as a function of boron concentration, with left/right panels for (1\;1\;0)/(1\;0\;0) planes and top/middle/bottom rows for substrate, bulk, and free-edge regions.}
 \label{fig:IP_Plots_E2}
\end{figure}

\begin{sidewaystable*}[]
\centering
    \caption{(1\;1\;0) inter-planar distance $d_{110}$ and lattice expansion $\Delta d_{110}/d_{110}$ as a function of dopant concentration for the substrate, bulk, and free-edge regions of the E2A, E2B, and E2C crystals.}
    \resizebox{\textheight}{!}{%
    \begin{tabular}{ccccccc}
    \multicolumn{1}{c|}{\multirow{2}{*}{B Concentration, $x$}} & \multicolumn{2}{c|}{Substrate Region} & \multicolumn{2}{c|}{Bulk Region} & \multicolumn{2}{c}{Free Edge} \\
    \cline{2-7} \multicolumn{1}{c|}{} & \multicolumn{1}{c|}{\rule{0pt}{2.5ex}$d_{110}$ (\unit{\angstrom})} & \multicolumn{1}{c|}{$\Delta d_{110}/d_{110}$ (\SI{e-3}{\angstrom})} & \multicolumn{1}{c|}{\rule{0pt}{2.5ex}$d_{110}$ (\unit{\angstrom})} & \multicolumn{1}{c|}{$\Delta d_{110}/d_{110}$ (\SI{e-3}{\angstrom})} & \multicolumn{1}{c|}{\rule{0pt}{2.5ex}$d_{110}$ (\unit{\angstrom})} & $\Delta d_{110}/d_{110}$ (\SI{e-3}{\angstrom}) \\ \hline
    \multicolumn{7}{c}{\rule{0pt}{2.5ex}\textbf{E2A}} \\ \hline
    \multicolumn{1}{c|}{0.00} & \multicolumn{1}{c|}{$1.2612\pm$\num{0.00018}} & \multicolumn{1}{c|}{$0.0822\pm0.1466$} & \multicolumn{1}{c|}{$1.2612\pm$\num{0.00013}} & \multicolumn{1}{c|}{$0.0556\pm0.1002$} & \multicolumn{1}{c|}{$1.2604\pm$\num{0.00035}} & $0.1127\pm0.2810$ \\
    \multicolumn{1}{c|}{0.01} & \multicolumn{1}{c|}{$1.2612\pm$\num{0.00021}} & \multicolumn{1}{c|}{$0.4727\pm0.1692$} & \multicolumn{1}{c|}{$1.2616\pm$\num{0.00014}} & \multicolumn{1}{c|}{$0.5094\pm0.1113$} & \multicolumn{1}{c|}{$1.2606\pm$\num{0.00038}} & $0.6503\pm0.2983$ \\
    \multicolumn{1}{c|}{0.02} & \multicolumn{1}{c|}{$1.2613\pm$\num{0.00021}} & \multicolumn{1}{c|}{$0.9241\pm0.1695$} & \multicolumn{1}{c|}{$1.2622\pm$\num{0.00016}} & \multicolumn{1}{c|}{$0.9925\pm0.1238$} & \multicolumn{1}{c|}{$1.2607\pm$\num{0.0004}} & $1.1576\pm0.3166$ \\
    \multicolumn{1}{c|}{0.03} & \multicolumn{1}{c|}{$1.2615\pm$\num{0.00023}} & \multicolumn{1}{c|}{$1.5064\pm0.1832$} & \multicolumn{1}{c|}{$1.2626\pm$\num{0.00017}} & \multicolumn{1}{c|}{$1.4343\pm0.1356$} & \multicolumn{1}{c|}{$1.2608\pm$\num{0.00041}} & $1.5837\pm0.3218$ \\
    \multicolumn{1}{c|}{0.04} & \multicolumn{1}{c|}{$1.2615\pm$\num{0.00023}} & \multicolumn{1}{c|}{$1.9545\pm0.1861$} & \multicolumn{1}{c|}{$1.2631\pm$\num{0.00018}} & \multicolumn{1}{c|}{$1.9092\pm0.1441$} & \multicolumn{1}{c|}{$1.2609\pm$\num{0.00043}} & $2.0597\pm0.3421$ \\
    \multicolumn{1}{c|}{0.05} & \multicolumn{1}{c|}{$1.2617\pm$\num{0.00026}} & \multicolumn{1}{c|}{$2.4820\pm0.2041$} & \multicolumn{1}{c|}{$1.2635\pm$\num{0.0002}} & \multicolumn{1}{c|}{$2.3203\pm0.1594$} & \multicolumn{1}{c|}{$1.2611\pm$\num{0.00046}} & $2.5071\pm0.3659$ \\ \toprule
    \multicolumn{7}{c}{\rule{0pt}{2.5ex}\textbf{E2B}} \\ \hline
    \multicolumn{1}{c|}{0.00} & \multicolumn{1}{c|}{$1.2611\pm$\num{0.0002}} & \multicolumn{1}{c|}{$0.0559\pm0.1553$} & \multicolumn{1}{c|}{$1.2612\pm$\num{0.00012}} & \multicolumn{1}{c|}{$0.0446\pm0.0965$} & \multicolumn{1}{c|}{$1.2604\pm$\num{0.00037}} & $0.1609\pm0.2914$ \\
    \multicolumn{1}{c|}{0.01} & \multicolumn{1}{c|}{$1.2612\pm$\num{0.0002}} & \multicolumn{1}{c|}{$0.4746\pm0.1576$} & \multicolumn{1}{c|}{$1.2617\pm$\num{0.00013}} & \multicolumn{1}{c|}{$0.5578\pm0.1029$} & \multicolumn{1}{c|}{$1.2605\pm$\num{0.00036}} & $0.6154\pm0.2893$ \\
    \multicolumn{1}{c|}{0.02} & \multicolumn{1}{c|}{$1.2614\pm$\num{0.00023}} & \multicolumn{1}{c|}{$0.9948\pm0.1811$} & \multicolumn{1}{c|}{$1.2622\pm$\num{0.00014}} & \multicolumn{1}{c|}{$1.0465\pm0.1128$} & \multicolumn{1}{c|}{$1.2607\pm$\num{0.0004}} & $1.1354\pm0.3165$ \\
    \multicolumn{1}{c|}{0.03} & \multicolumn{1}{c|}{$1.2614\pm$\num{0.00023}} & \multicolumn{1}{c|}{$1.4605\pm0.1786$} & \multicolumn{1}{c|}{$1.2627\pm$\num{0.00016}} & \multicolumn{1}{c|}{$1.5270\pm0.1248$} & \multicolumn{1}{c|}{$1.2608\pm$\num{0.00042}} & $1.6000\pm0.3343$ \\
    \multicolumn{1}{c|}{0.04} & \multicolumn{1}{c|}{$1.2615\pm$\num{0.00024}} & \multicolumn{1}{c|}{$1.9560\pm0.1893$} & \multicolumn{1}{c|}{$1.2632\pm$\num{0.00017}} & \multicolumn{1}{c|}{$2.0036\pm0.1348$} & \multicolumn{1}{c|}{$1.2609\pm$\num{0.00043}} & $2.0639\pm0.3408$ \\
    \multicolumn{1}{c|}{0.05} & \multicolumn{1}{c|}{$1.2617\pm$\num{0.00026}} & \multicolumn{1}{c|}{$2.5315\pm0.2099$} & \multicolumn{1}{c|}{$1.2637\pm$\num{0.00018}} & \multicolumn{1}{c|}{$2.4761\pm0.1436$} & \multicolumn{1}{c|}{$1.2610\pm$\num{0.00044}} & $2.4346\pm0.3475$ \\ \toprule
    \multicolumn{7}{c}{\rule{0pt}{2.5ex}\textbf{E2C}} \\ \hline
    \multicolumn{1}{c|}{0.00} & \multicolumn{1}{c|}{$1.2611\pm$\num{0.00019}} & \multicolumn{1}{c|}{$0.0566\pm0.1527$} & \multicolumn{1}{c|}{$1.2612\pm$\num{0.00011}} & \multicolumn{1}{c|}{$0.0821\pm0.0910$} & \multicolumn{1}{c|}{$1.2604\pm$\num{0.00036}} & $0.1468\pm0.2825$ \\
    \multicolumn{1}{c|}{0.01} & \multicolumn{1}{c|}{$1.2613\pm$\num{0.00021}} & \multicolumn{1}{c|}{$0.5234\pm0.1633$} & \multicolumn{1}{c|}{$1.2617\pm$\num{0.00012}} & \multicolumn{1}{c|}{$0.5685\pm0.0976$} & \multicolumn{1}{c|}{$1.2605\pm$\num{0.00037}} & $0.5750\pm0.2951$ \\
    \multicolumn{1}{c|}{0.02} & \multicolumn{1}{c|}{$1.2613\pm$\num{0.00021}} & \multicolumn{1}{c|}{$0.9153\pm0.1671$} & \multicolumn{1}{c|}{$1.2623\pm$\num{0.00014}} & \multicolumn{1}{c|}{$1.0895\pm0.1103$} & \multicolumn{1}{c|}{$1.2606\pm$\num{0.00039}} & $1.0936\pm0.3079$ \\
    \multicolumn{1}{c|}{0.03} & \multicolumn{1}{c|}{$1.2615\pm$\num{0.00022}} & \multicolumn{1}{c|}{$1.4960\pm0.1752$} & \multicolumn{1}{c|}{$1.2628\pm$\num{0.00015}} & \multicolumn{1}{c|}{$1.6356\pm0.1186$} & \multicolumn{1}{c|}{$1.2608\pm$\num{0.00043}} & $1.6506\pm0.3398$ \\
    \multicolumn{1}{c|}{0.04} & \multicolumn{1}{c|}{$1.2616\pm$\num{0.00025}} & \multicolumn{1}{c|}{$1.9725\pm0.1964$} & \multicolumn{1}{c|}{$1.2633\pm$\num{0.00017}} & \multicolumn{1}{c|}{$2.0899\pm0.1310$} & \multicolumn{1}{c|}{$1.2610\pm$\num{0.00044}} & $2.1586\pm0.3492$ \\
    \multicolumn{1}{c|}{0.05} & \multicolumn{1}{c|}{$1.2617\pm$\num{0.00027}} & \multicolumn{1}{c|}{$2.5177\pm0.2163$} & \multicolumn{1}{c|}{$1.2638\pm$\num{0.00018}} & \multicolumn{1}{c|}{$2.5738\pm0.1428$} & \multicolumn{1}{c|}{$1.2610\pm$\num{0.00047}} & $2.4183\pm0.3717$
    \end{tabular}
    }
    \label{tab:IP_table_E2_110}
\end{sidewaystable*}

\begin{sidewaystable*}[]
\centering
    \caption{(1\;0\;0) inter-planar distance $d_{100}$ and lattice expansion $\Delta d_{100}/d_{100}$ as a function of dopant concentration for the substrate, bulk, and free-edge regions of the E2A, E2B, and E2C crystals.}
    \resizebox{\textheight}{!}{%
    \begin{tabular}{ccccccc}
    \multicolumn{1}{c|}{\multirow{2}{*}{B Concentration, $x$}} & \multicolumn{2}{c|}{Substrate Region} & \multicolumn{2}{c|}{Bulk Region} & \multicolumn{2}{c}{Free Edge} \\
    \cline{2-7} \multicolumn{1}{c|}{} & \multicolumn{1}{c|}{\rule{0pt}{2.5ex}$d_{100}$ (\unit{\angstrom})} & \multicolumn{1}{c|}{$\Delta d_{100}/d_{100}$ (\SI{e-3}{\angstrom})} & \multicolumn{1}{c|}{\rule{0pt}{2.5ex}$d_{100}$ (\unit{\angstrom})} & \multicolumn{1}{c|}{$\Delta d_{100}/d_{100}$ (\SI{e-3}{\angstrom})} & \multicolumn{1}{c|}{\rule{0pt}{2.5ex}$d_{100}$ (\unit{\angstrom})} & $\Delta d_{100}/d_{100}$ (\SI{e-3}{\angstrom}) \\ \hline
    \multicolumn{7}{c}{\rule{0pt}{2.5ex}\textbf{E2A}} \\ \hline
    \multicolumn{1}{c|}{0.00} & \multicolumn{1}{c|}{$0.8918\pm$\num{8.5e-05}} & \multicolumn{1}{c|}{$0.0510\pm0.0955$} & \multicolumn{1}{c|}{$0.8918\pm$\num{9.5e-05}} & \multicolumn{1}{c|}{$0.0555\pm0.1060$} & \multicolumn{1}{c|}{$0.8919\pm$\num{8.6e-05}} & $0.1494\pm0.0963$ \\
    \multicolumn{1}{c|}{0.01} & \multicolumn{1}{c|}{$0.8922\pm$\num{8.9e-05}} & \multicolumn{1}{c|}{$0.4994\pm0.1003$} & \multicolumn{1}{c|}{$0.8922\pm$\num{0.0001}} & \multicolumn{1}{c|}{$0.5551\pm0.1156$} & \multicolumn{1}{c|}{$0.8924\pm$\num{9.4e-05}} & $0.6915\pm0.1056$ \\
    \multicolumn{1}{c|}{0.02} & \multicolumn{1}{c|}{$0.8926\pm$\num{9.6e-05}} & \multicolumn{1}{c|}{$0.9335\pm0.1078$} & \multicolumn{1}{c|}{$0.8927\pm$\num{0.00011}} & \multicolumn{1}{c|}{$1.0606\pm0.1240$} & \multicolumn{1}{c|}{$0.8928\pm$\num{0.0001}} & $1.1484\pm0.1116$ \\
    \multicolumn{1}{c|}{0.03} & \multicolumn{1}{c|}{$0.8931\pm$\num{0.0001}} & \multicolumn{1}{c|}{$1.5648\pm0.1157$} & \multicolumn{1}{c|}{$0.8931\pm$\num{0.00012}} & \multicolumn{1}{c|}{$1.5202\pm0.1293$} & \multicolumn{1}{c|}{$0.8932\pm$\num{0.0001}} & $1.5813\pm0.1167$ \\
    \multicolumn{1}{c|}{0.04} & \multicolumn{1}{c|}{$0.8935\pm$\num{0.00011}} & \multicolumn{1}{c|}{$2.0067\pm0.1225$} & \multicolumn{1}{c|}{$0.8935\pm$\num{0.00013}} & \multicolumn{1}{c|}{$2.0157\pm0.1476$} & \multicolumn{1}{c|}{$0.8936\pm$\num{0.00011}} & $2.0618\pm0.1191$ \\
    \multicolumn{1}{c|}{0.05} & \multicolumn{1}{c|}{$0.8940\pm$\num{0.00011}} & \multicolumn{1}{c|}{$2.4907\pm0.1275$} & \multicolumn{1}{c|}{$0.8939\pm$\num{0.0002}} & \multicolumn{1}{c|}{$2.4502\pm0.2294$} & \multicolumn{1}{c|}{$0.8940\pm$\num{0.00011}} & $2.5434\pm0.1287$ \\ \toprule
    \multicolumn{7}{c}{\rule{0pt}{2.5ex}\textbf{E2B}} \\ \hline
    \multicolumn{1}{c|}{0.00} & \multicolumn{1}{c|}{$0.8918\pm$\num{8.1e-05}} & \multicolumn{1}{c|}{$0.0773\pm0.0903$} & \multicolumn{1}{c|}{$0.8918\pm$\num{9.6e-05}} & \multicolumn{1}{c|}{$0.0593\pm0.1078$} & \multicolumn{1}{c|}{$0.8918\pm$\num{8.6e-05}} & $0.1030\pm0.0970$ \\
    \multicolumn{1}{c|}{0.01} & \multicolumn{1}{c|}{$0.8922\pm$\num{8.7e-05}} & \multicolumn{1}{c|}{$0.4946\pm0.0977$} & \multicolumn{1}{c|}{$0.8922\pm$\num{0.0001}} & \multicolumn{1}{c|}{$0.5394\pm0.1147$} & \multicolumn{1}{c|}{$0.8923\pm$\num{8.4e-05}} & $0.5618\pm0.0947$ \\
    \multicolumn{1}{c|}{0.02} & \multicolumn{1}{c|}{$0.8926\pm$\num{9e-05}} & \multicolumn{1}{c|}{$0.9750\pm0.1008$} & \multicolumn{1}{c|}{$0.8927\pm$\num{0.00011}} & \multicolumn{1}{c|}{$1.0346\pm0.1210$} & \multicolumn{1}{c|}{$0.8928\pm$\num{9.4e-05}} & $1.1654\pm0.1056$ \\
    \multicolumn{1}{c|}{0.03} & \multicolumn{1}{c|}{$0.8930\pm$\num{9.4e-05}} & \multicolumn{1}{c|}{$1.4380\pm0.1056$} & \multicolumn{1}{c|}{$0.8931\pm$\num{0.00012}} & \multicolumn{1}{c|}{$1.5526\pm0.1312$} & \multicolumn{1}{c|}{$0.8931\pm$\num{0.0001}} & $1.5479\pm0.1134$ \\
    \multicolumn{1}{c|}{0.04} & \multicolumn{1}{c|}{$0.8935\pm$\num{0.00011}} & \multicolumn{1}{c|}{$1.9887\pm0.1178$} & \multicolumn{1}{c|}{$0.8935\pm$\num{0.00015}} & \multicolumn{1}{c|}{$1.9921\pm0.1632$} & \multicolumn{1}{c|}{$0.8936\pm$\num{0.00011}} & $2.0627\pm0.1214$ \\
    \multicolumn{1}{c|}{0.05} & \multicolumn{1}{c|}{$0.8940\pm$\num{0.00011}} & \multicolumn{1}{c|}{$2.5362\pm0.1194$} & \multicolumn{1}{c|}{$0.8939\pm$\num{0.00018}} & \multicolumn{1}{c|}{$2.4451\pm0.1990$} & \multicolumn{1}{c|}{$0.8939\pm$\num{0.00011}} & $2.4260\pm0.1187$ \\ \toprule
    \multicolumn{7}{c}{\rule{0pt}{2.5ex}\textbf{E2C}} \\ \hline
    \multicolumn{1}{c|}{0.00} & \multicolumn{1}{c|}{$0.8918\pm$\num{8.6e-05}} & \multicolumn{1}{c|}{$0.0664\pm0.0961$} & \multicolumn{1}{c|}{$0.8918\pm$\num{9.7e-05}} & \multicolumn{1}{c|}{$0.0675\pm0.1084$} & \multicolumn{1}{c|}{$0.8919\pm$\num{8.2e-05}} & $0.1680\pm0.0923$ \\
    \multicolumn{1}{c|}{0.01} & \multicolumn{1}{c|}{$0.8922\pm$\num{8.8e-05}} & \multicolumn{1}{c|}{$0.4767\pm0.0987$} & \multicolumn{1}{c|}{$0.8922\pm$\num{0.0001}} & \multicolumn{1}{c|}{$0.5412\pm0.1153$} & \multicolumn{1}{c|}{$0.8923\pm$\num{8.7e-05}} & $0.5874\pm0.0974$ \\
    \multicolumn{1}{c|}{0.02} & \multicolumn{1}{c|}{$0.8926\pm$\num{9.3e-05}} & \multicolumn{1}{c|}{$0.9257\pm0.1040$} & \multicolumn{1}{c|}{$0.8927\pm$\num{0.00011}} & \multicolumn{1}{c|}{$1.0333\pm0.1231$} & \multicolumn{1}{c|}{$0.8927\pm$\num{8.9e-05}} & $1.0728\pm0.0998$ \\
    \multicolumn{1}{c|}{0.03} & \multicolumn{1}{c|}{$0.8931\pm$\num{9.9e-05}} & \multicolumn{1}{c|}{$1.4601\pm0.1109$} & \multicolumn{1}{c|}{$0.8931\pm$\num{0.00012}} & \multicolumn{1}{c|}{$1.5242\pm0.1297$} & \multicolumn{1}{c|}{$0.8933\pm$\num{0.0001}} & $1.7050\pm0.1134$ \\
    \multicolumn{1}{c|}{0.04} & \multicolumn{1}{c|}{$0.8934\pm$\num{9.4e-05}} & \multicolumn{1}{c|}{$1.8875\pm0.1060$} & \multicolumn{1}{c|}{$0.8935\pm$\num{0.00014}} & \multicolumn{1}{c|}{$1.9949\pm0.1585$} & \multicolumn{1}{c|}{$0.8937\pm$\num{0.0001}} & $2.1577\pm0.1162$ \\
    \multicolumn{1}{c|}{0.05} & \multicolumn{1}{c|}{$0.8940\pm$\num{0.0001}} & \multicolumn{1}{c|}{$2.5042\pm0.1161$} & \multicolumn{1}{c|}{$0.8940\pm$\num{0.00015}} & \multicolumn{1}{c|}{$2.4749\pm0.1682$} & \multicolumn{1}{c|}{$0.8939\pm$\num{0.00014}} & $2.3906\pm0.1610$  
    \end{tabular}
    }
    \label{tab:IP_table_E2_100}
\end{sidewaystable*}

\subsection{E3 Crystals}
\begin{figure}[H]
 \subfloat[]{\includegraphics[width=0.48\textwidth]{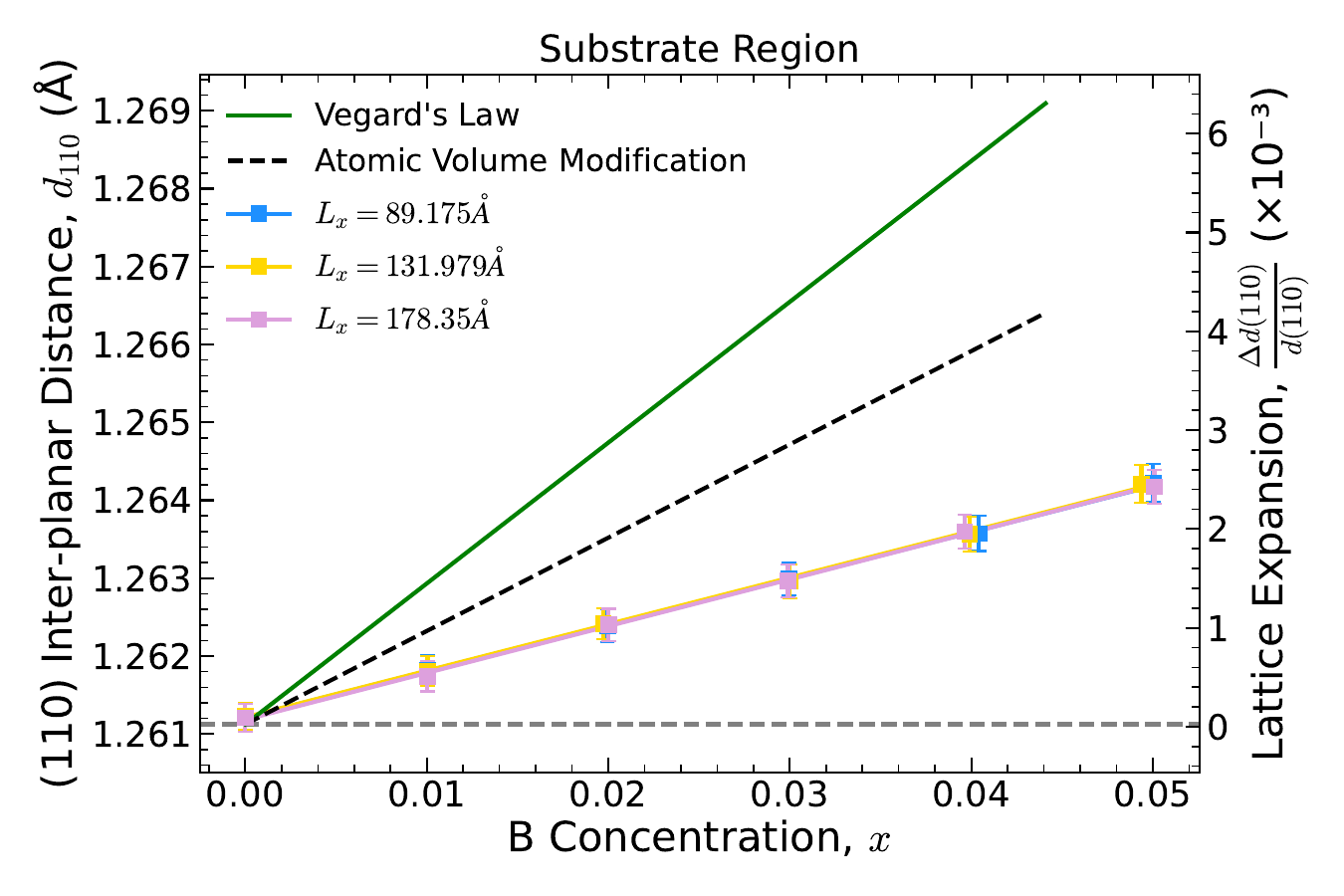}}\hfill
 \setcounter{subfigure}{3}
 \subfloat[]{\includegraphics[width=0.48\textwidth]{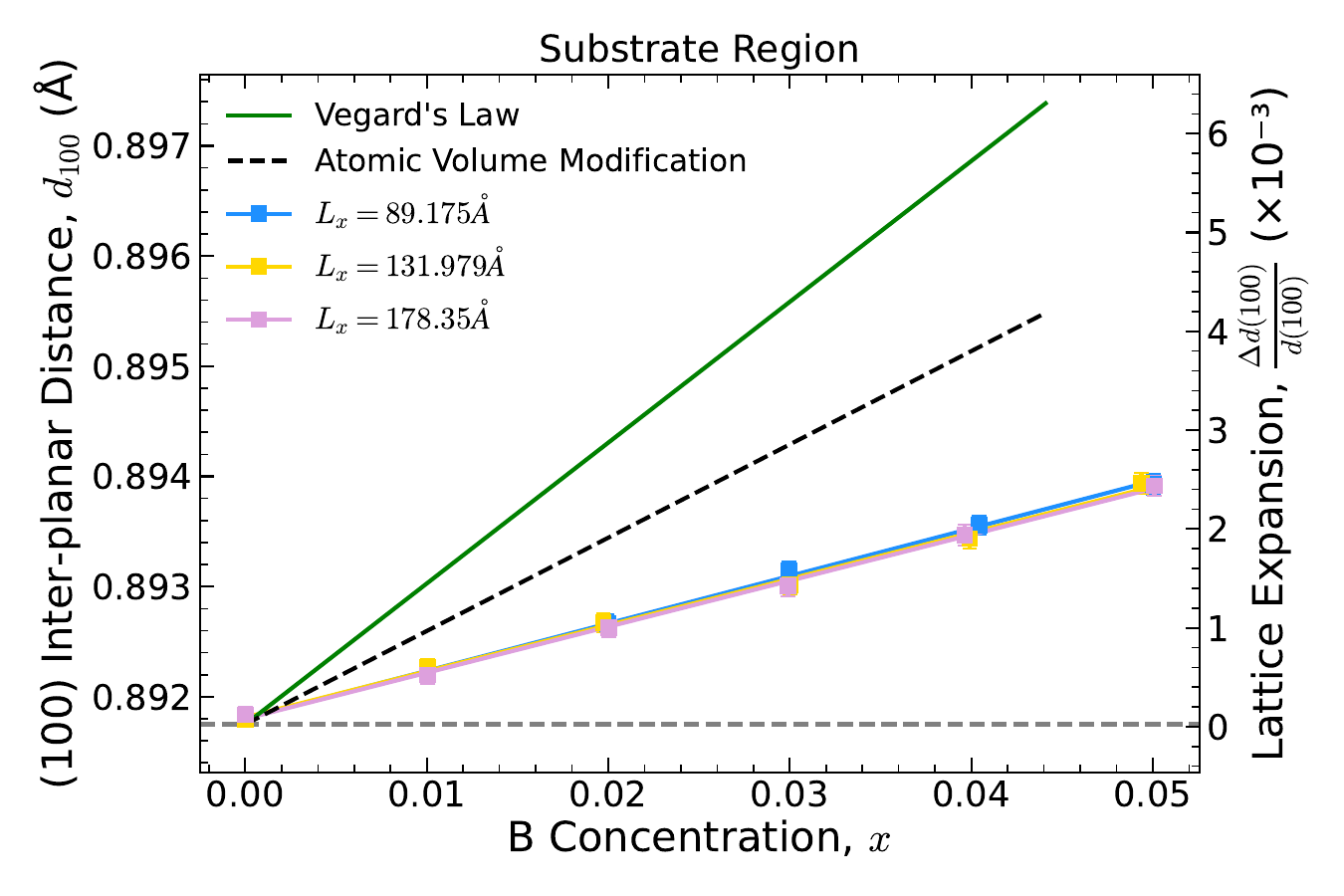}}\\[-2ex]
 \setcounter{subfigure}{1}
 \subfloat[]{\includegraphics[width=0.48\textwidth]{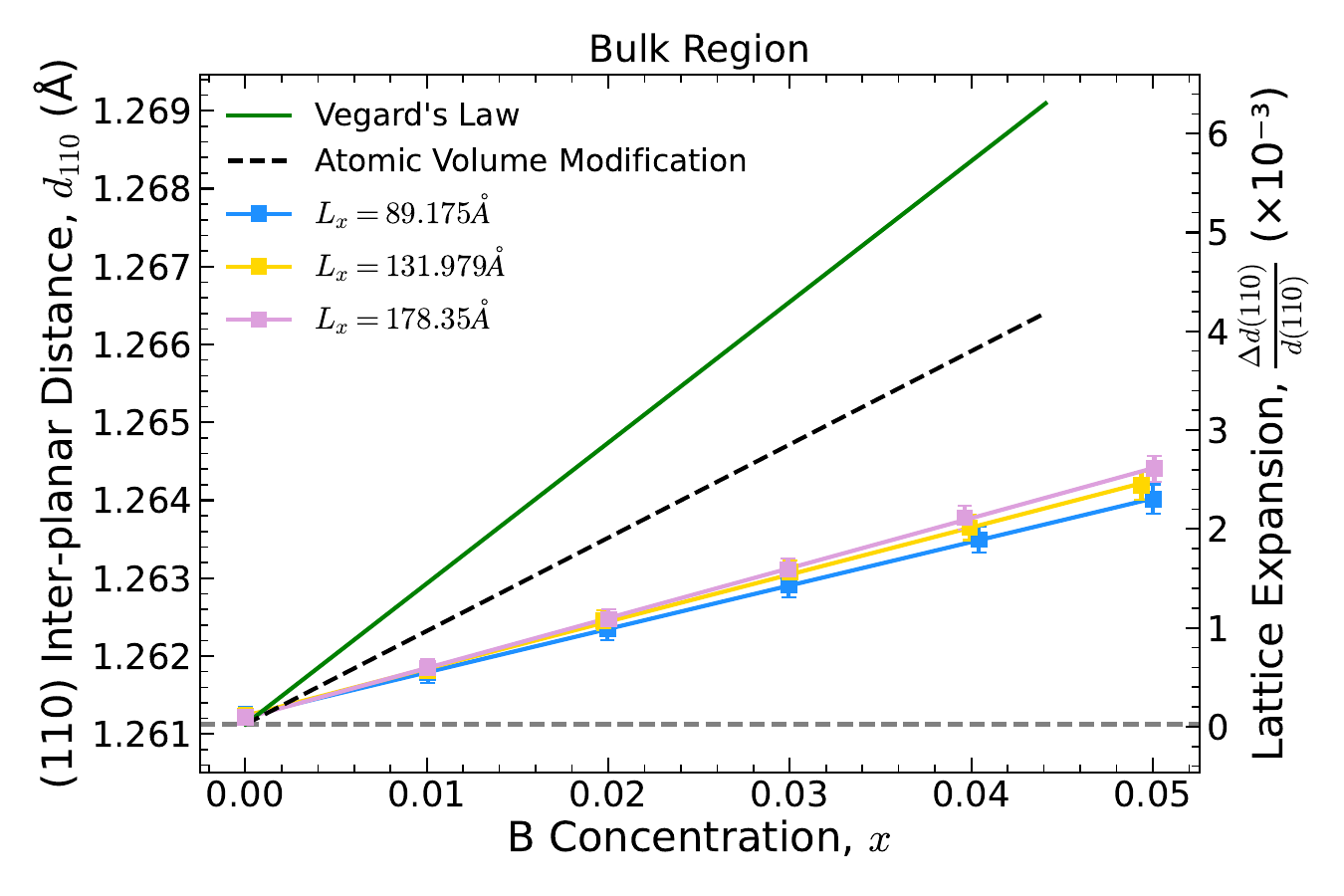}}\hfill
 \setcounter{subfigure}{4}
 \subfloat[]{\includegraphics[width=0.48\textwidth]{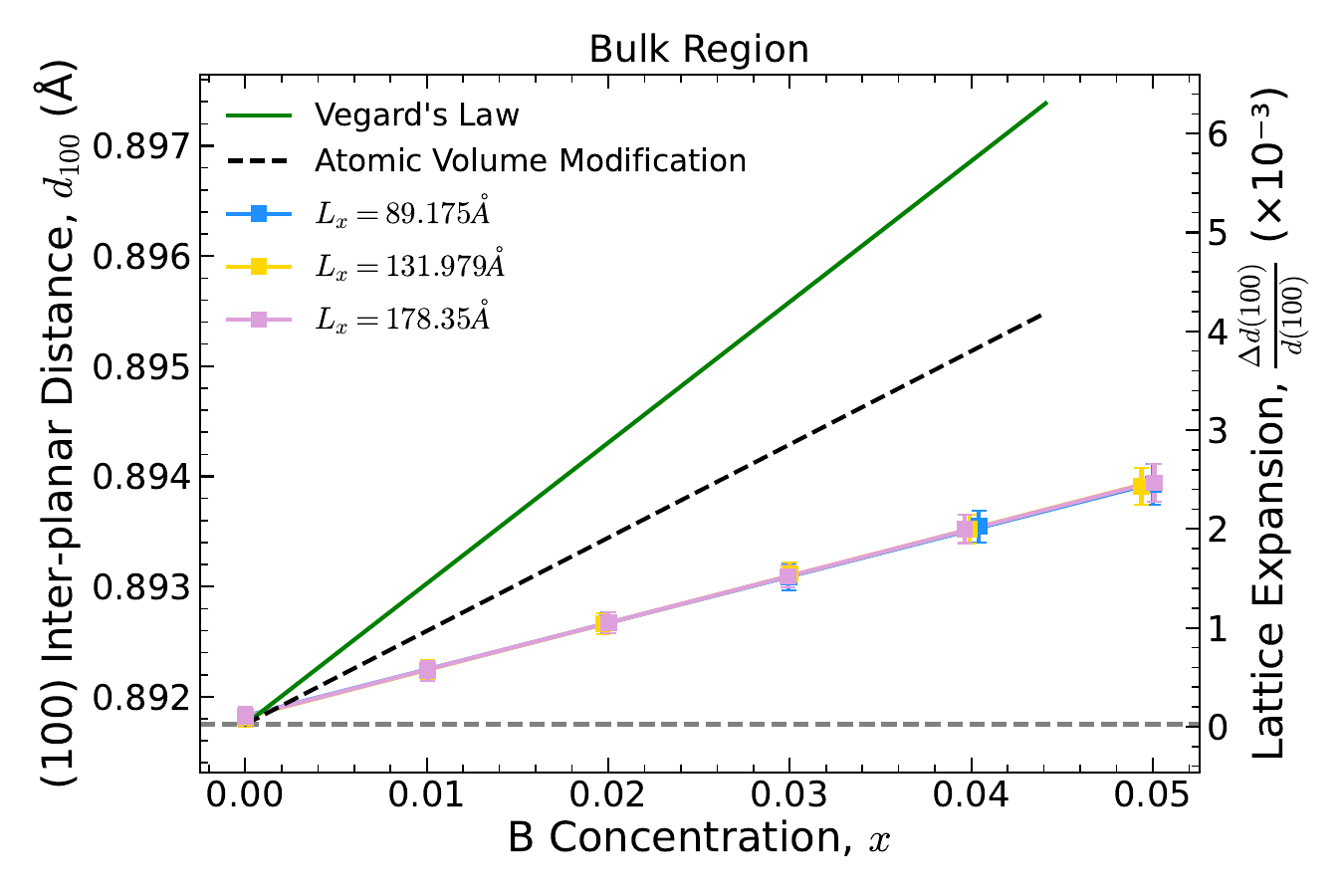}}\\[-2ex]
 \setcounter{subfigure}{2}
 \subfloat[]{\includegraphics[width=0.48\textwidth]{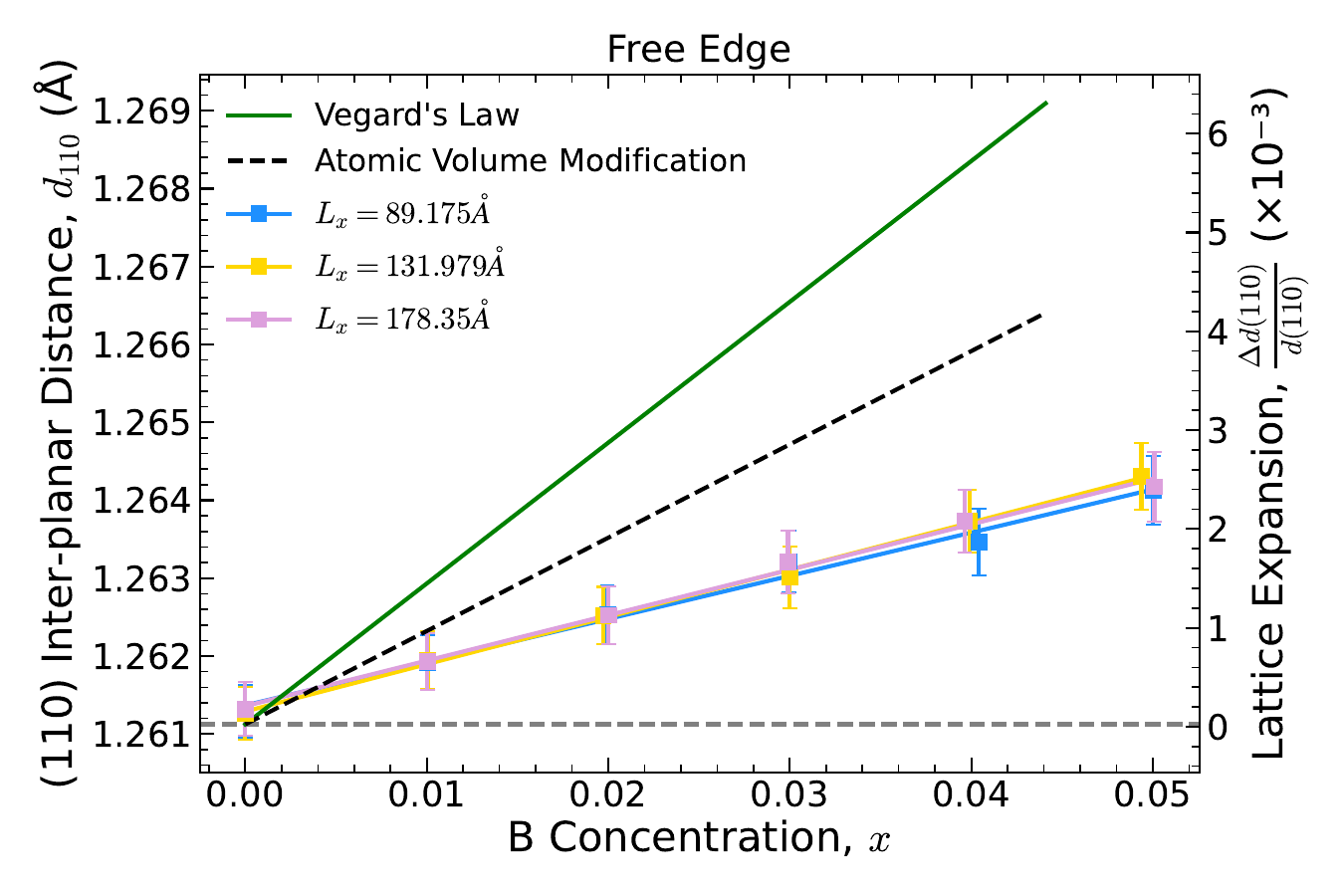}}\hfill
 \setcounter{subfigure}{5}
 \subfloat[]{\includegraphics[width=0.48\textwidth]{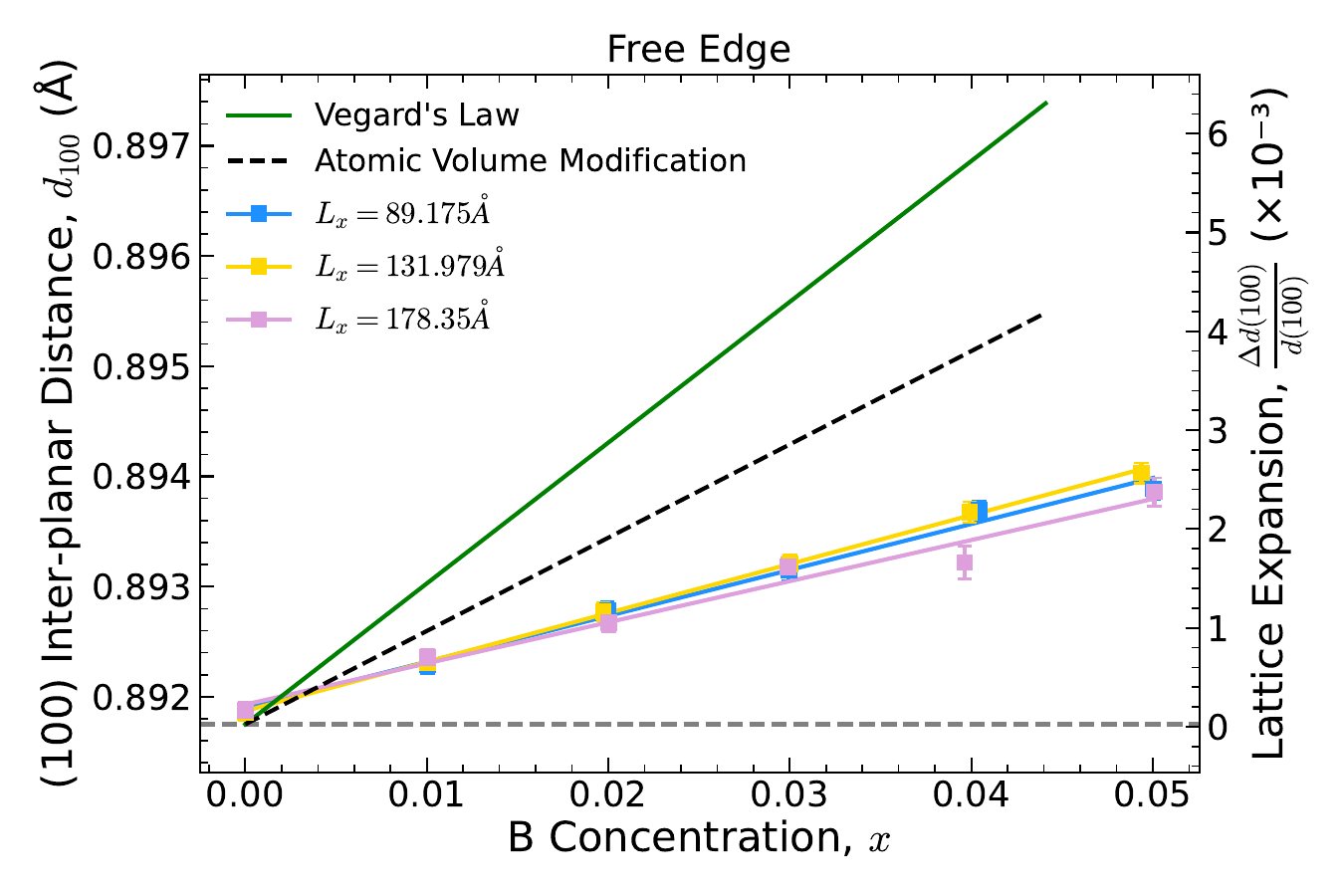}}\\[-2ex]
 \caption{Region-by-region inter-planar distances of E3-category crystals as a function of boron concentration, with left/right panels for (1\;1\;0)/(1\;0\;0) planes and top/middle/bottom rows for substrate, bulk, and free-edge regions.}
 \label{fig:IP_Plots_E3}
\end{figure}

\begin{sidewaystable*}[]
\centering
    \caption{(1\;1\;0) inter-planar distance $d_{110}$ and lattice expansion $\Delta d_{110}/d_{110}$ as a function of dopant concentration for the substrate, bulk, and free-edge regions of the E3A, E3B, and E3C crystals.}
    \resizebox{\textheight}{!}{%
    \begin{tabular}{ccccccc}
    \multicolumn{1}{c|}{\multirow{2}{*}{B Concentration, $x$}} & \multicolumn{2}{c|}{Substrate Region} & \multicolumn{2}{c|}{Bulk Region} & \multicolumn{2}{c}{Free Edge} \\
    \cline{2-7} \multicolumn{1}{c|}{} & \multicolumn{1}{c|}{\rule{0pt}{2.5ex}$d_{110}$ (\unit{\angstrom})} & \multicolumn{1}{c|}{$\Delta d_{110}/d_{110}$ (\SI{e-3}{\angstrom})} & \multicolumn{1}{c|}{\rule{0pt}{2.5ex}$d_{110}$ (\unit{\angstrom})} & \multicolumn{1}{c|}{$\Delta d_{110}/d_{110}$ (\SI{e-3}{\angstrom})} & \multicolumn{1}{c|}{\rule{0pt}{2.5ex}$d_{110}$ (\unit{\angstrom})} & $\Delta d_{110}/d_{110}$ (\SI{e-3}{\angstrom}) \\ \hline
    \multicolumn{7}{c}{\rule{0pt}{2.5ex}\textbf{E3A}} \\ \hline
    \multicolumn{1}{c|}{0.00} & \multicolumn{1}{c|}{$1.2611\pm$\num{0.00017}} & \multicolumn{1}{c|}{$0.0713\pm0.1346$} & \multicolumn{1}{c|}{$1.2612\pm$\num{0.00012}} & \multicolumn{1}{c|}{$0.0873\pm0.0918$} & \multicolumn{1}{c|}{$1.2605\pm$\num{0.00034}} & $0.1331\pm0.2716$ \\
    \multicolumn{1}{c|}{0.01} & \multicolumn{1}{c|}{$1.2613\pm$\num{0.00019}} & \multicolumn{1}{c|}{$0.5531\pm0.1492$} & \multicolumn{1}{c|}{$1.2616\pm$\num{0.00013}} & \multicolumn{1}{c|}{$0.5244\pm0.1016$} & \multicolumn{1}{c|}{$1.2606\pm$\num{0.00035}} & $0.6308\pm0.2796$ \\
    \multicolumn{1}{c|}{0.02} & \multicolumn{1}{c|}{$1.2613\pm$\num{0.00021}} & \multicolumn{1}{c|}{$1.0025\pm0.1675$} & \multicolumn{1}{c|}{$1.2621\pm$\num{0.00014}} & \multicolumn{1}{c|}{$0.9701\pm0.1115$} & \multicolumn{1}{c|}{$1.2607\pm$\num{0.00037}} & $1.1141\pm0.2948$ \\
    \multicolumn{1}{c|}{0.03} & \multicolumn{1}{c|}{$1.2614\pm$\num{0.00021}} & \multicolumn{1}{c|}{$1.4789\pm0.1644$} & \multicolumn{1}{c|}{$1.2625\pm$\num{0.00016}} & \multicolumn{1}{c|}{$1.4149\pm0.1234$} & \multicolumn{1}{c|}{$1.2608\pm$\num{0.0004}} & $1.6554\pm0.3135$ \\
    \multicolumn{1}{c|}{0.04} & \multicolumn{1}{c|}{$1.2615\pm$\num{0.00022}} & \multicolumn{1}{c|}{$1.9412\pm0.1777$} & \multicolumn{1}{c|}{$1.2630\pm$\num{0.00017}} & \multicolumn{1}{c|}{$1.8811\pm0.1359$} & \multicolumn{1}{c|}{$1.2607\pm$\num{0.00043}} & $1.8521\pm0.3407$ \\
    \multicolumn{1}{c|}{0.05} & \multicolumn{1}{c|}{$1.2616\pm$\num{0.00024}} & \multicolumn{1}{c|}{$2.4548\pm0.1927$} & \multicolumn{1}{c|}{$1.2634\pm$\num{0.00019}} & \multicolumn{1}{c|}{$2.2912\pm0.1508$} & \multicolumn{1}{c|}{$1.2609\pm$\num{0.00044}} & $2.3820\pm0.3510$ \\ \toprule
    \multicolumn{7}{c}{\rule{0pt}{2.5ex}\textbf{E3B}} \\ \hline
    \multicolumn{1}{c|}{0.00} & \multicolumn{1}{c|}{$1.2611\pm$\num{0.00017}} & \multicolumn{1}{c|}{$0.0823\pm0.1352$} & \multicolumn{1}{c|}{$1.2612\pm$\num{0.00011}} & \multicolumn{1}{c|}{$0.0786\pm0.0859$} & \multicolumn{1}{c|}{$1.2605\pm$\num{0.00034}} & $0.1130\pm0.2714$ \\
    \multicolumn{1}{c|}{0.01} & \multicolumn{1}{c|}{$1.2613\pm$\num{0.00019}} & \multicolumn{1}{c|}{$0.5431\pm0.1472$} & \multicolumn{1}{c|}{$1.2617\pm$\num{0.00012}} & \multicolumn{1}{c|}{$0.5457\pm0.0941$} & \multicolumn{1}{c|}{$1.2606\pm$\num{0.00037}} & $0.6500\pm0.2945$ \\
    \multicolumn{1}{c|}{0.02} & \multicolumn{1}{c|}{$1.2613\pm$\num{0.0002}} & \multicolumn{1}{c|}{$1.0266\pm0.1563$} & \multicolumn{1}{c|}{$1.2622\pm$\num{0.00013}} & \multicolumn{1}{c|}{$1.0572\pm0.1035$} & \multicolumn{1}{c|}{$1.2607\pm$\num{0.00037}} & $1.1072\pm0.2916$ \\
    \multicolumn{1}{c|}{0.03} & \multicolumn{1}{c|}{$1.2614\pm$\num{0.00022}} & \multicolumn{1}{c|}{$1.4543\pm0.1716$} & \multicolumn{1}{c|}{$1.2627\pm$\num{0.00014}} & \multicolumn{1}{c|}{$1.5497\pm0.1135$} & \multicolumn{1}{c|}{$1.2606\pm$\num{0.0004}} & $1.4954\pm0.3177$ \\
    \multicolumn{1}{c|}{0.04} & \multicolumn{1}{c|}{$1.2615\pm$\num{0.00023}} & \multicolumn{1}{c|}{$1.9376\pm0.1794$} & \multicolumn{1}{c|}{$1.2632\pm$\num{0.00016}} & \multicolumn{1}{c|}{$2.0023\pm0.1273$} & \multicolumn{1}{c|}{$1.2610\pm$\num{0.00041}} & $2.0670\pm0.3233$ \\
    \multicolumn{1}{c|}{0.05} & \multicolumn{1}{c|}{$1.2616\pm$\num{0.00024}} & \multicolumn{1}{c|}{$2.4455\pm0.1904$} & \multicolumn{1}{c|}{$1.2636\pm$\num{0.00018}} & \multicolumn{1}{c|}{$2.4300\pm0.1414$} & \multicolumn{1}{c|}{$1.2611\pm$\num{0.00043}} & $2.5208\pm0.3411$ \\ \toprule
    \multicolumn{7}{c}{\rule{0pt}{2.5ex}\textbf{E3C}} \\ \hline
    \multicolumn{1}{c|}{0.00} & \multicolumn{1}{c|}{$1.2611\pm$\num{0.00017}} & \multicolumn{1}{c|}{$0.0659\pm0.1380$} & \multicolumn{1}{c|}{$1.2612\pm$\num{0.0001}} & \multicolumn{1}{c|}{$0.0730\pm0.0818$} & \multicolumn{1}{c|}{$1.2605\pm$\num{0.00035}} & $0.1540\pm0.2743$ \\
    \multicolumn{1}{c|}{0.01} & \multicolumn{1}{c|}{$1.2612\pm$\num{0.00019}} & \multicolumn{1}{c|}{$0.4895\pm0.1544$} & \multicolumn{1}{c|}{$1.2617\pm$\num{0.00011}} & \multicolumn{1}{c|}{$0.5726\pm0.0877$} & \multicolumn{1}{c|}{$1.2606\pm$\num{0.00036}} & $0.6370\pm0.2878$ \\
    \multicolumn{1}{c|}{0.02} & \multicolumn{1}{c|}{$1.2613\pm$\num{0.00021}} & \multicolumn{1}{c|}{$1.0113\pm0.1648$} & \multicolumn{1}{c|}{$1.2622\pm$\num{0.00012}} & \multicolumn{1}{c|}{$1.0679\pm0.0987$} & \multicolumn{1}{c|}{$1.2607\pm$\num{0.00037}} & $1.1092\pm0.2949$ \\
    \multicolumn{1}{c|}{0.03} & \multicolumn{1}{c|}{$1.2614\pm$\num{0.00021}} & \multicolumn{1}{c|}{$1.4605\pm0.1697$} & \multicolumn{1}{c|}{$1.2627\pm$\num{0.00014}} & \multicolumn{1}{c|}{$1.5747\pm0.1103$} & \multicolumn{1}{c|}{$1.2608\pm$\num{0.0004}} & $1.6499\pm0.3211$ \\
    \multicolumn{1}{c|}{0.04} & \multicolumn{1}{c|}{$1.2615\pm$\num{0.00022}} & \multicolumn{1}{c|}{$1.9619\pm0.1752$} & \multicolumn{1}{c|}{$1.2633\pm$\num{0.00015}} & \multicolumn{1}{c|}{$2.1029\pm0.1195$} & \multicolumn{1}{c|}{$1.2610\pm$\num{0.0004}} & $2.0686\pm0.3202$ \\
    \multicolumn{1}{c|}{0.05} & \multicolumn{1}{c|}{$1.2615\pm$\num{0.00022}} & \multicolumn{1}{c|}{$2.4177\pm0.1730$} & \multicolumn{1}{c|}{$1.2638\pm$\num{0.00017}} & \multicolumn{1}{c|}{$2.6034\pm0.1322$} & \multicolumn{1}{c|}{$1.2610\pm$\num{0.00044}} & $2.4175\pm0.3527$
    \end{tabular}
    }
    \label{tab:IP_table_E3_110}
\end{sidewaystable*}

\begin{sidewaystable*}[]
\centering
    \caption{(1\;0\;0) inter-planar distance $d_{100}$ and lattice expansion $\Delta d_{100}/d_{100}$ as a function of dopant concentration for the substrate, bulk, and free-edge regions of the E3A, E3B, and E3C crystals.}
    \resizebox{\textheight}{!}{%
    \begin{tabular}{ccccccc}
    \multicolumn{1}{c|}{\multirow{2}{*}{B Concentration, $x$}} & \multicolumn{2}{c|}{Substrate Region} & \multicolumn{2}{c|}{Bulk Region} & \multicolumn{2}{c}{Free Edge} \\
    \cline{2-7} \multicolumn{1}{c|}{} & \multicolumn{1}{c|}{\rule{0pt}{2.5ex}$d_{100}$ (\unit{\angstrom})} & \multicolumn{1}{c|}{$\Delta d_{100}/d_{100}$ (\SI{e-3}{\angstrom})} & \multicolumn{1}{c|}{\rule{0pt}{2.5ex}$d_{100}$ (\unit{\angstrom})} & \multicolumn{1}{c|}{$\Delta d_{100}/d_{100}$ (\SI{e-3}{\angstrom})} & \multicolumn{1}{c|}{\rule{0pt}{2.5ex}$d_{100}$ (\unit{\angstrom})} & $\Delta d_{100}/d_{100}$ (\SI{e-3}{\angstrom}) \\ \hline
    \multicolumn{7}{c}{\rule{0pt}{2.5ex}\textbf{E3A}} \\ \hline
    \multicolumn{1}{c|}{0.00} & \multicolumn{1}{c|}{$0.8918\pm$\num{6.5e-05}} & \multicolumn{1}{c|}{$0.0678\pm0.0725$} & \multicolumn{1}{c|}{$0.8918\pm$\num{8.4e-05}} & \multicolumn{1}{c|}{$0.0876\pm0.0940$} & \multicolumn{1}{c|}{$0.8919\pm$\num{7.4e-05}} & $0.1405\pm0.0829$ \\
    \multicolumn{1}{c|}{0.01} & \multicolumn{1}{c|}{$0.8922\pm$\num{7.8e-05}} & \multicolumn{1}{c|}{$0.5022\pm0.0872$} & \multicolumn{1}{c|}{$0.8922\pm$\num{9.1e-05}} & \multicolumn{1}{c|}{$0.5470\pm0.1015$} & \multicolumn{1}{c|}{$0.8923\pm$\num{8.1e-05}} & $0.6062\pm0.0906$ \\
    \multicolumn{1}{c|}{0.02} & \multicolumn{1}{c|}{$0.8927\pm$\num{8.1e-05}} & \multicolumn{1}{c|}{$1.0262\pm0.0910$} & \multicolumn{1}{c|}{$0.8927\pm$\num{9.3e-05}} & \multicolumn{1}{c|}{$1.0366\pm0.1047$} & \multicolumn{1}{c|}{$0.8928\pm$\num{9e-05}} & $1.1532\pm0.1012$ \\
    \multicolumn{1}{c|}{0.03} & \multicolumn{1}{c|}{$0.8931\pm$\num{8.3e-05}} & \multicolumn{1}{c|}{$1.5665\pm0.0932$} & \multicolumn{1}{c|}{$0.8931\pm$\num{0.00012}} & \multicolumn{1}{c|}{$1.5017\pm0.1343$} & \multicolumn{1}{c|}{$0.8931\pm$\num{9.1e-05}} & $1.5662\pm0.1021$ \\
    \multicolumn{1}{c|}{0.04} & \multicolumn{1}{c|}{$0.8936\pm$\num{8.8e-05}} & \multicolumn{1}{c|}{$2.0250\pm0.0984$} & \multicolumn{1}{c|}{$0.8935\pm$\num{0.00014}} & \multicolumn{1}{c|}{$2.0133\pm0.1622$} & \multicolumn{1}{c|}{$0.8937\pm$\num{8.6e-05}} & $2.1793\pm0.0968$ \\
    \multicolumn{1}{c|}{0.05} & \multicolumn{1}{c|}{$0.8939\pm$\num{9.5e-05}} & \multicolumn{1}{c|}{$2.4403\pm0.1065$} & \multicolumn{1}{c|}{$0.8939\pm$\num{0.00019}} & \multicolumn{1}{c|}{$2.4418\pm0.2113$} & \multicolumn{1}{c|}{$0.8939\pm$\num{0.0001}} & $2.3968\pm0.1116$ \\ \toprule
    \multicolumn{7}{c}{\rule{0pt}{2.5ex}\textbf{E3B}} \\ \hline
    \multicolumn{1}{c|}{0.00} & \multicolumn{1}{c|}{$0.8918\pm$\num{6.5e-05}} & \multicolumn{1}{c|}{$0.0538\pm0.0731$} & \multicolumn{1}{c|}{$0.8918\pm$\num{8.3e-05}} & \multicolumn{1}{c|}{$0.0680\pm0.0926$} & \multicolumn{1}{c|}{$0.8919\pm$\num{7.2e-05}} & $0.1191\pm0.0812$ \\
    \multicolumn{1}{c|}{0.01} & \multicolumn{1}{c|}{$0.8923\pm$\num{7.5e-05}} & \multicolumn{1}{c|}{$0.5816\pm0.0836$} & \multicolumn{1}{c|}{$0.8922\pm$\num{8.9e-05}} & \multicolumn{1}{c|}{$0.5499\pm0.1001$} & \multicolumn{1}{c|}{$0.8923\pm$\num{8e-05}} & $0.6305\pm0.0893$ \\
    \multicolumn{1}{c|}{0.02} & \multicolumn{1}{c|}{$0.8927\pm$\num{8.4e-05}} & \multicolumn{1}{c|}{$1.0319\pm0.0945$} & \multicolumn{1}{c|}{$0.8927\pm$\num{9.6e-05}} & \multicolumn{1}{c|}{$1.0223\pm0.1076$} & \multicolumn{1}{c|}{$0.8928\pm$\num{8.8e-05}} & $1.1447\pm0.0990$ \\
    \multicolumn{1}{c|}{0.03} & \multicolumn{1}{c|}{$0.8930\pm$\num{8.5e-05}} & \multicolumn{1}{c|}{$1.4174\pm0.0957$} & \multicolumn{1}{c|}{$0.8931\pm$\num{0.0001}} & \multicolumn{1}{c|}{$1.5336\pm0.1127$} & \multicolumn{1}{c|}{$0.8932\pm$\num{8.3e-05}} & $1.6301\pm0.0933$ \\
    \multicolumn{1}{c|}{0.04} & \multicolumn{1}{c|}{$0.8934\pm$\num{9.2e-05}} & \multicolumn{1}{c|}{$1.8892\pm0.1034$} & \multicolumn{1}{c|}{$0.8935\pm$\num{0.00013}} & \multicolumn{1}{c|}{$1.9851\pm0.1454$} & \multicolumn{1}{c|}{$0.8937\pm$\num{9.3e-05}} & $2.1613\pm0.1047$ \\
    \multicolumn{1}{c|}{0.05} & \multicolumn{1}{c|}{$0.8939\pm$\num{9.7e-05}} & \multicolumn{1}{c|}{$2.4506\pm0.1090$} & \multicolumn{1}{c|}{$0.8939\pm$\num{0.00017}} & \multicolumn{1}{c|}{$2.4247\pm0.1868$} & \multicolumn{1}{c|}{$0.8940\pm$\num{9.4e-05}} & $2.5581\pm0.1058$ \\ \toprule
    \multicolumn{7}{c}{\rule{0pt}{2.5ex}\textbf{E3C}} \\ \hline
    \multicolumn{1}{c|}{0.00} & \multicolumn{1}{c|}{$0.8918\pm$\num{7.1e-05}} & \multicolumn{1}{c|}{$0.0979\pm0.0791$} & \multicolumn{1}{c|}{$0.8918\pm$\num{8.4e-05}} & \multicolumn{1}{c|}{$0.0833\pm0.0938$} & \multicolumn{1}{c|}{$0.8919\pm$\num{7.4e-05}} & $0.1406\pm0.0835$ \\
    \multicolumn{1}{c|}{0.01} & \multicolumn{1}{c|}{$0.8922\pm$\num{7.7e-05}} & \multicolumn{1}{c|}{$0.5018\pm0.0868$} & \multicolumn{1}{c|}{$0.8922\pm$\num{9e-05}} & \multicolumn{1}{c|}{$0.5457\pm0.1013$} & \multicolumn{1}{c|}{$0.8924\pm$\num{7.6e-05}} & $0.6810\pm0.0853$ \\
    \multicolumn{1}{c|}{0.02} & \multicolumn{1}{c|}{$0.8926\pm$\num{8.2e-05}} & \multicolumn{1}{c|}{$0.9823\pm0.0916$} & \multicolumn{1}{c|}{$0.8927\pm$\num{9.6e-05}} & \multicolumn{1}{c|}{$1.0362\pm0.1073$} & \multicolumn{1}{c|}{$0.8927\pm$\num{8e-05}} & $1.0325\pm0.0902$ \\
    \multicolumn{1}{c|}{0.03} & \multicolumn{1}{c|}{$0.8930\pm$\num{9.2e-05}} & \multicolumn{1}{c|}{$1.4098\pm0.1036$} & \multicolumn{1}{c|}{$0.8931\pm$\num{0.0001}} & \multicolumn{1}{c|}{$1.5040\pm0.1138$} & \multicolumn{1}{c|}{$0.8932\pm$\num{8.6e-05}} & $1.6044\pm0.0962$ \\
    \multicolumn{1}{c|}{0.04} & \multicolumn{1}{c|}{$0.8935\pm$\num{9.3e-05}} & \multicolumn{1}{c|}{$1.9265\pm0.1046$} & \multicolumn{1}{c|}{$0.8935\pm$\num{0.00013}} & \multicolumn{1}{c|}{$1.9877\pm0.1437$} & \multicolumn{1}{c|}{$0.8932\pm$\num{0.00015}} & $1.6471\pm0.1671$ \\
    \multicolumn{1}{c|}{0.05} & \multicolumn{1}{c|}{$0.8939\pm$\num{9.4e-05}} & \multicolumn{1}{c|}{$2.4272\pm0.1051$} & \multicolumn{1}{c|}{$0.8939\pm$\num{0.00017}} & \multicolumn{1}{c|}{$2.4581\pm0.1960$} & \multicolumn{1}{c|}{$0.8939\pm$\num{0.00013}} & $2.3654\pm0.1465$
    \end{tabular}
    }
    \label{tab:IP_table_E3_100}
\end{sidewaystable*}

\subsection{S Crystals}
\begin{figure}[H]
 \subfloat[]{\includegraphics[width=0.48\textwidth]{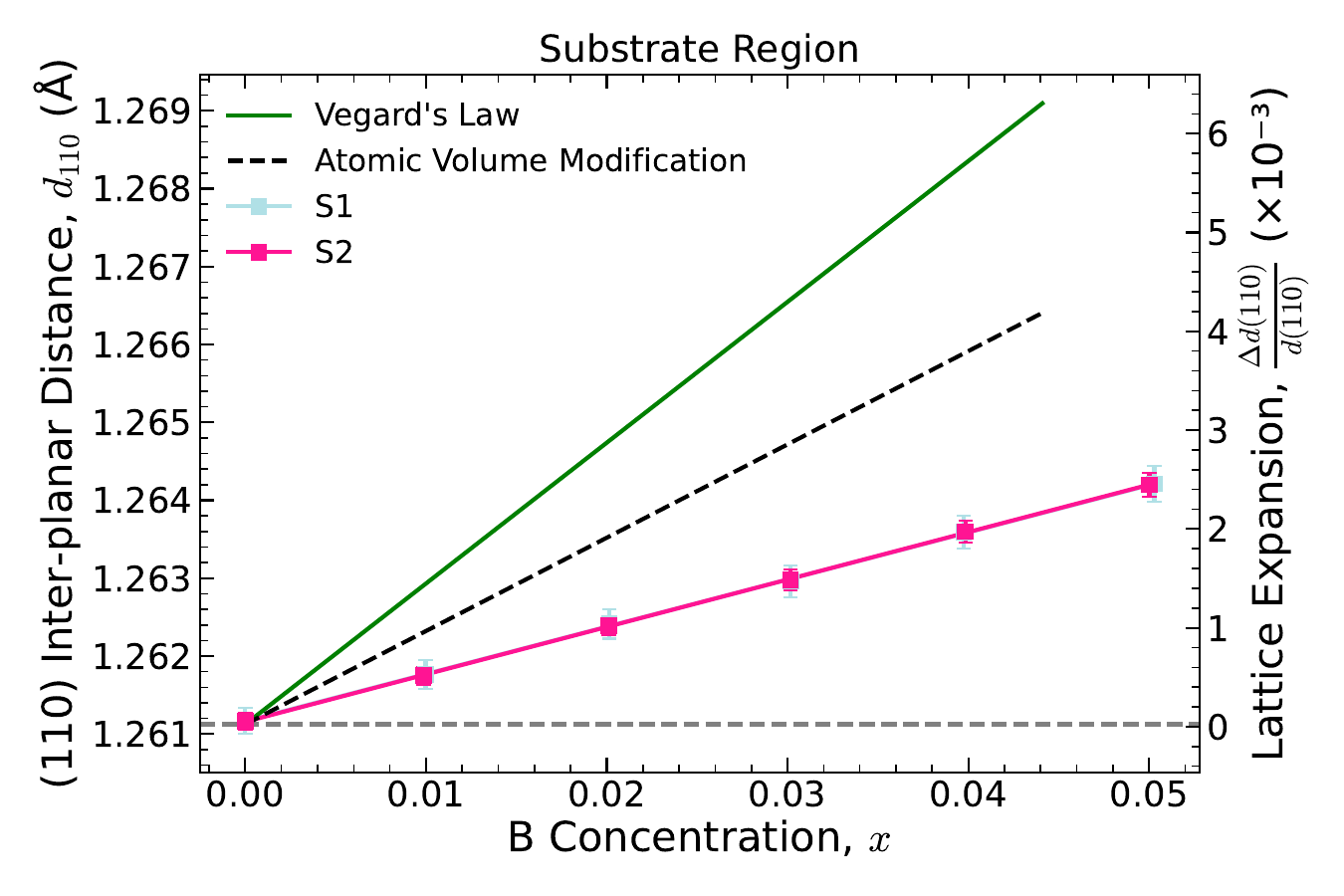}}\hfill
 \setcounter{subfigure}{3}
 \subfloat[]{\includegraphics[width=0.48\textwidth]{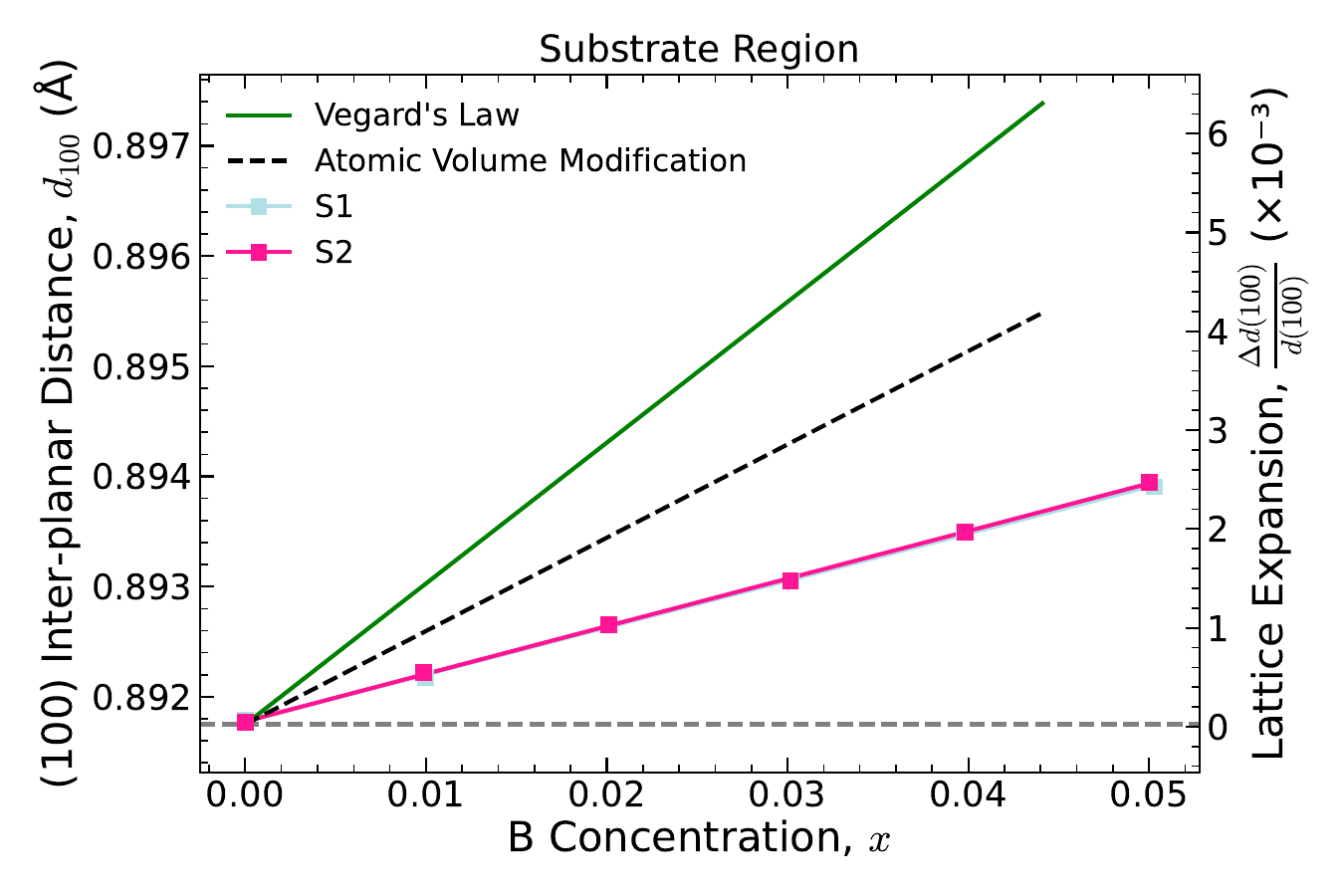}}\\[-2ex]
 \setcounter{subfigure}{1}
 \subfloat[]{\includegraphics[width=0.48\textwidth]{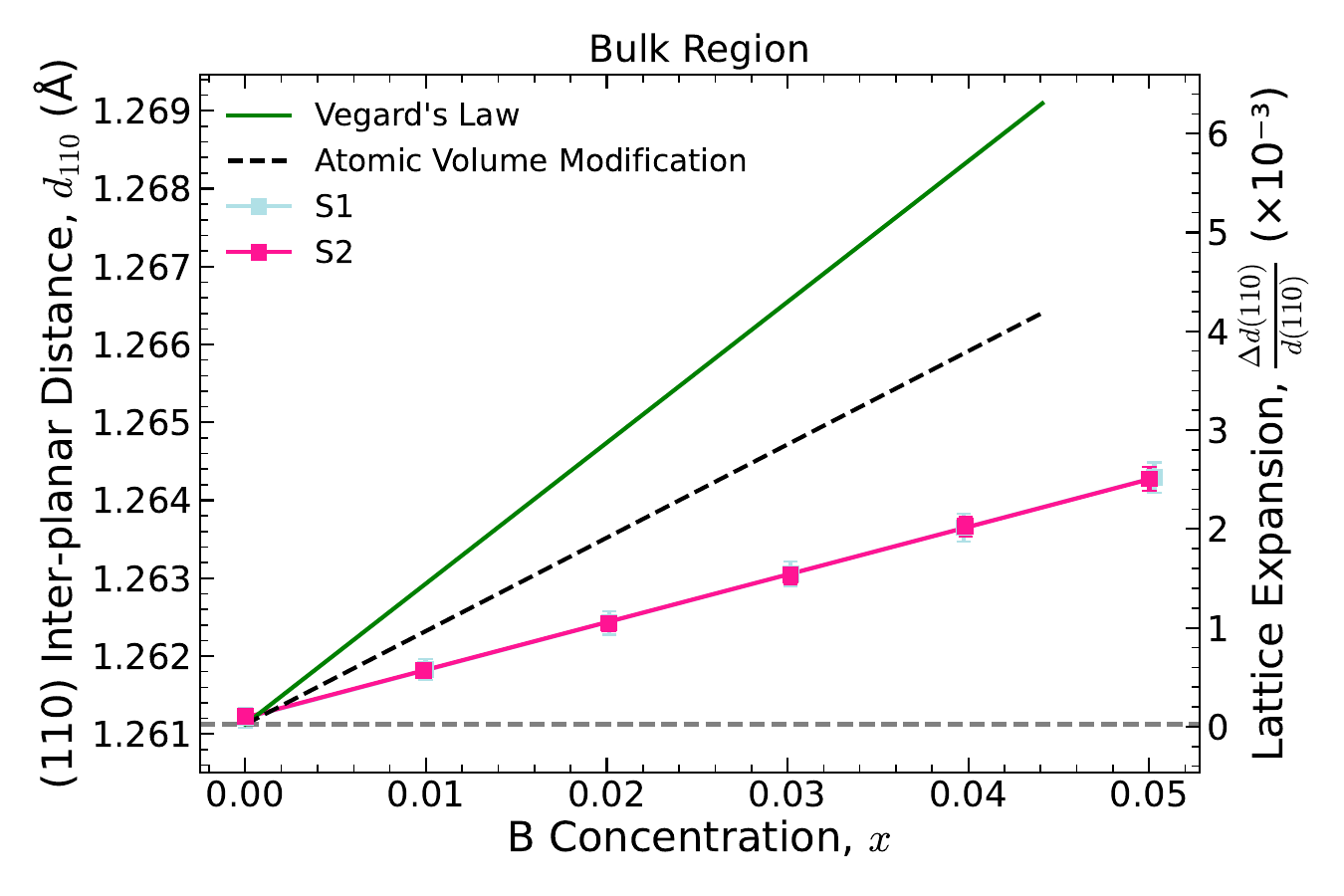}}\hfill
 \setcounter{subfigure}{4}
 \subfloat[]{\includegraphics[width=0.48\textwidth]{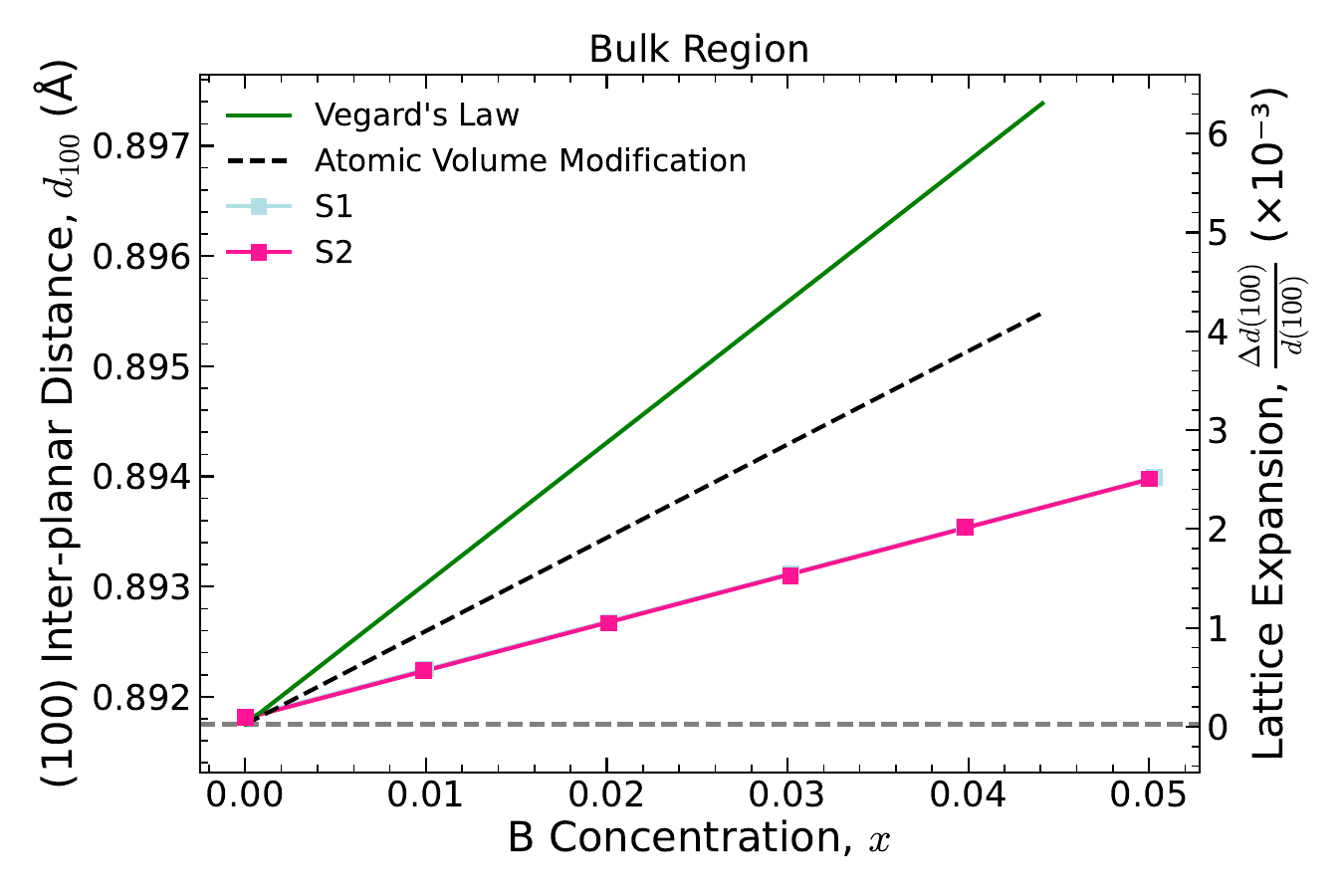}}\\[-2ex]
 \setcounter{subfigure}{2}
 \subfloat[]{\includegraphics[width=0.48\textwidth]{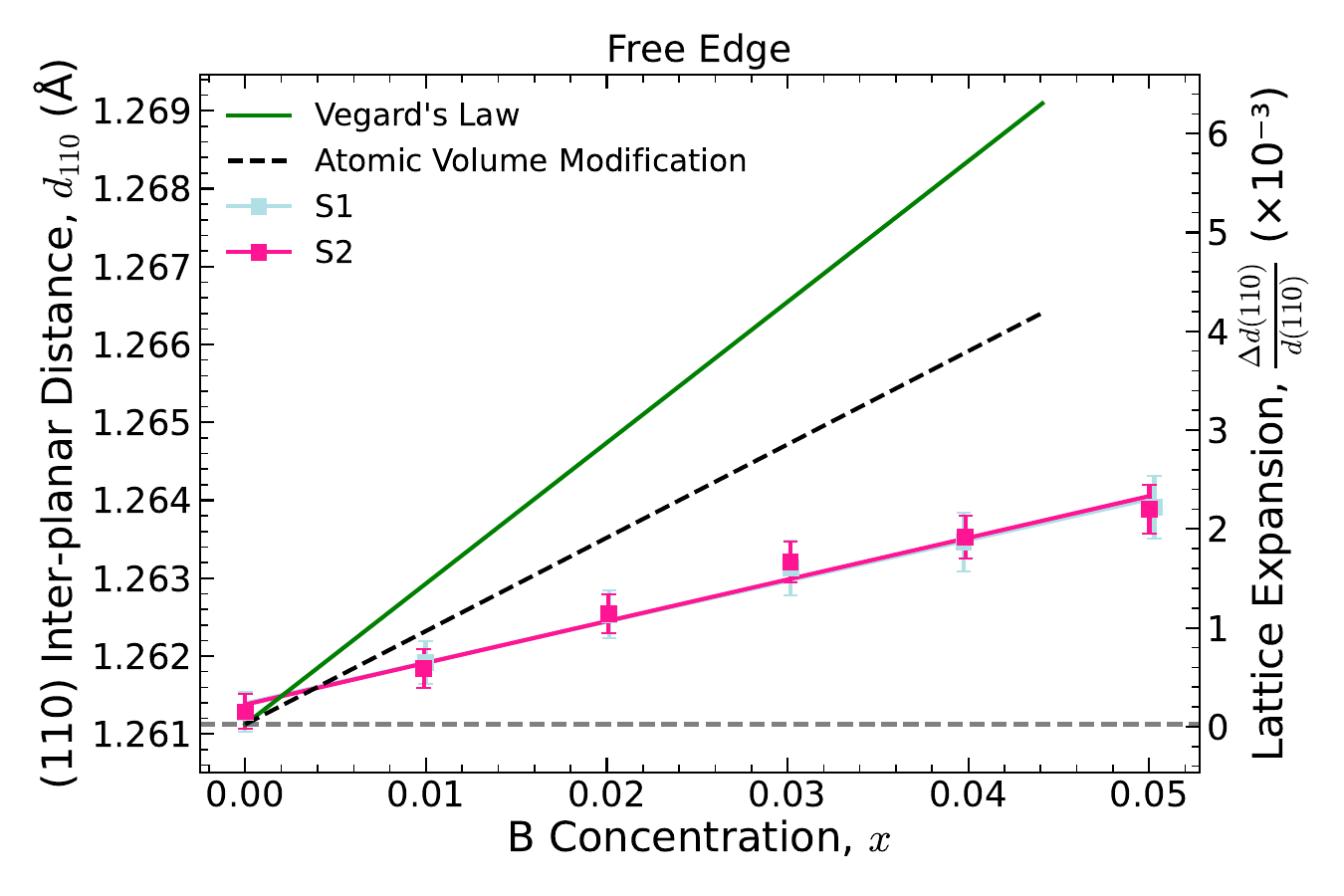}}\hfill
 \setcounter{subfigure}{5}
 \subfloat[]{\includegraphics[width=0.48\textwidth]{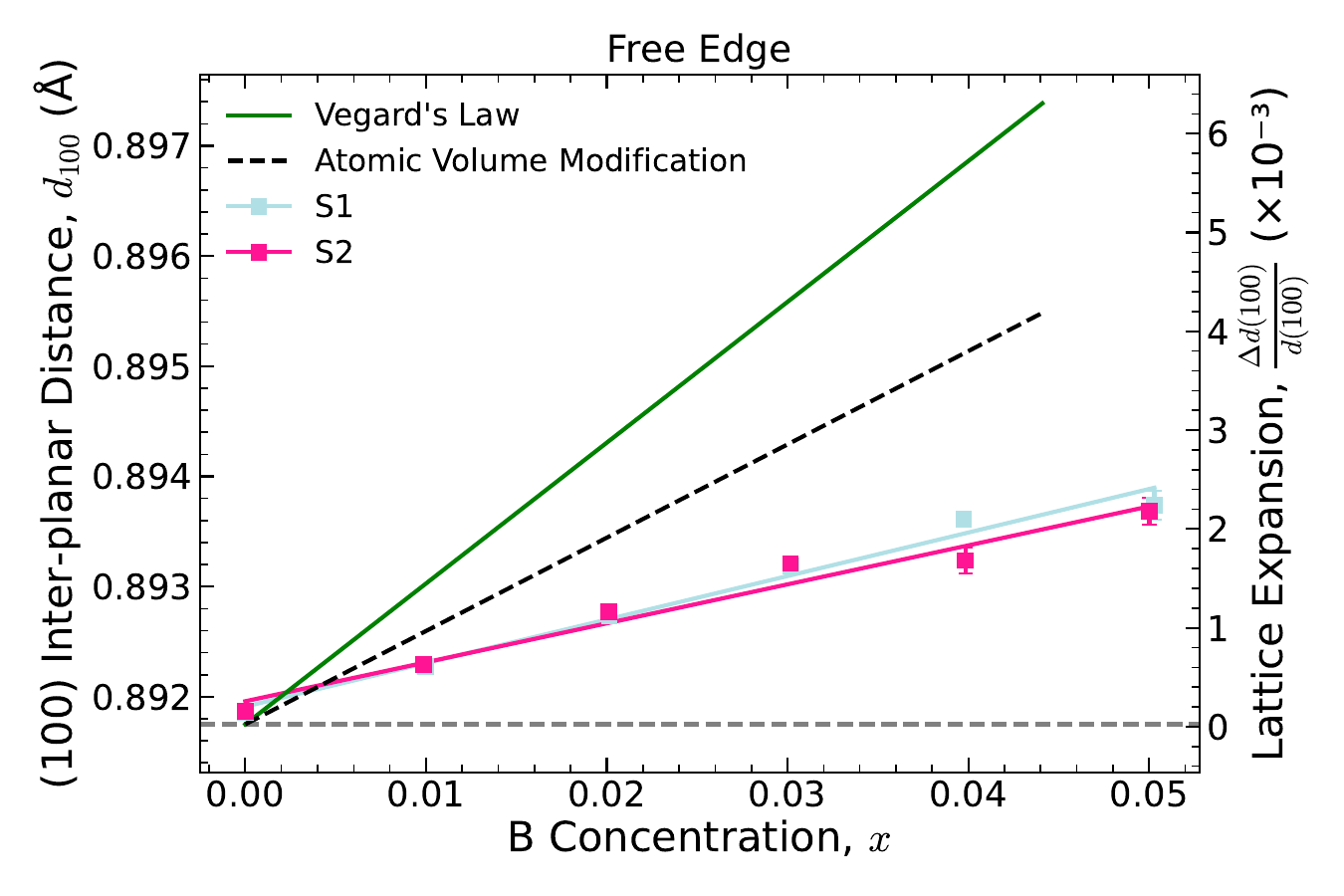}}\\[-2ex]
 \caption{Region-by-region inter-planar distances of S-category crystals as a function of boron concentration, with left/right panels for (1\;1\;0)/(1\;0\;0) planes and top/middle/bottom rows for substrate, bulk, and free-edge regions.}
 \label{fig:IP_Plots_S}
\end{figure}

\begin{sidewaystable*}[]
\centering
    \caption{(1\;1\;0) inter-planar distance $d_{110}$ and lattice expansion $\Delta d_{110}/d_{110}$ as a function of dopant concentration for the substrate, bulk, and free-edge regions of the S1 and S2 crystals.}
    \resizebox{\textheight}{!}{%
    \begin{tabular}{ccccccc}
    \multicolumn{1}{c|}{\multirow{2}{*}{B Concentration, $x$}} & \multicolumn{2}{c|}{Substrate Region} & \multicolumn{2}{c|}{Bulk Region} & \multicolumn{2}{c}{Free Edge} \\
    \cline{2-7} \multicolumn{1}{c|}{} & \multicolumn{1}{c|}{\rule{0pt}{2.5ex}$_{110}d$ (\unit{\angstrom})} & \multicolumn{1}{c|}{$\Delta d_{110}/d_{110}$ (\SI{e-3}{\angstrom})} & \multicolumn{1}{c|}{\rule{0pt}{2.5ex}$d_{110}$ (\unit{\angstrom})} & \multicolumn{1}{c|}{$\Delta d_{110}/d_{110}$ (\SI{e-3}{\angstrom})} & \multicolumn{1}{c|}{\rule{0pt}{2.5ex}$d_{110}$ (\unit{\angstrom})} & $\Delta d_{110}/d_{110}$ (\SI{e-3}{\angstrom}) \\ \hline
    \multicolumn{7}{c}{\rule{0pt}{2.5ex}\textbf{S1}} \\ \hline
    \multicolumn{1}{c|}{0.00} & \multicolumn{1}{c|}{$1.2611\pm$\num{0.00017}} & \multicolumn{1}{c|}{$0.0329\pm0.1331$} & \multicolumn{1}{c|}{$1.2611\pm$\num{0.00012}} & \multicolumn{1}{c|}{$0.0660\pm0.0987$} & \multicolumn{1}{c|}{$1.2609\pm$\num{0.00026}} & $0.1251\pm0.2047$ \\
    \multicolumn{1}{c|}{0.01} & \multicolumn{1}{c|}{$1.2612\pm$\num{0.00019}} & \multicolumn{1}{c|}{$0.5054\pm0.1474$} & \multicolumn{1}{c|}{$1.2613\pm$\num{0.00013}} & \multicolumn{1}{c|}{$0.5578\pm0.1051$} & \multicolumn{1}{c|}{$1.2609\pm$\num{0.00028}} & $0.6283\pm0.2216$ \\
    \multicolumn{1}{c|}{0.02} & \multicolumn{1}{c|}{$1.2612\pm$\num{0.00019}} & \multicolumn{1}{c|}{$1.0204\pm0.1490$} & \multicolumn{1}{c|}{$1.2614\pm$\num{0.00015}} & \multicolumn{1}{c|}{$1.0323\pm0.1201$} & \multicolumn{1}{c|}{$1.2610\pm$\num{0.00031}} & $1.1168\pm0.2443$ \\
    \multicolumn{1}{c|}{0.03} & \multicolumn{1}{c|}{$1.2612\pm$\num{0.0002}} & \multicolumn{1}{c|}{$1.4550\pm0.1594$} & \multicolumn{1}{c|}{$1.2616\pm$\num{0.00016}} & \multicolumn{1}{c|}{$1.5321\pm0.1245$} & \multicolumn{1}{c|}{$1.2609\pm$\num{0.00035}} & $1.5913\pm0.2787$ \\
    \multicolumn{1}{c|}{0.04} & \multicolumn{1}{c|}{$1.2613\pm$\num{0.00021}} & \multicolumn{1}{c|}{$1.9542\pm0.1677$} & \multicolumn{1}{c|}{$1.2617\pm$\num{0.00018}} & \multicolumn{1}{c|}{$2.0006\pm0.1399$} & \multicolumn{1}{c|}{$1.2610\pm$\num{0.00038}} & $1.8519\pm0.2982$ \\
    \multicolumn{1}{c|}{0.05} & \multicolumn{1}{c|}{$1.2613\pm$\num{0.00023}} & \multicolumn{1}{c|}{$2.4458\pm0.1849$} & \multicolumn{1}{c|}{$1.2618\pm$\num{0.00019}} & \multicolumn{1}{c|}{$2.5112\pm0.1543$} & \multicolumn{1}{c|}{$1.2610\pm$\num{0.00041}} & $2.2111\pm0.3213$ \\ \toprule
    \multicolumn{7}{c}{\rule{0pt}{2.5ex}\textbf{S2}} \\ \hline
    \multicolumn{1}{c|}{0.00} & \multicolumn{1}{c|}{$1.2611\pm$\num{0.00011}} & \multicolumn{1}{c|}{$0.0311\pm0.0854$} & \multicolumn{1}{c|}{$1.2611\pm$\num{8e-05}} & \multicolumn{1}{c|}{$0.0800\pm0.0638$} & \multicolumn{1}{c|}{$1.2609\pm$\num{0.00022}} & $0.1277\pm0.1773$ \\
    \multicolumn{1}{c|}{0.01} & \multicolumn{1}{c|}{$1.2612\pm$\num{0.00012}} & \multicolumn{1}{c|}{$0.4941\pm0.0927$} & \multicolumn{1}{c|}{$1.2613\pm$\num{8.6e-05}} & \multicolumn{1}{c|}{$0.5497\pm0.0685$} & \multicolumn{1}{c|}{$1.2608\pm$\num{0.00025}} & $0.5667\pm0.1997$ \\
    \multicolumn{1}{c|}{0.02} & \multicolumn{1}{c|}{$1.2612\pm$\num{0.00012}} & \multicolumn{1}{c|}{$1.0007\pm0.0977$} & \multicolumn{1}{c|}{$1.2614\pm$\num{9.9e-05}} & \multicolumn{1}{c|}{$1.0300\pm0.0784$} & \multicolumn{1}{c|}{$1.2610\pm$\num{0.00025}} & $1.1277\pm0.1961$ \\
    \multicolumn{1}{c|}{0.03} & \multicolumn{1}{c|}{$1.2613\pm$\num{0.00013}} & \multicolumn{1}{c|}{$1.4702\pm0.1059$} & \multicolumn{1}{c|}{$1.2615\pm$\num{0.00011}} & \multicolumn{1}{c|}{$1.5168\pm0.0879$} & \multicolumn{1}{c|}{$1.2610\pm$\num{0.00026}} & $1.6527\pm0.2039$ \\
    \multicolumn{1}{c|}{0.04} & \multicolumn{1}{c|}{$1.2613\pm$\num{0.00014}} & \multicolumn{1}{c|}{$1.9599\pm0.1144$} & \multicolumn{1}{c|}{$1.2617\pm$\num{0.00013}} & \multicolumn{1}{c|}{$2.0133\pm0.1037$} & \multicolumn{1}{c|}{$1.2611\pm$\num{0.00027}} & $1.9062\pm0.2160$ \\
    \multicolumn{1}{c|}{0.05} & \multicolumn{1}{c|}{$1.2613\pm$\num{0.00015}} & \multicolumn{1}{c|}{$2.4364\pm0.1227$} & \multicolumn{1}{c|}{$1.2618\pm$\num{0.00015}} & \multicolumn{1}{c|}{$2.4991\pm0.1189$} & \multicolumn{1}{c|}{$1.2610\pm$\num{0.00031}} & $2.1879\pm0.2485$
    \end{tabular}
    }
    \label{tab:IP_table_S_110}
\end{sidewaystable*}

\begin{sidewaystable*}[]
\centering
    \caption{(1\;0\;0) inter-planar distance $d_{100}$ and lattice expansion $\Delta d_{100}/d_{100}$ as a function of dopant concentration for the substrate, bulk, and free-edge regions of the S1 and S2 crystals.}
    \resizebox{\textheight}{!}{%
    \begin{tabular}{ccccccc}
    \multicolumn{1}{c|}{\multirow{2}{*}{B Concentration, $x$}} & \multicolumn{2}{c|}{Substrate Region} & \multicolumn{2}{c|}{Bulk Region} & \multicolumn{2}{c}{Free Edge} \\
    \cline{2-7} \multicolumn{1}{c|}{} & \multicolumn{1}{c|}{\rule{0pt}{2.5ex}$d_{100}$ (\unit{\angstrom})} & \multicolumn{1}{c|}{$\Delta d_{100}/d_{100}$ (\SI{e-3}{\angstrom})} & \multicolumn{1}{c|}{\rule{0pt}{2.5ex}$d_{100}$ (\unit{\angstrom})} & \multicolumn{1}{c|}{$\Delta d_{100}/d_{100}$ (\SI{e-3}{\angstrom})} & \multicolumn{1}{c|}{\rule{0pt}{2.5ex}$d_{100}$ (\unit{\angstrom})} & $\Delta d_{100}/d_{100}$ (\SI{e-3}{\angstrom}) \\ \hline
    \multicolumn{7}{c}{\rule{0pt}{2.5ex}\textbf{S1}} \\ \hline
    \multicolumn{1}{c|}{0.00} & \multicolumn{1}{c|}{$0.8918\pm$\num{4.4e-05}} & \multicolumn{1}{c|}{$0.0459\pm0.0493$} & \multicolumn{1}{c|}{$0.8918\pm$\num{5.1e-05}} & \multicolumn{1}{c|}{$0.0773\pm0.0571$} & \multicolumn{1}{c|}{$0.8919\pm$\num{4.8e-05}} & $0.1213\pm0.0538$ \\
    \multicolumn{1}{c|}{0.01} & \multicolumn{1}{c|}{$0.8922\pm$\num{5e-05}} & \multicolumn{1}{c|}{$0.4717\pm0.0560$} & \multicolumn{1}{c|}{$0.8923\pm$\num{5.5e-05}} & \multicolumn{1}{c|}{$0.5621\pm0.0614$} & \multicolumn{1}{c|}{$0.8923\pm$\num{5.5e-05}} & $0.5887\pm0.0617$ \\
    \multicolumn{1}{c|}{0.02} & \multicolumn{1}{c|}{$0.8926\pm$\num{4.9e-05}} & \multicolumn{1}{c|}{$1.0073\pm0.0544$} & \multicolumn{1}{c|}{$0.8927\pm$\num{5.8e-05}} & \multicolumn{1}{c|}{$1.0344\pm0.0655$} & \multicolumn{1}{c|}{$0.8927\pm$\num{5.3e-05}} & $1.1005\pm0.0590$ \\
    \multicolumn{1}{c|}{0.03} & \multicolumn{1}{c|}{$0.8931\pm$\num{5.5e-05}} & \multicolumn{1}{c|}{$1.4646\pm0.0612$} & \multicolumn{1}{c|}{$0.8931\pm$\num{6.2e-05}} & \multicolumn{1}{c|}{$1.5360\pm0.0699$} & \multicolumn{1}{c|}{$0.8932\pm$\num{5.6e-05}} & $1.6082\pm0.0628$ \\
    \multicolumn{1}{c|}{0.04} & \multicolumn{1}{c|}{$0.8935\pm$\num{5.1e-05}} & \multicolumn{1}{c|}{$1.9580\pm0.0566$} & \multicolumn{1}{c|}{$0.8935\pm$\num{6.2e-05}} & \multicolumn{1}{c|}{$2.0077\pm0.0700$} & \multicolumn{1}{c|}{$0.8936\pm$\num{5.9e-05}} & $2.0904\pm0.0663$ \\
    \multicolumn{1}{c|}{0.05} & \multicolumn{1}{c|}{$0.8939\pm$\num{6.4e-05}} & \multicolumn{1}{c|}{$2.4172\pm0.0713$} & \multicolumn{1}{c|}{$0.8940\pm$\num{6.9e-05}} & \multicolumn{1}{c|}{$2.5108\pm0.0773$} & \multicolumn{1}{c|}{$0.8937\pm$\num{0.00013}} & $2.2286\pm0.1493$ \\ \toprule
    \multicolumn{7}{c}{\rule{0pt}{2.5ex}\textbf{S2}} \\ \hline
    \multicolumn{1}{c|}{0.00} & \multicolumn{1}{c|}{$0.8918\pm$\num{2.9e-05}} & \multicolumn{1}{c|}{$0.0181\pm0.0322$} & \multicolumn{1}{c|}{$0.8918\pm$\num{3.2e-05}} & \multicolumn{1}{c|}{$0.0686\pm0.0354$} & \multicolumn{1}{c|}{$0.8919\pm$\num{3.3e-05}} & $0.1316\pm0.0374$ \\
    \multicolumn{1}{c|}{0.01} & \multicolumn{1}{c|}{$0.8922\pm$\num{3.1e-05}} & \multicolumn{1}{c|}{$0.5278\pm0.0350$} & \multicolumn{1}{c|}{$0.8922\pm$\num{3.4e-05}} & \multicolumn{1}{c|}{$0.5454\pm0.0377$} & \multicolumn{1}{c|}{$0.8923\pm$\num{3.8e-05}} & $0.6062\pm0.0430$ \\
    \multicolumn{1}{c|}{0.02} & \multicolumn{1}{c|}{$0.8927\pm$\num{3.3e-05}} & \multicolumn{1}{c|}{$1.0138\pm0.0368$} & \multicolumn{1}{c|}{$0.8927\pm$\num{3.7e-05}} & \multicolumn{1}{c|}{$1.0279\pm0.0415$} & \multicolumn{1}{c|}{$0.8928\pm$\num{3.7e-05}} & $1.1440\pm0.0413$ \\
    \multicolumn{1}{c|}{0.03} & \multicolumn{1}{c|}{$0.8931\pm$\num{3.2e-05}} & \multicolumn{1}{c|}{$1.4606\pm0.0355$} & \multicolumn{1}{c|}{$0.8931\pm$\num{3.9e-05}} & \multicolumn{1}{c|}{$1.5130\pm0.0439$} & \multicolumn{1}{c|}{$0.8932\pm$\num{3.6e-05}} & $1.6357\pm0.0398$ \\
    \multicolumn{1}{c|}{0.04} & \multicolumn{1}{c|}{$0.8935\pm$\num{3.6e-05}} & \multicolumn{1}{c|}{$1.9560\pm0.0406$} & \multicolumn{1}{c|}{$0.8935\pm$\num{4.1e-05}} & \multicolumn{1}{c|}{$2.0063\pm0.0462$} & \multicolumn{1}{c|}{$0.8932\pm$\num{0.00012}} & $1.6677\pm0.1365$ \\
    \multicolumn{1}{c|}{0.05} & \multicolumn{1}{c|}{$0.8939\pm$\num{4e-05}} & \multicolumn{1}{c|}{$2.4650\pm0.0449$} & \multicolumn{1}{c|}{$0.8940\pm$\num{4.4e-05}} & \multicolumn{1}{c|}{$2.4995\pm0.0498$} & \multicolumn{1}{c|}{$0.8937\pm$\num{0.00012}} & $2.1704\pm0.1340$
    \end{tabular}
    }
    \label{tab:IP_table_S_100}
\end{sidewaystable*}

\subsection{Summary of Inter-planar Distances Across Crystal Sizes}

\begin{figure}[H]
    \centering
    \includegraphics[width=\textwidth]{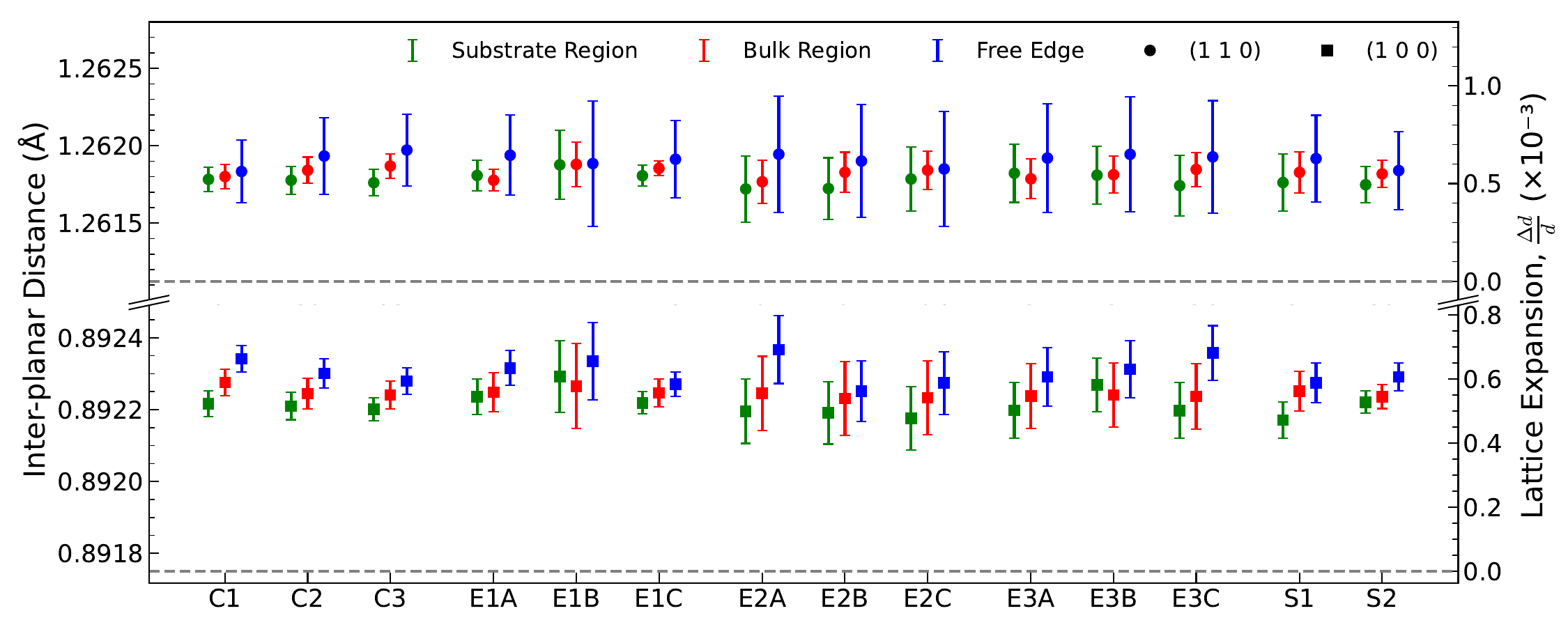}
    \caption{Calculated values of the (1\;1\;0) (circles) and (1\;0\;0) (squares) inter-planar distances for all crystal sizes and for each spatial region of the crystal: substrate (green), bulk (red), and free edge (blue).}
    \label{fig:Dist_Bars_All}
\end{figure}

\Cref{fig:Dist_Bars_All} shows the (1\;1\;0) (circles) and (1\;0\;0) inter-planar distances for a concentration of $x=0.01$ across the full range of crystal sizes and spatial regions.

\section{Statistical Analysis and Validation} \label{sec:Statistics}
The random replacement of carbon atoms in the diamond lattice with boron atoms differs fundamentally from the process of MPCVD. To ensure that this method of crystal generation is valid and yields physically meaningful results, a comprehensive statistical analysis was conducted. This analysis quantifies the accuracy of the achieved dopant concentrations relative to the targets, the reliability of structural measurements over a statistically significant sample, and the randomness of dopant distributions.

\subsection{Dopant Concentration Analysis}

\begin{figure}[t!]
    \centering
    \includegraphics[width=\columnwidth]{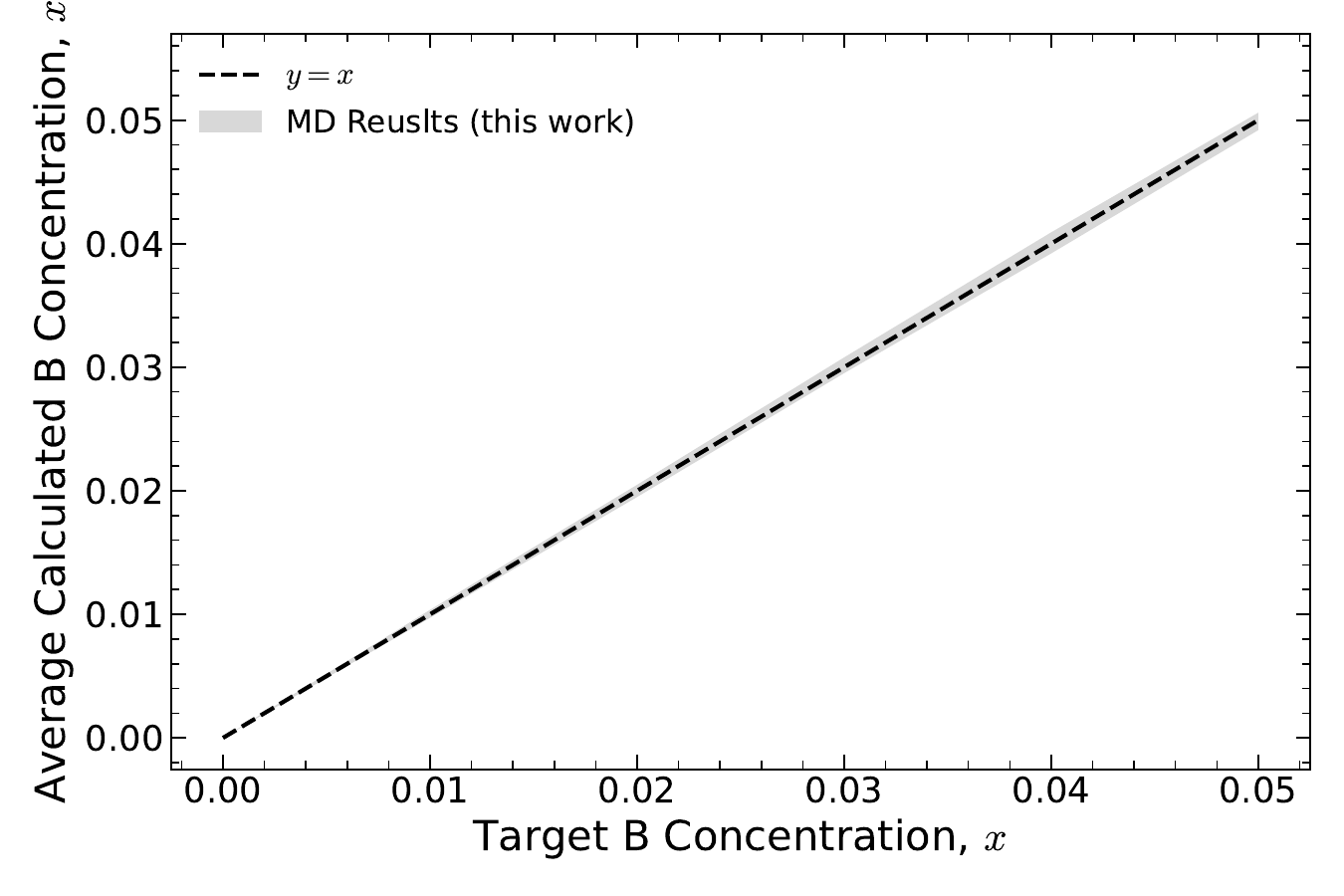}
    \caption{Plot of the calculated dopant concentration averaged across all simulations of a given crystal size as a function of the target dopant concentrations. The grey shaded region represents the full range of concentrations across all crystal sizes. The black dashed line corresponding to a perfect correlation between target and calculated concentration. In all cases the coefficient of determination $R^2=0.999$.}
    \label{fig:Dopant_Concentrations}
\end{figure}

To assess the accuracy of the doping procedure relative to the target concentration, the dopant concentration of each unique distribution of dopant atoms was calculated, and averaged over all crystals of a given size. These are shown in \Cref{fig:Dopant_Concentrations} in comparison to the targeted dopant concentration. A grey shaded region is used to represent the full range of concentrations calculated across all crystal sizes. A full breakdown of these calculated values is provided in the SI. Linear regression was applied for each crystal size to obtain the coefficient of determination ($R^2$). This quantifies how closely the calculated dopant concentrations match the target concentrations, with higher values indicating better agreement, and $R^2=1$ perfect agreement. In all cases an $R^2$ value of 0.999 was obtained, verifying a near-perfect correlation between the calculated and target dopant concentrations. This demonstrates that the crystal generation methodology reliably reproduces the intended dopant concentration. Small deviations are observed at higher concentrations, indicated by the widening of the shaded region, which may be attributed to the random selection process, and are on the same order ($\sim\SI{e-4}{}$) as the uncertainty on secondary-ion mass spectrometry (SIMS) measurements of dopant concentration in crystals grown via MPCVD for use in gamma-ray CLS \cite{supp:Hartmut_2025}, and much smaller than those presented in \citet{supp:Brazhkin_2006}.

\subsection{Convergence Analysis}

\begin{figure*}[t]
 \subfloat[]{\includegraphics[width=0.48\textwidth]{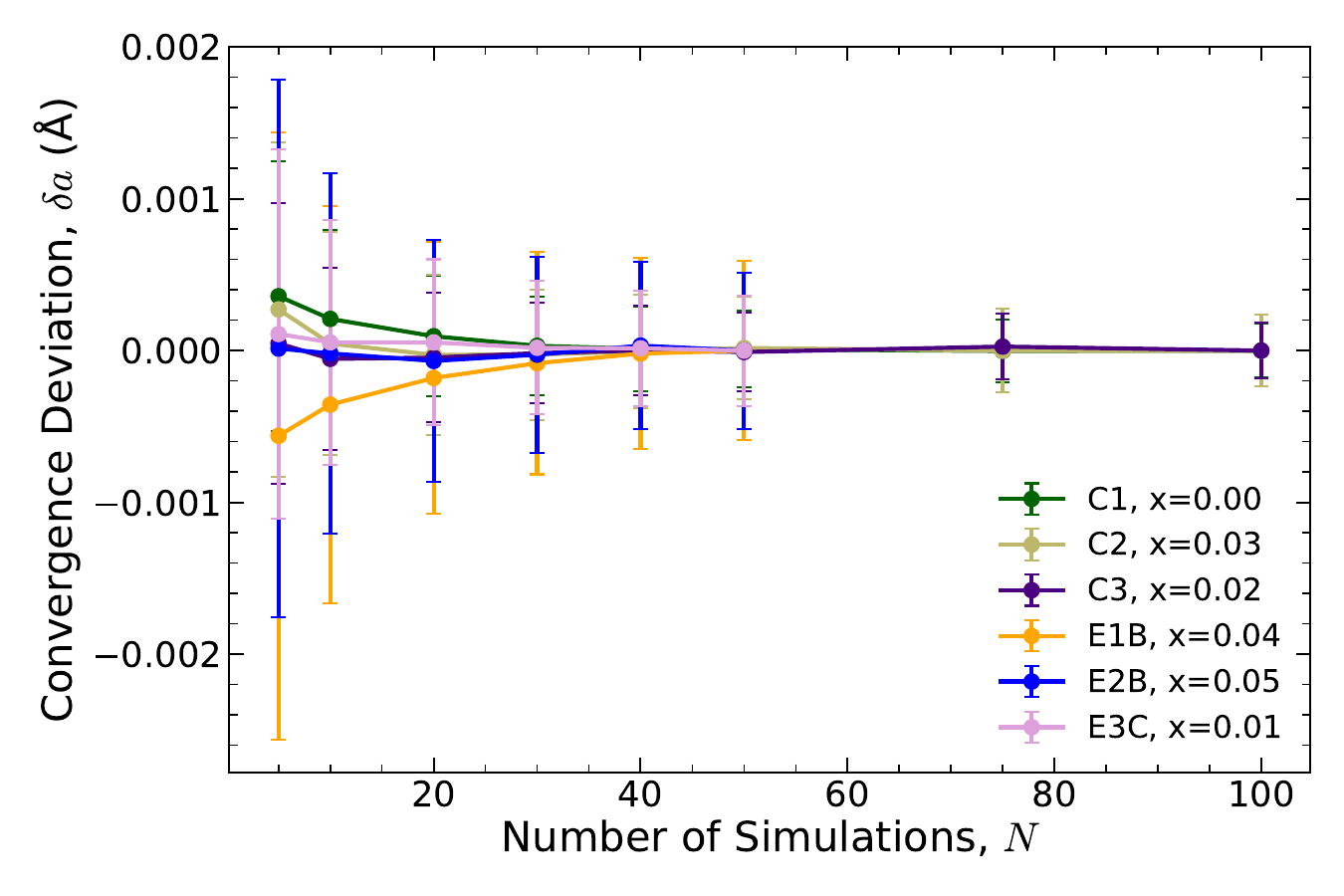}}\hfill
 \subfloat[]{\includegraphics[width=0.48\textwidth]{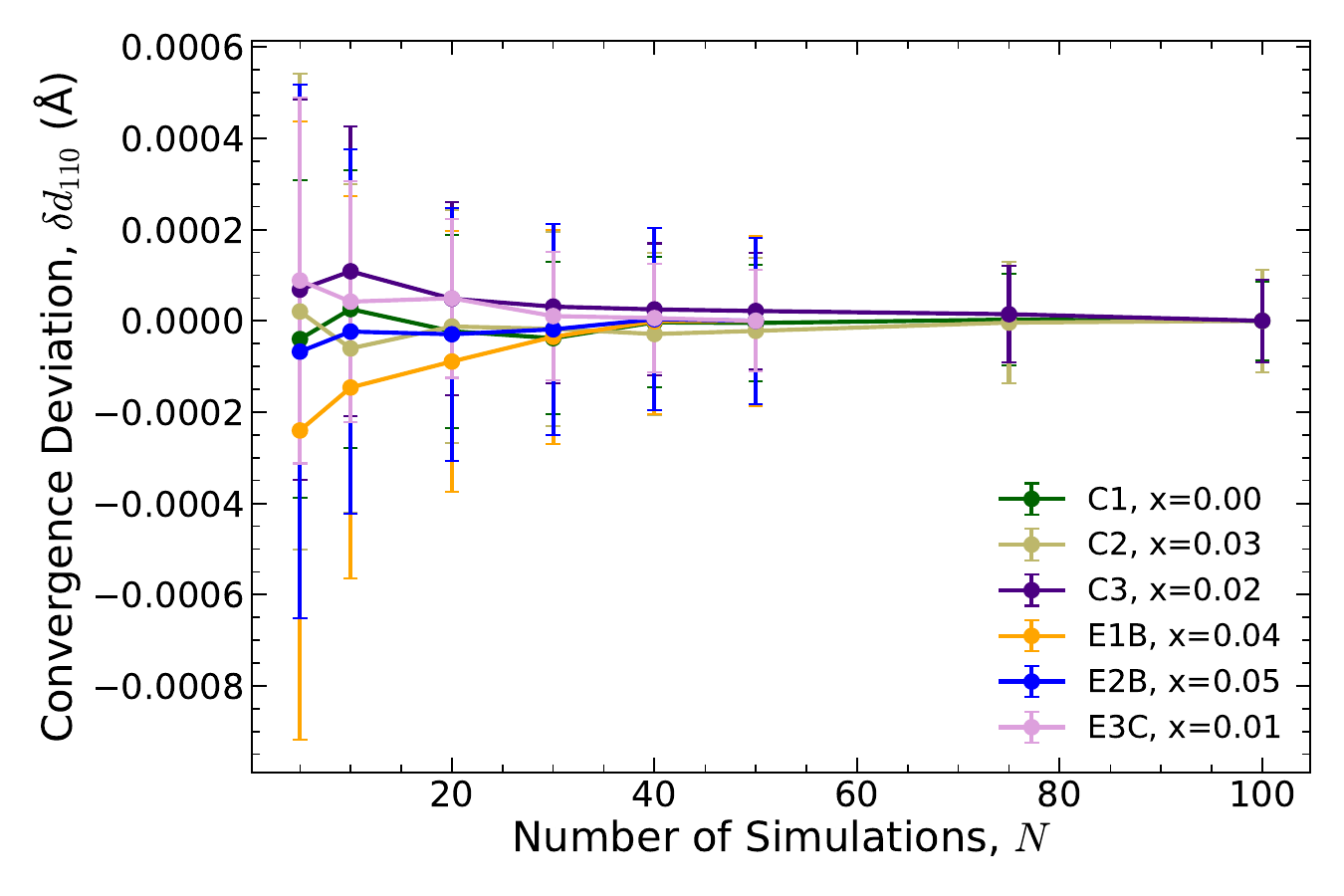}}\\[2ex]
 \subfloat[]{\includegraphics[width=0.48\textwidth]{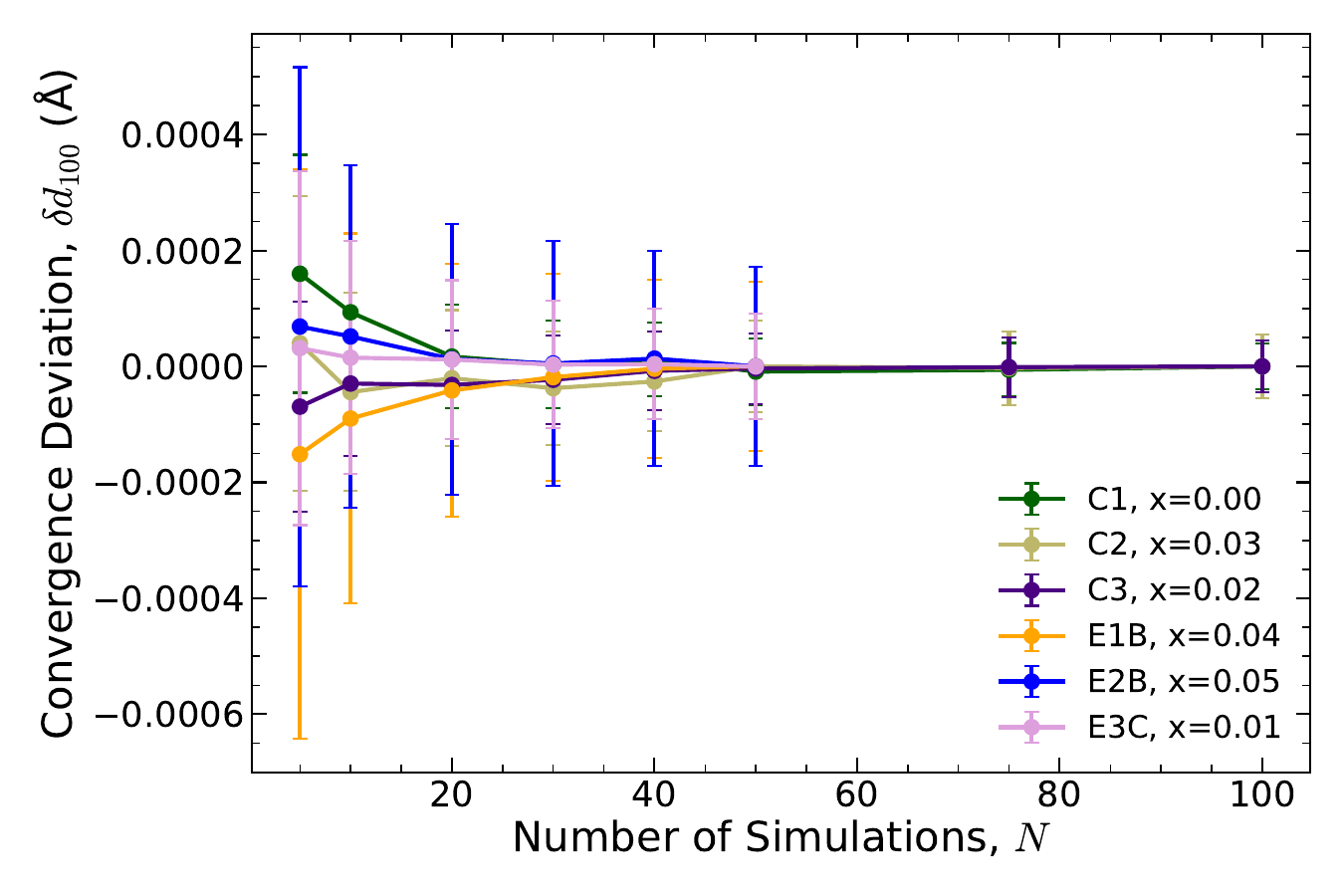}}
 \caption{Plots showing the convergence deviation (\Cref{eqn:Convergence_Deviation}) of the lattice constant \textbf{(a)}, (1\;1\;0) inter-planar distance \textbf{(b)}, and (1\;0\;0) inter-planar distance with increasing number of simulations $N$ for a representative sample of crystal sizes and dopant concentrations: C1 at $x=0.00$, C2 at $x=0.03$, C3 at $x=0.02$, E1B at $x=0.04$, E2B at $x=0.05$, and E3C at $x=0.01$.}
 \label{fig:Convergence_Plots}
\end{figure*}

To verify that a sufficient number of statistical repeats, i.e. unique crystal structures, have been conducted, the convergence of the ensemble averages lattice constant and inter-planar distances was calculated over an increasing number of simulations $N$ for a given crystal size and dopant concentration. This was conducted for $N=5,$ 10, 20, 30, 40, and 50 simulations for a representative sample of crystal sizes and dopant concentrations. These cases were chosen as they cover the full spectrum of crystal sizes and concentrations considered in this study, from the smallest crystal sizes at the smallest concentration, up to the largest at the highest concentration. For cases with more than 50 simulations available, recalculations were also performed with 75 simulations and with the maximum available. To quantify convergence, we use the convergence deviations $\delta a,\,\delta d_{110},$ and $\delta d_{100}$:

\begin{equation}
    \delta a = a_N - a_{N_{\text{max}}},
    \label{eqn:Convergence_Deviation}
\end{equation}

\noindent This is defined as the difference between the ensemble averages computed over $N$ simulations and that obtained using the maximum number of simulations. This was chosen instead of the absolute ensemble average, as variations in dopant concentration yield different mean values, which are distributed across the $Y$ axis and would obscure subtle convergence trends. The convergence deviation ensures that all metrics converge to the same point, allowing for direct comparison. These results are shown in Figures \ref{fig:Convergence_Plots}a for the lattice constant, \ref{fig:Convergence_Plots}b for the (1\;1\;0) inter-planar distance, and \ref{fig:Convergence_Plots}c for the (1\;0\;0) inter-planar distance. In all cases, the ensemble average rapidly converges, with deviations diminishing significantly by around 50 simulations. For systems where more than 50 simulations were performed, the differences between the ensemble averages computed using 50 and the maximum number of available simulations are minimal, typically on the order of $\sim\SI{1e-5}{\angstrom}$, and can be attributed to statistical fluctuations. This demonstrates that 50 simulations, which have been conducted in all cases, is sufficient to produce statistically meaningful results for the structural analysis.

\subsection{B$-$B Nearest Neighbour Distance}

\begin{figure}[t!]
    \centering
    \includegraphics[width=\columnwidth]{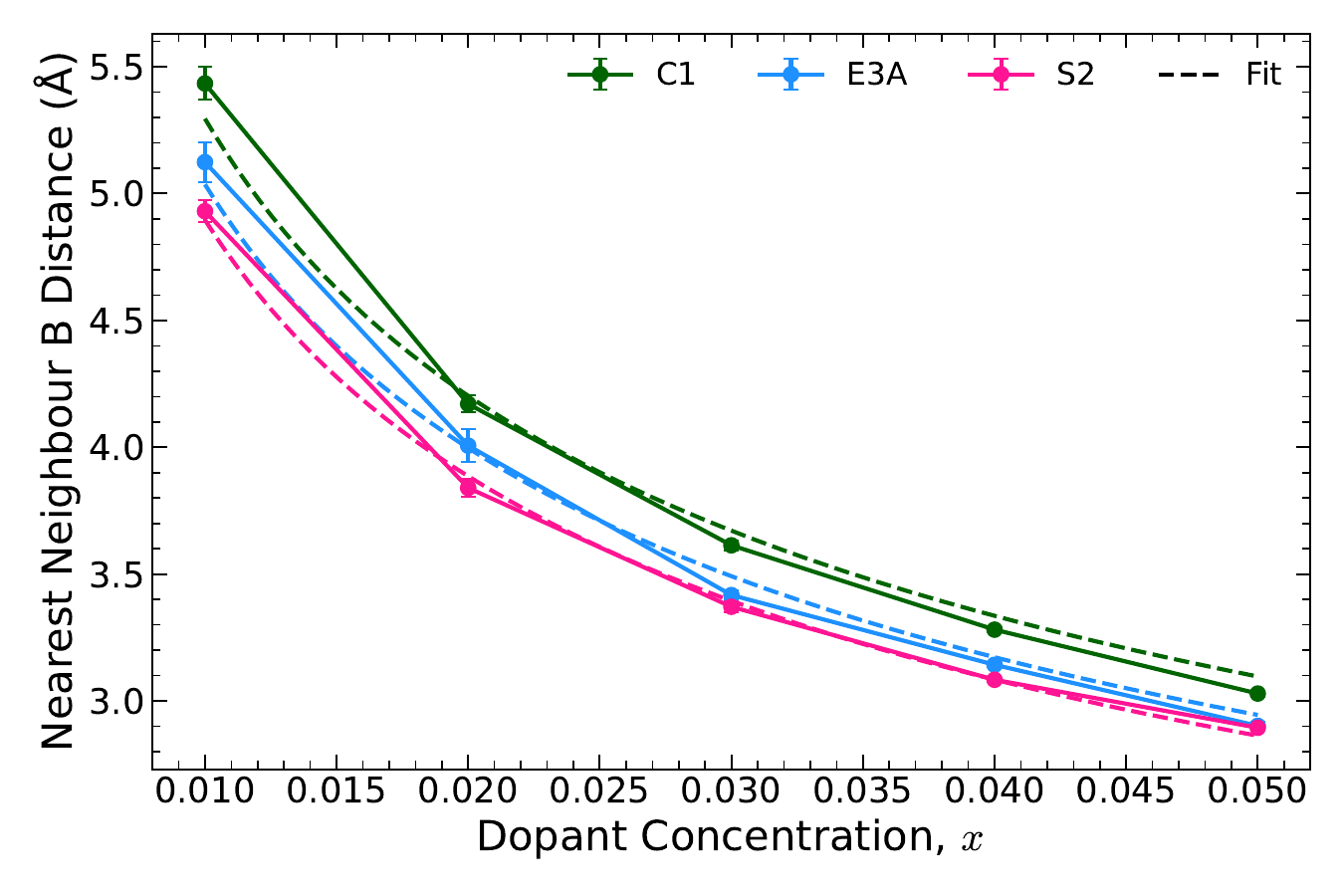}
    \caption{Plot of the average B$-$B nearest-neighbour distances as a function of dopant concentration for a representative example of crystal sizes. A fit to the relation $ A/x^{1/3}$ is also shown for each case.}
    \label{fig:Nearest_Neighbour}
\end{figure}

The average nearest-neighbour distance between boron dopant atoms $\langle d_{\text{B$-$B}}\rangle$ provides a quantitative measure of the spatial distribution of dopants within the crystal lattice, and can therefore be used to determine if the dopant atom distribution is indeed random. For each crystal, the shortest B$-$B separation was calculated and averaged across all simulations for a given dopant concentration and crystal size. In systems where dopant atoms are randomly distributed in three dimensions, it is expected that the average inter-dopant distance should scale as $\langle d_{\text{B$-$B}}\rangle\propto x^{-1/3}$ \cite{supp:Zhang_2011}. To test this expectation, the calculated B$-$B nearest-neighbour distances were fitted to $A/x^{1/3}$, where $A$ is a fitted parameter specific to each crystal size. A representative sample of these results can be seen in \Cref{fig:Nearest_Neighbour}, encompassing the full range of crystal sizes. The fitted curves closely reproduce the simulation results across all crystal sizes, confirming that the dopant atoms remain randomly distributed, without significant clustering. This is in agreement with the analysis of \ce{B2} clusters conducted in \Cref{sec:LC}.

Minor deviations from the expected $x^{-1/3}$ behaviour are seen in the case of the smallest crystal C1, particularly for a boron concentration of $x=0.01$. This can be attributed to statistical undersampling due to the small number of boron atoms, typically around 20 per simulation at this crystal size. This effect diminishes with increasing crystal size, as the number of dopant atoms increases and the average value becomes statistically more stable. At $x=0.01$, larger crystals consistently converge to an average B$-$B separation of approximately \SI{5}{\angstrom}, in agreement with the theoretical expectation.

\section{Poisson's ratio} \label{subsec:Poisson}

Here we present additional analysis on the determination of Poisson’s ratio for the anisotropic expansion of \CB~crystals. This was omitted from the main text for brevity, as further analysis in this section is required. Nevertheless, we believe that this analysis is important, and have therefore included it here. A second set of MD simulations was conducted starting from previously optimised structures. The simulation box dimensions in the $Y$ and $Z$ directions were reduced by one-tenth of the standard C–C bond length (\SI{0.154}{\angstrom}) to impose a controlled in-plane compression, simulating uniaxial stress along the $X$ direction while constraining the transverse directions. The $X$-direction box size was left unchanged to allow measurement of the axial response. All other simulation parameters were unchanged. Axial and transverse strains were calculated from the changes in crystal dimensions, and Poisson’s ratio was subsequently evaluated.

Poisson's ratio $\nu$ quantifies the extent to which a material deforms perpendicular to an applied stress. Poisson’s ratio therefore provides a metric for mechanical response and serves as a benchmark for the crystal generation methodology. The simulation protocol for these calculations is outlined in the main text. Poisson’s ratio is defined as the negative ratio of transverse to axial strain,, $\varepsilon_{\text{trans}}$ and $\varepsilon_{\text{axial}}$:

\begin{equation}
    \nu = -\frac{\varepsilon_{\text{trans}}}{\varepsilon_{\text{axial}}}
    \label{eqn:Poisson_Ratio}
\end{equation}

For the determination of Poisson’s ratio, the transverse and axial strains, $\varepsilon_{\text{trans}} = \varepsilon_{y}, \varepsilon_{z}$ and $\varepsilon_{\text{axial}} = \varepsilon_{x}$ were calculated from the changes in crystal dimensions $L_X$, $L_Y$, and $L_Z$ between the initial and final frames of the simulation:


\begin{equation}
    \varepsilon_\alpha = \frac{L_\alpha-L_{0\alpha}}{L_{0\alpha}}
\end{equation}

\noindent where $\alpha=(X,Y,Z)$, and $L_{0\alpha}$ and $L_{\alpha}$ are the crystal sizes at the start and end of the simulation, respectively. Directional Poisson’s ratios were then calculated as:

\begin{equation}
    \nu_{XY} = -\frac{\varepsilon_Y}{\varepsilon_X}, \, \nu_{XZ} = -\frac{\varepsilon_Z}{\varepsilon_X}
\end{equation}

\noindent and take their average to obtain the overall Poisson’s ratio:

\begin{equation}
    \nu = \frac{\nu_{xy} + \nu_{xz}}{2}
\end{equation}

For pure diamond, Poisson’s ratio is often cited as approximately 0.2; however, this value strongly depends strongly on factors such as crystal type (single vs. polycrystalline), crystallographic orientation, and grain structure \cite{supp:Klein_1993, supp:Mohr_2014}. Due to strong elastic anisotropy, diamond exhibits directional Poisson’s ratios in the range $0.00786 \leq \nu \leq 0.115$ \cite{supp:Hess_2012}, depending on the crystallographic orientation. For (1\;0\;0) single-crystal diamond, as studied in these simulations, reported values are typically close to $\nu \approx 0.10$; for example, $\nu = 0.1089$ \cite{supp:Klein_1993} and $\nu = 0.1031$ \cite{supp:Clerc_2005}.

\begin{figure}[t!]
    \centering
    \includegraphics[width=\columnwidth]{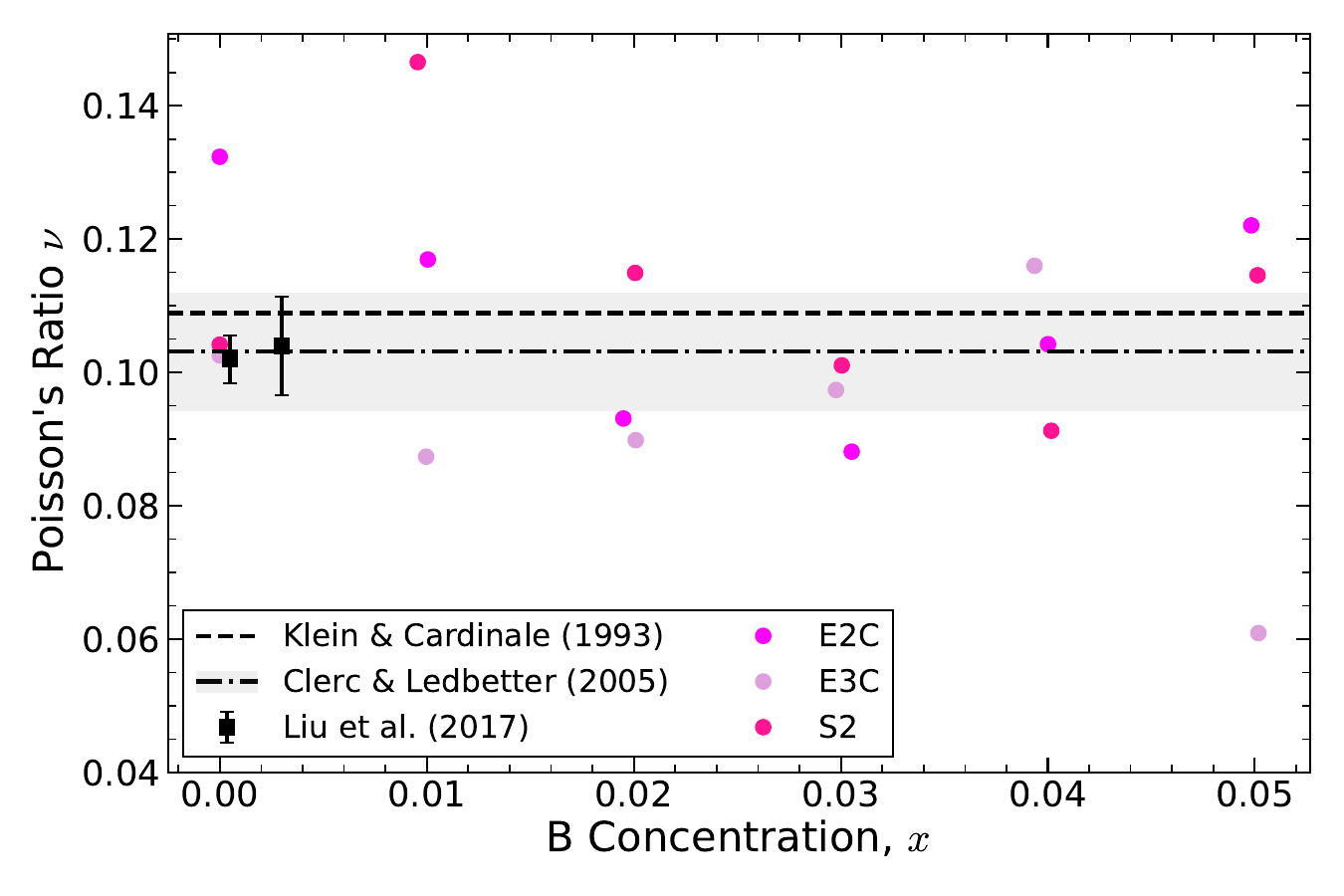}
    \caption{Plot of Poisson's ratio as a function of the boron dopant concentration for various crystal sizes. Black lines indicate the Poisson’s ratio values for single-crystal diamond reported by Klein and Cardinale (1993) \cite{supp:Klein_1993} and Clerc and Ledbetter (2005) \cite{supp:Clerc_2005}, and BDD crystal values reported by Liu \textit{et al.} \cite{supp:Liu_2017} shown as black squares.}
    \label{fig:Poisson_Ratio}
\end{figure}

\Cref{fig:Poisson_Ratio} shows the calculated Poisson's ratio over a subset of the larger non-C-category crystals, compared against literature data. These three crystal sizes were chosen due to their higher atom counts (see Table 2 in the main text); small crystals with few atomic planes along the axial ($X$) suffer from statistical undersampling and enhanced surface effects: when the number of atoms or atomic planes along the axial ($X$) direction suffer from statistical undersampling and enhanced surface effects. In such cases, surface relaxation and finite-size perturbations can bias the axial strain $\varepsilon_x$, leading to exaggerated Poisson ratios. For pure diamond ($x=0.00$), calculated values range from $0.08 \leq \nu \leq 0.14$, centered around $\nu\approx0.10$. The two largest crystals (E3C and S2), providing the best atomic statistics, show agreement with single-crystal [1\;0\;0] literature values \cite{supp:Klein_1993, supp:Clerc_2005, supp:Liu_2017}.

The scatter observed in the results can be attributed to both statistical and methodological effects. The determination of Poisson’s ratio (\Cref{eqn:Poisson_Ratio}) relies on the axial and transverse strains, which are very small $(\varepsilon\sim\SI{e-4}{})$. As such, even small thermal fluctuations can amplify the results. Averaging over the final \SI{10}{\pico\second} of each simulation was employed to mitigate this effect, but significant run-to-run variability can remain. Approximately 50 simulations were conducted for each crystal size; while this is sufficient for structural metrics such as inter-planar distances (analyzed in the next section), a larger number of repeats may be required due to the sensitivity of strain calculations. In addition, finite simulation box sizes can enhance surface effects, as previously described. Despite these factors, all results remain within the literature range for diamond $0.00786 \leq \nu \leq 0.115$ \cite{supp:Hess_2012}. The observed variability is therefore considered a natural consequence of molecular dynamics simulations and does not compromise the crystal generation methodology.

Poisson’s ratio was also evaluated for crystals doped with boron, which was not considered in Refs.~\cite{supp:Klein_1993, supp:Clerc_2005} but is addressed by Liu \textit{et al.} \cite{supp:Liu_2017}. While Poisson’s ratio is not explicitly reported in that work, values of the elastic coefficients $C_{11}$ and $C_{12}$ are provided for single-crystal diamond doped with boron at concentrations from \SI{50}{\text{ppm}} to \SI{3000}{\text{ppm}} (i.e. $x\approx0.05\%$ and 0.3\%, respectively). In general they observed an overall elastic softening of the crystal with increasing boron content. Poisson's ratio can be inferred from the elastic coefficients $C_{11}$ and $C_{12}$ using \cite{supp:Nye_1985}:

\begin{equation}
    \nu = \frac{C_{12}}{C_{11} + C_{12}}
\end{equation}

Using the values of $C_{11}$, and $C_{12}$ reported by Liu \textit{et al.} \cite{supp:Liu_2017}, Poisson’s ratio was found to increase slightly from 0.102 at $\sim0.05\%$ to 0.104 at $\sim0.3\%$. This concentration range is much smaller than that considered in the present work; however, these results suggest that Poisson’s ratio is only weakly affected by boron doping at low concentrations. Simulations show consistent behaviour across all dopant concentrations, with a possible increase in scatter at higher concentrations. Due to the limited experimental data at higher dopant levels, a larger ensemble of MD repeats would be required to resolve concentration-dependent trends definitively.

\end{document}